\documentclass[useAMS,usenatbib]{mn2e}
\usepackage{booktabs}
\usepackage{natbib}
\usepackage{graphicx,multirow,rotating,pdflscape,footnote}
\usepackage[fleqn]{amsmath}
\usepackage{times}
\usepackage{url}
\usepackage{lastpage}
\usepackage[a4paper,left=1.5cm,right=1.5cm,top=1.5cm,bottom=1.5cm]{geometry}

\title[A search for ionized jets towards MYSOs]{A search for ionized jets towards massive young stellar objects}
\author[S. J. D. Purser et al.]{S. J. D. Purser,$^{1}$\thanks{E-mail:
pysjp@leeds.ac.uk} S. L. Lumsden,$^{1}$ M. G. Hoare,$^{1}$ J. S. Urquhart,$^{2,3}$ N. Cunningham,$^{1,4}$
\newauthor C. R. Purcell,$^{5}$ K. J. Brooks,$^{6}$ G. Garay,$^{7}$ A. E. G\'{u}zman$^{7}$ and M. A. Voronkov$^{6}$\\
$^{1}$School of Physics and Astronomy, University of Leeds, Leeds LS2 9JT, UK\\
$^{2}$Max-Planck-Institute f\"ur Radioastronomie, Auf dem H\"ugel 69, D-53121 Bonn, Germany\\
$^{3}$Centre for Astrophysics and Planetary Science, The University of Kent, Canterbury, Kent CT2 7NH, UK\\
$^{4}$National Radio Astronomy Observatory, PO Box 2, Green Bank, WV 24944, USA\\
$^{5}$Sydney Institute for Astronomy (SIfA), School of Physics, The University of Sydney, NSW 2006, Australia\\
$^{6}$Australia Telescope National Facility, CSIRO Astronomy and Space Science, PO Box 76, Epping, NSW 1710, Australia\\
$^{7}$Departamento de Astronom\'{i}a, Universidad de Chile, Casilla 36-D, Santiago, Chile}

\begin{document}
\date{}
\pagerange{\pageref{firstpage}--\pageref{lastpage}} \pubyear{2016}
\maketitle

\label{firstpage}
\begin{abstract}
Radio continuum observations using the Australia telescope compact array at 5.5, 9.0, 17.0 and 22.8 GHz have detected free-free emission associated with 45 of 49 massive young stellar objects and HII regions.  Of these, 26 sources are classified as ionized jets (12 of which are candidates), 2 as ambiguous jets or disc winds, 1 as a disc-wind, 14 as HII regions and 2 were unable to be categorised. Classification as ionized jets is based upon morphology, radio flux and spectral index, in conjunction with previous observational results at other wavelengths. Radio-luminosity and momentum are found to scale with bolometric luminosity in the same way as low-mass jets, indicating a common mechanism for jet production across all masses. In 13 of the jets, we see associated non-thermal/optically-thin lobes resulting from shocks either internal to the jet and/or at working surfaces. Ten jets display non-thermal (synchrotron emission) spectra in their lobes, with an average spectral index of $\alpha = -0.55$ consistent with Fermi acceleration in shocks. This shows that magnetic fields are present, in agreement with models of jet formation incorporating magnetic fields. Since the production of collimated radio jets is associated with accretion processes, the results presented in this paper support the picture of disc-mediated accretion for the formation of massive stars with an upper-limit on the jet phase lasting approximately $6.5\times10^4\,\mathrm{yr}$. Typical mass loss rates in the jet are found to be 1.4$\times10^{-5}\,\mathrm{M}_\odot\mathrm{yr^{-1}}$ with associated momentum rates of the order $(1\--2)\times10^{-2}\,\mathrm{M}_\odot\mathrm{km\,s^{-1}\,yr}^{-1}$.
\end{abstract}


\section{Introduction}
\label{sec:intro}
The formation of massive stars still presents a problem in astrophysics today. With their powerful outflows and radiative outputs, they play a significant role in local (cloud/clump scale) as well as galactic evolution, thus understanding massive young stellar objects (MYSOs) and their formation is of critical importance. Current theories on the exact mechanisms of formation include the incorporation of accretion through discs \citep{McKeeTan2002}, similar to the low-mass paradigms of \citet{shu1987a}, as well as a competitive accretion scenario whereby the size of the reservoir of accreted material onto a MYSO's disc stems from the gravitational potential in which it resides \citep[][]{Bonnell2001a}. Due to sampling/observational constraints (a by-product of the comparative rarity and extreme distances of massive stars), only a few cases of accretion discs around massive stars have been observed \citep[e.g.Cep A HW2, HH 80-81;][respectively]{patel2005,gonzalezHHjet2012}, most notably the disc imaged through near-infrared interferometry by \citet{Kraus2010} around the massive young stellar object, IRAS 13481$\--$6124 (G310.0135+00.3892 in this paper). With the advent of ALMA however, we expect this type of observation to become more common. One such example is that of the MYSO G345.4938+01.4677 \citep{Guzman2014} which showed evidence in SO$_2$ molecular lines of  a rotating disc structure perpendicularly oriented to the previously established ionized jet \citep{Guzman2010}.



For this work, we turn our attention towards the larger scale jets of ionized material moving away from, and perpendicular to, the disc. As well as carrying away excess angular momentum (avoiding terminal `spin-up' of the protostar), they also provide a source of momentum to drive/entrain large-scale molecular outflows, almost ubiquitously associated with sites of massive star formation \citep{Beuther2002}. With their detection, they provide indirect evidence for the accretion process itself, while measurement of jet velocities, momenta and mass loss rates \citep{eisloffel2000} can provide an insight into the exact accretion mechanisms in operation. In turn this may aid in differentiating between competitive/Bondi-Hoyle accretion \citep{Bonnell2001a} and turbulent core accretion models \citep{McKeeTan2002}. 

Ideas put forward to explain both the production and collimation of ionized jets from massive stars are based upon those of their low-mass counterparts. Mechanisms have been suggested, which can be categorised as either being driven purely in the disc, or by disc/stellar wind interactions. 

A pure disc-wind model was produced by \citet{BlandfordPayne1982}, who suggested that a poloidal magnetic field originating from the accretion disc itself, would act as a launching mechanism for ionized material travelling inwards in the disc. This material would travel along the field lines and away from the disc, accurately adhering to the analogy of `beads on a wire'. As the accretion disc rotates, these field lines become more twisted (due to magnetic coupling) further from the disc and eventually collimation is induced by `hoop stresses' from the field upon the ionized material. Simulations by \citet{Seifried2012} showed that massive protostellar jets were likely driven by this mechanism, with varying hoop stresses (enhanced by weaker magnetic fields, and suppressed by sub-Keplerian disc motions in the presence of strong fields) in the disc affecting the degree of collimation. This result predicted that collimation should relate to the evolutionary stage of the MYSO, with lower degrees of collimation for younger MYSOs. The jet in W75N VLA 2 shows an example whereby the collimation is evolving with time, with collimation increasing with time \citep{CarrascoGonzalez2015}.

On the other hand, \citet{Shu1994} proposed a model involving both stellar, and disc magnetic fields, referred to as the `X-wind model'. This is so named because the launching radius for the jet in the disc is located at the X-shaped lines of equipotential (balancing points of gravitational and centrifugal forces) located near the truncated (by strong stellar magnetic fields), inner surface of the disc itself. With the inward flow of material in the accretion disc, magnetic coupling bows the field lines inwards. Consequently, this inflowing material follows these field lines up and away from the disc. Magneto-centrifugal processes accelerate the gas supersonically in a direction parallel to the poloidal field lines, resulting in an outflowing wind. Twisted magnetic fields further away from the disc's mid-plane focus the ionized material, resulting in a collimated, ionized jet. When applied to massive star formation, the X-wind model presents a problem however. Stellar magnetic fields are intrinsically related to the convective nature of their protostars. For low-mass protostars, these convective models adhere well to reality with the onset of deuterium burning \citep{Stahler1988}. Simulations of the protostellar evolution of MYSOs, \citep[e.g.][]{hosokawa2009} show them to evolve through radiative as well as convective stages, so this phase is much more limited in time.

The observation by \citet{CarrascoGonzalezMagJet2010} of polarised emission from an ionized jet associated with the HH 80-81 object, revealed possible poloidal field lines, suggesting a magnetic mechanism for jet formation around MYSOs. However, this type of result is not common in the literature, and the formation of jets around massive protostars is still an open question.

Observationally, much information can be gleaned from measurement of the spectral index of a jet's free-free radio emission. A paper by \citet{Reynolds1986} derived the following relationships between a jet's physical parameters:

\begin{equation} \label{eq:reynolds}
\alpha=2+\frac{2.1}{q_\tau}\left(1+\epsilon + q_T\right)
\end{equation}

\begin{equation} \label{eq:reynoldsAlphaBeta}
\alpha = \frac{4\beta-6.2}{2\beta-1}
\end{equation}

\noindent Where $\alpha$ is the spectral index ($S_\nu\propto\nu^{\alpha}$), $\epsilon$ represents the degree of collimation and $q_T$ and $q_\tau$ are essentially the powers to which the temperature and optical depth in the jet vary with distance from the base of the jet. For equation \ref{eq:reynoldsAlphaBeta}, $\mathrm{n}_\mathrm{e}\propto\mathrm{r}^{-\beta}$ where $\mathrm{n}_\mathrm{e}$ is the number density of electrons, and $\mathrm{r}$ is the distance from the MYSO).

For a standard (i.e. not accelerating or recombining), bi-conical jet, a spectral index of $\sim0.6$ is expected, however spectral indices of $-0.1\rightarrow1.6$ are found when considering a variety of jet scenarios. Although the derivation of the spectral index cannot be used to infer temperature or density profiles along the jet, we can deduce information about the jet's geometry (collimation) and processes likely ongoing within the flow (such as further acceleration or recombination for example), as summarised in Table 1 of \citet{Reynolds1986}. As well as the spectral index, we can measure the behaviour of the deconvolved major axis of the jet with frequency (which is dependent upon the relationship between electron density and radial distance from the powering source), i.e. $\theta_\mathrm{maj}\propto\nu^\gamma$, where $\theta_\mathrm{maj}$ is the deconvolved major axis and $\gamma$ is the measured index. For a conical wind ($\alpha=0.6$), we expect $\gamma=-0.7$ as can be seen from equations \ref{eq:reynoldsAlphaBeta} and \ref{eq:reynoldsGammaBeta}, demonstrating the connection between these two variables.

\begin{equation} \label{eq:reynoldsGammaBeta}
\gamma = -\frac{2.1}{2\beta-1}
\end{equation}

Specific examples of radio jets in the literature, associated with sites of high mass star formation, are those of HH 80-81  (G010.8411$-$02.5919) and Cepheus HW-2. These MYSOs harbour thermal jets found to have proper motions of $\sim500\,\mathrm{km\,s}^{-1}$, and $\sim480\,\mathrm{km\,s}^{-1}$ \citep[][respectively]{Marti1998,Curiel2006}, which are much greater than the velocities typically seen in low-mass cases of $\sim200\,\mathrm{km\,s}^{-1}$ \citep{Arce2007}. Interestingly, even larger velocities were reported for other parts within the HH 80-81 jet of $600\--1400\,\mathrm{km\,s}^{-1}$ in an earlier study \citep{Marti1995}. Cepheus HW-2 consists of a simple elongated structure with a spectral index of $\alpha=0.69\pm0.06$, indicative of a conical bipolar jet \citep{Rodriguez1994}. Conversely HH 80-81 consists of a more complex picture whereby a central, thermal source ($\alpha\sim0.2$) and two non-thermal lobes ($\alpha\sim-0.3$) align on a NE-SW orientation \citep{Marti1993}. These two examples show the range of complexities which radio observations of ionized jets reveal.

Systematic searches for ionized jets are few and far between, and conducted with sporadic success. \citet{GibbHoare2007} performed radio observations from 5 to 43 GHz towards 5 MYSOs with spectral indices ranging from 0.2 to 0.8, all consistent with ionized jets/winds \citep{Reynolds1986}. Although no jets were definitively identified, two sources (S106-IR and S140-IRS1), whose radio emission was perpendicular to their associated molecular outflows, were classified as equatorial disc winds. Another survey by \citet{Guzman2012}, detected two radio jets (G343.1261-00.0623 and G345.4938+01.4677 in this paper) from a sample of 33 (8 of which were observed) IR-luminous objects taken, in part, from the RMS survey (see \S\ref{sec:RMS}). An issue with this survey's sampling methodology, was selection criteria based upon positive radio spectral indices and low radio to IR luminosities, the radio data for which was taken from \citet{Walsh1998} and \citet{Urquhart2007ATCA}. This resulted in contamination of the sample by unresolved, optically-thick HII regions with low radio fluxes, which were under-luminous compared to radio fluxes expected from the bolometric luminosities (for an optically thin HII region). The survey's $4\sigma$ detection limits were $\sim0.8\,\mathrm{mJy}$, at 1.4 and 2.4 GHz and $\sim0.4\,\mathrm{mJy}$, at 4.8 and 8.6 GHz. Since we expect radio-weak, thermal spectra from the ionized jets, at those frequencies emission from such objects may have remained undetected. Most recently \citet{Moscadelli2016} selected a sub-sample of MYSOs (ZAMS type B3-O7) from the Bar and Spiral Structure Legacy (BeSSeL) survey, which displayed water maser activity, and previous compact ($<1\arcsec$), and weak (which they define as $<50\,\mathrm{mJy}$) radio emission. Of the 40 objects in the sample, 11 were observed with the VLA at C, K and Ku-bands ($\theta_\mathrm{res}\sim0.2\arcsec$), 10 displayed spectral indices typical of thermal jets, 5 of which were confirmed to have collimated (opening angles, $\theta_0$, of between 10 and 30$\degr$) outflows via maser kinematics. Spectral indices ranged between $0.1\leq \alpha \leq 1.1$, with inferred momentum rates of between $10^{-3}\--1\,\mathrm{M_\odot kms^{-1}yr^{-1}}$. Interestingly, lobes of elongated, non-thermal emission were observed towards two of the sample, suggestive of both particle-acceleration and the presence of magnetic fields.

With these previous ionized jet surveys in mind, our own RMS survey-derived selection of targets, in conjunction with the targets of \citet{Guzman2012}, was used to create an unbiased sample of MYSOs which could harbour ionized jets. 

In this paper, we introduce the methodology of our sample selection (\S\ref{sec:Sampling}), and explain the technicalities of its observation in \S\ref{sec:Obs}. Section \ref{sec:results} contains a discussion of cleaned and imaged radio data (Figure \ref{fig:ContourPlots} in the online version) and their relation to previous observations within the literature. This is done for a selection of specific examples representative of the sample as a whole. An overall, statistical approach is then employed in \S\ref{sec:Analysis}, from which implications to current jet and accretion models are discussed. These findings are summarised in \S\ref{sec:Conclusions}.

\section{A sample of MYSOs}
\label{sec:Sampling}

\subsection{The RMS survey}
\label{sec:RMS}

The \underline{R}ed \underline{M}SX \underline{S}ource survey \citep{Lumsden2013} is a systematic search for MYSOs using previous surveys as well as mm, radio and infrared follow-up observations to aid in the classification of objects catalogued by the Midcourse Space eXperiment \citep[MSX; ][]{Price2001MSX}. This extensive, ordered approach was necessary considering the relative sparsity of forming massive stars in the Galaxy. 

In order to extract genuine MYSO candidates from the MSX database, contaminant objects such as planetary nebulae, old dusty stars and resolved ultra-compact HII regions had to be removed. Since most of the bolometric luminosity of a MYSO lies at infrared wavelengths (due to dust reprocessing), a comparison of fluxes at a variety of IR-wavelengths (based on known MYSOs) was employed. The exact sampling methodologies used can be found in \citet{Lumsden2002}. 

VLA \citep{Urquhart2009VLA} and ATCA \citep{Urquhart2007ATCA} observations at 5 GHz ($\theta_{res}\sim1\arcsec$) helped to remove compact UCHIIs, while mm observations by MOPRA ($^{13}$CO$(1-0)$) and JCMT ($^{13}$CO$(2-1)$) allowed the removal of close, low-mass YSOs by the determination of kinematic distances and therefore absolute luminosities \citep{Urquhart2007CO}. The relative strength of the CO lines helped eliminate dusty, evolved stars (stronger in young stars) while those evolved stars lying along the line of sight to $\mathrm{H}_2$ molecular clouds were picked out on the basis of their distinct near-IR absorption spectra. 

The resulting RMS database\footnote{\url{http://rms.leeds.ac.uk/cgi-bin/public/RMS_DATABASE.cgi}} catalogues $\sim 900$ embedded, young massive stars \citep{RMS2014} of which both HII regions and MYSOs are included. 

\subsection{Sample selection}
\label{sec:sampleselection}

In total, we have a sample of 49 individual objects which are listed in Table \ref{tab:targets}. Of these, 34 were selected to form a distance-limited ($\mathrm{d}<7\mathrm{kpc}$) sample of MYSOs. An even spread in the luminosities of the objects was required for unbiased sampling in terms of bolometric luminosity and consequently 6 of these objects fall in the range $700\,\mathrm{L}_\odot\leq\mathrm{L}_\star<3000\,\mathrm{L}_\odot$, 11 in the range $3000\,\mathrm{L}_\odot\leq\mathrm{L}_\star\leq10000\,\mathrm{L}_\odot$, 9 in the range $10000\,\mathrm{L}_\odot\leq\mathrm{L}_\star\leq30000\,\mathrm{L}_\odot$ and 8 with $\mathrm{L}_\star>30000\,\mathrm{L}_\odot$. It is worth noting that derived bolometric luminosities have changed for some of the sample since the observations. The previous observations of \citet{Urquhart2007ATCA} have demonstrated them all to be radio weak with $\mathrm{S}_\nu<2\,\mathrm{mJy}$. These sources are at elevations conducive to more circular synthetic apertures, whereby $\delta<-34\degr$. Targets with declinations greater than the quoted cut-off have been included due to their scientific value, as well as filler objects for allocated times when the other sources were not fully risen above the horizon (i.e. elevation $< 20\degr$). A further 8 objects were added to complete the observations (at 17 and 22.8 GHz) which \citet{Guzman2012} performed from 1.4 to 8.6 GHz with the ATCA. One of the Guzman sample (G317.4298-00.5612) was observed at 5 and 9 GHz to check for variability. Four of the objects were added on the basis of being associated to 4.5 $\micron$ excesses, or extended green objects \citep[EGOs][]{Cyganowski2008}, a signature of shocks possibly caused by jet activity \citep{SmithRosen2005}. The latter two samples were not distance-limited and also were used to fill in coverage in right ascension.

\section{Observations and data reduction}

\subsection{Radio observations with the ATCA}
\label{sec:Obs}
All radio observations were made using the Australia Telescope Compact Array \citep[ATCA; See chapter 1, \S1 of][for an overview]{FraterBrooks1992} in the 6A and 6B configurations, over a period of 12 days, spread over 4 years. A total of 49 individual, compact objects were observed at 4 different frequency bands (centred on 5.5, 9.0, 17.0 and 22.8 GHz). These frequencies were observed using a bandwidth of 2048 MHz (XX, YY, XY, and YX polarizations) split evenly either side of the  central frequencies \citep[an overview of the CABB correlator's capabilities can be found in][]{Wilson2011}. This bandwidth was split into channels of width 1 MHz (minimising the averaging out effect of RFI over broad channels and giving the ability to easily recognise and flag it). From this point on the frequency bands observed shall be referred to as 5.5, 9, 17 and 22.8 GHz bands.

The highest frequency band has been also chosen to incorporate the $\mathrm{H}_2\mathrm{O}(6\rightarrow5)$ transition at 22.235 GHz, in order to observe masing activity and allow more accurate, phase-only self-calibration from the relevant channels. 

Throughout the radio bands, radio recombination lines should also be observable in the strongest sources, giving some degree of kinematic information for the ionized gas. Table \ref{tab:RRLs} shows the radio recombination lines falling within the observed bands, and their effective, velocity resolutions for a channel width of 1 MHz.

\begin{table}
\centering
\begin{tabular}{lcc}
\hline 
\textbf{Band}   & \textbf{RRLs} & \boldmath$\Delta\mathrm{v}$      \\ 
\textbf{(GHz)}  &               & \boldmath$(\mathrm{km\,s^{-1}})$ \\ 
\hline 
5.5  & $\mathrm{H100\alpha-H112\alpha}$ & 54.5  \\ 
9    & $\mathrm{H87\alpha-H93\alpha}$   & 33.3  \\ 
17   & $\mathrm{H71\alpha-H73\alpha}$   & 17.6  \\ 
22.8 & $\mathrm{H65\alpha,\,H66\alpha}$ & 13.1  \\ 
\hline 
\end{tabular} 
\caption{A table of the observable radio recombination lines for each observed band, and the associated resolution per 1 MHz channel (at the centre of each band).}
\label{tab:RRLs}
\end{table}

The range of scales the instrument was sensitive to were $2.0\--16.5\arcsec$, $1.2\--10.1\arcsec$, $0.6\--5.4\arcsec$ and $0.5\--4.0\arcsec$ over the $5.5$, $9$, $17$ and $22.8$ GHz frequency bands respectively. These correspond to a maximum baseline length of 6 km and a minimum of $\sim300$ m.

Individual scan times on flux and phase calibrators were 5 minutes and 90 seconds respectively and on target sources were between 5 and 20 minutes dependent on atmospheric conditions and observing frequencies. The average total integration time on each science target, $\tau_\mathrm{int}$, was $\sim75$ minutes. Observations at 5.5 and 9 GHz were all conducted during February 2013, with a typical RMS noise level calculated to be 17 and 20 $\mathrm{\mu Jy/beam}$ respectively. The upper two frequency bands were observed during the months of February 2010, April 2011, September 2011 and January 2012, giving a range of conditions particularly affecting the 22.8 GHz band which is more sensitive to atmospheric water vapour content (greater during the southern hemisphere's summer months). Theoretical RMS noise levels therefore range between $40\--41$ and $73\--106$ $\mathrm{\mu Jy/beam}$ for the 17 and 22.8 GHz bands respectively, during the months of September (lower noise) and January. As a note, the typical values for the RMS noise were calculated assuming the removal of 200 channels (100 at each edge of the bandwidth), RFI interference, typical weather during the relevant months, $\delta\sim-50\degr$ and a robustness of 0.

\subsection{Data reduction}
\label{sec:datareduction}
For calibrating and cleaning the data, the Multichannel Image Reconstruction Image Analysis and Display (MIRIAD) software package \citep{miriad} was used. For the measurement of various parameters, the Common Astronomy Software Applications (CASA) was implemented \citep{CASARef}.

\begin{table*}
\centering
\caption{Target sources observed by ATCA between 12/02/2010 and 28/02/2013, their bolometric luminosities and distances \citep{Lumsden2013}. Sample selection is indicated by the superscripted symbol on each source's name, whereby $\star$ indicates a filler object of scientific value, $\dagger$ indicates an object which is part of the \citet{Guzman2012} sample and $\ddagger$ indicates an MYSO associated to an EGO \citep{Cyganowski2008}. Any source without a symbol is part of the distance limited sample. The footnote marks for the distances indicate how they were calculated: a) kinematic distance determined using the source velocity; b) spectrophotometric distance; c) maser parallax distance; d) distances taken from the literature. The footnote marking for the bolometric luminosities show the relevant calculation methods which were: a) spectral energy distribution (SED) fit to the available infrared fluxes (e.g., 2MASS, MSX) including HiGAL (Mottram et al., in prep.); b) SED fit to infrared and either MIPSGAL 70$\micron$ or IRAS Galaxy ATLAS 60$\micron$ fluxes (\citet{Mottram2010}, \citet{Mottram2011a}); c) MSX 21$\micron$ band flux using a scaling relationship determined from a comparison with sources where SED fits have been possible \citep{Mottram2011a}.  Also included are the epochs during which their data was recorded by the ATCA for the 5.5/9 GHz setup and 17/22.8 GHz setup. The epoch notation corresponds to the following dates:  E1, 12/02/2010-13/02/2010; E2, 16/04/2011; E3, 10/09/2011-11/09/2011; E4, 05/01/2012-07/01/2012; E5, 25/02/2013-28/02/2013. Date ranges include all dates in between those provided. Hyphenated table cells denote the source not being observed at the relevant frequencies. The final column denotes each object's final classification, whereby HII regions follow the accepted acronyms, N/D indicates a non-detection, U/K indicates where no robust classification could be made, DW indicates a disc-wind, while Jet/DW indicates an ambiguous classification and the presence of an (L) and/or (C) after `jet' indicates whether lobes were present and/or the object was only classified as a jet candidate, respectively.}

\renewcommand{\thefootnote}{\alph{footnote}}

\begin{tabular}{lcccccccc}
\toprule
\textbf{Object Name} & \textbf{Associated Object} & \textbf{R.A. (J2000)} & \textbf{Dec. (J2000)} & \textbf{D} & $\mathbf{L_{Bol}}$ & \textbf{5.5 / 9} & \textbf{17 / 22.8} & \textbf{Class.} \\
 &  &  &  & \textbf{(kpc)} & \boldmath$(\mathrm{L}_\odot)$ & \textbf{(GHz)} & \textbf{(GHz)} & \\
\hline
G010.8411--02.5919$^\star$ & IRAS 18162--2048 & $18^\mathrm{h}19^\mathrm{m}12.09^\mathrm{s}$ & $-20\degr47\arcmin30.90\arcsec$ & 1.9\footnotemark[1] & $2.40\times 10^4$\footnotemark[2] & -- & E2 & Jet\\
G012.9090--00.2607$^\star$ & IRAS 18117--1753 & $18^\mathrm{h}14^\mathrm{m}39.56^\mathrm{s}$ & $-17\degr52\arcmin02.30\arcsec$ & 2.4\footnotemark[3] & $3.20\times 10^4$\footnotemark[1] & -- & E3 & Jet (C)\\
G014.9958--00.6732$^\star$ & -- & $18^\mathrm{h}20^\mathrm{m}19.47^\mathrm{s}$ & $-16\degr13\arcmin29.80\arcsec$ & 2.0\footnotemark[3] & $1.30\times 10^4$\footnotemark[3] & -- & E3 & Jet (C)\\
G251.2337--01.9535 & IRAS 08008--3423 & $08^\mathrm{h}02^\mathrm{m}42.97^\mathrm{s}$ & $-34\degr31\arcmin48.70\arcsec$ & 4.6\footnotemark[1] & $8.20\times 10^3$\footnotemark[1] & E5 & E4 & Jet (C)\\
G254.0491--00.5615 & IRAS 08140--3559 & $08^\mathrm{h}15^\mathrm{m}57.12^\mathrm{s}$ & $-36\degr08\arcmin06.80\arcsec$ & 3.0\footnotemark[1] & $1.70\times 10^3$\footnotemark[1] & E5 & E1 & Jet\\
G254.0548--00.0961 & IRAS 08159--3543 & $08^\mathrm{h}17^\mathrm{m}52.62^\mathrm{s}$ & $-35\degr52\arcmin47.60\arcsec$ & 2.8\footnotemark[1] & $1.90\times 10^3$\footnotemark[1] & E5 & E1 & Jet (L)\\
G263.2283+01.5712 & IRAS 08513--4201 & $08^\mathrm{h}53^\mathrm{m}09.46^\mathrm{s}$ & $-42\degr13\arcmin07.60\arcsec$ & 0.7\footnotemark[4] & $1.20\times 10^3$\footnotemark[2] & E5 & E1 & Jet\\
G263.7434+00.1161 & IRAS 08470--4321 & $08^\mathrm{h}48^\mathrm{m}48.64^\mathrm{s}$ & $-43\degr32\arcmin29.00\arcsec$ & 0.7\footnotemark[4] & $1.20\times 10^3$\footnotemark[1] & E5 & E1 & Jet (L)\\
G263.7759--00.4281 & IRAS 08448--4343 & $08^\mathrm{h}46^\mathrm{m}34.84^\mathrm{s}$ & $-43\degr54\arcmin29.80\arcsec$ & 0.7\footnotemark[4] & $1.30\times 10^3$\footnotemark[1] & E5 & E1 & Jet (L)\\
G265.1438+01.4548 & RCW 36 & $08^\mathrm{h}59^\mathrm{m}27.40^\mathrm{s}$ & $-43\degr45\arcmin03.70\arcsec$ & 0.7\footnotemark[4] & $7.19\times 10^2$\footnotemark[3] & -- & E1,E4 & Jet/DW\\
G268.3957--00.4842 & IRAS 09017--4716 & $09^\mathrm{h}03^\mathrm{m}25.08^\mathrm{s}$ & $-47\degr28\arcmin27.50\arcsec$ & 0.7\footnotemark[4] & $3.00\times 10^3$\footnotemark[1] & -- & E1 & N/D\\
G274.0649--01.1460A & -- & $09^\mathrm{h}24^\mathrm{m}42.54^\mathrm{s}$ & $-52\degr01\arcmin50.60\arcsec$ & 5.7\footnotemark[1] & $6.00\times 10^3$\footnotemark[1] & E5 & E4 & HCHII\\
G282.2988--00.7769 & IRAS 10082--5647 & $10^\mathrm{h}10^\mathrm{m}00.32^\mathrm{s}$ & $-57\degr02\arcmin07.30\arcsec$ & 3.7\footnotemark[1] & $4.00\times 10^3$\footnotemark[2] & -- & E1 & N/D\\
G283.9146--01.0485 & -- & $10^\mathrm{h}18^\mathrm{m}49.96^\mathrm{s}$ & $-58\degr10\arcmin11.30\arcsec$ & 4.9\footnotemark[1] & $6.00\times 10^3$\footnotemark[1] & -- & E1 & N/D\\
G284.2438--01.1302 & IRAS 10187-5810 & $10^\mathrm{h}20^\mathrm{m}35.17^\mathrm{s}$ & $-58\degr25\arcmin07.10\arcsec$ & 4.2\footnotemark[1] & $4.40\times 10^3$\footnotemark[1] & E5 & E1 & CHII\\
G286.2086+00.1694 & IRAS 10365--5803 & $10^\mathrm{h}38^\mathrm{m}32.70^\mathrm{s}$ & $-58\degr19\arcmin14.30\arcsec$ & 2.3\footnotemark[1] & $8.00\times 10^3$\footnotemark[1] & -- & E1 & Jet (L,C)\\
G287.3716+00.6444 & IRAS 10460--5811 & $10^\mathrm{h}48^\mathrm{m}04.55^\mathrm{s}$ & $-58\degr27\arcmin01.50\arcsec$ & 4.5\footnotemark[1] & $1.80\times 10^4$\footnotemark[2] & E5 & E1 & CHII\\
G298.2620+00.7394 & IRAS 12091--6129 & $12^\mathrm{h}11^\mathrm{m}47.68^\mathrm{s}$ & $-61\degr46\arcmin18.80\arcsec$ & 4.0\footnotemark[1] & $1.50\times 10^4$\footnotemark[1] & E5 & E3 & DW\\
G300.9674+01.1499$^\dagger$ & IRAS 12320--6122 & $12^\mathrm{h}34^\mathrm{m}53.22^\mathrm{s}$ & $-61\degr39\arcmin40.00\arcsec$ & 4.3\footnotemark[1] & $4.50\times 10^4$\footnotemark[3] & -- & E4 & UCHII\\
G301.1364--00.2249$^\dagger$ & IRAS 12326--6245 & $12^\mathrm{h}35^\mathrm{m}35.13^\mathrm{s}$ & $-63\degr02\arcmin31.60\arcsec$ & 4.3\footnotemark[1] & $3.80\times 10^5$\footnotemark[1] & -- & E3 & HCHII\\
G305.1940--00.0051$^\ddagger$ & -- & $13^\mathrm{h}11^\mathrm{m}14.44^\mathrm{s}$ & $-62\degr47\arcmin25.50\arcsec$ & 4.0\footnotemark[2] & $7.00\times 10^3$\footnotemark[1] & E5 & -- & N/D\\
G305.2017+00.2072A & IRAS 13079--6218 & $13^\mathrm{h}11^\mathrm{m}10.45^\mathrm{s}$ & $-62\degr34\arcmin38.60\arcsec$ & 4.0\footnotemark[2] & $3.00\times 10^4$\footnotemark[2] & -- & E4 & U/K\\
G305.5610+00.0124 & IRAS 13111--6228 & $13^\mathrm{h}14^\mathrm{m}26.36^\mathrm{s}$ & $-62\degr44\arcmin30.40\arcsec$ & 4.0\footnotemark[2] & $1.20\times 10^4$\footnotemark[1] & E5 & E4 & HCHII\\
G308.9176+00.1231A & IRAS 13395--6153 & $13^\mathrm{h}43^\mathrm{m}01.70^\mathrm{s}$ & $-62\degr08\arcmin51.20\arcsec$ & 5.3\footnotemark[1] & $9.00\times 10^4$\footnotemark[1] & -- & E4 & UCHII\\
G310.0135+00.3892 & IRAS 13481--6124 & $13^\mathrm{h}51^\mathrm{m}37.85^\mathrm{s}$ & $-61\degr39\arcmin07.50\arcsec$ & 3.2\footnotemark[1] & $6.70\times 10^4$\footnotemark[1] & E5 & E3 & Jet (L)\\
G310.1420+00.7583A & IRAS 13484--6100 & $13^\mathrm{h}51^\mathrm{m}58.27^\mathrm{s}$ & $-61\degr15\arcmin41.70\arcsec$ & 5.4\footnotemark[1] & $8.00\times 10^3$\footnotemark[1] & E5 & E4 & Jet (L)\\
G313.7654--00.8620$^\ddagger$ & IRAS 14212--6131 & $14^\mathrm{h}25^\mathrm{m}01.53^\mathrm{s}$ & $-61\degr44\arcmin57.60\arcsec$ & 7.8\footnotemark[1] & $6.10\times 10^4$\footnotemark[1] & E5 & -- & Jet (L)\\
G317.4298--00.5612$^\dagger$ & IRAS 14477--5947 & $14^\mathrm{h}51^\mathrm{m}37.59^\mathrm{s}$ & $-60\degr00\arcmin19.40\arcsec$ & 14.2\footnotemark[1] & $3.50\times 10^5$\footnotemark[1] & E5 & -- & HCHII\\
G317.8908--00.0578$^\dagger$ & IRAS 14492--5908 & $14^\mathrm{h}53^\mathrm{m}06.16^\mathrm{s}$ & $-59\degr20\arcmin59.40\arcsec$ & 13.6\footnotemark[1] & $6.30\times 10^4$\footnotemark[1] & -- & E2 & UCHII\\
G318.9480--00.1969A & -- & $15^\mathrm{h}00^\mathrm{m}55.31^\mathrm{s}$ & $-58\degr58\arcmin52.60\arcsec$ & 2.4\footnotemark[1] & $1.00\times 10^4$\footnotemark[1] & E5 & E3 & Jet/DW\\
G326.6618+00.5207 & IRAS 15412--5359 & $15^\mathrm{h}45^\mathrm{m}02.84^\mathrm{s}$ & $-54\degr09\arcmin03.00\arcsec$ & 1.8\footnotemark[2] & $1.40\times 10^4$\footnotemark[1] & E5 & E2 & Jet (C)\\
G327.1192+00.5103 & IRAS 15437--5343 & $15^\mathrm{h}47^\mathrm{m}32.80^\mathrm{s}$ & $-53\degr52\arcmin39.30\arcsec$ & 4.9\footnotemark[1] & $3.70\times 10^4$\footnotemark[1] & E5 & E2 & Jet\\
G331.3576+01.0626 & IRAS 16026--5035 & $16^\mathrm{h}06^\mathrm{m}25.78^\mathrm{s}$ & $-50\degr43\arcmin22.00\arcsec$ & 4.5\footnotemark[1] & $1.80\times 10^4$\footnotemark[1] & E5 & E3 & Jet (C)\\
G331.5414--00.0675$^\ddagger$ & -- & $16^\mathrm{h}12^\mathrm{m}09.00^\mathrm{s}$ & $-51\degr25\arcmin47.00\arcsec$ & 5.0\footnotemark[1] & $6.00\times 10^4$\footnotemark[1] & E5 & -- & UCHII\\
G332.0939--00.4206 & IRAS 16124--5110 & $16^\mathrm{h}16^\mathrm{m}16.46^\mathrm{s}$ & $-51\degr18\arcmin25.20\arcsec$ & 3.6\footnotemark[2] & $9.30\times 10^4$\footnotemark[1] & E5 & E2 & Jet (L)\\
G332.8256--00.5498A & IRAS 16164--5046 & $16^\mathrm{h}20^\mathrm{m}11.06^\mathrm{s}$ & $-50\degr53\arcmin16.20\arcsec$ & 3.6\footnotemark[2] & $1.30\times 10^5$\footnotemark[1] & E5 & E2 & UCHII\\
G332.9868--00.4871 & RCW 106 & $16^\mathrm{h}20^\mathrm{m}37.81^\mathrm{s}$ & $-50\degr43\arcmin49.60\arcsec$ & 3.6\footnotemark[2] & $1.80\times 10^4$\footnotemark[1] & E5 & E4 & Jet (C)\\
G337.8442--00.3748$^\dagger$ & IRAS 16367--4701 & $16^\mathrm{h}40^\mathrm{m}26.68^\mathrm{s}$ & $-47\degr07\arcmin13.10\arcsec$ & 3.0\footnotemark[1] & $3.70\times 10^4$\footnotemark[1] & -- & E2 & HCHII\\
G338.9196+00.5495 & -- & $16^\mathrm{h}40^\mathrm{m}34.04^\mathrm{s}$ & $-45\degr42\arcmin07.90\arcsec$ & 4.2\footnotemark[1] & $3.20\times 10^4$\footnotemark[2] & E5 & E3 & Jet (C)\\
G339.6221--00.1209 & IRAS 16424--4531 & $16^\mathrm{h}46^\mathrm{m}05.99^\mathrm{s}$ & $-45\degr36\arcmin43.90\arcsec$ & 2.8\footnotemark[1] & $1.90\times 10^4$\footnotemark[1] & E5 & -- & Jet (L,C)\\
G339.8838--01.2588 & IRAS 16484--4603 & $16^\mathrm{h}52^\mathrm{m}04.66^\mathrm{s}$ & $-46\degr08\arcmin33.60\arcsec$ & 2.7\footnotemark[1] & $6.40\times 10^4$\footnotemark[2] & E5 & E2 & Jet (L)\\
G340.0543--00.2437A & IRAS 16445--4516 & $16^\mathrm{h}48^\mathrm{m}13.69^\mathrm{s}$ & $-45\degr21\arcmin42.50\arcsec$ & 3.8\footnotemark[1] & $2.90\times 10^4$\footnotemark[1] & E5 & -- & Jet (C)\\
G340.2480--00.3725$^\dagger$ & IRAS 16458--4512 & $16^\mathrm{h}49^\mathrm{m}29.97^\mathrm{s}$ & $-45\degr17\arcmin44.40\arcsec$ & 3.7\footnotemark[1] & $3.40\times 10^4$\footnotemark[1] & -- & E2 & UCHII\\
G343.1261--00.0623$^\dagger$ & IRAS 16547--4247 & $16^\mathrm{h}58^\mathrm{m}17.20^\mathrm{s}$ & $-42\degr52\arcmin07.10\arcsec$ & 2.8\footnotemark[1] & $1.90\times 10^4$\footnotemark[1] & -- & E2, E4 & Jet (L)\\
G343.5213--00.5171 & -- & $17^\mathrm{h}01^\mathrm{m}34.04^\mathrm{s}$ & $-42\degr50\arcmin19.70\arcsec$ & 3.2\footnotemark[1] & $6.70\times 10^3$\footnotemark[1] & E5 & E4 & Jet (L,C)\\
G345.4938+01.4677$^\dagger$ & IRAS 16562--3959 & $16^\mathrm{h}59^\mathrm{m}41.61^\mathrm{s}$ & $-40\degr03\arcmin43.30\arcsec$ & 2.4\footnotemark[2] & $1.50\times 10^5$\footnotemark[2] & -- & E2 & Jet (L)\\
G345.5043+00.3480 & IRAS 17008--4040 & $17^\mathrm{h}04^\mathrm{m}22.87^\mathrm{s}$ & $-40\degr44\arcmin23.50\arcsec$ & 2.0\footnotemark[1] & $1.00\times 10^5$\footnotemark[1] & E5 & E4 & Jet (C)\\
G345.9561+00.6123 & IRAS 17012--4009 & $17^\mathrm{h}04^\mathrm{m}43.00^\mathrm{s}$ & $-40\degr13\arcmin13.40\arcsec$ & 2.5\footnotemark[1] & $2.40\times 10^3$\footnotemark[1] & E5 & E3 & U/K\\
G348.6972--01.0263$^\ddagger$ & RCW 121 & $17^\mathrm{h}19^\mathrm{m}58.91^\mathrm{s}$ & $-38\degr58\arcmin14.80\arcsec$ & 2.8\footnotemark[2] & $1.30\times 10^5$\footnotemark[1] & E5 & -- & UCHII\\

\bottomrule
\end{tabular}
\label{tab:targets}
\end{table*}

Multi frequency synthesis was used to image the visibility datasets, and at 5.5, 9 and 17 GHz, the multi-frequency CLEAN algorithm (MFCLEAN in miriad) was used to account for the variation in intensity distribution with frequency, since the fractional bandwidth is $>$12.5\% \citep{Sault1994}. In Fourier transforming the data, the visibilities were weighted according to $T_\mathrm{sys}$ in order to down-weight data taken at times of increased atmospheric instability and thus improve the signal-to-noise ratio of the resultant images. This acted to improve the accuracy of recorded fluxes, at the cost of more elongated restoring beams.

Phase only self-calibration was performed on any strong point source (extended sources give erroneous solutions for the limited baseline numbers of the ATCA) in the field of view, and solutions were applied only if they converged over the course of multiple iterations (in terms of the rms deviations on the gain phases). Time intervals over which to produce calibration solutions took account of the calibrating source's flux and the frequency of the observations (since the atmosphere is more stable at lower frequencies). At 22.8 GHz, if masing activity was present for the 22.23508 GHz water line, the strongest channel was imaged and used as a model for phase-only self-calibration. Simultaneous amplitude and phase self calibration was not performed on any of the datasets since there are not enough constraints with 15 baselines to solve for both the real and imaginary parts of the gains.

\subsection{Measurement of fluxes and spectral indices}
\label{sec:measurement}

For the clean images of our observed fields, both resolved and unresolved emission had to be measured. Thus, integrated radio flux measurements on sources in the field/pointing centre were performed in one of two ways, dependent on this morphology. For simple (i.e. Gaussian-like, both resolved and compact) morphologies, the CASA task, IMFIT, was used to measure the deconvolved dimensions, positions and fluxes. IMFIT works by the fitting of one/several Gaussian profile(s) to the image plane (from a roughly guessed model), whose properties estimate those of the source. Residual maps of the Gaussian model(s) subtracted from the real emission were consequently inspected to make sure that no components were unaccounted for (i.e. the residual map is `noise-like'). For extended and non-Gaussian morphologies the fluxes were integrated over the 3$\sigma$ flux area, positions were taken as the peak emission's location and deconvolved dimensions were measured using equation \ref{eq:deconvolveddims} below. 

\begin{equation} \label{eq:deconvolveddims}
\theta_\mathrm{G} = \sqrt{\theta_\mathrm{source}^2-\theta_\mathrm{beam}^2}
\end{equation}

\noindent where $\theta_\mathrm{G}$ is the deconvolved size of the source, $\theta_\mathrm{source}$ is the measured FWHM of the source and $\theta_\mathrm{beam}$ is the FWHM of the restoring beam.

The image rms deviation was measured in portions of the field without emission, close to the source (so residual artefacts from the cleaning process, such as side lobes, were included in the noise estimation).

The measurement of spectral indices for each source was performed using least squares fitting algorithms. Errors on the fluxes included those provided by the noise/Gaussian fit parameters in addition to a generally accepted 10$\%$ error on the absolute flux calibration of the ATCA. In the case where only two frequencies are present, the spectral index is estimated from the difference between each flux, whose associated error is estimated by combining the fluxes' errors (plus the previously mentioned 10$\%$) in quadrature. Similar techniques are used for the derivation of the power-law index associated to the variation of major axis length with frequency, $\gamma$.

As a note, thermal sources are defined as those possessing a spectral index of $\alpha \geq -0.1$ and are assumed to be dominated by thermal Bremsstrahlung, while non-thermal sources have $\alpha < -0.1$ and are assumed to be dominated by synchrotron emission. We define the optically thin and thick regimes for free-free emission as $-0.1 \leq \alpha \leq 1$ and $1 < \alpha \leq 2$ respectively.

\begin{figure}
\includegraphics[width=84mm]{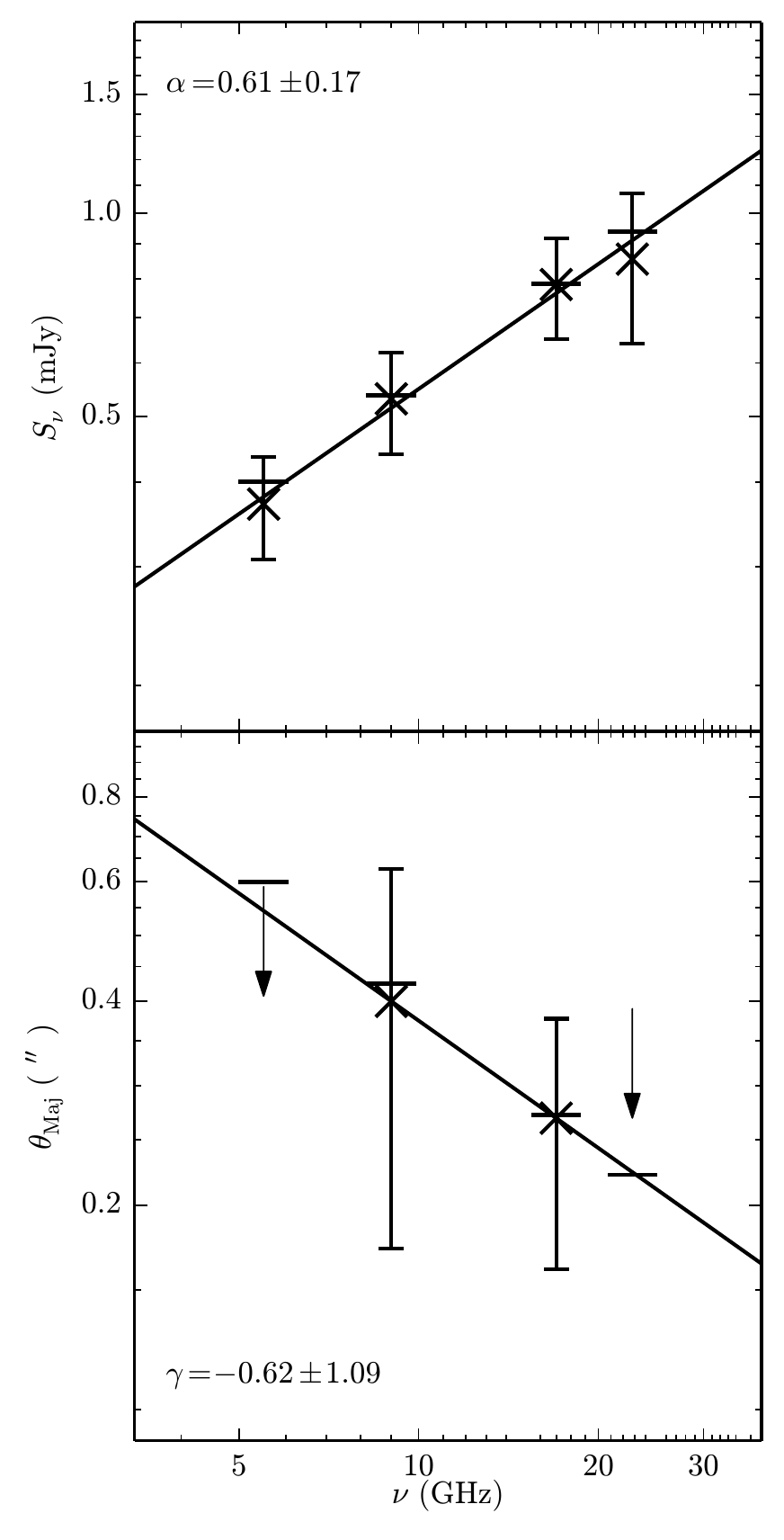}
\caption{The plotted fluxes (top) and deconvolved major axes (bottom) for both the model (wide horizontal lines) and those recovered through deconvolution with a Gaussian (crosses), for a synthetic dataset.  The best fits to the measured fluxes and major axes are shown as black lines. Error bars on the recovered fluxes include a 10$\%$ absolute flux-scale uncertainty expected from observations using the ATCA. Upper limits on the recovered major axes are also shown. Model parameters used were, $\alpha=0.6$, $\gamma=-0.7$, $\theta_\mathrm{Maj}=0.6\arcsec$ (major axis at 5.5 GHz) and $\theta_\mathrm{Min}=0.21\arcsec$ (minor axis at 5.5 GHz).}
\label{fig:ResolveOutTest}
\end{figure}

\subsection{Flux recovery with the 6km configuration}
\label{sec:fluxmodelling}

As a check, modelling was performed on a synthetic dataset in order to calculate how much flux is resolved out at each observing frequency. This is an expected effect of interferometry, since the shortest baseline (in units of wavelength) increases in size with increasing frequency and consequentially, the array's scale-sensitivity drops at larger spatial scales. 

The synthetic dataset was created with the use of the miriad task, UVGEN, and mirrors the real observations. Total time on target (positioned at the pointing/phase centres) for the fake observations was 90 minutes, split into five separate 18 minute scans with visibility integration times of 5 seconds. An elevation limit of 30$\degr$ was imposed (equivalent to a range in hour angle of $-5^\mathrm{h}\!\!\leq$H.A.$\leq4^\mathrm{h}$), and the source declination used ($-60\degr$) is conservative (for the sample) in terms of the maximum elevations for scans ($<60\degr$). Typical values of the system temperature of 30K, 35K, 45K and 100K were used for the 5.5, 9, 17 and 22.8 GHz bands respectively. For the model we used a source with a spectral index $\alpha=0.6$, a major axis index $\gamma=-0.7$, and a total flux (at 5.5 GHz) of 0.4 mJy. This is representative of a typical ionized wind or jet. The major and minor axes at 5.5 GHz were defined as 0.60$\arcsec$ (position angle of $0\degr$) and 0.21$\arcsec$ respectively, corresponding to an opening angle in the jet of $20\degr$. Restoring beams were oriented at a position angle of $175\degr$, with major axes approximately 30$\%$ larger than minor axes.

In Figure \ref{fig:ResolveOutTest}, we see the model fluxes, the recovered fluxes (through deconvolution with a Gaussian as discussed in \S\ref{sec:measurement}) and the derived spectral index/fit. Flux recovered varied between 91\% (at 22.8 GHz) and $>99$\% (at 17 GHz). A spectral index of 0.61$\pm0.17$ was derived, with a derived value for $\gamma$ of $\sim-0.6$ (physical sizes could not be deconvolved at 5.5 and 22.8 GHz). This test of the array's performance towards the types of object this paper focusses on, shows that the ATCA effectively recovers the flux and spectral index of a typical ionized jet/wind. The deconvolved sizes were not recovered at both the lower (due to SNR) and upper (due to SNR and the small spatial scales involved) frequencies, while errors were approaching 100$\%$ of the deconvolved size at 9 GHz. This shows that for low-SNR objects, the deconvolved dimensions may be unreliable, even though the recovered value for $\gamma$ in this case is reasonably close to the model's value. Further to this point, the position angles recovered at 9 and 17 GHz ($145\degr\pm46\degr$ and $126\degr\pm10\degr$ respectively), were not in agreement with the model. The major implication of this result for analysis of low radio SNR sources would be upon the credibility of deconvolved position angles, which should not be taken as absolute. Extended sources, such as UCHII regions, will likely not mirror these results.

\subsection{Extragalactic radio sources}
\label{sec:extragalactic}

A study by \citet{Jackson2004} examined the theoretical, radio, AGN population as a function of flux density and angular density. It considered and predicted FRI, FRII and star-forming galaxy populations on the celestial sphere for three flux-density limits (10 $\mathrm{\mu Jy}$ being the highest) at a frequency of 1.4 GHz. From this work, we extrapolate the predicted  populations to a conservative 3$\sigma$ limit (60 $\mu\mathrm{Jy}$ at 5.5/9 GHz, 90 $\mu\mathrm{Jy}$ at 17 GHz and 300 $\mu\mathrm{Jy}$ at 22.8 GHz) of our study, and further reduce the populations at each frequency in accordance with the observed average spectral index of radio galaxies \citep[$\bar{\alpha}=-0.65$][]{Bornancini2010}. In Table \ref{tab:AGNs}, we have summarised how many radio galaxies we expect to observe for 3 fields of view.

\begin{table}
\centering
\caption{Predicted radio galaxy populations, interpolated from the study by \citet{Jackson2004} using simple power laws for each population. For each frequency, different flux density limits are used (60 $\mu\mathrm{Jy}$ at 5.5/9 GHz, 90 $\mu\mathrm{Jy}$ at 17 GHz and 300 $\mu\mathrm{Jy}$ at 22.8 GHz) typical of our observations.}
\begin{tabular}{ccccc}
\hline 
\textbf{Type} & \boldmath$\nu$ & \multicolumn{3}{c}{\textbf{Field of View}}\\
 & \boldmath$\mathrm{(GHz)}$ & \boldmath$(1\degr)^2$ & \boldmath$(6\arcsec)^2$ & \boldmath$(10\arcmin)^2$ \\ 
\hline 
\multirow{ 4}{*}{FRI} & 5.5 & 222 & 2.2 & 0.002 \\ 
 
 & 9.0 & 161 & 1.6 & 0.001 \\ 
 
 & 17.0 & 88 & 0.881 & 0.001 \\ 
 
 & 22.8 & 41 & 0.414 & $<0.001$ \\ 
\hline 
\multirow{ 4}{*}{FRII} & 5.5 & 22 & 0.225 & $<0.001$ \\ 
 
 & 9.0 & 16 & 0.163 & $<0.001$ \\ 
 
 & 17.0 & 9 & 0.108 & $<0.001$ \\ 
 
 & 22.8 & 9 & 0.089 & $<0.001$ \\ 
\hline 
\multirow{ 4}{*}{Star-forming} & 5.5 & 1061 & 11 & 0.008 \\ 
 
 & 9.0 & 770 & 7.7 & 0.006 \\ 
 
 & 17.0 & 391 & 3.9 & 0.003 \\ 
 
 & 22.8 & 147 & 1.5 & 0.001 \\ 
\hline 
\multirow{ 4}{*}{Total} & 5.5 & 1305 & 13 & 0.010 \\ 
 
 & 9.0 & 947 & 9.5 & 0.007 \\ 
 
 & 17.0 & 488 & 4.9 & 0.004 \\ 
 
 & 22.8 & 197 & 2.0 & 0.001 \\ 
\hline 
\end{tabular}
\label{tab:AGNs}
\end{table}

This estimate shows that even at 5.5 GHz we only expect to observe one radio galaxy, within 5 arcseconds of the pointing centre, for every 50 fields of view. Moreover the star-forming galaxy population has been overestimated, since a power-law is a rough approximation which overestimates the population. We also assumed point-like profiles, however this is not valid for all and likely our observations will resolve many of the closer radio galaxies out, especially at the higher frequencies. 

Another study by \citet{Anglada1998} provides extragalactic source number density estimates (their equation A2), which can be extrapolated to other frequencies (using the average spectral index value quoted above). This alternate approach produces number densities $\sim5$ times lower than those calculated above accross all 4 frequencies. On this basis, we do not expect this type of object to be a concern for our analysis.

\section{Results}
\label{sec:results}

We detect emission in 45 of the 49 fields observed. The maps of radio flux are shown in Figure \ref{fig:ContourPlots} (online version only) at each observed frequency. For the ionized jets identified (and other associated emission), deconvolved positions are given in Table \ref{tab:JetsLobesPositions}, fluxes/spectral indices are given in Table \ref{tab:Jets+LobeFluxes} and sizes/derived values for $\gamma$ are in Table \ref{tab:Jets+LobeSizes} (these tables are in the online version only). Fluxes, sizes and other derived properties of radio sources identified as HII regions are given in Table \ref{tab:HIIProperties} (online version only). Any sources that were not part of the sample, but identified within the primary beam of the observations at 5.5 and 9 GHz, have both their fluxes (primary-beam corrected) and positions listed in Table \ref{tab:ExtraSources} (online version).

\subsection{Classification of the radio emission}
\label{sec:classification}

A high degree of similarity between the different types of object (especially at current resolutions) is, by and large, the main problem and cannot be solved by observations of only one type. In summary, morphology, radio/bolometric flux comparison, radio-spectral features and correlation with previous observations at different wavelengths form the bulk of this process. 

An initial inspection of the radio emission's morphology is usually enough to remove extended HII regions from the sample. For the more compact radio sources (some of which may be embedded in extended emission), a comparison of the measured radio fluxes with those expected from the bolometric luminosities \citep[inferred from the models of][]{Davies2011} aids in the separation of HCHIIs and other types of objects. For this study, a cutoff of 20$\%$ of the expected radio flux (at the highest observed frequency, thus limiting optical depth effects) was used for identification. A potential pitfall at this point originates from the optical depths towards the prospective HII regions. If still in the optically thick regime (typical for HCHIIs at the frequencies in this work), the radio flux may be underestimated leading to a false classification as a jet candidate further down the line. Therefore, to avoid an ambiguous classification as a jet/HII region, if strong Br$\gamma$ emission is seen (from NIR spectra of Lumsden et al., in prep.), an automatic classification as a HII region is enacted. The 1.644$\micron$ [FeII] transition is also of interest as it is indicative of J-type shocks with shock velocities $>50\,\mathrm{km\,s^{-1}}$ \citep{HollenbachMcKee1989}. These shocks arise from the interaction of the jets with ambient material and therefore the presence of this line within the spectra is taken as evidence of ionized jets or winds.

From the radio flux maps, measurement of both $\alpha$ and $\gamma$ enables the differentiation between various jet scenarios in accordance with the models of \citet{Reynolds1986} (i.e. equations \ref{eq:reynoldsAlphaBeta} and \ref{eq:reynoldsGammaBeta}). The deconvolved position angle for the major axis of the jet's emission is taken to be the jet's propagation axis, which can then be compared to position angles for molecular ouflows present in the literature. In cases whereby the difference between these two position angles is $<45\degr$ (i.e. parallel), this is taken as enough evidence to warrant classification as a jet. Should this alignment be perpendicular whilst $\alpha\sim0.6$ and $\gamma\sim-0.7$, a disc wind scenario is deduced. Another indicator in the literature for the outflow axis (to which position angles can be compared) is 2.12$\micron$ H$_2$ emission \citep[such as in G310.0135+00.3892,][]{Caratti2015} which is indicative of collisional excitation of molecular hydrogen through shocks attributable to protostellar outflows and/or large Lyman continuum fluxes \citep{Wolfire1991}. This emission can arise in both C and J-type shocks, unlike the [FeII] line.

Optically-thin \citep[$\alpha\sim-0.1$, e.g. G345.4938+01.4677,][]{Guzman2010} and/or non-thermal lobes of emission \citep[$\alpha<-0.1$, e.g. G343.1261--00.0623,][]{Garay2003-G343.1261}, may be present where the jet shock-ionizes its environment through collisional interaction at working surfaces. The alignment angle of these lobes with the central, thermal jet defines the jet's axis for which comparison with observations in the literature can be conducted. As with the deconvolved position angles of the thermal jet component, should the derived outflow axis be parallel to molecular outflows and/or 2.12$\micron$ H$_2$ emission, this is taken as evidence of a jet and classification as a jet with lobes is made. If perpendicular, classification as a candidate jet with lobes is instead decided.

It is important to emphasize at this point that classification is not an absolute procedure and each case is assessed individually. Although the process presented above provides a robust algorithm in the majority of cases, proper consideration is still required especially in low signal to noise examples where full information cannot be recovered. 

\subsection{Object results}
\label{sec:objresults}
We report the identification of 28 sources with the characteristics expected of ionized jets, two of which can be identified as either disc-winds or ionized jets (G265.1438+01.4548 and G318.9480-00.1969A). Twelve of these jets are given the classification of jet candidate. One source (G298.2620+00.7394) is identified as a disc wind. Of the original sample, 14 HII regions (5 HCHII, 7 UCHII and two compact HII regions) have been identified, with 6 other UCHIIs identified in the fields of view for the images. The conditions for classification of the UCHII regions are based on those of \citet{WoodChurchwell1989}. It is also important to note that in cases where there is no clear cut distinction, based upon radius and emission measure, the measured radius of the HII region takes precedence in the classification process.

Four of the sources were not detected at any observing frequency (G268.3957-00.4842, G282.2988-00.7769, G283.9146-01.0485, and G305.1940-00.0051). Figure \ref{fig:DistLum} shows 1 (which was observed at 9 GHz) of the 4 non-detections as a lower-limit, which lies above the flux expected from an ionized jet. 

Two objects evaded classification, G305.2017+00.2072A and G345.9561+00.6123. G305.2017+00.2072A imaging displayed evidence of extended, resolved out emission towards the object complicating its deconvolution. Removing shorter baselines did not help to improve the images significantly. Although a faint, compact component was present at 17 GHz coincident with both the MSX point source position and a 6.7 GHz methanol maser, it was not detected at 22.8 GHz and thus no spectral index could be derived. With a bolometric luminosity of $2.4\times10^3$L$_\odot$, G345.9561+00.6123 is one of the less luminous MYSOs in the sample. For the radio emission from an optically thin HII region, a value for S$_\mathrm{9GHz}$D$^2$ (see section \ref{sec:RadioLumBolLum}) of 1.6 mJy$\cdot$kpc$^2$ is expected for L$_\mathrm{Bol.}=2400\,\mathrm{L}_\odot$, the measured value being $(0.56\pm0.13)$ mJy$\cdot$kpc$^2$ at 9 GHz and $(1.75\pm0.44)$ mJy$\cdot$kpc$^2$ at 17 GHz. Either this may represent an optically thin (at 17 GHz or above), small, emergent HII region or an ionized jet with a large range of possible spectral indices, hence why it remains unclassified.

In summary, the objects reported here have had their classifications tabulated in table \ref{tab:detections}.

\begin{table}
\centering
\caption{A summary of the numbers and types of objects within the sample, as well as their incidence rate with water masers in the 22.8 GHz band (which takes account of which sources were observed at 22.8 GHz and which were not).}
\begin{tabular}{ccccc}
\hline 
\textbf{Type} & \textbf{Sub-type} & \textbf{Count} & \textbf{Masers?} & \textbf{(\%)}\\ 
\hline 
Ionized Jet    & with lobes       & 10 & 7 & 78 \\ 
•              & w/o lobes        & 4  & 2 & 50 \\
Ionized Jet (C)& with lobes       & 3  & 1 & 50 \\ 
•              & w/o lobes        & 9  & 4 & 50 \\
Jet/Disc Wind  & -                & 2  & 1 & 50\\ 
Disc wind      & -                & 1  & 0 & 0 \\ 
HII region     & HCHII (embedded) & 3  & 1 & 50 \\ 
•              & HCHII (isolated) & 2  & 1 & 50 \\ 
•              & UCHII            & 7  & 5 & 100 \\ 
•              & Compact HII      & 2  & 1 & 50 \\ 
Unknown        & -                & 2  & 0 & 0 \\ 
Non-detection  & -                & 4  & - & - \\ 
\hline 
\end{tabular} 
\label{tab:detections}
\end{table}


\section{Analysis and Discussion}
\label{sec:Analysis}
The analyses presented in this section make use of the Kendall-tau rank correlation test \citep{KendallTau} in order to examine the degree of correlation between various observables. It is used on the basis that it is a non-parametric test, making no assumptions of underlying distributions. Results for each of the correlations presented throughout this section are summarised in Table \ref{tab:tauvalues}. For consistency, we initially calculate the Kendall-tau correlation coefficient for measured flux (not distance-flux, $\mathrm{S}\cdot\mathrm{D^2}$) with distance to be $\tau=0.18$ with a p-value of 0.38, showing an absence of a Malmquist bias in the distance-limited sample. For fitting the data, we assume a simple power-law and use the method of least squares.


\begin{table}
\centering
\caption{A summary values for the Kendall-Tau coefficient of correlation, the associated p-value, for both the distance-limited (D/L) sample of detected jets and all jets detected (the second column). Presence of (J) next to the sample type denotes jet classifications only, while the absence denotes jets and candidate jets. The size of each sample is tabulated in the third column. $\tau_\mathrm{p}$ represents the partial Kendall-Tau correlation coefficient and the associated p-value ($\mathrm{p_{p}}$), whilst controlling for distance.}
\begin{tabular}{llccccc}
\hline
\boldmath$(\mathrm{x},\,\mathrm{y})$ & \textbf{S} & \boldmath$\mathrm{n}$ &\boldmath$\tau$ & \boldmath$\mathrm{p}$ & \boldmath$\tau_\mathrm{p}$ & \boldmath$\mathrm{p_{p}}$ \\ 
\hline
\rule{0pt}{2ex}($\mathrm{L_{Bol}}$      & D/L (J)  & $10$ & $0.674$ & $0.007$ & $0.479$ & $0.023$ \\
~~~vs.                                  & D/L      & $22$ & $0.418$ & $0.006$ & $0.294$ & $0.062$ \\
$\,\mathrm{S_\nu D^2}$)                 & All (J)  & $14$ & $0.552$ & $0.006$ & $0.479$ & $0.023$ \\ 
                                        & All      & $28$ & $0.423$ & $0.002$ & $0.351$ & $0.010$ \\
 
\rule{0pt}{4ex}($\mathrm{M_{Cl.}}$      & D/L (J)  & $05$ & $0.800$ & $0.050$ & $0.802$ & $0.102$ \\
~~~vs.                                  & D/L      & $14$ & $0.626$ & $0.002$ & $0.548$ & $0.009$ \\
$\,\mathrm{S_\nu D^2}$)                 & All (J)  & $08$ & $0.643$ & $0.026$ & $0.645$ & $0.042$ \\ 
                                        & All      & $17$ & $0.603$ & $<0.001$& $0.551$ & $0.039$ \\ 
\rule{0pt}{4ex}($\mathrm{M_{Cl.}}$

                                        & D/L (J)  & $05$ & $0.400$ & $0.327$ & $0.535$ & $0.276$ \\ 
~~~vs.                                  & D/L      & $14$ & $0.473$ & $0.019$ & $0.476$ & $0.023$ \\
$\,\dot{\mathrm{M}}_\mathrm{Jet}$)      & All (J)  & $08$ & $0.357$ & $0.216$ & $0.382$ & $0.228$ \\ 
                                        & All      & $17$ & $0.485$ & $0.007$ & $0.486$ & $0.009$ \\
                                         
\rule{0pt}{4ex}($\mathrm{L_{Bol}}$      & D/L (J)  & $10$ & $0.809$ & $0.001$ & $0.742$ & $0.005$ \\
~~~vs.                                  & D/L      & $21$ & $0.600$ & $<0.001$& $0.550$ & $<0.001$ \\
$\,\mathrm{F_{Jet}}$)                   & All (J)  & $14$ & $0.818$ & $<0.001$& $0.796$ & $<0.001$ \\ 
                                        & All      & $25$ & $0.617$ & $<0.001$& $0.586$ & $<0.001$ \\ 
\hline 
\end{tabular} 
\label{tab:tauvalues}
\end{table}

\subsection{Radio luminosity vs. bolometric luminosity}
\label{sec:RadioLumBolLum}
Using the parameters of stellar atmospheres reported in \citet{Davies2011} (for L$_\mathrm{Bol.}>10^3\mathrm{L}_\odot$), it is possible to infer both the spectral type, and UV photon flux, of a ZAMS star with a specific bolometric luminosity. Since the Lyman flux is intrinsically associated with the amount of material which a young protostar can ionize, we can infer the (optically thin) radio continuum flux expected from an optically thin HII region produced by such a UV flux. Therefore, using this expected radio flux from the bolometric luminosities of the objects,  can help to distinguish between HII regions and less evolved phases of massive star formation (see \S\ref{sec:classification}). The specific radio luminosity expected for ionized jets and winds is not as easy to constrain, due to a lack of theoretical models which incorporate all the variables likely influencing the radio luminosity of jets. Therefore, in order to provide us with an idea of the fluxes expected of these objects, we look towards the empirical relations of radio luminosity against bolometric luminosity, S$_\nu\mathrm{D}^2\simeq10^{-2.1}\mathrm{L}_\mathrm{Bol} \,^{0.6}$, derived by \citet{Anglada1995}. As this relation is derived from a low-mass sample with a maximum bolometric luminosity of $\sim500\mathrm{L}_\odot$ it serves as a comparison for the fluxes we expect to see for thermal jets from MYSOs. Should the mechanism for ionization/jet production be different to the models of shock-ionization \citep{Curiel1987}, we should expect to see a deviation towards the higher masses of the MYSOs.

\begin{figure}
\includegraphics[width=84mm]{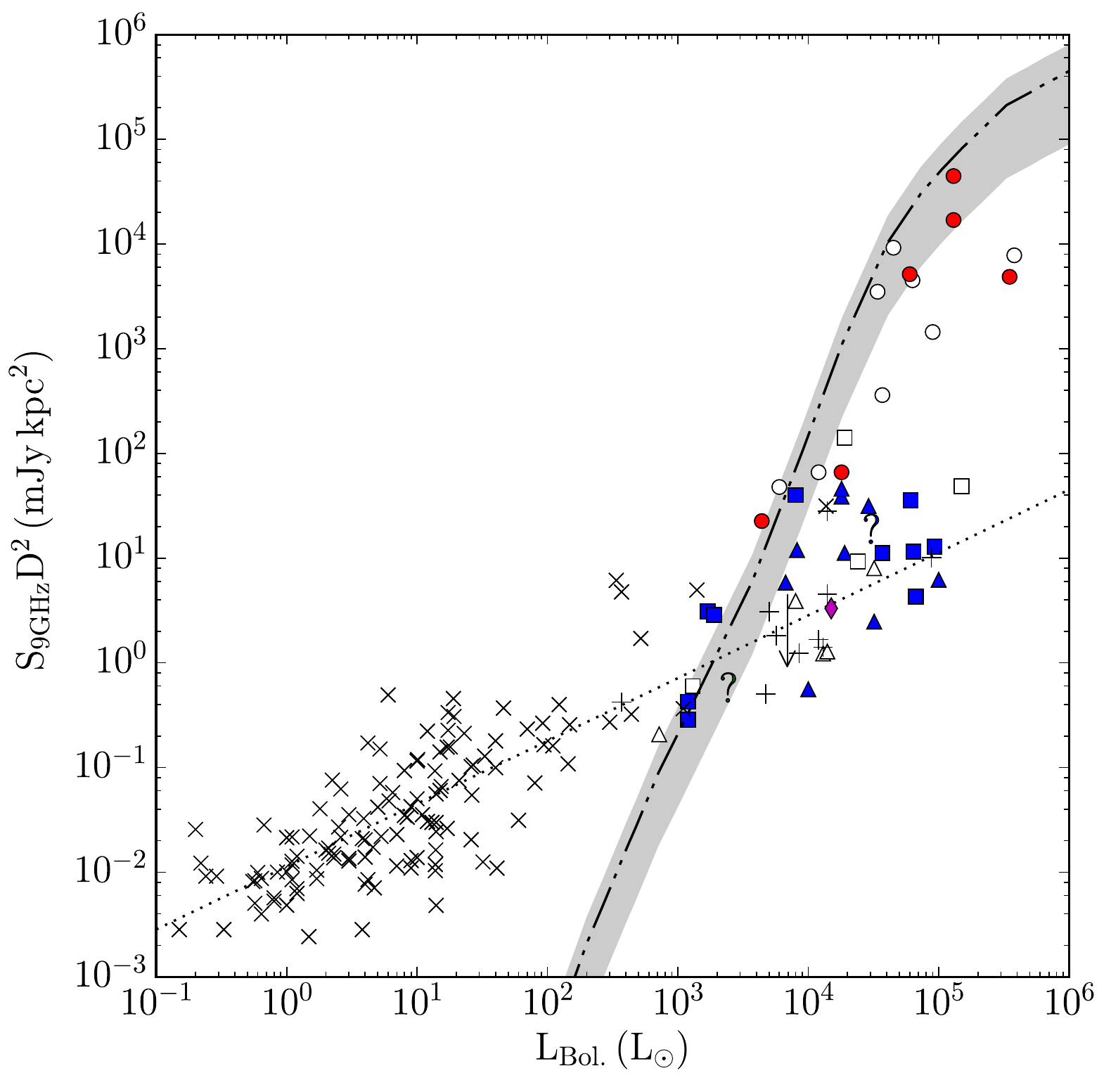}
\caption{A plot of the 9 GHz distance-luminosities against bolometric luminosity for all detected objects (apart from non-thermal lobes) in the sample. Circles (red) represent HII regions, squares (blue) represent ionized jets, diamonds (magenta) represent disc-winds, triangles (blue) represent jet candidates and `$\times$' markers mark low-mass cases normalized to 9 GHz \citep[assuming $\alpha=0.6$][]{Anglada1995,Furuya2003,AMI2011a,AMI2012}. Filled (coloured online) symbols show cases where the flux has been directly measured at 9GHz, while hollow symbols show cases where the flux has been interpolated from measured spectral indices and higher frequency observations. The `+' symbols represent the sample of high-mass jets from \citet{Moscadelli2016}. Lower limits ($<3\sigma$) of non-detections are shown. The dot-dashed line represents the ionized flux expected from the Lyman continuum, taken from stellar models by \citet{Davies2011} for L$_\mathrm{Bol.}>10^3\mathrm{L}_\odot$, and \citet{Thompson1984} for L$_\mathrm{Bol.}\leq10^3\mathrm{L}_\odot$. Grey shading represents the area where the radio-luminosity is between $20\--180\%$ of that expected from the bolometric luminosity (for a HII region). The dotted line represents the empirical relation derived for the low-mass case \citep{Anglada1995}, normalized to 9GHz (assuming $\alpha=0.6$).}
\label{fig:DistLum}
\end{figure}

Figure \ref{fig:DistLum} shows structure attributable to the clear divide between the bright, HII regions near the expected Lyman-continuum line and the jet-like objects which are not as bright as expected from their bolometric luminosities. The HII regions which occupy radio luminosities similar to those of the jets are either resolved out extended regions (higher bolometric luminosities), or are small HCHII regions around lower bolometric luminosity objects. Those HII regions which have lower than expected radio fluxes are usually still in the optically thick regime. It is worth noting that we also expect to see lower radio luminosities for the HII regions than expected from the models of \citet{Davies2011}, due to significant dust absorption of Lyman continuum photons \citep{Wood1988}.

The jets themselves can be fitted (distance-limited sample) with a power-law with an index of 0.63$\pm$0.21 ($\tau=0.67$, p-value$=0.007$). Fitting all jets/candidate (28 sources) yields an index value of 0.75$\pm$0.17. We must add that fitting the low-mass jet sample of \citet{Anglada1995} with the same algorithms as employed upon our high-mass sample yields a higher coefficient of $0.67\pm0.10$ ($\tau=0.55$, p-value$<0.001$) than that recorded by \citet{Anglada1995}. When fitting both the high-mass and low-mass sample of jets together, we arrive at a coefficient value of 0.64$\pm$0.04 ($\tau=0.73$, p-value$\simeq0$). These results imply that similar processes explain the ionization within the high-mass jets, as with their lower mass counterparts \citep[i.e. the shock-ionization models of ][]{Curiel1987}. Some jets however have higher fluxes similar to that expected from HII regions (such as G310.1420+00.7583A). This may be explained by a small ionized core at small radii around the MYSO \citep[which is part of the models of][]{Reynolds1986}, existing with the collimated jet contemporaneously, although higher resolution observations would be required to determine this (sub-milliarcsecond for $1\-5$ AU at a distance of 3 kpc). In such cases, these objects perhaps may represent transition objects between jet and HII phases.

Of the 34 MYSOs from which the distance-limited sample is comprised, we detect a total of 22 objects displaying the characteristics of ionized jets. Of these, 10 (3 of which are candidates) are associated with radio lobes, 10 without (7 of which are candidates) and 2 are ambiguously categorized as jet or disc winds. The mean (and its standard error) bolometric luminosities of isolated jets and jets with lobes are $(2.6\pm0.9)\times 10^4 \mathrm{L_\odot}$ and $(2.7\pm1.0)\times 10^4 \mathrm{L_\odot}$ with medians of $1.8\times 10^4\mathrm{L_\odot}$ and $8.0\times 10^3\mathrm{L_\odot}$ respectively, showing only small differences between the two populations in this respect.

\subsection{9GHz Luminosity vs. Clump Mass}
\label{sec:RadioLumClumpMass}
An ATLASGAL survey by \citet{Urquhart2014}, recorded the $870\micron$ fluxes and sizes of a sample of clumps, 35 of which coincide with sources in this survey. We define the term `clump' as being the substructure within giant molecular clouds (GMCs) which ultimately evolve into stellar clusters. In turn, they harbour the smaller substructure of `cores', which are defined as the evolutionary precursor of single (which may or may not display multiplicity) stellar systems \citep{zinnecker2007a}. Of those 35 sources, 21 have been classified as exhibiting jet-like characteristics. The clump mass was calculated using the referenced dataset in conjunction with the equations of \citet{Hildebrand1983}. 

\begin{figure}
\includegraphics[width=84mm]{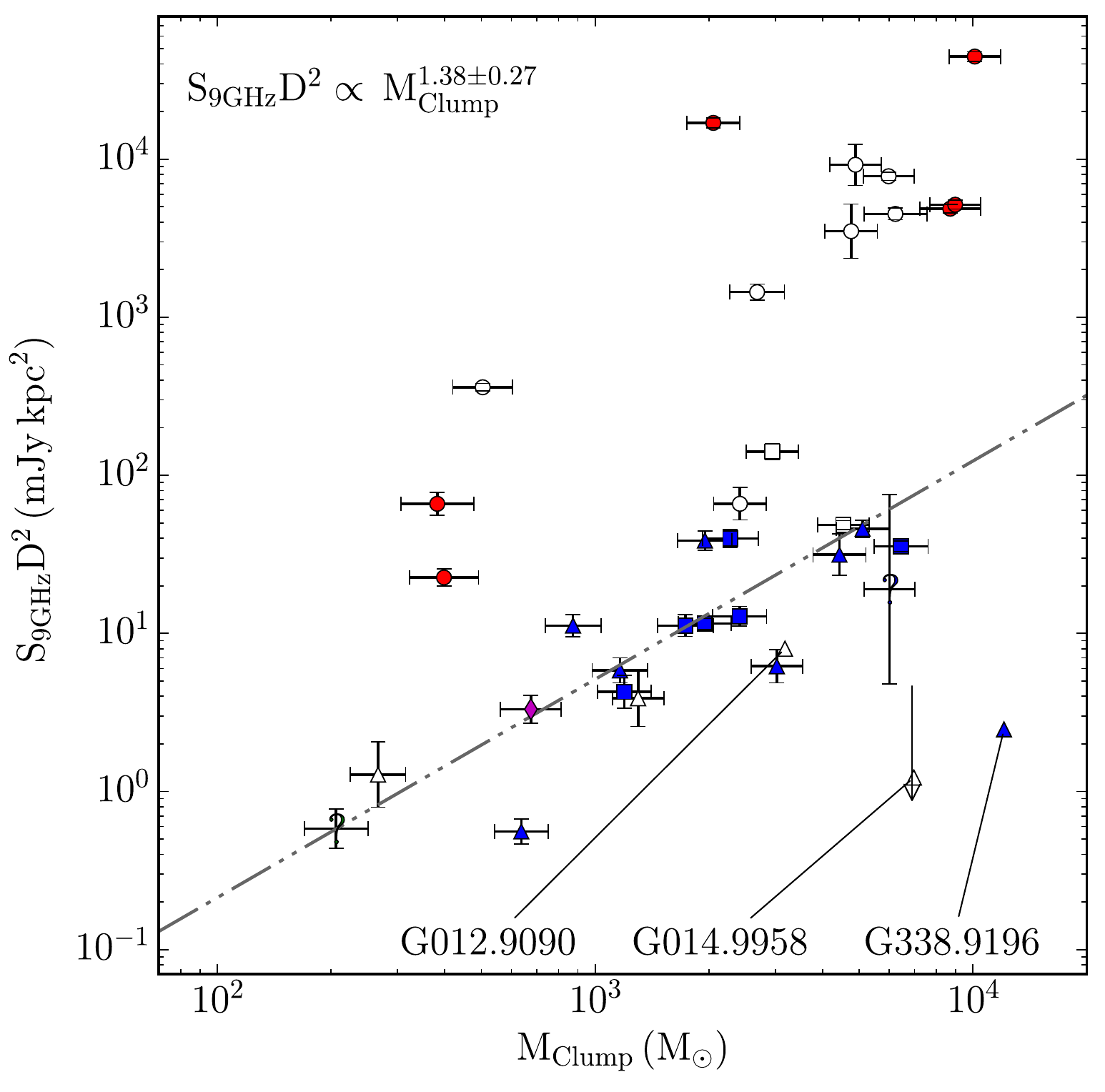}
\caption{A plot of the 9GHz distance fluxes against associated clump masses for HII regions and objects displaying properties associated with either ionized jets or disc winds. The coefficient to the fitted power-law fit is displayed in the top left corner and is represented as the dot-dashed line. Symbols have the same meanings as in Figure \ref{fig:DistLum}.}
\label{fig:FluxVsClumpMass}
\end{figure}

In Figure \ref{fig:FluxVsClumpMass} the radio luminosities at 9GHz have been plotted against the calculated clump masses and a relation of S$_\nu$D$^2\propto\,$M$_\mathrm{Clump}^{1.38\pm0.27}$ is derived for all jets and candidates (1.33$\pm$0.29 for the jets and candidates from the distance-limited sample). There appears to be two outliers toward the heavier end of the clump masses. G014.9958--00.6732 suffered from elongated beams and noisier data as a result of its observed times. G338.9196+00.5495 is known to be in an extremely active star formation region in general, as well as being located at the edge of its natal clump (see \S\ref{sec:JetProps}).

Neglecting the two outliers as well as G012.9090--00.2607 (for the same justification as for G014.9958--00.6732), we calculate the correlation coefficient between the clump masses and distance fluxes at 9GHz (for the distance-limited sample of jets and candidates, neglecting G012.9090--00.2607, G014.9958--00.6732 and G338.9196+00.5495) to be $\tau=0.63$ with a p-value of $0.002$. This indicates a high likelihood of a positive correlation between the two quantities and justifies the empirical power-law, fitted in Figure \ref{fig:FluxVsClumpMass}. For the partial Kendall correlation coefficient for distance-flux with clump mass (controlling for distance) we calculate a value of $\tau=0.55$ with a p-value of $0.009$. This statistical result again supports the correlation between jet luminosity and clump mass, as well as the fact that the most massive stars form in the most massive clumps.

As a further note, the two objects classified as unknown are consistent with the fitted power-law for the jets and candidates. This provides evidence that these objects may be ionized jets, a classification which higher signal-to-noise observations will help to definitively clarify.

\subsection{Radio luminosity vs. IR colours}
\label{sec:RadioLumIRCol}
\begin{figure}
\includegraphics[width=84mm]{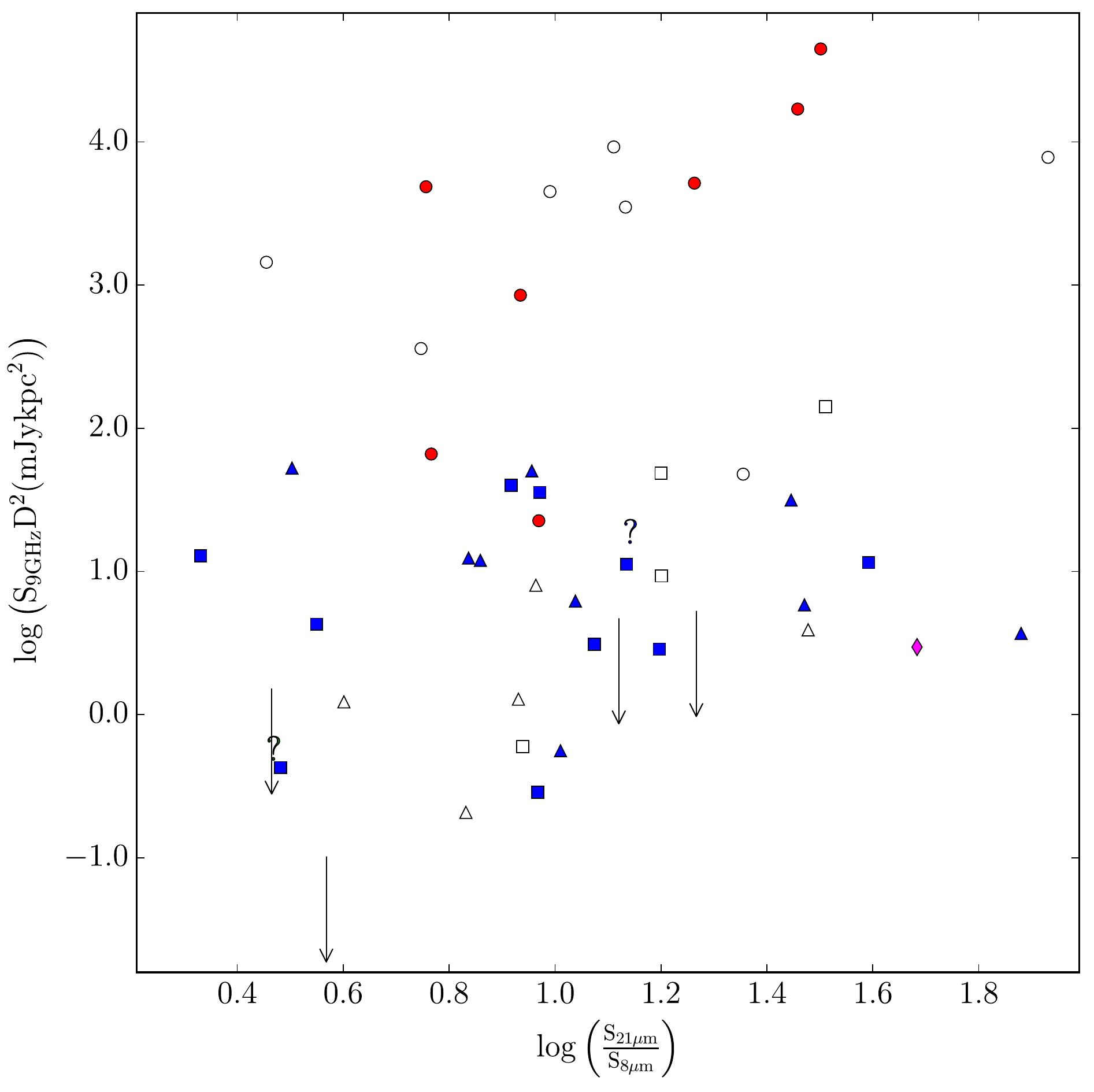}
\caption{A plot of the logarithm of the 9GHz fluxes against the MSX 8$\micron$/21$\micron$ colour for the nearest associated MSX point source, for the HII regions and ionized jets detected. Symbols follow the conventions of Figure \ref{fig:DistLum}}
\label{fig:RadioLumVsMSXCols}
\end{figure}

As discussed in \S\ref{sec:RMS}, the reddening of sources in infrared surveys can be an accurate indicator of youth since the greater the degree of reddening, the more embedded a source will be in its environment. Extrapolating this reasoning to the MYSO phase, the radio luminosity at 9 GHz has been plotted against the MSX (21$\micron$-8$\micron$) colour in Figure \ref{fig:RadioLumVsMSXCols}.
	
%

While the HII regions appear to show higher radio luminosities with the more reddened mid-infrared colours, the jets show no correlation, holding similar fluxes over the sampled colour-space. This may indicate that the jet phase occupies a relatively large portion of an MYSO's lifetime. Non-detections occupy the same range of colours as the detected jets themselves. Considering the average bolometric luminosities for the detections and non-detections of the sample are 3.7$\times10^4$L$_\odot$ and 3.7$\times10^3$L$_\odot$ respectively, the signal to noise ratio may be insufficient to detect any free-free emission, should it exist coincident with the MSX sources with no detected radio flux. It is also interesting to note that the sources exhibiting disc-wind  properties occupy a wide range of colours. However, of these, the unambiguous disc wind (G298.2620+00.7394) holds one of the reddest colours of the jets/disc-winds in the sample. Definite conclusions about whether disc winds represent separate evolutionary stages in MYSO evolution would require a larger sample of such objects than that presented in this work.

\subsection{Coincidence with maser and line emission}
With regards to the water maser line at 22.23508 GHz, we detect emission within $10\arcsec$ of 22 of the 42 sources observed in the 22.8 GHz band. Two more of the 42 sources have a detection of water maser activity in previous studies \citep{Forster1999}, making a total of 24 of 42 sources. The exact breakdown in terms of object type is summarised in Table \ref{tab:detections}.


In comparison with the survey by \citet{Urquhart2009}, we find a detection rate of 52\% compared to their rate of 27\%. This is likely due to the SNR ($\sim50$ times lower) of this survey since it is continuum oriented (i.e. 1 MHz channels), compared to the previous line oriented study ($4\sigma\sim 1\mathrm{Jy}$).  Given the (spectrally) low-resolution nature of continuum observations, no kinematic information could be derived.

\subsection{Ionized jets and their general properties}
\label{sec:JetProps}
Using equation \ref{eq:jetmassloss} \citep{Reynolds1986}, it is possible to infer a mass loss rate in the jet: 

\begin{equation}
\dot{M}_\mathrm{Jet}=\frac{9.38\times10^{-6}\,v_8 \, \mu \, S_\mathrm{mJy}^{3/4} \, d_\mathrm{kpc}^{3/2} \, \theta_0^{3/4}}{x_0 \, \nu_{10}^{\alpha} \, \nu_{m10}^{0.45-\frac{3\alpha}{4}}\,T_4^{0.075}\,(\sin i)^{1/4}\,F^{3/4}}
\label{eq:jetmassloss}
\end{equation}

\noindent where $\alpha$ is the derived spectral index, $\mu$ is the average particle mass (as a fraction of the proton mass), $x_{o}$ is the ionization fraction, $v_\mathrm{8}$ is the terminal velocity of the jet (assumed to be $500\,\mathrm{km\,s}^{-1}$ from proper motion studies in the literature), $\mathrm{S}_{\nu}$ is the integrated flux density, D is the distance to the source, $\nu_{\mathrm{m10}}$ is the turnover frequency, $i$ is the inclination angle (assumed to be $\sim39^\mathrm{o}$), $T$ is the electron temperature and  $\theta_0$ is 
the opening angle at the base of the jet, in radians ($\theta_0=2\tan^{-1}(\theta_\mathrm{min}/\theta_\mathrm{maj})$). F is given below in equation \ref{eq:Fmassloss}:

\begin{equation}
F\equiv F(q_\tau,\,\alpha) \equiv \frac{4.41}{q_\tau(\alpha-2)(\alpha+0.1)}
\label{eq:Fmassloss}
\end{equation}
\noindent where $q_\tau$ is the power-law coefficient with which the opacity falls with distance along the jet propagation axis.

Equations \ref{eq:jetmassloss} and \ref{eq:Fmassloss} necessitate us to make assumptions for the non-observable parameters contained within them. We assume a higher ionization fraction than in the low-mass case \citep[$\sim0.1$][]{Hartigan1994} of 0.2, a typical jet velocity of 500 km s$^{-1}$, (as seen in \S\ref{sec:intro} for other examples of ionized jets around MYSOs) and an electron temperature of $10000\,\mathrm{K}$. For the inclination angle, a value of $39.54\degr$ is adopted since the average value for $\sin(i)$ (for uniformly distributed, random inclinations) is found to be $\sim0.64$. These values are used for all of the jets in the sample. The turnover frequency is assumed to be 50 GHz, apart from for the jets whereby $\alpha<0$, in which case we are already in the optically thin regime and the lowest observed frequency is used. The highest (i.e most resolving) frequency measurements of the minor/major axes are used in the calculation of the opening angle, for which we find a range between $16\--80\degr$ and a median value of $41\degr$, higher than the range of opening angles found by \citet{Moscadelli2016} towards their sample of MYSOs ($10\--30\degr$), and much greater than typical low-mass cases \citep[$5\--10\degr$][]{Tsinganos2009}. The widest opening angle was found towards G313.7654-00.8620, which represents a morphologically complex MYSO in the radio, which may have affected the deconvolution (reflected in opening angle errors of $42\degr$). It is important to note at this point that in cases where the opening angle could not be directly measured, we adopted a value of $(20\pm15)\degr$ in line with typical values within the literature.

A more complex approach is used when assuming values for $q_\tau$. Since (from Equation \ref{eq:reynolds}) it is dependent on values for $\epsilon$ and $q_\mathrm{T}$, we assume one of three models. In the case of $0.4\leq\alpha<0.8$, we assume $\epsilon=1$ and $q_\mathrm{T}=0$, the ``standard" spherical model. Where $\alpha$ is measured to be less than 0.4, we assume $\epsilon=2/3$ and $q_\mathrm{T}=0$, the ``standard" collimated model. Finally, in the instances where $\alpha\geq0.8$, we assign values of $\epsilon=1$ and $q_\mathrm{T}=-0.5$, a conical, recombining jet model.

\begin{figure}
\includegraphics[width=84mm]{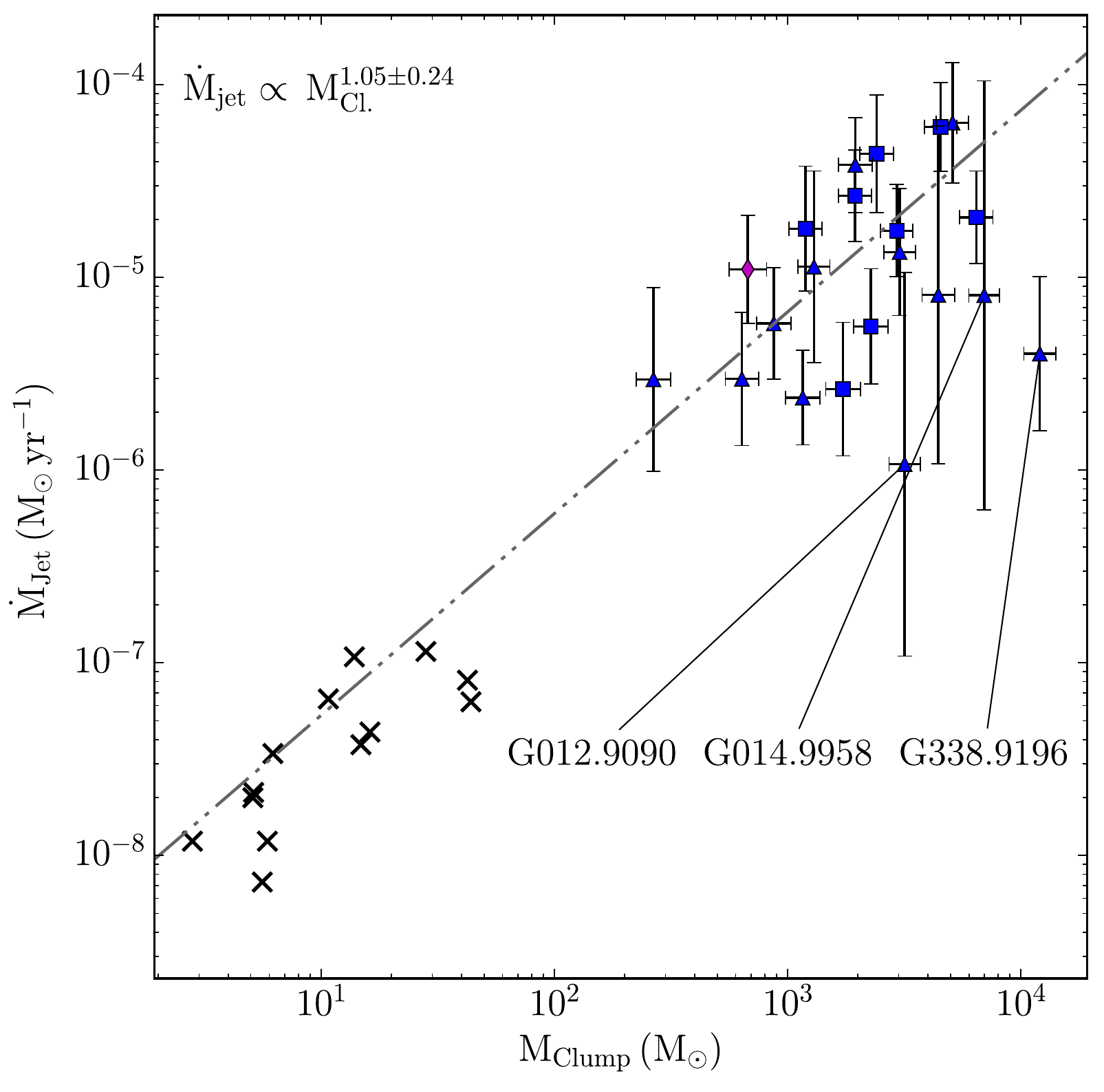}
\caption{A plot of the inferred mass loss rates against ATLASGAL derived clump masses \citep{Urquhart2014} in the jet for all the sources in our sample displaying jet-like characteristics. The dot-dashed line represents the power-law fit to the data, with the power-law index indicated in the top left of the plot. Data for low-mass jets from \citet{AMI2012} is represented by `$\times$' symbols.}
\label{fig:JetMassLossVsClumpMass}
\end{figure}

Derived mass loss rates range between $(0.3-64)\times10^{-6}\mathrm{M}_\odot\mathrm{yr}^{-1}$ for all jets/candidates. We calculate a corresponding average mass loss rate of $(1.5\pm0.3)\times10^{-5}\mathrm{M}_\odot\mathrm{yr}^{-1}$, with a median value of $8\times10^{-6}\mathrm{M}_\odot\mathrm{yr}^{-1}$. Plotting the derived mass loss rates against the clump masses (Figure \ref{fig:JetMassLossVsClumpMass}) shows a power law with an index of $1.05\pm0.24$. For this correlation we calculate $\tau=0.485$ with an associated p-value of $\sim0.007$. For those objects comprising the distance limited sample, we calculated $\tau = 0.473$ and a p-value of 0.019. As shown in table \ref{tab:tauvalues}, these values do not appreciably change for the partial correlation coefficients, whilst controlling for distance. This confirms that the mass loss rates in the jets are related to the clump masses, without a distance related bias.

The same three sources were neglected from the correlation tests (G012.9090--00.2607, G014.9958--00.6732 and G338.9196+00.5495) as in \S\ref{sec:RadioLumClumpMass}. W33A (G012.9090--00.2067) and the Kleinmann-Wright \citep{Kleinmann1973} object (G014.9958.00-6732) both lie at declinations conducive to elongated beam shapes, thus adversely affecting their deconvolved sizes. Indeed, W33A was found to have a deconvolved size of $(0.53\arcsec\pm0.14)\times(0.27\arcsec\pm0.11)$ at 43GHz by \citet{VanDerTak2005}, giving an opening angle at the base of the jet of 0.94 rads, in contrast to our derived value of 0.18 rads. This would in turn derive $\dot{\mathrm{M}}_\mathrm{jet}=(1.7\pm1.0)\times10^{-5}\mathrm{M}_\odot\mathrm{yr}^{-1}$ (our derived value is $\sim1\times10^{-6}\mathrm{M}_\odot\mathrm{yr}^{-1}$). W33A is also known to be composed of at least 3 separate sources at very high resolutions, Q1, Q2 and Q3 (Q1 represents the main jet). This complicates the analysis, requiring higher resolutions observations (at all frequencies) in order to separate and determine the natures of each source. 

The KW object is positioned at the edges of two ATLASGAL (R$_\mathrm{eff}\sim60\arcsec$ and R$_\mathrm{eff}\sim90\arcsec$) clumps, the membership to which is uncertain. If the accretion, and hence jet mass loss rates, are intrinsically tied to the reservoir of material from the clump, this may explain (in combination with the elongated beam) its current (relatively low) mass loss rate of $\sim4\times10^{-6}\mathrm{M}_\odot\mathrm{yr}^{-1}$.

G338.9196+00.5495 images had low signal to noise ratios, showed the presence of 5 HCHII/UCHIIs within 1$\arcmin$ of the source, and displayed partially resolved out extended emission at the lowest frequencies. The busy nature of the environment would affect the reservoir of available material to the MYSO and hence affect jet outflow rates, as well as the quality of the images and deconvolved dimensions of the object itself (which is unresolved at 17 and 22.8 GHz). ATLASGAL images of G338.9196+00.5495 also show its parental clump to be comprised of two, non-resolved clumps, with the source offset on the southern edge of the smaller, thus the clump's mass may be overestimated. Considering that G338.9196+00.5495 is also an outlier in Figure \ref{fig:FluxVsClumpMass}, it is likely that the clump mass is poorly constrained as discussed above.

As a check with the rest of the data, all sources apart from the KW object, G338.9196+00.5495 and G345.9561+00.6123 (which is still of uncertain classification) are positioned within the inner 25\% of their parental clumps \citep[c.f.][who found 90\% of embedded objects were separated by less than 0.5pc from their associated clump's peak flux position]{Urquhart2014}.

\begin{figure}
\includegraphics[width=84mm]{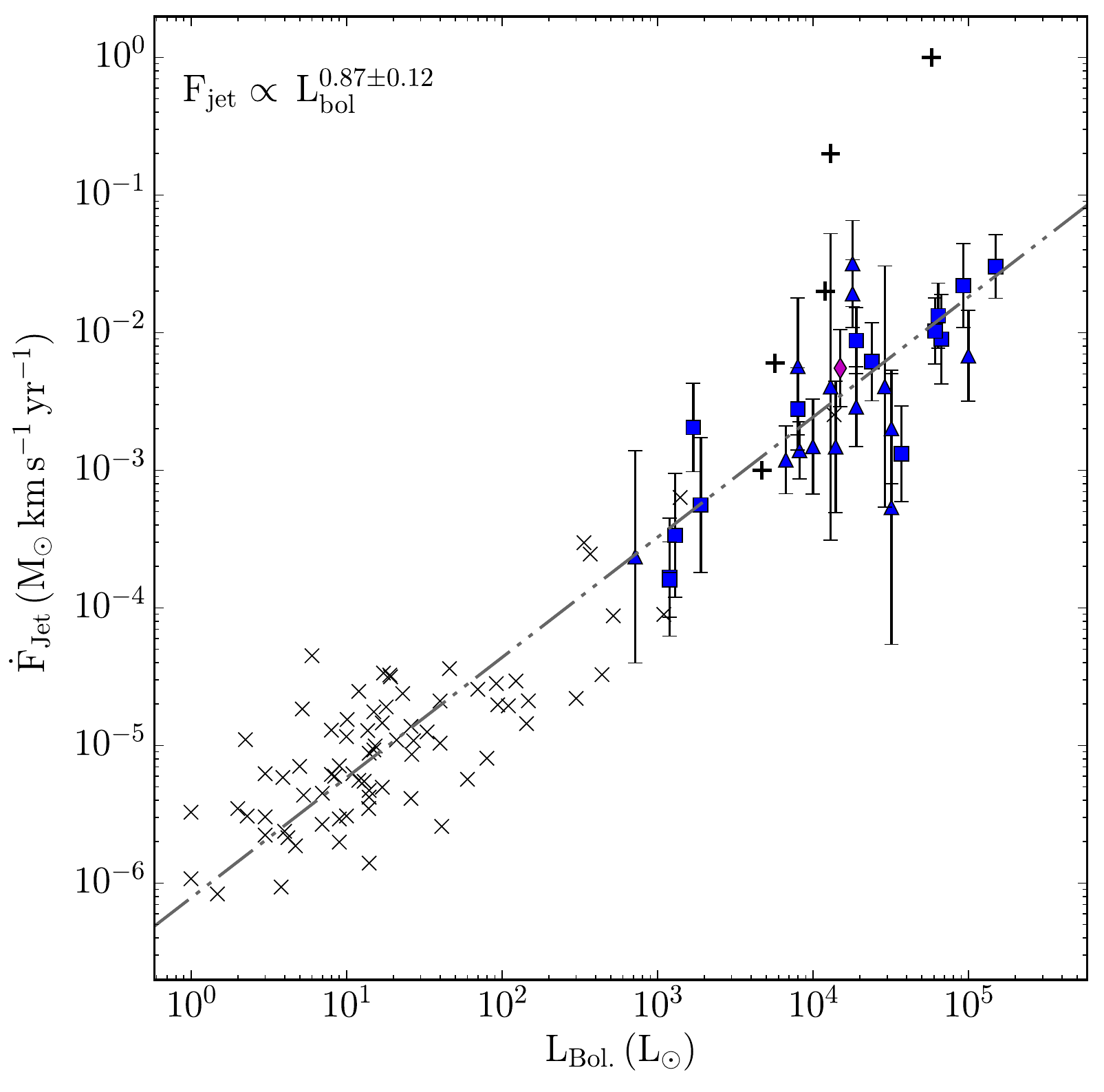}
\caption{A plot of the inferred momentum rate against bolometric luminosity for the sources in our sample displaying jet-like characteristics. The dot-dashed line represents the power-law fit to the data, with the power-law index indicated in the top left of the plot. Data for low-mass jets from \citet{AMI2012} is represented by `$\times$' symbols, assuming a velocity of $200\,\mathrm{km\,s^{-1
}}$ unless otherwise indicated in the literature. The `$+$' symbol indicates data for the high mass sample from \citet{Moscadelli2016}.}
\label{fig:JetMomentaVsL_Bol}
\end{figure}

Inferred momentum rates (assuming a terminal velocity for the jet of $500\mathrm{km\,s}^{-1}$) against the bolometric luminosities of the powering sources, are plotted in Figure \ref{fig:JetMomentaVsL_Bol}. We find the relation presented in Equation \ref{eq:JetMomentaLbolRelationOurs} for those sources comprising the distance limited sample (derived values change little between jet and jet/candidate sub-samples). Comparing this to the CO outflow momentum rate/bolometric luminosity relations found by \citet{CabritBertout1992} (Equation \ref{eq:JetMomentaLbolRelationCabrits}) and \citet{Maud2015CO} (Equation \ref{eq:JetMomentaLbolRelationLukes}):

\begin{equation}
\log_{10}\mathrm{F}_\mathrm{jet}=(-6.11\pm0.49)+(0.87\pm0.12)\log_{10}\mathrm{L}_\mathrm{bol}
\label{eq:JetMomentaLbolRelationOurs} 
\end{equation}
\begin{equation}
\log_{10}\mathrm{F}_\mathrm{CO}=(-4.36\pm0.12)+(0.69\pm0.05)\log_{10}\mathrm{L}_\mathrm{bol}
\label{eq:JetMomentaLbolRelationCabrits} 
\end{equation}
\begin{equation}
\log_{10}\mathrm{F}_\mathrm{CO}=(-4.60\pm0.46)+(0.61\pm0.11)\log_{10}\mathrm{L}_\mathrm{bol}
\label{eq:JetMomentaLbolRelationLukes} 
\end{equation}

Within errors, the momentum rate of ionized jets appears to depend more heavily on the bolometric luminosities than the molecular outflows. However the incidence rate of massive molecular outflows in the (6kpc distance-limited) MYSO sample of \citet{Maud2015CO} was found to be 66$\%$, similar to the incidence rate of jets and candidates for our distance-limited sample of 65$\%$. The power-law coefficient for the relation between the molecular mass outflow rate and clump mass was found to be $0.55\pm0.10$. In comparison to our fitted value of $1.06\pm0.24$ for the jet outflow rate, this indicates a heavier dependence of the outflow rates on the clump mass for ionized jets, than for the molecular outflows.

\begin{figure}
\includegraphics[width=84mm]{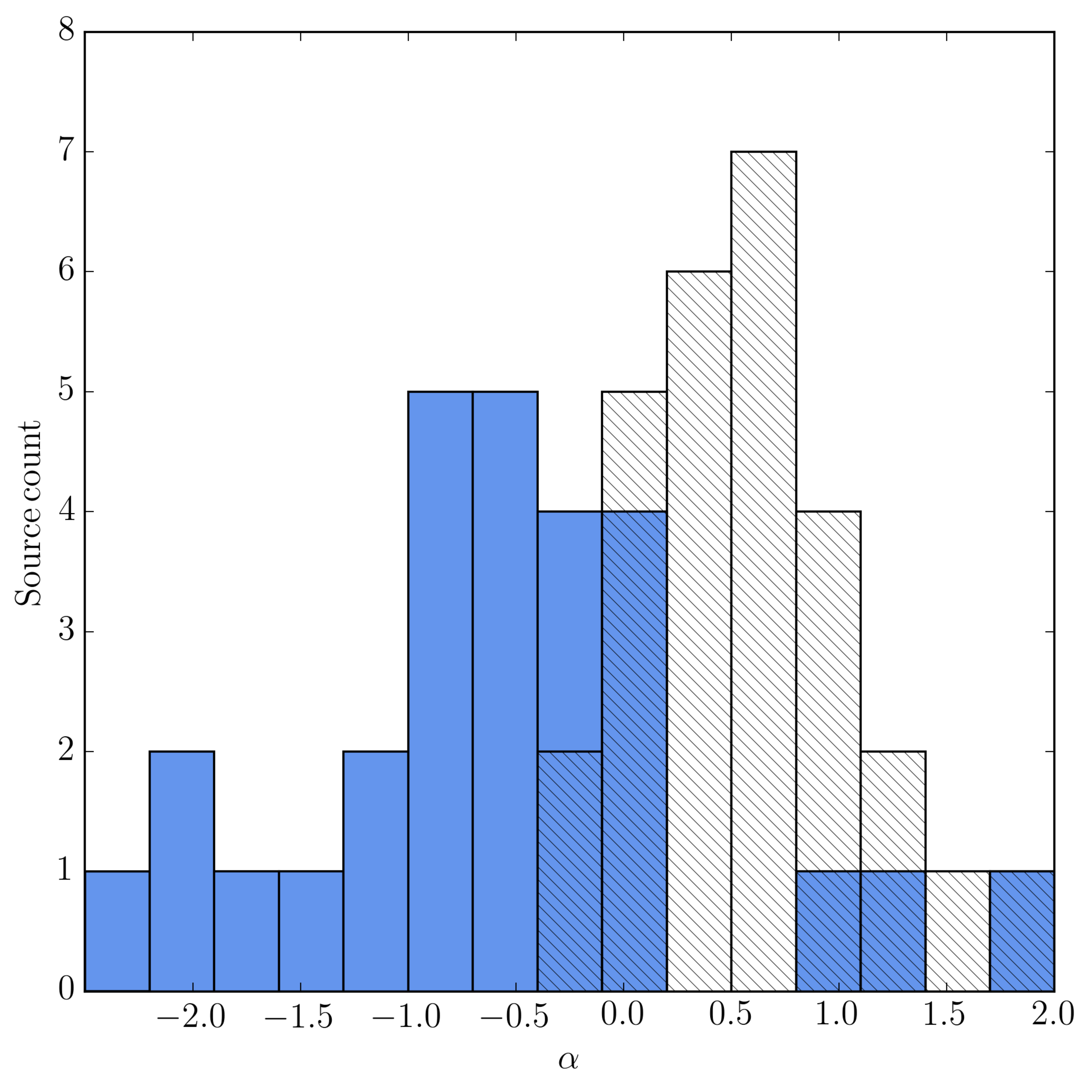}
\caption{A histogram of all the spectral indices derived for those sources determined to exhibit jet-like characteristics (hatched). The solid blue bars are for the associated lobes. Binning starts at $\alpha=-2.5$ and bin widths are set at 0.3 due to the sample size.}
\label{fig:SpecIndHist}
\end{figure}

In Figure \ref{fig:SpecIndHist}, the histogram of the spectral indices derived for all jets in our sample is presented. This shows a peak in the jet's spectral indices of $\sim0.6$ as per the standard, conical jet model. All detected jets exhibit spectral indices expected from the models of \citet{Reynolds1986} of between -0.1 and 1.6 within errors. As for the lobes of emission associated to the thermal jets, we derive non-thermal spectral indices for lobes associated to 10 of the 13 jets. An average spectral index in the non-thermal lobes of $-0.55$ is also calculated showing that synchrotron emission is the dominant emission mechanism in these cases and therefore magnetic fields are present in the environments of ionized jets from MYSOs. This result also agrees with radio lobes being produced via Fermi acceleration at shocks, whereby we expect $\alpha = (1-p)/2$, where $p$ is the power-law coefficient for the electron energy distribution ($p=2$ in head-on shocks). Four of the jets show associated lobes with thermal spectral indices, G310.0135+00.3892 (SW), G310.1420+00.7583A (B), G313.7654-00.8620 (B2) and G332.0939-00.4206 (E). For all four cases, apart from  G313.7654-00.8620, the spectra of the components are not well fitted by simple power-laws suggesting these components may represent extended emission (i.e. resolved out at higher frequencies), the natures of which are indeterminate. G313.7654-00.8620 was observed at the lower two frequencies only, and so the presence of this effect is unknown, raising another possibility of multiplicity. For the SW component of G310.0135+00.3892, the alignment between the central radio source and this component is parallel to derived outflow directions in the literature. Therefore, it is likely this represents thermal emission from the jet material however more sensitive observations with higher signal-to-noise are required to conclude this definitively.

\begin{figure}
\includegraphics[width=84mm]{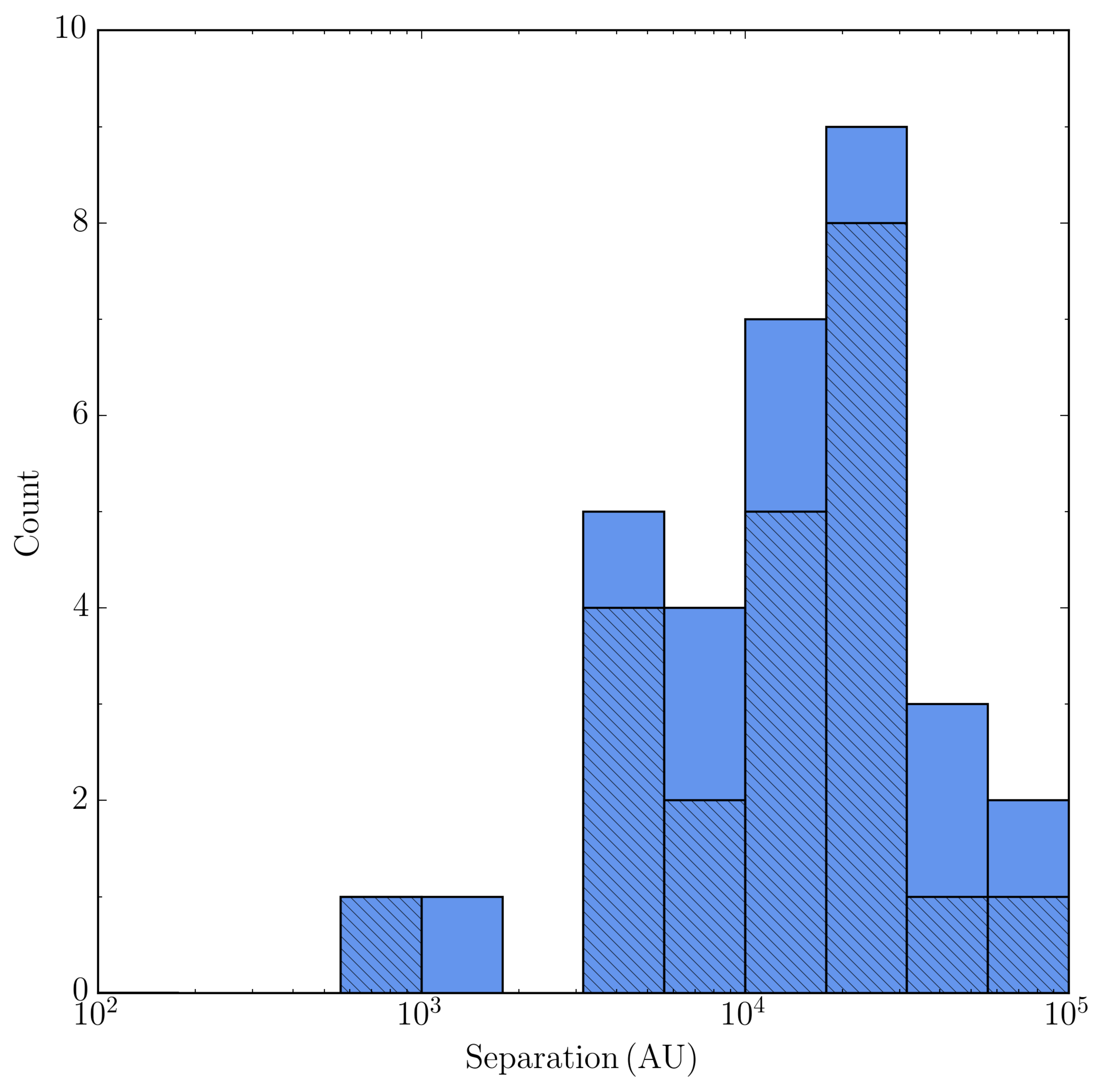}
\caption{A logarithmic histogram of the spatial separations between all lobes (solid bars), and lobes where $\alpha<0$ (hatched bars), and their powering central jets.}
\label{fig:LobeSep}
\end{figure}

For those jets identified which had associated lobes of emission, the separation of these lobes from the central jet was calculated (assuming an inclination of 90$\degr$) and in Figure \ref{fig:LobeSep}, the histogram of logarithmic lobe separations is plotted. A peak in the separations at $\sim10^4\mathrm{AU}$ from the central object is evident, with a mean value of $2.0\times10^4\mathrm{AU}$ (and a median value of $1.3\times10^4\mathrm{AU}$). The minimum separation calculated ($\sim600\mathrm{AU}$) was for the NW component of G263.7759--00.4281, while the maximum ($\sim90000\mathrm{AU}$) was for component D of G313.7654--00.8620. If these lobes represent internal shocks within the jet as a result of periodic ejection coupled with non-constant ejection velocities we can calculate the typical time between these ejections. Assuming an ejection velocity of $500\mathrm{km\,s^{-1}}$ for the first ejection, which increases by 10\% for the second, and using the typical value for lobe separations of $1.3\times10^4\mathrm{AU}$, it can be inferred that the time between these ejections is typically $\sim10\mathrm{years}$.

It is important to note that the population of small separation lobes is under-sampled and we cannot reliably comment on lobe separations for distances $< 4000\mathrm{AU}$, due to the minimum resolvable scales of the observations. In order to investigate the spectral consequences of this resolution limit, we analysed the spectral indices for the combined fluxes of north and south lobes, for both G263.7434+00.1161 and G310.0135+00.3892 (due to their relative morphological simplicities). Derived values for the spectral index of the combined emission were found to be $\alpha=0.09\pm0.13$ and $\alpha=0.42\pm0.17$ for G263.7434+00.1161 and G310.0135+00.3892 (for the thermal jets alone, $\alpha=0.39\pm0.18$ and $\alpha=1.27\pm0.18$) respectively, with the combined fluxes being fitted well with a simple power law. This indicates that any unresolved, optically thin lobes which are inseparable from their powering, thermal jets, would flatten the jet's spectrum without strongly distorting its power law profile. This effect could act to decrease the measured spectral index for jets, whose flux incorporates those of optically thin or non-thermal lobes, offering another explanation for a slightly larger proportion of jets with spectral indices $<0.6$ (see Figure \ref{fig:SpecIndHist}).




\subsection{HII regions and their general properties}
\begin{figure}
\includegraphics[width=84mm]{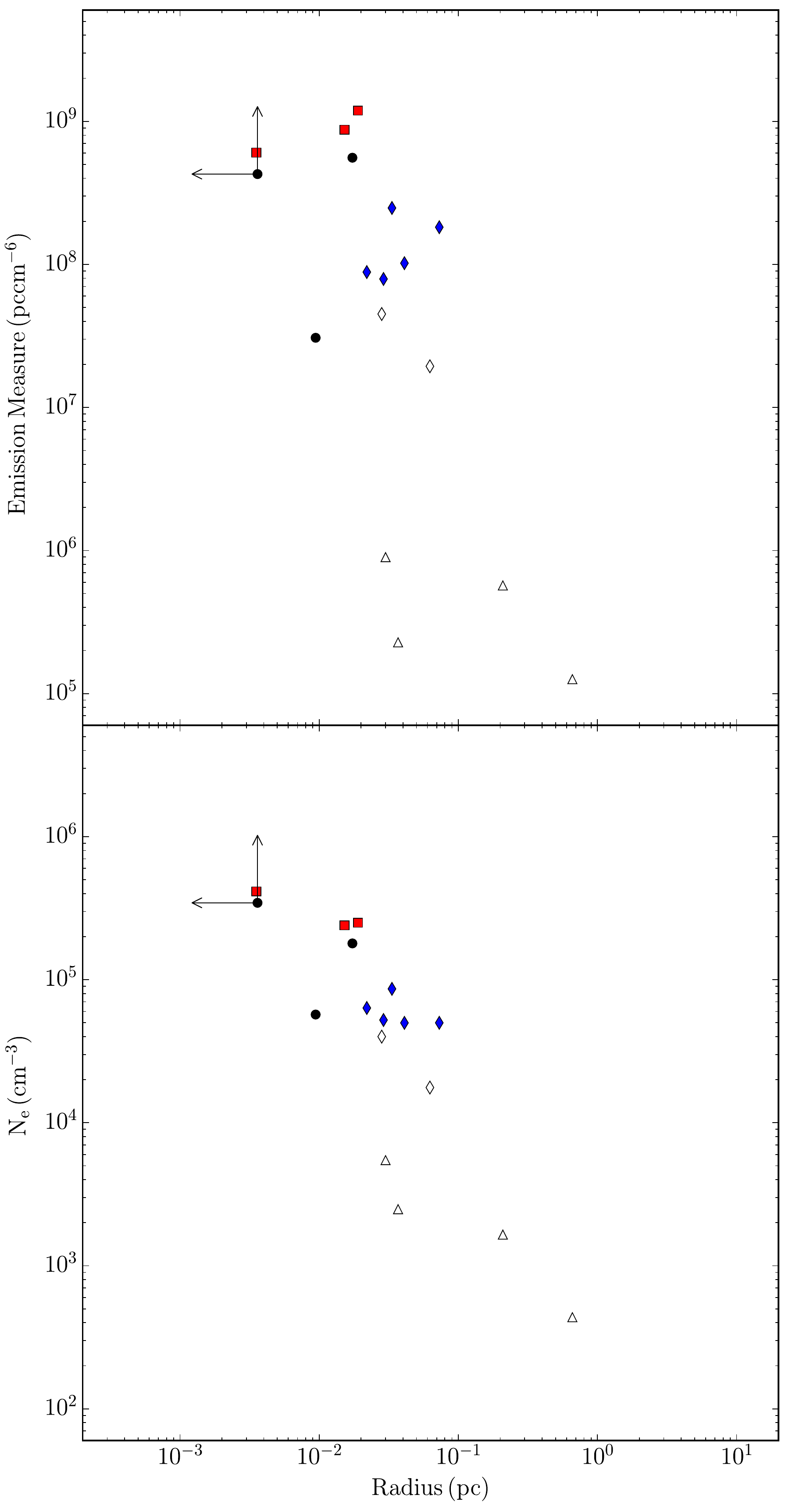}
\caption{A plot of the calculated emission measures against radius for the HII regions detected in our sample (top) and the corresponding electron densities inferred against radius (bottom). Circles (black) represent the compact HCHII `core' seen in 3 HII regions, squares (red) represent other HCHII regions, diamonds (blue) represent UCHII regions, and triangles represent compact and classical HII regions. Any symbols in white suffer from resolving out effects due to their scale.}
\label{fig:EMvsRadiiHIIs}
\end{figure}

From the HII regions that were included in the overall sample, we have calculated both the emission measures and physical radii (deconvolved). The three HCHII regions displaying a compact core have been separated from their extended background emission, with both components analysed separately. Both emission measures and radii are  plotted in Figure \ref{fig:EMvsRadiiHIIs}. 

For the HCHII regions in our sample, we find an average emission measure of 6.51$\times10^8\mathrm{pc\,cm}^{-6}$ and corresponding electron density of 2.28$\times10^5$cm$^{-3}$. For the ultra-compact HII regions detected, we find values of 1.09$\times10^8\mathrm{pc\,cm}^{-6}$ and 5.13$\times10^4$cm$^{-3}$ for the average emission measure and electron density. Average radii were calculated to be 0.013pc and 0.041pc, while average bolometric luminosities were calculated to be $(1.6\pm0.4)\times10^5$L$_\odot$ and $(7.9\pm1.1)\times10^4$L$_\odot$, for HCHIIs and UCHIIs respectively \citep[assuming a $34\%$ error in the bolometric luminosities, as in][]{Mottram2011a}. Although HII regions are not the focus of this paper, the derived properties provide a consistency check for typical values reported in the literature \citep{Mezger1967,WoodChurchwell1989,Hoare2007PRPL}.

\section{Conclusions}
\label{sec:Conclusions}
In this work, we present the results of radio observations made towards 49 massive young stellar objects potentially harbouring ionized jets, 34 of which form a distance-limited sample . Our goal is to establish how common this phenomenon is for the formation process of massive stars, the relation to possible accretion processes/models and the physical parameters typical of the jets themselves. In summary, our main results and conclusions are:
\begin{enumerate}
\item From the distance-limited sample of 34 MYSOs we detect 22 ionized jets and jet candidates 	showing this to be a common phenomenon in forming massive stars. Disc based accretion processes are therefore the dominant mechanism of accretion (up to masses of at least 25M$_\odot$).
\item Ionized jets are more common than disc winds, by at least an order of magnitude. This suggests that either ionized jets are generally brighter than disc winds (if they are concurrent), or that a distinct disc-wind phase exists and is relatively short in comparison.  
\item Assuming a MYSO phase lifetime of $\sim10^5$yr \citep{McKeeTan2002}, a jet-phase lasting $(2.9-6.5)\times10^4\,\mathrm{yr}$ is implied from the incidence rate of jets and jet candidates. Since some objects potentially represent a transition phase between ionized jet and HII region, this time range is placed at the later end of an MYSO's lifetime, implying accretion doesn't completely halt after the initial production of an HII region.
\item The luminosity and momenta of our sample of massive, ionized jets scale with bolometric luminosity in the same way as low-mass jets, supporting the idea that these jets are produced in the same way over YSO mass ranges upto at least 25$\mathrm{M_\odot}$.
\item Typical jet mass-loss rates and momenta rates are $1.4\times10^{-5}$M$_\odot$yr$^{-1}$ and $0.7\times10^{-2}$M$_\odot\mathrm{km\,s}^{-1}$yr$^{-1}$ respectively, assuming an average ionization fraction of 20$\%$ and jet velocity of 500$\mathrm{km\,s}^{-1}$. This suggests an average mass of $\sim1\--2\mathrm{M}_\odot$ (corresponding to $\sim10^{48}\mathrm{ergs}$) is lost through the ionized jet mechanism, to the ISM, over the course of an MYSOs jet-phase.
\item Jet outflow rates (and hence accretion rates) are closely related to the mass of the molecular clump from which they form, which agrees with the idea that the most massive stars form in the most massive clumps. This is also supported by the fact that $88\%$ of the jets are found in the inner 25$\%$ of their associated clump.
\item Synchrotron emission is commonplace (present in $\sim75\%$ of the jets displaying associated lobes) and as a result, this lends support to the idea of magnetic collimation in ionized jets around MYSOs, as in the low mass case.
\end{enumerate}

\section*{Acknowledgments}
We thank the referee for constructive comments. SJDP gratefully acknowledges the studentship funded by the Science and Technology Facilities Council of the United Kingdom (STFC). AEG acknowledges support from FONCEDYT grant 3150570. The National Radio Astronomy Observatory is a facility of the National Science Foundation operated under cooperative agreement by Associated Universities, Inc. This paper has made use of information from the RMS survey database at http://www.ast.leeds.ac.uk/RMS which was constructed with support from the Science and Technology Facilities Council of the United Kingdom.

\footnotesize{
	\bibliographystyle{mn2e.bst}
	\bibliography{./Paper_Short}
}

\clearpage

\onecolumn

\appendix

\section{Tables}

\begin{table}
\centering
\caption{The deconvolved positions of the MYSOs displaying characteristics of jets and/or disc-winds, as well as their associated lobes at all 4 observing frequencies. In cases where positions are given without errors, we fixed the positions to allow deconvolution solutions to converge (generally from the highest frequency where deconvolution with no fixed parameters is possible).} 
\begin{scriptsize}
\begin{tabular}{lcccccccccc}
\hline
\textbf{Name} & \textbf{Comp.} & \textbf{Type} & \multicolumn{4}{c}{\textbf{5.5 GHz}} & \multicolumn{4}{c}{\textbf{9GHz}} \\
 & & & \textbf{R.A. (J2000)} & \boldmath$\Delta\alpha$ & \textbf{Dec.(J2000)} & \boldmath$\Delta\delta$ & \textbf{R.A. (J2000)} & \boldmath$\Delta\alpha$ & \textbf{Dec. (J2000)} & \boldmath$\Delta\delta$ \\
\hline
G251.2337-01.9535 &  & Jet (C) & $08^\mathrm{h}02^\mathrm{m}42.98^\mathrm{s}$ & 0.030 & $-34\degr31\arcmin49.15\arcsec$ & 0.082 & $08^\mathrm{h}02^\mathrm{m}42.98^\mathrm{s}$ & 0.021 & $-34\degr31\arcmin49.12\arcsec$ & 0.051 \\
G254.0491-00.5615 &  & Jet & $08^\mathrm{h}15^\mathrm{m}57.12^\mathrm{s}$ & 0.077 & $-36\degr08\arcmin07.16\arcsec$ & 0.196 & $08^\mathrm{h}15^\mathrm{m}57.13^\mathrm{s}$ & 0.040 & $-36\degr08\arcmin06.90\arcsec$ & 0.085 \\
G254.0548-00.0961 & C & Jet & $08^\mathrm{h}17^\mathrm{m}52.55^\mathrm{s}$ & 0.044 & $-35\degr52\arcmin47.64\arcsec$ & 0.115 & $08^\mathrm{h}17^\mathrm{m}52.56^\mathrm{s}$ & 0.029 & $-35\degr52\arcmin47.63\arcsec$ & 0.067 \\
 & W & Lobe & $08^\mathrm{h}17^\mathrm{m}52.31^\mathrm{s}$ & 0.131 & $-35\degr52\arcmin51.62\arcsec$ & 0.340 & $08^\mathrm{h}17^\mathrm{m}52.27^\mathrm{s}$ & 0.130 & $-35\degr52\arcmin50.96\arcsec$ & 0.298 \\
G263.2283+01.5712 &  & Jet & $08^\mathrm{h}53^\mathrm{m}09.45^\mathrm{s}$ & 0.034 & $-42\degr13\arcmin07.72\arcsec$ & 0.061 & $08^\mathrm{h}53^\mathrm{m}09.46^\mathrm{s}$ & 0.016 & $-42\degr13\arcmin07.85\arcsec$ & 0.029 \\
G263.7434+00.1161 & N & Lobe & $08^\mathrm{h}48^\mathrm{m}48.47^\mathrm{s}$ & 0.033 & $-43\degr32\arcmin24.02\arcsec$ & 0.057 & $08^\mathrm{h}48^\mathrm{m}48.47^\mathrm{s}$ & 0.026 & $-43\degr32\arcmin24.01\arcsec$ & 0.042 \\
 & S & Jet & $08^\mathrm{h}48^\mathrm{m}48.65^\mathrm{s}$ & 0.035 & $-43\degr32\arcmin28.58\arcsec$ & 0.060 & $08^\mathrm{h}48^\mathrm{m}48.66^\mathrm{s}$ & 0.023 & $-43\degr32\arcmin28.70\arcsec$ & 0.036 \\
G298.2620+00.7394 &  & DW & $12^\mathrm{h}11^\mathrm{m}47.67^\mathrm{s}$ & 0.123 & $-61\degr46\arcmin20.86\arcsec$ & 0.085 & $12^\mathrm{h}11^\mathrm{m}47.67^\mathrm{s}$ & 0.080 & $-61\degr46\arcmin20.98\arcsec$ & 0.052 \\
G310.0135+00.3892 & N & Lobe & $13^\mathrm{h}51^\mathrm{m}37.84^\mathrm{s}$ & 0.078 & $-61\degr39\arcmin06.30\arcsec$ & 0.207 & $13^\mathrm{h}51^\mathrm{m}37.85^\mathrm{s}$ & 0.019 & $-61\degr39\arcmin06.34\arcsec$ & 0.048 \\
 & S & Jet & $13^\mathrm{h}51^\mathrm{m}37.85^\mathrm{s}$ & 0.223 & $-61\degr39\arcmin07.94\arcsec$ & 0.597 & $13^\mathrm{h}51^\mathrm{m}37.86^\mathrm{s}$ & 0.053 & $-61\degr39\arcmin07.80\arcsec$ & 0.076 \\
 & SW & Lobe & $13^\mathrm{h}51^\mathrm{m}37.66^\mathrm{s}$ & 0.735 & $-61\degr39\arcmin08.72\arcsec$ & 0.441 & $13^\mathrm{h}51^\mathrm{m}37.73^\mathrm{s}$ & 0.182 & $-61\degr39\arcmin09.02\arcsec$ & 0.254 \\
G310.1420+00.7583 & A1 & Jet & $13^\mathrm{h}51^\mathrm{m}58.39^\mathrm{s}$ & - & $-61\degr15\arcmin41.19\arcsec$ & - & $13^\mathrm{h}51^\mathrm{m}58.38^\mathrm{s}$ & - & $-61\degr15\arcmin41.22\arcsec$ & - \\
 & A2 & Lobe & $13^\mathrm{h}51^\mathrm{m}58.29^\mathrm{s}$ & - & $-61\degr15\arcmin41.40\arcsec$ & - & $13^\mathrm{h}51^\mathrm{m}58.28^\mathrm{s}$ & - & $-61\degr15\arcmin41.37\arcsec$ & - \\
 & A3 & Lobe & $13^\mathrm{h}51^\mathrm{m}58.17^\mathrm{s}$ & - & $-61\degr15\arcmin41.75\arcsec$ & - & $13^\mathrm{h}51^\mathrm{m}58.15^\mathrm{s}$ & - & $-61\degr15\arcmin41.83\arcsec$ & - \\
 & A4 & Lobe & $13^\mathrm{h}51^\mathrm{m}57.94^\mathrm{s}$ & - & $-61\degr15\arcmin42.20\arcsec$ & - & $13^\mathrm{h}51^\mathrm{m}57.95^\mathrm{s}$ & - & $-61\degr15\arcmin42.26\arcsec$ & - \\
 & B & Lobe & $13^\mathrm{h}51^\mathrm{m}58.70^\mathrm{s}$ & 0.069 & $-61\degr15\arcmin40.90\arcsec$ & 0.069 & $13^\mathrm{h}51^\mathrm{m}58.73^\mathrm{s}$ & - & $-61\degr15\arcmin40.58\arcsec$ & - \\
 & C & Lobe & $13^\mathrm{h}51^\mathrm{m}59.10^\mathrm{s}$ & 0.016 & $-61\degr15\arcmin40.00\arcsec$ & 0.021 & $13^\mathrm{h}51^\mathrm{m}59.07^\mathrm{s}$ & - & $-61\degr15\arcmin40.02\arcsec$ & - \\
 & D & Lobe & $13^\mathrm{h}51^\mathrm{m}59.54^\mathrm{s}$ & 0.009 & $-61\degr15\arcmin40.10\arcsec$ & 0.011 & $13^\mathrm{h}51^\mathrm{m}59.55^\mathrm{s}$ & - & $-61\degr15\arcmin40.00\arcsec$ & - \\
G313.7654-00.8620 & A1 & Jet & $14^\mathrm{h}25^\mathrm{m}01.59^\mathrm{s}$ & 0.021 & $-61\degr44\arcmin57.71\arcsec$ & 0.026 & $14^\mathrm{h}25^\mathrm{m}01.59^\mathrm{s}$ & 0.016 & $-61\degr44\arcmin57.77\arcsec$ & 0.019 \\
 & A2 & Lobe & $14^\mathrm{h}25^\mathrm{m}01.82^\mathrm{s}$ & - & $-61\degr44\arcmin58.96\arcsec$ & - & $14^\mathrm{h}25^\mathrm{m}01.82^\mathrm{s}$ & 0.112 & $-61\degr44\arcmin58.96\arcsec$ & 0.134 \\
 & B1 & Lobe & $14^\mathrm{h}25^\mathrm{m}01.10^\mathrm{s}$ & 0.050 & $-61\degr44\arcmin56.24\arcsec$ & 0.061 & $14^\mathrm{h}25^\mathrm{m}01.10^\mathrm{s}$ & 0.061 & $-61\degr44\arcmin55.93\arcsec$ & 0.073 \\
 & B2 & Lobe & $14^\mathrm{h}25^\mathrm{m}01.20^\mathrm{s}$ & - & $-61\degr44\arcmin57.92\arcsec$ & - & $14^\mathrm{h}25^\mathrm{m}01.20^\mathrm{s}$ & 0.080 & $-61\degr44\arcmin57.92\arcsec$ & 0.096 \\
 & C & Lobe & $14^\mathrm{h}25^\mathrm{m}02.23^\mathrm{s}$ & 0.031 & $-61\degr45\arcmin01.36\arcsec$ & 0.038 & $14^\mathrm{h}25^\mathrm{m}02.24^\mathrm{s}$ & 0.029 & $-61\degr45\arcmin01.34\arcsec$ & 0.035 \\
 & D & Lobe & $14^\mathrm{h}25^\mathrm{m}00.06^\mathrm{s}$ & 0.084 & $-61\degr44\arcmin53.17\arcsec$ & 0.103 & $14^\mathrm{h}25^\mathrm{m}00.06^\mathrm{s}$ & 0.061 & $-61\degr44\arcmin53.35\arcsec$ & 0.073 \\
G318.9480-00.1969 &  & Jet/DW & $15^\mathrm{h}00^\mathrm{m}55.22^\mathrm{s}$ & 0.143 & $-58\degr58\arcmin53.02\arcsec$ & 0.203 & $15^\mathrm{h}00^\mathrm{m}55.28^\mathrm{s}$ & 0.066 & $-58\degr58\arcmin52.84\arcsec$ & 0.091 \\
G326.6618+00.5207 &  & Jet (C) & - & - & - & - & - & - & - & - \\
G327.1192+00.5103 &  & Jet & $15^\mathrm{h}47^\mathrm{m}32.78^\mathrm{s}$ & 0.057 & $-53\degr52\arcmin39.69\arcsec$ & 0.073 & $15^\mathrm{h}47^\mathrm{m}32.79^\mathrm{s}$ & 0.054 & $-53\degr52\arcmin39.68\arcsec$ & 0.071 \\
G331.3576+01.0626 &  & Jet (C) & $16^\mathrm{h}06^\mathrm{m}25.79^\mathrm{s}$ & 0.063 & $-50\degr43\arcmin22.46\arcsec$ & 0.085 & $16^\mathrm{h}06^\mathrm{m}25.82^\mathrm{s}$ & 0.029 & $-50\degr45\arcmin02.82\arcsec$ & 0.040 \\
G332.0939-00.4206 & E & Lobe & $16^\mathrm{h}16^\mathrm{m}16.73^\mathrm{s}$ & 0.064 & $-51\degr18\arcmin26.68\arcsec$ & 0.100 & $16^\mathrm{h}16^\mathrm{m}16.75^\mathrm{s}$ & 0.055 & $-51\degr18\arcmin26.60\arcsec$ & 0.080 \\
 & W & Jet & $16^\mathrm{h}16^\mathrm{m}16.46^\mathrm{s}$ & 0.032 & $-51\degr18\arcmin25.21\arcsec$ & 0.051 & $16^\mathrm{h}16^\mathrm{m}16.48^\mathrm{s}$ & 0.025 & $-51\degr18\arcmin25.35\arcsec$ & 0.036 \\
G332.9868-00.4871 &  & Jet (C) & $16^\mathrm{h}20^\mathrm{m}37.78^\mathrm{s}$ & 0.038 & $-50\degr43\arcmin49.47\arcsec$ & 0.041 & $16^\mathrm{h}20^\mathrm{m}37.79^\mathrm{s}$ & 0.013 & $-50\degr43\arcmin49.51\arcsec$ & 0.019 \\
G338.9196+00.5495 &  & Jet (C) & $16^\mathrm{h}40^\mathrm{m}34.06^\mathrm{s}$ & 0.124 & $-45\degr42\arcmin07.21\arcsec$ & 0.240 & $16^\mathrm{h}40^\mathrm{m}34.03^\mathrm{s}$ & 0.035 & $-45\degr42\arcmin07.61\arcsec$ & 0.071 \\
G339.6221-00.1209 & E & Jet (C) & $16^\mathrm{h}46^\mathrm{m}06.02^\mathrm{s}$ & 0.022 & $-45\degr36\arcmin43.63\arcsec$ & 0.049 & $16^\mathrm{h}46^\mathrm{m}06.03^\mathrm{s}$ & 0.014 & $-45\degr36\arcmin43.75\arcsec$ & 0.026 \\
 & W & Lobe & $16^\mathrm{h}46^\mathrm{m}05.64^\mathrm{s}$ & 0.139 & $-45\degr36\arcmin45.52\arcsec$ & 0.347 & $16^\mathrm{h}46^\mathrm{m}05.65^\mathrm{s}$ & 0.088 & $-45\degr36\arcmin45.42\arcsec$ & 0.166 \\
G339.8838-01.2588 & C & Jet & $16^\mathrm{h}52^\mathrm{m}04.69^\mathrm{s}$ & 0.056 & $-46\degr08\arcmin34.22\arcsec$ & 0.085 & $16^\mathrm{h}52^\mathrm{m}04.67^\mathrm{s}$ & 0.016 & $-46\degr08\arcmin34.16\arcsec$ & 0.035 \\
 & NE & Jet & $16^\mathrm{h}52^\mathrm{m}04.94^\mathrm{s}$ & 0.014 & $-46\degr08\arcmin31.21\arcsec$ & 0.021 & $16^\mathrm{h}52^\mathrm{m}04.94^\mathrm{s}$ & 0.009 & $-46\degr08\arcmin31.22\arcsec$ & 0.019 \\
 & SW & Jet & $16^\mathrm{h}52^\mathrm{m}04.17^\mathrm{s}$ & 0.043 & $-46\degr08\arcmin40.97\arcsec$ & 0.065 & $16^\mathrm{h}52^\mathrm{m}04.20^\mathrm{s}$ & 0.049 & $-46\degr08\arcmin41.47\arcsec$ & 0.107 \\
G340.0543-00.2437 & D & Jet (C) & $16^\mathrm{h}48^\mathrm{m}13.84^\mathrm{s}$ & 0.060 & $-45\degr21\arcmin46.11\arcsec$ & 0.110 & $16^\mathrm{h}48^\mathrm{m}13.82^\mathrm{s}$ & 0.029 & $-45\degr21\arcmin45.88\arcsec$ & 0.045 \\
G343.5213-00.5171 & E & Lobe & $17^\mathrm{h}01^\mathrm{m}34.11^\mathrm{s}$ & 0.348 & $-42\degr50\arcmin19.62\arcsec$ & 0.336 & $17^\mathrm{h}01^\mathrm{m}34.10^\mathrm{s}$ & 0.186 & $-42\degr50\arcmin19.62\arcsec$ & 0.321 \\
 & W & Jet (C) & $17^\mathrm{h}01^\mathrm{m}33.94^\mathrm{s}$ & 0.087 & $-42\degr50\arcmin19.53\arcsec$ & 0.097 & $17^\mathrm{h}01^\mathrm{m}33.94^\mathrm{s}$ & 0.047 & $-42\degr50\arcmin19.54\arcsec$ & 0.076 \\
G345.5043+00.3480 &  & Jet (C) & $17^\mathrm{h}04^\mathrm{m}22.91^\mathrm{s}$ & 0.070 & $-40\degr44\arcmin23.08\arcsec$ & 0.116 & $17^\mathrm{h}04^\mathrm{m}22.90^\mathrm{s}$ & 0.061 & $-40\degr44\arcmin22.86\arcsec$ & 0.100 \\
\hline
\end{tabular}
\end{scriptsize}
\label{tab:JetsLobesPositions}
\end{table}
\begin{table}
\centering
\contcaption{}
\begin{scriptsize}
\begin{tabular}{lcccccccccc}
\hline
\textbf{Name} & \textbf{Comp.} & \textbf{Type} & \multicolumn{4}{c}{\textbf{17 GHz}} & \multicolumn{4}{c}{\textbf{22.8 GHz}} \\
 & & & \textbf{R.A. (J2000)} & \boldmath$\Delta\alpha$ & \textbf{Dec.(J2000)} & \boldmath$\Delta\delta$ & \textbf{R.A. (J2000)} & \boldmath$\Delta\alpha$ & \textbf{Dec. (J2000)} & \boldmath$\Delta\delta$ \\
\hline
G010.8411-02.5919 &  & Jet & $18^\mathrm{h}19^\mathrm{m}12.10^\mathrm{s}$ & 0.005 & $-20\degr47\arcmin30.99\arcsec$ & 0.029 & $18^\mathrm{h}19^\mathrm{m}12.09^\mathrm{s}$ & 0.006 & $-20\degr47\arcmin31.01\arcsec$ & 0.039 \\
G012.9090-00.2607 &  & Jet (C) & $18^\mathrm{h}14^\mathrm{m}39.52^\mathrm{s}$ & 0.024 & $-17\degr51\arcmin59.69\arcsec$ & 0.108 & $18^\mathrm{h}14^\mathrm{m}39.51^\mathrm{s}$ & 0.017 & $-17\degr51\arcmin59.87\arcsec$ & 0.068 \\
G014.9958-00.6732 &  & Jet (C) & $18^\mathrm{h}20^\mathrm{m}19.48^\mathrm{s}$ & 0.024 & $-16\degr13\arcmin29.77\arcsec$ & 0.077 & $18^\mathrm{h}20^\mathrm{m}19.48^\mathrm{s}$ & 0.010 & $-16\degr13\arcmin29.95\arcsec$ & 0.030 \\
G251.2337-01.9535 &  & Jet (C) & $08^\mathrm{h}02^\mathrm{m}42.97^\mathrm{s}$ & 0.017 & $-34\degr31\arcmin48.83\arcsec$ & 0.036 & $08^\mathrm{h}02^\mathrm{m}42.96^\mathrm{s}$ & 0.079 & $-34\degr31\arcmin48.47\arcsec$ & 0.260 \\
G254.0491-00.5615 &  & Jet & $08^\mathrm{h}15^\mathrm{m}57.12^\mathrm{s}$ & 0.063 & $-36\degr08\arcmin06.03\arcsec$ & 0.161 & - & - & - & - \\
G254.0548-00.0961 & C & Jet & $08^\mathrm{h}17^\mathrm{m}52.55^\mathrm{s}$ & 0.041 & $-35\degr52\arcmin47.61\arcsec$ & 0.075 & $08^\mathrm{h}17^\mathrm{m}52.54^\mathrm{s}$ & 0.051 & $-35\degr52\arcmin47.74\arcsec$ & 0.156 \\
G263.2283+01.5712 &  & Jet & $08^\mathrm{h}53^\mathrm{m}09.45^\mathrm{s}$ & 0.028 & $-42\degr13\arcmin08.21\arcsec$ & 0.039 & $08^\mathrm{h}53^\mathrm{m}09.45^\mathrm{s}$ & 0.021 & $-42\degr13\arcmin08.09\arcsec$ & 0.041 \\
G263.7434+00.1161 & N & Lobe & $08^\mathrm{h}48^\mathrm{m}48.48^\mathrm{s}$ & 0.020 & $-43\degr32\arcmin23.61\arcsec$ & 0.046 & - & - & - & - \\
 & S & Jet & $08^\mathrm{h}48^\mathrm{m}48.65^\mathrm{s}$ & 0.011 & $-43\degr32\arcmin28.42\arcsec$ & 0.025 & $08^\mathrm{h}48^\mathrm{m}48.65^\mathrm{s}$ & 0.021 & $-43\degr32\arcmin28.42\arcsec$ & 0.046 \\
G263.7759-00.4281 &  & Jet & $08^\mathrm{h}46^\mathrm{m}34.85^\mathrm{s}$ & 0.008 & $-43\degr54\arcmin30.24\arcsec$ & 0.013 & $08^\mathrm{h}46^\mathrm{m}34.85^\mathrm{s}$ & 0.008 & $-43\degr54\arcmin30.33\arcsec$ & 0.013 \\
 & NW & Lobe & $08^\mathrm{h}46^\mathrm{m}34.80^\mathrm{s}$ & 0.014 & $-43\degr54\arcmin29.35\arcsec$ & 0.021 & $08^\mathrm{h}46^\mathrm{m}34.82^\mathrm{s}$ & 0.045 & $-43\degr54\arcmin29.69\arcsec$ & 0.076 \\
G265.1438+01.4548 &  & Jet/DW & $08^\mathrm{h}59^\mathrm{m}27.40^\mathrm{s}$ & 0.022 & $-43\degr45\arcmin03.75\arcsec$ & 0.038 & $08^\mathrm{h}59^\mathrm{m}27.40^\mathrm{s}$ & 0.002 & $-43\degr45\arcmin03.75\arcsec$ & 0.073 \\
G286.2086+00.1694 & B & Jet (C) & $10^\mathrm{h}38^\mathrm{m}32.15^\mathrm{s}$ & 0.039 & $-58\degr19\arcmin08.60\arcsec$ & 0.084 & $10^\mathrm{h}38^\mathrm{m}32.16^\mathrm{s}$ & 0.033 & $-58\degr19\arcmin08.58\arcsec$ & 0.044 \\
 & B2 & Lobe & $10^\mathrm{h}38^\mathrm{m}32.23^\mathrm{s}$ & 0.083 & $-58\degr19\arcmin08.62\arcsec$ & 0.044 & - & - & - & - \\
G298.2620+00.7394 &  & DW & $12^\mathrm{h}11^\mathrm{m}47.69^\mathrm{s}$ & 0.042 & $-61\degr46\arcmin21.08\arcsec$ & 0.043 & $12^\mathrm{h}11^\mathrm{m}47.69^\mathrm{s}$ & 0.042 & $-61\degr46\arcmin20.97\arcsec$ & 0.047 \\
G310.0135+00.3892 & N & Lobe & $13^\mathrm{h}51^\mathrm{m}37.85^\mathrm{s}$ & 0.020 & $-61\degr39\arcmin06.25\arcsec$ & 0.017 & $13^\mathrm{h}51^\mathrm{m}37.85^\mathrm{s}$ & 0.045 & $-61\degr39\arcmin06.25\arcsec$ & 0.039 \\
 & S & Jet & $13^\mathrm{h}51^\mathrm{m}37.85^\mathrm{s}$ & 0.016 & $-61\degr39\arcmin07.69\arcsec$ & 0.014 & $13^\mathrm{h}51^\mathrm{m}37.86^\mathrm{s}$ & 0.012 & $-61\degr39\arcmin07.81\arcsec$ & 0.014 \\
G310.1420+00.7583 & A1 & Jet & $13^\mathrm{h}51^\mathrm{m}58.39^\mathrm{s}$ & - & $-61\degr15\arcmin41.17\arcsec$ & - & $13^\mathrm{h}51^\mathrm{m}58.39^\mathrm{s}$ & 0.009 & $-61\degr15\arcmin41.18\arcsec$ & 0.017 \\
 & A2 & Lobe & $13^\mathrm{h}51^\mathrm{m}58.28^\mathrm{s}$ & - & $-61\degr15\arcmin41.35\arcsec$ & - & $13^\mathrm{h}51^\mathrm{m}58.28^\mathrm{s}$ & 0.019 & $-61\degr15\arcmin41.35\arcsec$ & 0.037 \\
 & A3 & Lobe & $13^\mathrm{h}51^\mathrm{m}58.16^\mathrm{s}$ & - & $-61\degr15\arcmin41.61\arcsec$ & - & $13^\mathrm{h}51^\mathrm{m}58.16^\mathrm{s}$ & 0.026 & $-61\degr15\arcmin41.72\arcsec$ & 0.050 \\
 & A4 & Lobe & $13^\mathrm{h}51^\mathrm{m}57.94^\mathrm{s}$ & - & $-61\degr15\arcmin42.14\arcsec$ & - & $13^\mathrm{h}51^\mathrm{m}57.95^\mathrm{s}$ & 0.054 & $-61\degr15\arcmin42.10\arcsec$ & 0.104 \\
 & B & Lobe & $13^\mathrm{h}51^\mathrm{m}58.69^\mathrm{s}$ & - & $-61\degr15\arcmin40.48\arcsec$ & - & - & - & - & - \\
 & C & Lobe & $13^\mathrm{h}51^\mathrm{m}59.07^\mathrm{s}$ & 0.036 & $-61\degr15\arcmin39.95\arcsec$ & 0.089 & - & - & - & - \\
 & D & Lobe & $13^\mathrm{h}51^\mathrm{m}59.57^\mathrm{s}$ & - & $-61\degr15\arcmin40.02\arcsec$ & - & $13^\mathrm{h}51^\mathrm{m}59.57^\mathrm{s}$ & 0.052 & $-61\degr15\arcmin39.82\arcsec$ & 0.101 \\
G318.9480-00.1969 &  & Jet/DW & $15^\mathrm{h}00^\mathrm{m}55.28^\mathrm{s}$ & 0.063 & $-58\degr58\arcmin52.72\arcsec$ & 0.050 & $15^\mathrm{h}00^\mathrm{m}55.29^\mathrm{s}$ & 0.050 & $-58\degr58\arcmin52.80\arcsec$ & 0.044 \\
G326.6618+00.5207 &  & Jet (C) & $15^\mathrm{h}45^\mathrm{m}02.84^\mathrm{s}$ & 0.017 & $-54\degr09\arcmin03.38\arcsec$ & 0.036 & $15^\mathrm{h}45^\mathrm{m}02.83^\mathrm{s}$ & 0.017 & $-54\degr09\arcmin03.43\arcsec$ & 0.040 \\
G327.1192+00.5103 &  & Jet & $15^\mathrm{h}47^\mathrm{m}32.80^\mathrm{s}$ & 0.026 & $-53\degr52\arcmin39.68\arcsec$ & 0.053 & $15^\mathrm{h}47^\mathrm{m}32.80^\mathrm{s}$ & 0.024 & $-53\degr52\arcmin39.68\arcsec$ & 0.054 \\
G331.3576+01.0626 &  & Jet (C) & $16^\mathrm{h}06^\mathrm{m}25.76^\mathrm{s}$ & 0.012 & $-50\degr43\arcmin22.47\arcsec$ & 0.010 & $16^\mathrm{h}06^\mathrm{m}25.78^\mathrm{s}$ & 0.017 & $-50\degr43\arcmin22.55\arcsec$ & 0.024 \\
G332.0939-00.4206 & E & Lobe & - & - & - & - & - & - & - & - \\
 & W & Jet & $16^\mathrm{h}16^\mathrm{m}16.44^\mathrm{s}$ & 0.002 & $-51\degr18\arcmin24.80\arcsec$ & 0.004 & $16^\mathrm{h}16^\mathrm{m}16.44^\mathrm{s}$ & 0.004 & $-51\degr18\arcmin24.83\arcsec$ & 0.009 \\
G332.9868-00.4871 &  & Jet (C) & $16^\mathrm{h}20^\mathrm{m}37.80^\mathrm{s}$ & 0.002 & $-50\degr43\arcmin49.86\arcsec$ & 0.004 & $16^\mathrm{h}20^\mathrm{m}37.80^\mathrm{s}$ & 0.004 & $-50\degr43\arcmin49.84\arcsec$ & 0.007 \\
G338.9196+00.5495 &  & Jet (C) & $16^\mathrm{h}40^\mathrm{m}34.01^\mathrm{s}$ & 0.039 & $-45\degr42\arcmin07.46\arcsec$ & 0.052 & $16^\mathrm{h}40^\mathrm{m}34.01^\mathrm{s}$ & 0.022 & $-45\degr42\arcmin07.47\arcsec$ & 0.029 \\
G339.8838-01.2588 & C & Jet & $16^\mathrm{h}52^\mathrm{m}04.67^\mathrm{s}$ & 0.004 & $-46\degr08\arcmin34.07\arcsec$ & 0.010 & $16^\mathrm{h}52^\mathrm{m}04.67^\mathrm{s}$ & 0.004 & $-46\degr08\arcmin34.07\arcsec$ & 0.010 \\
 & NE & Jet & $16^\mathrm{h}52^\mathrm{m}04.94^\mathrm{s}$ & 0.007 & $-46\degr08\arcmin31.10\arcsec$ & 0.018 & $16^\mathrm{h}52^\mathrm{m}04.94^\mathrm{s}$ & 0.011 & $-46\degr08\arcmin31.03\arcsec$ & 0.030 \\
 & SW & Jet & $16^\mathrm{h}52^\mathrm{m}04.15^\mathrm{s}$ & 0.015 & $-46\degr08\arcmin39.97\arcsec$ & 0.037 & $16^\mathrm{h}52^\mathrm{m}04.15^\mathrm{s}$ & 0.027 & $-46\degr08\arcmin39.95\arcsec$ & 0.072 \\
G343.1261-00.0623 & C & Lobe & $16^\mathrm{h}58^\mathrm{m}17.67^\mathrm{s}$ & 0.025 & $-42\degr52\arcmin12.60\arcsec$ & 0.065 & $16^\mathrm{h}58^\mathrm{m}17.67^\mathrm{s}$ & 0.015 & $-42\degr52\arcmin12.57\arcsec$ & 0.044 \\
 & Central & Jet & $16^\mathrm{h}58^\mathrm{m}17.21^\mathrm{s}$ & 0.006 & $-42\degr52\arcmin07.16\arcsec$ & 0.013 & $16^\mathrm{h}58^\mathrm{m}17.21^\mathrm{s}$ & 0.005 & $-42\degr52\arcmin07.15\arcsec$ & 0.014 \\
 & N1 & Lobe & $16^\mathrm{h}58^\mathrm{m}16.82^\mathrm{s}$ & 0.019 & $-42\degr51\arcmin57.00\arcsec$ & 0.051 & $16^\mathrm{h}58^\mathrm{m}16.81^\mathrm{s}$ & 0.030 & $-42\degr51\arcmin57.02\arcsec$ & 0.087 \\
 & N2 & Lobe & $16^\mathrm{h}58^\mathrm{m}16.88^\mathrm{s}$ & 0.019 & $-42\degr51\arcmin58.41\arcsec$ & 0.051 & $16^\mathrm{h}58^\mathrm{m}16.87^\mathrm{s}$ & 0.040 & $-42\degr51\arcmin58.39\arcsec$ & 0.117 \\
 & N4 & Lobe & $16^\mathrm{h}58^\mathrm{m}16.99^\mathrm{s}$ & 0.013 & $-42\degr52\arcmin03.64\arcsec$ & 0.035 & $16^\mathrm{h}58^\mathrm{m}16.99^\mathrm{s}$ & 0.012 & $-42\degr52\arcmin03.67\arcsec$ & 0.035 \\
 & S1 & Lobe & $16^\mathrm{h}58^\mathrm{m}17.46^\mathrm{s}$ & 0.005 & $-42\degr52\arcmin16.48\arcsec$ & 0.014 & $16^\mathrm{h}58^\mathrm{m}17.46^\mathrm{s}$ & 0.005 & $-42\degr52\arcmin16.44\arcsec$ & 0.013 \\
 & S3 & Lobe & $16^\mathrm{h}58^\mathrm{m}17.25^\mathrm{s}$ & 0.071 & $-42\degr52\arcmin08.55\arcsec$ & 0.095 & $16^\mathrm{h}58^\mathrm{m}17.24^\mathrm{s}$ & 0.062 & $-42\degr52\arcmin08.77\arcsec$ & 0.128 \\
G343.5213-00.5171 & E & Lobe & - & - & - & - & - & - & - & - \\
 & W & Jet (C) & $17^\mathrm{h}01^\mathrm{m}33.94^\mathrm{s}$ & 0.104 & $-42\degr50\arcmin19.27\arcsec$ & 0.159 & $17^\mathrm{h}01^\mathrm{m}33.93^\mathrm{s}$ & 0.044 & $-42\degr50\arcmin20.02\arcsec$ & 0.128 \\
G345.4938+01.4677 & C & Jet & $16^\mathrm{h}59^\mathrm{m}41.63^\mathrm{s}$ & 0.002 & $-40\degr03\arcmin43.73\arcsec$ & 0.005 & $16^\mathrm{h}59^\mathrm{m}41.63^\mathrm{s}$ & 0.001 & $-40\degr03\arcmin43.74\arcsec$ & 0.004 \\
 & IE & Lobe & $16^\mathrm{h}59^\mathrm{m}41.88^\mathrm{s}$ & 0.060 & $-40\degr03\arcmin44.28\arcsec$ & 0.060 & $16^\mathrm{h}59^\mathrm{m}41.88^\mathrm{s}$ & 0.040 & $-40\degr03\arcmin44.11\arcsec$ & 0.040 \\
 & IW & Lobe & $16^\mathrm{h}59^\mathrm{m}41.37^\mathrm{s}$ & 0.060 & $-40\degr03\arcmin42.95\arcsec$ & 0.060 & $16^\mathrm{h}59^\mathrm{m}41.39^\mathrm{s}$ & 0.040 & $-40\degr03\arcmin43.38\arcsec$ & 0.040 \\
 & OE & Lobe & $16^\mathrm{h}59^\mathrm{m}44.11^\mathrm{s}$ & 0.060 & $-40\degr03\arcmin51.89\arcsec$ & 0.060 & - & - & - & - \\
 & OW & Lobe & $16^\mathrm{h}59^\mathrm{m}39.77^\mathrm{s}$ & 0.060 & $-40\degr03\arcmin41.66\arcsec$ & 0.060 & - & - & - & - \\
G345.5043+00.3480 &  & Jet (C) & $17^\mathrm{h}04^\mathrm{m}22.91^\mathrm{s}$ & 0.010 & $-40\degr44\arcmin22.81\arcsec$ & 0.018 & $17^\mathrm{h}04^\mathrm{m}22.91^\mathrm{s}$ & 0.006 & $-40\degr44\arcmin22.89\arcsec$ & 0.015 \\
\hline
\end{tabular}
\end{scriptsize}
\end{table}

\begin{table}
\centering
\caption{Fluxes and spectral indices of the MYSOs displaying the characteristics of jets and/or disc-winds in our sample, as well as their associated lobes. Hypenated entries mean simply that particular source was not observed at that frequency and upper-limits to the flux are given where available.}
\label{tab:Jets+LobeFluxes}
\begin{scriptsize}
\begin{tabular}{ccccccccccc}
\hline
\multicolumn{3}{c}{\textbf{Source}} & \multicolumn{4}{c}{\textbf{Integrated Flux Density (mJy)}} & $\mathbf{\alpha}$\\[5pt]\textbf{Name} & \textbf{Component} & \textbf{Type} &\textbf{5.5 GHz} & \textbf{9.0 GHz} & \textbf{17.0 GHz} & \textbf{22.8 GHz} & \\
\hline
G010.8411-02.5919 & & Jet &       -  &       -  &     4.75$\pm$0.14 &     6.29$\pm$0.11 & 0.96$\pm$0.16 & \\
[5pt]
G012.9090-00.2607 & & Jet (C) &       -  &       -  &     1.49$\pm$0.15 &     1.54$\pm$0.08 & 0.11$\pm$0.52 & \\
[5pt]
G014.9958-00.6732 & & Jet (C) &       -  &       -  &     0.91$\pm$0.10 &     1.50$\pm$0.09 & 1.71$\pm$0.58 & \\
[5pt]
G251.2337-01.9535 & & Jet (C) &     0.51$\pm$0.02 &     0.56$\pm$0.02 &     0.54$\pm$0.03 &     0.64$\pm$0.12 & 0.08$\pm$0.06 & \\
[5pt]
G254.0491-00.5615 & & Jet &     0.30$\pm$0.03 &     0.34$\pm$0.03 &     0.58$\pm$0.10 & $<$    0.31  & 0.56$\pm$0.19 & \\
[5pt]
G254.0548-00.0961 &C & Jet &     0.27$\pm$0.08 &     0.36$\pm$0.09 &     0.35$\pm$0.04 &     0.40$\pm$0.05 & 0.22$\pm$0.19 & \\
 &W & Lobe &     0.11$\pm$0.02 &     0.04$\pm$0.01 & $<$    0.09  & $<$    0.32  & -2.09$\pm$0.92 & \\
[5pt]
G263.2283+01.5712 & & Jet &     0.44$\pm$0.02 &     0.58$\pm$0.02 &     0.70$\pm$0.06 &     0.84$\pm$0.09 & 0.19$\pm$0.24 & \\
[5pt]
G263.7434+00.1161 &S & Jet &     0.73$\pm$0.03 &     0.87$\pm$0.04 &     1.01$\pm$0.04 &     1.43$\pm$0.16 & 0.39$\pm$0.18 & \\
 &N & Lobe &     0.58$\pm$0.02 &     0.56$\pm$0.03 &     0.44$\pm$0.03 & $<$    0.37  & -0.45$\pm$0.31 & \\
[5pt]
G263.7759-00.4281 & & Jet &       -  &       -  &     1.70$\pm$0.05 &     1.98$\pm$0.17 & 0.52$\pm$0.39 & \\
 &NW & Lobe &       -  &       -  &     2.16$\pm$0.10 &     1.72$\pm$0.32 & -0.78$\pm$0.80 & \\
[5pt]
G265.1438+01.4548 & & Jet/DW &       -  &       -  &     0.60$\pm$0.05 &     0.70$\pm$0.10 & 0.54$\pm$0.79 & \\
[5pt]
G286.2086+00.1694 & B& Jet (C) &       -  &       -  &     1.20$\pm$0.10 &     1.51$\pm$0.24 & 0.78$\pm$0.61 & \\
 &B2 & Lobe &       -  &       -  &     0.34$\pm$0.04 &     $<0.32$ & $<-0.27$ & \\
[5pt]
G298.2620+00.7394 & & DW &     0.21$\pm$0.04 &     0.21$\pm$0.03 &     0.33$\pm$0.05 &     0.70$\pm$0.08 & 0.79$\pm$0.20 & \\
[5pt]
G305.5610+00.0124 &C & HCHII &       -  &       -  &     4.18$\pm$0.24 &     4.21$\pm$0.16 & 0.02$\pm$0.33 & \\
[5pt]
G310.0135+00.3892 &S & Jet &     0.27$\pm$0.16 &     0.42$\pm$0.08 &     0.91$\pm$0.07 &     1.35$\pm$0.07 & 1.27$\pm$0.18 & \\
 &N & Lobe &     0.81$\pm$0.10 &     0.82$\pm$0.07 &     0.74$\pm$0.07 &     0.54$\pm$0.14 & -0.17$\pm$0.12 & \\
 &SW & Lobe &     0.12$\pm$0.08 &     0.16$\pm$0.07 & - & - & 0.70$\pm$1.66 & \\ 
[5pt]
G310.1420+00.7583A &A1 & Jet &     1.00$\pm$0.57 &     1.37$\pm$0.11 &     1.97$\pm$0.14 &     1.60$\pm$0.08 & 0.18$\pm$0.09 & \\
 &A2 & Lobe &     2.00$\pm$0.30 &     1.97$\pm$0.21 &     1.38$\pm$0.16 &     1.49$\pm$0.16 & -0.28$\pm$0.11 & \\
 &A3 & Lobe &     1.92$\pm$0.19 &     1.40$\pm$0.10 &     1.03$\pm$0.26 &     0.63$\pm$0.09 & -0.76$\pm$0.12 & \\
 &A4 & Lobe &     1.08$\pm$0.08 &     0.93$\pm$0.12 &     0.51$\pm$0.24 &     0.34$\pm$0.10 & -0.70$\pm$0.16 & \\
 &B & Lobe &     0.26$\pm$0.02 &     0.39$\pm$0.35 &     0.85$\pm$0.03 & $<$    0.22  & 1.04$\pm$0.07 & \\
 &C & Lobe &     1.40$\pm$0.03 &     1.05$\pm$0.56 &     0.21$\pm$0.03 & $<$    0.22  & -1.69$\pm$0.14 & \\
 &D & Lobe &     2.28$\pm$0.03 &     1.93$\pm$0.67 &     1.29$\pm$0.55 &     1.19$\pm$0.36 & -0.46$\pm$0.18 & \\
[5pt]
G313.7654-00.8620 &A1 & Jet &     0.50$\pm$0.01 &     0.58$\pm$0.02 &       -  &       -  & 0.32$\pm$0.11 & \\
 &A2 & Lobe &     0.20$\pm$0.04 &     0.14$\pm$0.03 &       -  &       -  & -0.68$\pm$0.86 & \\
 &B1 & Lobe &     0.45$\pm$0.03 &     0.28$\pm$0.03 &       -  &       -  & -0.96$\pm$0.34 & \\
 &B2 & Lobe &     0.05$\pm$0.01 &     0.12$\pm$0.02 &       -  &       -  & 1.90$\pm$0.77 & \\
&C & Lobe &     0.41$\pm$0.01 &     0.41$\pm$0.02 &       -  &       -  & 0.04$\pm$0.18 & \\
 &D & Lobe &     0.15$\pm$0.01 &     0.12$\pm$0.01 &       -  &       -  & -0.32$\pm$0.42 & \\
[5pt]
G318.9480-00.1969A & & Jet/DW &     0.12$\pm$0.03 &     0.10$\pm$0.01 &     0.14$\pm$0.02 &     0.33$\pm$0.06 & 0.69$\pm$0.22 & \\
[5pt]
G326.6618+00.5207 & & Jet (C) & $<$    0.26  & $<$    0.17  &     0.76$\pm$0.07 &     1.02$\pm$0.10 & 1.02$\pm$0.63 & \\
[5pt]
G327.1192+00.5103 & & Jet &     0.33$\pm$0.02 &     0.47$\pm$0.05 &     0.35$\pm$0.04 &     0.56$\pm$0.08 & 0.16$\pm$0.09 & \\
[5pt]
G331.3576+01.0626 & & Jet (C) &     1.55$\pm$0.19 &     1.91$\pm$0.17 &     2.70$\pm$0.14 &     2.94$\pm$0.24 & 0.48$\pm$0.09 & \\
[5pt]
G332.0939-00.4206 & E & Lobe &     0.32$\pm$0.04 &     0.47$\pm$0.09 &     $<0.14$ &     $<0.34$ & 1.24$\pm$0.65 & \\
 & W & Jet &     0.57$\pm$0.03 &     0.99$\pm$0.09 &     4.25$\pm$0.07 &     5.00$\pm$0.11 & 1.49$\pm$0.03 & \\
[5pt]
G332.9868-00.4871 &  & Jet (C) &     1.96$\pm$0.21 &     3.54$\pm$0.26 &     9.30$\pm$0.42 &     10.84$\pm$0.59 & 1.27$\pm$0.14 & \\
[5pt]
G338.9196+00.5495 & & Jet (C) &     $<0.14$ &     0.14$\pm$0.04 &     0.25$\pm$0.06 &     0.33$\pm$0.06 & 0.96$\pm$0.48 & \\
[5pt]
G339.6221-00.1209 &E & Jet (C) &     1.56$\pm$0.10 &     1.43$\pm$0.16 &       -  &       -  & -0.17$\pm$0.16 & \\
 &W & Lobe &     0.22$\pm$0.05 &     0.13$\pm$0.03 &       -  &       -  & -1.05$\pm$0.88 & \\
[5pt]
G339.8838-01.2588 &C & Jet &     0.88$\pm$0.06 &     1.58$\pm$0.05 &     2.44$\pm$0.05 &     3.01$\pm$0.07 & 0.77$\pm$0.04 & \\
 &NE & Lobe &     2.29$\pm$0.04 &     2.36$\pm$0.04 &     1.86$\pm$0.06 &     1.14$\pm$0.08 & -0.39$\pm$0.07 & \\
 &SW & Lobe &     1.40$\pm$0.08 &     1.50$\pm$0.16 &     0.71$\pm$0.05 &     0.43$\pm$0.07 & -0.72$\pm$0.11 & \\
[5pt]
G340.0543-00.2437D &D & Jet (C) &     1.80$\pm$0.37 &     2.18$\pm$0.55 &       -  &       -  & 0.39$\pm$0.67 & \\
[5pt]
G343.1261-00.0623 &Central & Jet &       -  &       -  &    16.80$\pm$0.51 &    16.30$\pm$0.59 & -0.10$\pm$0.23 & \\
 &C  & Lobe &       -  &       -  &     1.49$\pm$0.17 &     1.00$\pm$0.09 & -1.37$\pm$0.71 & \\
 &N1 & Lobe &       -  &       -  &     1.07$\pm$0.09 &     0.52$\pm$0.09 & -2.48$\pm$0.92 & \\
 &N2 & Lobe &       -  &       -  &     1.58$\pm$0.14 &     0.85$\pm$0.20 & -2.12$\pm$1.14 & \\
 &N4 & Lobe &       -  &       -  &     1.80$\pm$0.11 &     1.48$\pm$0.11 & -0.67$\pm$0.46 & \\
 &S1 & Lobe &       -  &       -  &     4.72$\pm$0.11 &     4.13$\pm$0.11 & -0.45$\pm$0.17 & \\
 &S3 & Lobe &       -  &       -  &     0.96$\pm$0.32 &     0.67$\pm$0.38 & -1.21$\pm$3.41 & \\
[5pt]
G343.5213-00.5171 & W & Jet (C) &     0.54$\pm$0.08 &     0.57$\pm$0.07 &     0.46$\pm$0.15 &     0.52$\pm$0.09 & -0.05$\pm$0.15 & \\
 &E  & Lobe & $0.14\pm0.08$  & $0.14\pm0.08$ &  $<0.09$ &  $<0.20$ & 0.04$\pm$1.56 & \\
[5pt]
G345.4938+01.4677 &C & Jet &       -  &       -  &    13.90$\pm$0.11 &    17.50$\pm$0.13 & 0.79$\pm$0.05 & \\
 &IE & Lobe &       -  &       -  &     1.93$\pm$0.13 &     1.97$\pm$0.50 & 0.07$\pm$1.11 & \\
 &IW & Lobe &       -  &       -  &     2.18$\pm$0.16 &     2.16$\pm$0.65 & -0.03$\pm$1.31 & \\
 &OE & Lobe &       -  &       -  &     0.97$\pm$0.07 & $<$    0.44  &   -  & \\
 &OW & Lobe &       -  &       -  &     1.84$\pm$0.07 & $<$    0.44  &   -  & \\
[5pt]
G345.5043+00.3480 & & Jet (C) &     0.82$\pm$0.12 &     1.55$\pm$0.30 &     2.93$\pm$0.17 &     3.24$\pm$0.18 & 0.92$\pm$0.10 & \\

\hline
\end{tabular}
\end{scriptsize}
\end{table}

\begin{landscape}
\begin{table}
\centering
\caption{A table of deconvolved, physical dimensions of MYSOs in the sample displaying jet-like and/or disc-wind characteristics. The fitted power-law index is derived along with its associated error. Hypenated entries denote where either the source was not observed, or the source was point-like after deconvolution.}
\begin{scriptsize}
\begin{tabular}{cccccccccccccccc}
\hline
\multicolumn{3}{c}{\textbf{Source}} & \multicolumn{4}{c}{$\mathbf{\theta_\mathrm{maj}\,(\arcsec)}$} & \multicolumn{4}{c}{$\mathbf{\theta_\mathrm{min}\,(\arcsec)}$} & \multicolumn{4}{c}{$\mathbf{\theta_\mathrm{P.A.}(\degr)}$} & $\mathbf{\gamma}$ \\

\textbf{Name} & \textbf{Comp.} & \textbf{Type} &\textbf{5.5 GHz} & \textbf{9.0 GHz} & \textbf{17.0 GHz} & \textbf{22.8 GHz} &\textbf{5.5 GHz} & \textbf{9.0 GHz} & \textbf{17.0 GHz} & \textbf{22.8 GHz} & \textbf{5.5 GHz} & \textbf{9.0 GHz} & \textbf{17.0 GHz} & \textbf{22.8 GHz} & \\
\hline

G010.8411-02.5919 & & Jet &   -  &   -  & 0.95$\pm$0.24 & 0.60$\pm$1.70 &   -  &   -  & 0.19$\pm$0.07 & 0.10$\pm$0.22 &   -  &   -  &  1.4$\pm$20.3 &  7.2$\pm$ 9.8 & -1.45$\pm$0.63 \\
[5pt]
G012.9090-00.2607 & & Jet (C) &   -  &   -  & - & - &   -  &   -  & - & - &   -  &   -  &  - & - & - \\
[5pt]
G014.9958-00.6732 & & Jet (C) &   -  &   -  & 0.34$\pm$0.04 &   -   &   -  &   -  & 0.05$\pm$0.16 &   -   &   -  &   -  &  9.8$\pm$ 3.7 &   -  &   -   \\
[5pt]
G251.2337-01.9535 & & Jet (C) &   -   &   -   & 0.41$\pm$0.08 &   -   &   -   &   -   & 0.26$\pm$0.05 &   -   &   -  &   -  & 154.9$\pm$20.0 &   -  &   -  \\
[5pt]
G254.0491-00.5615 & & Jet & 2.22$\pm$0.18 & 1.22$\pm$0.08 & 2.34$\pm$0.92 &   -  & 0.17$\pm$0.39 & 0.42$\pm$0.17 & 0.27$\pm$0.17 &   -  & 169.7$\pm$ 4.9 &  3.0$\pm$ 3.0 & 164.9$\pm$ 4.0 &   -  & -0.93$\pm$0.19 \\
[5pt]
G254.0548-00.0961 &C & Jet & 0.94$\pm$0.12 &   -  &   -   &   -  & 0.34$\pm$0.30 &   -  &   -   &   -  & 19.4$\pm$ 3.5 &   -  &   -  &   -  &   -  \\
 &W & Lobe & 2.95$\pm$0.76 &   -  &   -  &   -  & 1.91$\pm$1.15 &   -  &   -  &   -  & 78.0$\pm$165.0 &   -  &   -  &   -  &   -  \\
[5pt]
G263.2283+01.5712 & & Jet &   -   & 0.68$\pm$1.16 & 0.66$\pm$0.09 &   -  &   -   & 0.17$\pm$0.84 & 0.46$\pm$0.16 &   -  &   -  & 167.0$\pm$16.0 & 147.0$\pm$37.0 &   -  &   -   \\
[5pt]
G263.7434+00.1161 &S & Jet & 2.11$\pm$0.36 & 1.92$\pm$0.20 & 0.94$\pm$0.14 & 0.54$\pm$0.23 & 0.67$\pm$0.40 & 0.50$\pm$0.15 & 0.21$\pm$0.04 & 0.19$\pm$0.11 &  6.6$\pm$10.0 &  4.7$\pm$ 4.2 &  2.5$\pm$ 2.5 & 23.0$\pm$51.0 & -0.84$\pm$0.16 \\
 &N & Lobe &   -   & 1.23$\pm$0.26 &   -  &   -  &   -   & 0.36$\pm$0.24 &   -  &   -  &   -  & 173.3$\pm$ 9.9 &   -  &   -  &   -  \\
[5pt]
G263.7759-00.4281 & & Jet &   -  &   -  &   -   &   -  &   -  &   -  &   -   &   -   &   -  &   -  &   -  &   -  &   -  \\
 &NW & Lobe &   -  &   -  & 1.64$\pm$0.04 &   -   &   -  &   -  & 0.34$\pm$0.12 &   -   &   -  &   -  & 141.1$\pm$ 1.5 &   -  &   -   \\
[5pt]
G265.1438+01.4548 & & Jet/DW &   -  &   -  & 0.62$\pm$0.19 &   -  &   -  &   -  & 0.19$\pm$0.17 &   -  &   -  &   -  & 51.0$\pm$27.0 &   -  &   -  \\
[5pt]
G286.2086+00.1694 &B & Jet (C) &   -  &   -  & 1.08$\pm$0.05 & 0.90$\pm$0.36 &   -  &   -  & 0.65$\pm$0.08 & 0.26$\pm$0.15 &   -  &   -  & 17.2$\pm$ 2.7 & 49.0$\pm$45.0 & -0.59$\pm$1.36 \\
 &B2 & Lobe &   -  &   -  &   -   &   -  &   -  &   -  &   -   &   -  &   -  &   -  &   -  &   -  &   -  \\
[5pt]
G298.2620+00.7394 & & DW & 2.02$\pm$0.74 &   -   & 0.80$\pm$0.33 & 0.43$\pm$0.13 & 1.05$\pm$0.57 &   -   & 0.26$\pm$0.19 & 0.23$\pm$0.13 & 78$\pm$65 &   -  & 106$\pm$30 & 103$\pm$32 & -1.05$\pm$0.33 \\
[5pt]
G310.0135+00.3892 &S & Jet &   -  &   -   & 0.17$\pm$0.03 &   -   &   -  &   -   & 0.06$\pm$0.04 &   -   &   -  &   -  & 113.8$\pm$ 3.3 &   -  &   -   \\
 &N & Lobe &   -  &   -  &   -  &   -   &   -  &   -  &   -  &   -   &   -  &   -  &   -  &   -  &   -  \\
 &SW & Lobe &   -  &   -  &   -  &   -   &   -  &   -  &   -  &   -   &   -  &   -  &   -  &   -  &   -  \\ 
[5pt]
G310.1420+00.7583A &A1 & Jet &   -  & 0.73$\pm$0.15 &   -  & 0.34$\pm$0.02 &   -  & 0.13$\pm$0.03 &   -  & 0.09$\pm$0.04 &   -  & 86.1$\pm$17.2 &   -  &  2.4$\pm$ 0.8 & -0.84$\pm$0.28 \\
 &A2 & Lobe &   -  & 0.85$\pm$0.17 & 0.55$\pm$0.11 & 0.73$\pm$0.06 &   -  & 0.38$\pm$0.08 & 0.13$\pm$0.03 & 0.39$\pm$0.13 &   -  & 85.4$\pm$17.1 & -36.5$\pm$ 7.3 & 25.0$\pm$10.0 & -0.06$\pm$0.26 \\
 &A3 & Lobe & 1.31$\pm$0.26 & 0.94$\pm$0.19 &   -  &   -  & 0.80$\pm$0.16 & 0.28$\pm$0.06 &   -  &   -  & 85.1$\pm$17.0 & 80.6$\pm$16.1 &   -  &   -  & -0.69$\pm$0.82 \\
 &A4 & Lobe & 1.74$\pm$0.35 & 1.10$\pm$0.22 &   -  &   -  & 0.75$\pm$0.15 & 0.59$\pm$0.12 &   -  &   -  & -75.5$\pm$15.1 & 83.4$\pm$16.7 &   -  &   -  & -0.93$\pm$0.82 \\
 &B & Lobe &   -  & 1.70$\pm$0.34 & 4.27$\pm$0.85 &   -  &   -  & 0.27$\pm$0.05 & 0.26$\pm$0.05 &   -  &   -  & -85.2$\pm$17.0 & 83.0$\pm$16.6 &   -  & 1.44$\pm$0.64 \\
 &C & Lobe & 2.17$\pm$0.06 & 1.38$\pm$0.28 &   -   &   -  & 0.62$\pm$0.16 & 0.75$\pm$0.15 &   -   &   -  & 70.4$\pm$ 1.9 & 60.4$\pm$12.1 &   -  &   -  & -0.92$\pm$0.41 \\
 &D & Lobe & 1.38$\pm$0.04 & 1.38$\pm$0.28 & 1.28$\pm$0.26 & 1.40$\pm$0.20 & 1.10$\pm$0.06 & 1.05$\pm$0.21 & 0.95$\pm$0.19 & 0.23$\pm$0.11 & 105.5$\pm$ 7.8 & -76.7$\pm$15.3 & -46.0$\pm$ 9.2 & 81.6$\pm$ 5.6 & -0.01$\pm$0.09 \\
[5pt]
G313.7654-00.8620 &A1 & Jet &   -  & 0.47$\pm$0.14 &   -  &   -  &   -  & 0.40$\pm$0.27 &   -  &   -  &   -  & 106.0$\pm$159.0 &   -  &   -  &   -  \\
 &A2 & Lobe & 3.49$\pm$0.41 & 1.78$\pm$0.23 &   -  &   -  & 1.95$\pm$0.63 & 0.42$\pm$0.26 &   -  &   -  & 24.2$\pm$ 7.6 & 168.8$\pm$ 5.1 &   -  &   -  & -1.37$\pm$0.50 \\
 &B1 & Lobe & 2.23$\pm$0.16 & 1.52$\pm$0.20 &   -  &   -  & 1.31$\pm$0.24 & 0.80$\pm$0.35 &   -  &   -  & 137.5$\pm$ 8.3 & 128.0$\pm$13.0 &   -  &   -  & -0.78$\pm$0.41 \\
 &B2 & Lobe &   -  &   -   &   -  &   -  &   -  &   -   &   -  &   -  &   -  &   -  &   -  &   -  &   -  \\
 &C & Lobe & 0.86$\pm$1.16 & 0.81$\pm$0.14 &   -  &   -  & 0.19$\pm$1.40 & 0.72$\pm$0.18 &   -  &   -  & 123.0$\pm$72.0 & 98.0$\pm$170.0 &   -  &   -  &   -  \\
 &D & Lobe & 0.80$\pm$0.17 &   -  &   -  &   -  & 0.41$\pm$0.21 &   -  &   -  &   -  &  5.8$\pm$ 4.5 &   -  &   -  &   -  &   -  \\
[5pt]
G318.9480-00.1969A & & Jet/DW &   -   &   -  &   -  & 0.54$\pm$0.09 &   -   &   -  &   -  & 0.25$\pm$0.18 &   -  &   -  &   -  & 77.0$\pm$143.0 &   -   \\
[5pt]
\hline
\end{tabular}
\end{scriptsize}
\label{tab:Jets+LobeSizes}
\end{table}
\end{landscape}
\begin{landscape}
\begin{table}
\centering
\contcaption{}
\begin{scriptsize}
\begin{tabular}{cccccccccccccccc}
\hline
\multicolumn{3}{c}{\textbf{Source}} & \multicolumn{4}{c}{$\mathbf{\theta_\mathrm{maj}\,(\arcsec)}$} & \multicolumn{4}{c}{$\mathbf{\theta_\mathrm{min}\,(\arcsec)}$} & \multicolumn{4}{c}{$\mathbf{\theta_\mathrm{P.A.}(\degr)}$} & $\mathbf{\gamma}$ \\
\textbf{Name} & \textbf{Comp.} & \textbf{Type} &\textbf{5.5 GHz} & \textbf{9.0 GHz} & \textbf{17.0 GHz} & \textbf{22.8 GHz} &\textbf{5.5 GHz} & \textbf{9.0 GHz} & \textbf{17.0 GHz} & \textbf{22.8 GHz} & \textbf{5.5 GHz} & \textbf{9.0 GHz} & \textbf{17.0 GHz} & \textbf{22.8 GHz} & \\
\hline
G326.6618+00.5207 & & Jet (C) &   -  &   -  & 0.52$\pm$0.23 & 0.51$\pm$0.21 &   -  &   -  & 0.24$\pm$0.11 & 0.10$\pm$0.14 &   -  &   -  & 180.0$\pm$52.0 & 153.0$\pm$27.0 & -0.07$\pm$2.42 \\
[5pt]
G327.1192+00.5103 & & Jet &   -  & 1.68$\pm$0.35 &   -  &   -  &   -  & 0.53$\pm$0.46 &   -  &   -  &   -  & 157.0$\pm$36.0 &   -  &   -  &   -  \\
[5pt]
G331.3576+01.0626 & & Jet (C) &   -   & $<0.56$ &   -   & 0.28$\pm$0.11 &   -   & $<0.17$ &   -   & 0.23$\pm$0.16 &   -  & - &   -  & 119.0$\pm$143.0 & - \\
[5pt]
G332.0939-00.4206 &W & Jet &   -   & - & - &   -  &   -   & - & - &   -  &   -  &  - & - &   -  & - \\
 &E & Lobe & - &   -   &   -  &   -  & - &   -   &   -  &   -  & - &   -  &   -  &   -  & - \\
[5pt]
G332.9868-00.4871 & & Jet (C) &   -  &   -  &   -   &   -   &   -  &   -  &   -   &   -   &   -  &   -  &   -  &   -  &   -  \\
[5pt]
G338.9196+00.5495 & & Jet (C) & - & - & $<0.76$ &   -  & - & - &  $<0.14$ &   -  & - & - &   -  &   -  & - \\
[5pt]
G339.6221-00.1209 &E & Jet (C) & 1.48$\pm$0.28 & 1.25$\pm$0.23 &   -  &   -  & 0.22$\pm$0.12 & 0.20$\pm$0.11 &   -  &   -  & 7.5$\pm$ 5.3 &  177.6$\pm$4.5 &   -  &   -  & -0.33$\pm$0.52 \\
 &W & Lobe &   -  &   -  &   -  &   -  &   -  &   -  &   -  &   -  &   -  &   -  &   -  &   -  &   -  \\
[5pt]
G339.8838-01.2588 &C & Jet & 1.87$\pm$0.30 & 1.13$\pm$0.08 &   -   &   -   & 1.55$\pm$0.38 & 0.62$\pm$0.18 &   -   &   -   & 108.0$\pm$157.0 & 160.0$\pm$11.0 &   -  &   -  & -1.02$\pm$0.36 \\
 &NE & Lobe & 1.15$\pm$0.06 & 0.47$\pm$0.12 & 0.59$\pm$0.04 & 0.30$\pm$0.60 & 0.73$\pm$0.11 & 0.38$\pm$0.20 & 0.30$\pm$0.09 & 0.12$\pm$0.22 & 160.5$\pm$ 6.4 & 36.0$\pm$50.0 &  5.8$\pm$ 9.0 & 146.0$\pm$124.0 & -0.59$\pm$0.07 \\
 &SW & Lobe & 3.07$\pm$0.11 & 5.33$\pm$0.14 &   -   &   -  & 1.47$\pm$0.20 & 0.85$\pm$0.53 &   -   &   -  & 171.0$\pm$ 2.3 & 160.1$\pm$ 3.2 &   -  &   -  & 1.12$\pm$0.13 \\
[5pt]
G340.0543-00.2437D &D & Jet (C) &   -  &   -  &   -  &   -  &   -  &   -  &   -  &   -  &   -  &   -  &   -  &   -  & -1.43$\pm$0.66 \\
[5pt]
G343.1261-00.0623 &Central & Jet &   -  &   -  & 0.97$\pm$0.01 & 1.01$\pm$0.20 &   -  &   -  & 0.16$\pm$0.05 & 0.15$\pm$0.03 &   -  &   -  & 160.1$\pm$ 0.8 & 161.6$\pm$20.0 & 0.16$\pm$0.72 \\
 &C & Lobe &   -  &   -  &   -  & 0.69$\pm$0.27 &   -  &   -  &   -  & 0.18$\pm$0.20 &   -  &   -  &   -  & 160.0$\pm$20.0 & -0.61$\pm$1.82 \\
 &N1 & Lobe &   -  &   -  & 0.57$\pm$0.59 &   -  &   -  &   -  & 0.15$\pm$0.30 &   -  &   -  &   -  & 132.0$\pm$46.0 &   -  &   -  \\
 &N2 & Lobe &   -  &   -  & 0.84$\pm$0.07 & 1.10$\pm$0.19 &   -  &   -  & 0.62$\pm$0.15 & 0.32$\pm$0.41 &   -  &   -  &  9.3$\pm$30.9 & 132.0$\pm$48.0 & 0.91$\pm$0.90 \\
 &N4 & Lobe &   -  &   -  & 0.65$\pm$0.07 & 0.42$\pm$0.09 &   -  &   -  & 0.35$\pm$0.18 & 0.31$\pm$0.15 &   -  &   -  & 171.0$\pm$18.0 & 44.0$\pm$34.0 & -1.48$\pm$1.05 \\
 &S1 & Lobe &   -  &   -  & 0.64$\pm$0.03 & 0.70$\pm$0.02 &   -  &   -  & 0.35$\pm$0.06 & 0.27$\pm$0.07 &   -  &   -  & 23.1$\pm$ 6.0 & 36.4$\pm$ 3.8 & 0.31$\pm$0.26 \\
 &S3 & Lobe &   -  &   -  &   -  &   -  &   -  &   -  &   -  &   -  &   -  &   -  &   -  &   -  &   -  \\
[5pt]
G343.5213-00.5171 &W & Jet (C) &   -  &   -  & 1.73$\pm$0.35 & 1.20$\pm$0.33 &   -  &   -  & 0.41$\pm$0.08 & 0.63$\pm$0.50 &   -  &   -  & -33.0$\pm$30.0 & 167.0$\pm$35.0 & -1.25$\pm$1.65 \\
 &E & Lobe &   -  &   -  &   -  &   -  &   -  &   -  &   -  &   -  &   -  &   -  &   -  &   -  &   -  \\
[5pt]
G345.4938+01.4677 &C & Jet &   -  &   -  & 0.24$\pm$0.02 & 0.32$\pm$0.01 &   -  &   -  & 0.12$\pm$0.06 & 0.13$\pm$0.02 &   -  &   -  & 136.0$\pm$12.0 & 164.1$\pm$ 2.7 & 0.96$\pm$0.35 \\
 &IE & Lobe &   -  &   -  & 2.31$\pm$0.17 &   -   &   -  &   -  & 1.25$\pm$0.29 &   -   &   -  &   -  & 145.5$\pm$ 7.5 &   -  &   -   \\
 &IW & Lobe &   -  &   -  & 1.15$\pm$0.07 & 1.46$\pm$0.13 &   -  &   -  & 0.90$\pm$0.15 & 0.86$\pm$0.33 &   -  &   -  & 14.0$\pm$27.0 & 14.0$\pm$14.0 & 0.82$\pm$0.52 \\
 &OE & Lobe &   -  &   -  & 2.97$\pm$0.13 &   -  &   -  &   -  & 0.95$\pm$0.37 &   -  &   -  &   -  & 175.2$\pm$ 5.7 &   -  &   -  \\
 &OW & Lobe &   -  &   -  & 3.20$\pm$0.11 &   -  &   -  &   -  & 1.51$\pm$0.29 &   -  &   -  &   -  & 175.2$\pm$ 3.6 &   -  &   -  \\
[5pt]
G345.5043+00.3480 & & Jet (C) &   -   &   -  & 0.32$\pm$0.07 & $<0.37$ &   -   &   -  & 0.12$\pm$0.09 & $<0.04$ &   -  &   -  & 145$\pm$15 & - & - \\
\hline
\end{tabular}
\end{scriptsize}
\end{table}
\end{landscape}

\begin{table}
\centering
\caption{The deconvolved positions for sources of unknown classification.}
\begin{tabular}{lcccccccc}
\hline
\textbf{Name} & \multicolumn{4}{c}{\textbf{5.5 GHz}} & \multicolumn{4}{c}{\textbf{9GHz}} \\
 & \textbf{R.A. (J2000)} & \boldmath$\Delta\alpha$ & \textbf{Dec.(J2000)} & \boldmath$\Delta\delta$ & \textbf{R.A. (J2000)} & \boldmath$\Delta\alpha$ & \textbf{Dec. (J2000)} & \boldmath$\Delta\delta$ \\
\hline
G305.2017+00.2072 & - & - & - & - & - & -  & - & - \\
G345.9561+00.6123 & - & - & - & - & $17^\mathrm{h}04^\mathrm{m}42.99^\mathrm{s}$ & 0.151 & $-40\degr13\arcmin14.82\arcsec$ & 0.404 \\
\hline
 & \multicolumn{4}{c}{\textbf{17 GHz}} & \multicolumn{4}{c}{\textbf{22.8 GHz}} \\
\hline
G305.2017+00.2072 & $13^\mathrm{h}11^\mathrm{m}10.52^\mathrm{s}$ & 0.045 & $-62\degr34\arcmin38.70\arcsec$ & 0.054 & $13^\mathrm{h}11^\mathrm{m}10.51^\mathrm{s}$ & 0.058 & $-62\degr34\arcmin38.70\arcsec$ & 0.114 \\
G345.9561+00.6123 & $17^\mathrm{h}04^\mathrm{m}43.00^\mathrm{s}$ & 0.134 & $-40\degr13\arcmin13.82\arcsec$ & 0.22 & - & - & - & - \\
\hline
\end{tabular}
\label{tab:UKsPositions}
\end{table}

\begin{table}
\centering
\caption{Fluxes and spectral indices for sources of unknown classification.}
\label{tab:UKsFluxes}
\begin{tabular}{ccccccccccc}
\hline
\textbf{Source} & \multicolumn{4}{c}{\textbf{Integrated Flux Density (mJy)}} & $\mathbf{\alpha}$\\[5pt]
 &\textbf{5.5 GHz} & \textbf{9.0 GHz} & \textbf{17.0 GHz} & \textbf{22.8 GHz} & \\
\hline
G305.2017+00.2072 &   &       -  &  0.61$\pm$0.16 &     0.44$\pm$0.15 & -1.11$\pm$1.46 & \\
G345.9561+00.6123 &  $<0.07$ &  0.09$\pm$0.03 &     0.28$\pm$0.07 &   $<0.14$ & 1.78$\pm$0.89 & \\
\hline
\end{tabular}
\end{table}

\begin{table}
\centering
\caption{Deconvolved sizes and position angles for sources of unknown classification. The 5.5 and 22.8 GHz bands are not listed since they were either not observed or no dimensions could be deconvolved.}
\begin{tabular}{ccccccccccc}
\hline
\textbf{Source} & \multicolumn{2}{c}{$\mathbf{\theta_\mathrm{maj}\,(\arcsec)}$} & \multicolumn{2}{c}{$\mathbf{\theta_\mathrm{min}\,(\arcsec)}$} & \multicolumn{2}{c}{$\mathbf{\theta_\mathrm{P.A.}(\degr)}$} & $\mathbf{\gamma}$ \\
 & \textbf{9.0 GHz} & \textbf{17.0 GHz} & \textbf{9.0 GHz} & \textbf{17.0 GHz} & \textbf{9.0 GHz} & \textbf{17.0 GHz} & \\
\hline
G305.2017+00.2072 &  -  & 0.64$\pm$0.13 &  -  & 0.38$\pm$0.19 &  -  & 145.0$\pm$19.0  & - \\
G345.9561+00.6123 &  1.57$\pm$1.98  & 2.70$\pm$1.51 &   0.39$\pm$0.36  & 0.95$\pm$0.48& 174$\pm$10 & 168$\pm$50 & 0.85$\pm$2.17 \\
\hline
\end{tabular}
\label{tab:UKsSizes}
\end{table}

\begin{table}
\centering
\caption{A table of the positions, fluxes and deconvolved dimensions of the HII regions in our sample. Fluxes, positions and dimensions were derived using the IMFIT task in CASA. Asterisked sources are extended and are poorly fitted by the IMFIT task, and their positions represent that of the peak flux at each frequency, their fluxes are those integrated over the 3$\sigma$ areas of the clean maps and dimensions are their deconvolved sizes as calculated using equation \ref{eq:deconvolveddims}.}
\begin{tabular}{lccccccccc}
\hline
\textbf{Name} & \textbf{Morph.} & \textbf{Freq.} & \textbf{R.A. (J2000)} & \boldmath$\Delta\alpha$ & \textbf{Dec. (J2000)} & \boldmath$\Delta\delta$ & \boldmath$\mathrm{S}_\mathrm{\nu}$ & \boldmath$\theta_\mathrm{Maj.}$ &\boldmath$\theta_\mathrm{Min.}$\\
 & & \textbf{(GHz)} & & (\boldmath$\arcsec$) & & (\boldmath$\arcsec$) & \textbf{(mJy)}& (\boldmath$\arcsec$) & (\boldmath$\arcsec$)\\
\hline
G274.0649-01.1460A & Bipolar & 17.0 & $09^\mathrm{h}24^\mathrm{m}42.53^\mathrm{s}$ & 0.021 & $-52\degr01\arcmin51.00\arcsec$ & 0.027 & $2.5\pm0.2$ & $0.84\pm0.06$ & $0.08\pm0.44$ \\
 &  & 22.8 & $09^\mathrm{h}24^\mathrm{m}42.52^\mathrm{s}$ & 0.011 & $-52\degr01\arcmin50.71\arcsec$ & 0.020 & $3.1\pm0.2$ & $0.49\pm0.02$ & $0.11\pm0.04$ \\
G284.2438-01.1302 & Spherical & 5.5 & $10^\mathrm{h}20^\mathrm{m}35.17^\mathrm{s}$ & 0.028 & $-58\degr25\arcmin07.06\arcsec$ & 0.021 & $1.6\pm0.1$ & $2.05\pm0.08$ & $1.51\pm0.08$ \\
 &  & 9.0 & $10^\mathrm{h}20^\mathrm{m}35.24^\mathrm{s}$ & 0.119 & $-58\degr25\arcmin07.54\arcsec$ & 0.065 & $1.3\pm0.1$ & $2.54\pm0.22$ & $1.36\pm0.66$ \\
 &  & 17.0 & $10^\mathrm{h}20^\mathrm{m}35.13^\mathrm{s}$ & 0.035 & $-58\degr25\arcmin06.88\arcsec$ & 0.053 & $0.9\pm0.1$ & $1.48\pm0.08$ & $0.83\pm0.13$ \\
 &  & 22.8 & $10^\mathrm{h}20^\mathrm{m}35.18^\mathrm{s}$ & 0.079 & $-58\degr25\arcmin06.57\arcsec$ & 0.129 & $0.5\pm0.2$ & $<2.15$ & $<0.38$ \\
G287.3716+00.6444 * & Spherical & 5.5 & $10^\mathrm{h}48^\mathrm{m}04.68^\mathrm{s}$ & 0.018 & $-58\degr27\arcmin02.22\arcsec$ & 0.016 & $3.7\pm0.1$ & $1.63\pm0.10$ & $1.16\pm0.14$ \\
 &  & 9.0 & $10^\mathrm{h}48^\mathrm{m}04.68^\mathrm{s}$ & 0.017 & $-58\degr27\arcmin02.23\arcsec$ & 0.014 & $3.1\pm0.1$ & $1.35\pm0.05$ & $1.21\pm0.06$ \\
 &  & 17.0 & $10^\mathrm{h}48^\mathrm{m}04.70^\mathrm{s}$ & 0.050 & $-58\degr27\arcmin01.90\arcsec$ & 0.050 & $1.7\pm0.1$ & $1.54\pm0.52$ & $1.07\pm0.49$ \\
 &  & 22.8 & $10^\mathrm{h}48^\mathrm{m}04.71^\mathrm{s}$ & 0.055 & $-58\degr27\arcmin02.15\arcsec$ & 0.072 & $1.7\pm0.4$ & $1.09\pm0.16$ & $0.56\pm0.19$ \\
G300.9674+01.1499 * & Spherical & 17.0 & $12^\mathrm{h}34^\mathrm{m}53.24^\mathrm{s}$ & 0.050 & $-61\degr39\arcmin39.98\arcsec$ & 0.050 & $519.0\pm38.7$ & $0.57\pm0.01$ & $0.51\pm0.01$ \\
 &  & 22.8 & $12^\mathrm{h}34^\mathrm{m}53.24^\mathrm{s}$ & 0.030 & $-61\degr39\arcmin39.88\arcsec$ & 0.030 & $528.0\pm27.6$ & $0.57\pm0.01$ & $0.55\pm0.01$ \\
G301.1364-00.2249 & Bipolar & 17.0 & $12^\mathrm{h}35^\mathrm{m}35.10^\mathrm{s}$ & 0.002 & $-63\degr02\arcmin31.89\arcsec$ & 0.001 & $812.0\pm4.9$ & $0.67\pm0.01$ & $0.36\pm0.01$ \\
 &  & 22.8 & $12^\mathrm{h}35^\mathrm{m}35.09^\mathrm{s}$ & 0.001 & $-63\degr02\arcmin31.88\arcsec$ & 0.001 & $1100.0\pm5.4$ & $0.63\pm0.01$ & $0.36\pm0.01$ \\
G305.5610+00.0124 & Core/Halo & 17.0 & $13^\mathrm{h}14^\mathrm{m}26.39^\mathrm{s}$ & 0.010 & $-62\degr44\arcmin30.30\arcsec$ & 0.012 & $4.2\pm0.2$ & $0.42\pm0.03$ & $0.11\pm0.22$ \\
 &  & 22.8 & $13^\mathrm{h}14^\mathrm{m}26.39^\mathrm{s}$ & 0.005 & $-62\degr44\arcmin30.29\arcsec$ & 0.006 & $4.2\pm0.2$ & $<0.23$ & $<0.05$ \\
G308.9176+00.1231A * & Irregular & 17.0 & $13^\mathrm{h}43^\mathrm{m}01.73^\mathrm{s}$ & 0.040 & $-62\degr08\arcmin55.28\arcsec$ & 0.040 & $158.0\pm2.3$ & $3.19\pm0.03$ & $1.27\pm0.03$ \\
 &  & 22.8 & $13^\mathrm{h}43^\mathrm{m}01.72^\mathrm{s}$ & 0.040 & $-62\degr08\arcmin55.12\arcsec$ & 0.040 & $80.5\pm1.2$ & $3.49\pm0.08$ & $1.35\pm0.06$ \\
G317.4298-00.5612 & Bipolar & 5.5 & $14^\mathrm{h}51^\mathrm{m}37.63^\mathrm{s}$ & 0.002 & $-60\degr00\arcmin20.68\arcsec$ & 0.003 & $15.2\pm0.3$ & $0.77\pm0.02$ & $0.41\pm0.03$ \\
 &  & 9.0 & $14^\mathrm{h}51^\mathrm{m}37.62^\mathrm{s}$ & 0.001 & $-60\degr00\arcmin20.66\arcsec$ & 0.001 & $24.1\pm0.1$ & $0.45\pm0.01$ & $0.27\pm0.01$ \\
G317.8908-00.0578 & Spherical & 17.0 & $14^\mathrm{h}53^\mathrm{m}06.27^\mathrm{s}$ & 0.002 & $-59\degr21\arcmin00.00\arcsec$ & 0.005 & $15.7\pm0.2$ & $0.47\pm0.01$ & $0.38\pm0.02$ \\
 &  & 22.8 & $14^\mathrm{h}53^\mathrm{m}06.27^\mathrm{s}$ & 0.003 & $-59\degr21\arcmin00.28\arcsec$ & 0.005 & $12.8\pm0.2$ & $0.44\pm0.01$ & $0.27\pm0.02$ \\
G331.5414-00.0675 & Spherical & 5.5 & $16^\mathrm{h}12^\mathrm{m}09.02^\mathrm{s}$ & 0.019 & $-51\degr25\arcmin47.65\arcsec$ & 0.017 & $169.0\pm4.3$ & $1.08\pm0.07$ & $0.74\pm0.09$ \\
 &  & 9.0 & $16^\mathrm{h}12^\mathrm{m}09.01^\mathrm{s}$ & 0.003 & $-51\degr25\arcmin47.81\arcsec$ & 0.003 & $206.0\pm0.8$ & $0.77\pm0.01$ & $0.43\pm0.02$ \\
G332.8256-00.5498A * & Cometary & 5.5 & $16^\mathrm{h}20^\mathrm{m}11.06^\mathrm{s}$ & 0.200 & $-50\degr53\arcmin16.20\arcsec$ & 0.200 & $2490.0\pm88.6$ & $4.37\pm0.14$ & $3.67\pm0.08$ \\
 &  & 9.0 & $16^\mathrm{h}20^\mathrm{m}11.06^\mathrm{s}$ & 0.100 & $-50\degr53\arcmin13.80\arcsec$ & 0.100 & $3440.0\pm67.0$ & $3.42\pm0.12$ & $3.17\pm0.04$ \\
 &  & 17.0 & $16^\mathrm{h}20^\mathrm{m}11.04^\mathrm{s}$ & 0.040 & $-50\degr53\arcmin13.56\arcsec$ & 0.040 & $4090.0\pm67.6$ & $4.63\pm0.28$ & $2.98\pm0.14$ \\
 &  & 22.8 & $16^\mathrm{h}20^\mathrm{m}11.06^\mathrm{s}$ & 0.040 & $-50\degr53\arcmin13.48\arcsec$ & 0.040 & $4320.0\pm57.7$ & $3.19\pm0.06$ & $2.84\pm0.06$ \\
G337.8442-00.3748 & Spherical & 17.0 & $16^\mathrm{h}40^\mathrm{m}26.67^\mathrm{s}$ & 0.001 & $-47\degr07\arcmin13.11\arcsec$ & 0.002 & $42.7\pm0.1$ & $0.23\pm0.01$ & $0.18\pm0.01$ \\
 &  & 22.8 & $16^\mathrm{h}40^\mathrm{m}26.66^\mathrm{s}$ & 0.001 & $-47\degr07\arcmin13.02\arcsec$ & 0.003 & $43.9\pm0.3$ & $0.23\pm0.01$ & $0.19\pm0.01$ \\
G340.2480-00.3725 * & Spherical & 17.0 & $16^\mathrm{h}49^\mathrm{m}30.06^\mathrm{s}$ & 0.060 & $-45\degr17\arcmin44.75\arcsec$ & 0.060 & $210.0\pm19.3$ & $0.70\pm0.01$ & $0.63\pm0.02$ \\
 &  & 22.8 & $16^\mathrm{h}49^\mathrm{m}30.06^\mathrm{s}$ & 0.040 & $-45\degr17\arcmin44.73\arcsec$ & 0.040 & $442.0\pm40.7$ & $0.65\pm0.01$ & $0.43\pm0.01$ \\
G348.6972-01.0263 * & Spherical & 5.5 & $17^\mathrm{h}19^\mathrm{m}58.91^\mathrm{s}$ & 0.200 & $-38\degr58\arcmin15.22\arcsec$ & 0.200 & $1830.0\pm156.0$ & $4.75\pm0.06$ & $3.40\pm0.03$ \\
 &  & 9.0 & $17^\mathrm{h}19^\mathrm{m}58.88^\mathrm{s}$ & 0.150 & $-38\degr58\arcmin15.10\arcsec$ & 0.150 & $2020.0\pm154.0$ & $3.63\pm0.03$ & $2.90\pm0.04$ \\
\hline
\end{tabular}
\label{tab:HIIProperties}
\end{table}

\begin{table}
\centering
\caption{A table of derived properties for the jets and ambiguous jet/disc wind sources. The 4th column shows whether the distance-luminosities have been extrapolated from data at other frequencies with derived spectral indices. The 6th column shows the infrared flux (from MSX) ratio between 21$\micron$ and 8$\micron$, expressed logarithmically. Calculated ass loss rate is displayed in the final column. In the cases where an opening angle could not be calculated, we adopt a value of $\theta=40\pm20\degr$.}
\begin{scriptsize}
\begin{tabular}{lllcccc}
\hline
\textbf{Name} & \textbf{Type} & \boldmath$\mathrm{S}_\mathrm{9GHz}\mathrm{D}^2$ & \textbf{Extrap.?} & \boldmath$\mathrm{M}_\mathrm{Clump}$ & \boldmath$\log\frac{\mathrm{S}_\mathrm{21\micron}}{\mathrm{S}_\mathrm{8\micron}}$ & \boldmath$\dot{\mathrm{M}}_\mathrm{Jet}$\\
 & & \boldmath$(\mathrm{mJy\,kpc}^2)$ & & \boldmath$(\mathrm{M}_\odot)$ &  & \boldmath$(\times10^{-6}\,\mathrm{M}_\odot\mathrm{km\,s}^{-1}\mathrm{yr}^{-1})$\\
\hline
 & & & & \textbf{Jets} & &\\
\hline
G010.8411--02.5919 & Jet & $(9.30\pm0.62)\times10^{0}$ & Y & - & 1.20 & $12.30\pm8.04$\\
G012.9090--00.2607 & Jet(C) & $(8.01\pm1.73)\times10^{0}$ & Y & $(3.18\pm0.49)\times10^{3}$ & 0.96 & $1.07\pm2.46$\\
G014.9958--00.6732 & Jet(C) & $(1.22\pm0.30)\times10^{0}$ & Y & $(6.98\pm1.05)\times10^{3}$ & 0.60 & $8.08\pm20.72$\\
G251.2337--01.9535 & Jet(C) & $(1.19\pm0.05)\times10^{1}$ & N & - & 0.86 & $2.79\pm1.33$\\
G254.0491--00.5615 & Jet & $(1.69\pm0.15)\times10^{-1}$ & N & - & 1.07 & $4.10\pm3.04$\\
G254.0548--00.0961C & Jet(L) & $(2.86\pm0.71)\times10^{0}$ & N & - & 1.20 & $1.11\pm1.26$\\
G263.2283+01.5712 & Jet & $(2.86\pm0.09)\times10^{-1}$ & N & - & 0.97 & $0.33\pm0.32$\\
G263.7434+00.1161S & Jet(L) & $(4.26\pm0.18)\times10^{-1}$ & N & - & 0.48 & $0.32\pm0.20$\\
G263.7759--00.4281 & Jet(L) & $(5.98\pm1.07)\times10^{-1}$ & Y & - & 0.94 & $0.67\pm0.70$ \\
G286.2086+00.1694B & Jet(L,C) & $(3.90\pm1.45)\times10^{0}$ & Y & $(1.30\pm0.20)\times10^{3}$ & 1.48 & $11.37\pm13.01$\\
G310.0135+00.3892S & Jet(L) & $(4.27\pm0.81)\times10^{0}$ & N & $(1.19\pm0.19)\times10^{3}$ & 0.55 & $17.92\pm13.38$\\
G310.1420+00.7583A & Jet(L) & $(3.99\pm0.32)\times10^{1}$ & N & $(2.28\pm0.39)\times10^{3}$ & 0.92 & $5.57\pm3.84$\\
G313.7654--00.8620 & Jet(L) & $(3.56\pm0.10)\times10^{1}$ & N & $(6.45\pm1.06)\times10^{3}$ & 0.97 & $20.53\pm11.39$\\
G326.6618+00.5207 & Jet(C) & $(1.28\pm0.35)\times10^{0}$ & Y & $(2.66\pm0.45)\times10^{2}$ & 0.93 & $2.96\pm3.25$\\
G327.1192+00.5103 & Jet & $(1.12\pm0.12)\times10^{1}$ & N & $(1.73\pm0.30)\times10^{3}$ & 1.13 & $2.64\pm2.10$\\
G331.3576+01.0626 & Jet(C) & $(5.26\pm0.47)\times10^{1}$ & N & $(1.95\pm0.32)\times10^{3}$ & 0.50 & $38.38\pm21.78$\\
G332.0939--00.4206W & Jet(L) & $(1.59\pm0.08)\times10^{1}$ & N & $(2.41\pm0.40)\times10^{3}$ & 0.33 &  $43.89\pm30.77$\\
G332.9868--00.4871 & Jet(C) & $(5.03\pm0.12)\times10^{1}$ & N & $(5.10\pm0.81)\times10^{3}$ & 0.96 & $63.54\pm45.69$ \\
G338.9196+00.5495 & Jet(C) & $(3.69\pm0.60)\times10^{0}$ & N & $(1.21\pm0.19)\times10^{4}$ & 1.88 & $4.02\pm3.71$ \\
G339.6221--00.1209E & Jet(L) & $(1.24\pm0.04)\times10^{1}$ & N & $(8.74\pm1.48)\times10^{2}$ & 0.84 & $5.78\pm3.85$\\
G339.8838--01.2588C & Jet(L) & $(1.15\pm0.04)\times10^{1}$ & N & $(1.95\pm0.32)\times10^{3}$ & 1.59 & $26.53\pm14.47$\\
G340.0543--00.2437D & Jet(C) & $(3.15\pm0.79)\times10^{1}$ & N & $(4.43\pm0.71)\times10^{3}$ & 1.45 & $8.12\pm16.38$ \\
G343.1261--00.0623 & Jet(L) & $(1.42\pm0.10)\times10^{2}$ & Y & $(2.94\pm0.47)\times10^{3}$ & 1.51 & $17.50\pm9.66$\\
G343.5213--00.5171W & Jet(L,C) & $(5.84\pm0.77)\times10^{0}$ & N & $(1.16\pm0.20)\times10^{3}$ & 1.47 & $2.38\pm1.34$ \\
G345.4938+01.4677C & Jet(L) & $(4.84\pm0.10)\times10^{1}$ & Y & $(4.54\pm0.71)\times10^{3}$ & 1.20 & $60.39\pm32.10$ \\
G345.5043+00.3480 & Jet(C) & $(6.20\pm1.20)\times10^{0}$ & N & $(3.03\pm0.48)\times10^{3}$ & 1.04 & $13.56\pm10.28$ \\
\hline
 & & & & \textbf{Jets/DWs} & & \\
\hline
G265.1438+01.4548 & Jet/DW & $(2.07\pm0.73)\times10^{-1}$ & Y & - & 0.83 & $0.47\pm0.84$ \\
G298.2620+00.7394 & DW & $(2.96\pm0.29)\times10^{0}$ & N & $(6.76\pm1.25)\times10^{2}$ & 1.68 & $11.02\pm7.090$\\
G318.9480--00.1969A & Jet/DW & $(5.59\pm0.75)\times10^{-1}$ & N & $(6.38\pm1.05)\times10^{2}$ & 1.01 & $2.98\pm2.37$\\
\hline
\end{tabular}
\end{scriptsize}
\label{tab:MassLossAndEMs}
\end{table}

\begin{table}
\centering
\caption{A table of derived properties for HII regions. The final column shows how the emission measures for the HII regions have been derived. If the row's value is `model, this indicates the data across all frequencies has been fitted with a model of a homogeneous HII region, and an emission measure has been inferred from the fit. If a frequency is shown, this is the frequency used to calculate the emission measure, assuming the ionized gas is not opaque to its own emission. If the type is given as `HCHII core', this denotes that the properties derived are of the compact component located within the extended emission.}
\begin{scriptsize}
\begin{tabular}{lcccccccc}
\hline
\textbf{Name} & \textbf{Type} & \boldmath$\mathrm{S}_\mathrm{9GHz}\mathrm{D}^2$ & \textbf{Extrap.?} & \boldmath$\mathrm{M}_\mathrm{Clump}$ & \boldmath$\log\frac{\mathrm{S}_\mathrm{21\micron}}{\mathrm{S}_\mathrm{8\micron}}$ & \boldmath$\mathrm{Radius}$ & \boldmath$\mathrm{EM}$ & \textbf{Method}\\
 & & \boldmath$(\mathrm{mJy\,kpc}^2)$ & & \boldmath$(\mathrm{M}_\odot)$ & & \boldmath$(\mathrm{pc})$ & \boldmath$(\mathrm{pc\,cm}^{-6})$ & \\
\hline
G274.0649--01.1460A     & CHII  & $(4.77\pm0.96)\times10^{1}$ & Y & - & 1.36 & $0.209\pm0.010$ & $(5.70\pm0.60)\times10^{7}$ & 5.5 GHz \\ \vspace{-1mm}
G274.0649--01.1460      & HCHII & $(2.15\pm0.05)\times10^{3}$ & N & - & 1.36 & $0.009\pm0.001$ & $(3.07\pm0.40)\times10^{7}$ & 22.8 GHz \\
& Core & & & & & & & \\
G284.2438--01.1302      & CHII  & $(2.26\pm0.16)\times10^{1}$ & N & - & 0.97 & $0.037\pm0.002$ & $(2.28\pm0.26)\times10^{5}$ & 5.5 GHz \\
G287.3716+00.6444       & CHII  & $(6.60\pm0.77)\times10^{1}$ & N & $(3.82\pm0.85)\times10^{2}$ & 0.77 & $0.030\pm0.002$ &$(8.99\pm1.89)\times10^{5}$ & 5.5 GHz \\
G300.9674+01.1499       & UCHII & $(9.23\pm1.67)\times10^{3}$ & Y & $(4.89\pm0.76)\times10^{3}$ & 1.11 & $0.041\pm0.005$ &$(1.02\pm0.32)\times10^{8}$ & Model \\
G301.1364--00.2249N     & UCHII & $(3.02\pm0.26)\times10^{3}$ & Y & $(5.98\pm0.93)\times10^{3}$ & 1.93 & $0.029\pm0.002$ &$(7.91\pm1.29)\times10^{7}$ & Model \\
G301.1364--00.2249S     & HCHII & $(7.81\pm0.13)\times10^{3}$ & Y & $(5.98\pm0.93)\times10^{3}$ & 1.93 & $0.019\pm0.001$ &$(1.19\pm0.17)\times10^{9}$ & Model \\ \vspace{-1mm}
G305.5610+00.0124C      & HCHII & $(6.61\pm0.92)\times10^{1}$ & Y & $(2.42\pm0.39)\times10^{3}$ & 0.93 & $<0.004$ & $>4.28\times10^{8}$ & 22.8 GHz \\
& Core & & & & & & & \\
G308.9176+00.1231A      & UCHII & $(1.44\pm0.08)\times10^{3}$ & Y & $(2.68\pm0.45)\times10^{3}$ & 0.46 & $0.063\pm0.001$ &$(1.94\pm0.20)\times10^{7}$ & 17 GHz \\ \vspace{-1mm}
G317.4298--00.5612      & HCHII & $(4.86\pm0.01)\times10^{3}$ & N & $(8.71\pm1.60)\times10^{3}$ & 0.76 & $0.017\pm0.001$ &$(5.57\pm0.59)\times10^{8}$ & 9GHz \\ 
& Core & & & & & & & \\ \vspace{-1mm}
G317.4298--00.5612 (SE) & Classical & $(4.86\pm0.01)\times10^{3}$ & N & $(8.71\pm1.60)\times10^{3}$ & 0.76 & $0.599\pm0.004$ & $(5.57\pm0.58)\times10^{8}$ & 5.5 GHz \\ 
& HII & & & & & & & \\
G317.8908--00.0578      & UCHII & $(4.50\pm0.18)\times10^{3}$ & Y & $(6.23\pm1.19)\times10^{3}$ & 0.99 & $0.028\pm0.001$ &$(4.50\pm0.47)\times10^{7}$ & 17 GHz \\
G331.5414--00.0675      & UCHII & $(5.15\pm0.02)\times10^{3}$ & N & $(8.97\pm1.39)\times10^{3}$ & 1.26 & $0.015\pm0.001$ &$(8.74\pm1.17)\times10^{8}$ & 9GHz \\
G332.8256--00.5498A     & UCHII & $(4.46\pm0.09)\times10^{4}$ & N & $(1.01\pm0.16)\times10^{4}$ & 1.50 & $0.073\pm0.009$ &$(1.82\pm0.53)\times10^{8}$ & Model \\
G337.8442--00.3748      & HCHII & $(3.60\pm0.05)\times10^{2}$ & Y & $(5.04\pm0.92)\times10^{2}$ & 0.75 & $0.004\pm0.001$ &$(6.05\pm0.53)\times10^{8}$ & Model \\
G340.2480--00.3725      & UCHII & $(3.51\pm0.86)\times10^{3}$ & Y & $(4.76\pm0.76)\times10^{3}$ & 1.13 & $0.022\pm0.003$ &$(8.84\pm3.59)\times10^{7}$ & Model \\
G348.6972--01.0263      & UCHII & $(1.70\pm0.04)\times10^{4}$ & N & $(2.06\pm0.33)\times10^{3}$ & 1.46 & $0.035\pm0.002$ &$(2.48\pm0.30)\times10^{8}$ & 9GHz \\
\hline
\end{tabular}
\end{scriptsize}
\label{tab:HIIEMs}
\end{table}

\begin{table}
\centering
\caption{A table of properties for the additional sources detected in the primary beams (to the $20\%$ level) of both 5.5 and 9.0 GHz images (except for G254.0819-00.5614A/B which were observed at 17.0 and 22.8 GHz). Names of the sources are either inferred from their galactic positions, or are the names of the sources given in the RMS survey (if present in the database). The second column shows the morphology for the sources, which determined how fluxes and positions (at 9.0 GHz) were deduced. In the case of the Gaussian sources, position and fluxes were determined through the deconvolution of the source, by a Gaussian, in the image plane. For the extended sources, fluxes were determined by integrating over the 3$\sigma$ areas, while the positions are for the peak pixel in the source's flux distribution. All images were corrected for the primary beam at each frequency.}
\begin{scriptsize}
\begin{tabular}{lcccccccc}
\hline
\textbf{Name} & \textbf{Morph.} & \textbf{R.A. (J2000)} & \boldmath$\Delta\alpha$ & \textbf{Dec. (J2000)} & \boldmath$\Delta\delta$ & \boldmath$\mathrm{S}_\mathrm{5.5 GHz}$ & \boldmath$\mathrm{S}_\mathrm{9.0 GHz}$ & \boldmath$\alpha$\\
 & & & (\boldmath$\arcsec$) & & (\boldmath$\arcsec$) & \textbf{(mJy)}& \textbf{(mJy)} & \\
\hline
G251.2301-01.9918 & Gaussian & $08^\mathrm{h}02^\mathrm{m}33.12^\mathrm{s}$ & 0.011 & $-34\degr32\arcmin50.95\arcsec$ & 0.027 & $2.28\pm0.03$ & $1.44\pm0.05$ & $-0.9\pm0.2$ \\
G251.2414-01.9315 & Gaussian & $08^\mathrm{h}02^\mathrm{m}49.78^\mathrm{s}$ & 0.041 & $-34\degr31\arcmin30.35\arcsec$ & 0.104 & $0.38\pm0.02$ & $0.22\pm0.04$ & $-1.1\pm0.9$ \\
G254.0311-00.5793 & Gaussian & $08^\mathrm{h}15^\mathrm{m}49.78^\mathrm{s}$ & 0.058 & $-36\degr07\arcmin49.93\arcsec$ & 0.131 & $0.25\pm0.04$ & $0.20\pm0.02$ & $-0.5\pm0.8$ \\
G254.0364-00.5363 & Gaussian & $08^\mathrm{h}16^\mathrm{m}01.26^\mathrm{s}$ & 0.091 & $-36\degr06\arcmin39.52\arcsec$ & 0.220 & $0.20\pm0.03$ & $<0.09$ & $<-1.6$ \\
G254.0819-00.5614A & Gaussian & $08^\mathrm{h}16^\mathrm{m}02.61^\mathrm{s}$ & 0.015 & $-36\degr09\arcmin45.76\arcsec$ & 0.035 & $0.86\pm0.02$ & $1.00\pm0.03$ & $0.3\pm0.2$ \\
G254.0819-00.5614B & Gaussian & $08^\mathrm{h}16^\mathrm{m}02.76^\mathrm{s}$ & 0.060 & $-36\degr09\arcmin46.82\arcsec$ & 0.145 & $0.49\pm0.05$ & $0.71\pm0.16$ & $0.8\pm1.1$ \\
G254.1113-00.1012 & Gaussian & $08^\mathrm{h}18^\mathrm{m}00.71^\mathrm{s}$ & 0.008 & $-35\degr55\arcmin46.24\arcsec$ & 0.018 & $8.17\pm0.04$ & $4.59\pm0.07$ & $-1.2\pm0.1$ \\
G263.7696+00.1347 & Gaussian & $08^\mathrm{h}48^\mathrm{m}58.97^\mathrm{s}$ & 0.003 & $-43\degr32\arcmin59.94\arcsec$ & 0.005 & $14.70\pm0.08$ & $11.00\pm0.06$ & $-0.6\pm0.1$ \\
G286.2074+00.1698A\footnote{Where the fluxes are for the 17.0 and 22.8 GHz data respectively, and position is for the 22.8 GHz data.} & Gaussian & $10^\mathrm{h}38^\mathrm{m}32.16^\mathrm{s}$ & 0.023 & $-58\degr19\arcmin08.52\arcsec$ & 0.028 & $1.27\pm0.10$ & $1.89\pm0.32$ & $1.4\pm1.5$ \\
G286.2074+00.1698B\footnotemark[1] & Gaussian & $10^\mathrm{h}38^\mathrm{m}32.32^\mathrm{s}$ & 0.035 & $-58\degr19\arcmin08.75\arcsec$ & 0.043 & $0.35\pm0.04$ & $<0.31$ & $<-0.4$ \\
G287.3277+00.6608A & Gaussian & $10^\mathrm{h}47^\mathrm{m}50.17^\mathrm{s}$ & 0.093 & $-58\degr24\arcmin56.27\arcsec$ & 0.076 & $1.35\pm0.05$ & $0.99\pm0.13$ & $-0.6\pm0.6$ \\
G287.3277+00.6608B & Gaussian & $10^\mathrm{h}47^\mathrm{m}50.60^\mathrm{s}$ & 0.138 & $-58\degr24\arcmin53.73\arcsec$ & 0.112 & $1.06\pm0.06$ & $0.77\pm0.15$ & $-0.7\pm0.9$ \\
G287.4170+00.6473 & Gaussian & $10^\mathrm{h}48^\mathrm{m}23.77^\mathrm{s}$ & 0.053 & $-58\degr28\arcmin06.08\arcsec$ & 0.043 & $0.76\pm0.04$ & $0.58\pm0.04$ & $-0.6\pm0.4$ \\
G298.2481+00.7261 & Gaussian & $12^\mathrm{h}11^\mathrm{m}39.59^\mathrm{s}$ & 0.096 & $-61\degr47\arcmin00.04\arcsec$ & 0.066 & $0.21\pm0.02$ & $0.18\pm0.04$ & $-0.2\pm1.0$ \\
G298.2718+00.7520 & Gaussian & $12^\mathrm{h}11^\mathrm{m}53.49^\mathrm{s}$ & 0.022 & $-61\degr45\arcmin41.17\arcsec$ & 0.015 & $0.94\pm0.02$ & $0.84\pm0.02$ & $-0.2\pm0.2$ \\
G298.2744+00.7082 & Gaussian & $12^\mathrm{h}11^\mathrm{m}51.42^\mathrm{s}$ & 0.023 & $-61\degr48\arcmin18.26\arcsec$ & 0.016 & $1.00\pm0.02$ & $0.89\pm0.04$ & $-0.2\pm0.2$ \\
G305.3474-01.9987 & Gaussian & $13^\mathrm{h}14^\mathrm{m}07.01^\mathrm{s}$ & 0.100 & $-64\degr45\arcmin50.63\arcsec$ & 0.100 & $9.71\pm0.12$ & $4.97\pm0.04$ & $-1.4\pm0.1$ \\
G305.5367+00.0136 & Gaussian & $13^\mathrm{h}14^\mathrm{m}13.06^\mathrm{s}$ & 0.023 & $-62\degr44\arcmin34.53\arcsec$ & 0.024 & $1.35\pm0.06$ & $1.41\pm0.06$ & $0.1\pm0.3$ \\
G305.5528-00.0109 & Extended & $13^\mathrm{h}14^\mathrm{m}22.81^\mathrm{s}$ & 0.100 & $-62\degr46\arcmin00.17\arcsec$ & 0.100 & $22.05\pm0.80$ & $16.90\pm0.26$ & $-0.6\pm0.2$ \\
G309.9977+00.3960 & Gaussian & $13^\mathrm{h}51^\mathrm{m}29.28^\mathrm{s}$ & 0.053 & $-61\degr38\arcmin57.17\arcsec$ & 0.057 & $0.28\pm0.02$ & $0.20\pm0.02$ & $-0.7\pm0.5$ \\
G310.1380+00.7641 & Gaussian & $13^\mathrm{h}51^\mathrm{m}55.08^\mathrm{s}$ & 0.019 & $-61\degr15\arcmin31.16\arcsec$ & 0.024 & $0.43\pm0.02$ & $0.39\pm0.02$ & $-0.2\pm0.2$ \\
G310.1501+00.7313 & Gaussian & $13^\mathrm{h}52^\mathrm{m}04.74^\mathrm{s}$ & 0.056 & $-61\degr17\arcmin15.69\arcsec$ & 0.071 & $0.24\pm0.02$ & $0.14\pm0.02$ & $-1.1\pm0.6$ \\
G313.7293-00.8291 & Gaussian & $14^\mathrm{h}24^\mathrm{m}37.80^\mathrm{s}$ & 0.078 & $-61\degr43\arcmin53.37\arcsec$ & 0.094 & $0.27\pm0.02$ & $0.20\pm0.03$ & $-0.6\pm0.8$ \\
G313.7365-00.9279 & Gaussian & $14^\mathrm{h}24^\mathrm{m}58.93^\mathrm{s}$ & 0.063 & $-61\degr49\arcmin17.22\arcsec$ & 0.076 & $0.95\pm0.02$ & $0.94\pm0.10$ & $-0.0\pm0.5$ \\
G313.7514-00.8978 & Gaussian & $14^\mathrm{h}25^\mathrm{m}00.58^\mathrm{s}$ & 0.060 & $-61\degr47\arcmin17.03\arcsec$ & 0.072 & $0.36\pm0.03$ & $0.31\pm0.03$ & $-0.3\pm0.7$ \\
G313.7551-00.8195 & Gaussian & $14^\mathrm{h}24^\mathrm{m}48.30^\mathrm{s}$ & 0.087 & $-61\degr42\arcmin48.42\arcsec$ & 0.104 & $0.28\pm0.03$ & $0.20\pm0.03$ & $-0.6\pm0.9$ \\
G313.7628-00.9283 & Gaussian & $14^\mathrm{h}25^\mathrm{m}11.48^\mathrm{s}$ & 0.010 & $-61\degr48\arcmin45.14\arcsec$ & 0.013 & $3.35\pm0.03$ & $4.51\pm0.08$ & $0.6\pm0.1$ \\
G317.4713-00.6210 & Gaussian & $14^\mathrm{h}52^\mathrm{m}08.82^\mathrm{s}$ & 0.046 & $-60\degr02\arcmin25.59\arcsec$ & 0.068 & $2.47\pm0.03$ & $1.31\pm0.12$ & $-1.3\pm0.4$ \\
G317.4745-00.5843 & Gaussian & $14^\mathrm{h}52^\mathrm{m}02.32^\mathrm{s}$ & 0.072 & $-60\degr00\arcmin22.27\arcsec$ & 0.107 & $0.44\pm0.03$ & $0.63\pm0.09$ & $0.7\pm0.8$ \\
G318.9238-00.1639 & Gaussian & $15^\mathrm{h}00^\mathrm{m}38.40^\mathrm{s}$ & 0.006 & $-58\degr57\arcmin51.90\arcsec$ & 0.009 & $18.80\pm0.24$ & $7.07\pm0.09$ & $-2.0\pm0.1$ \\
G318.9490-00.2321 & Gaussian & $15^\mathrm{h}01^\mathrm{m}03.84^\mathrm{s}$ & 0.008 & $-59\degr00\arcmin44.71\arcsec$ & 0.011 & $3.03\pm0.03$ & $1.70\pm0.03$ & $-1.2\pm0.1$ \\
G327.1052+00.5612 & Gaussian & $15^\mathrm{h}47^\mathrm{m}15.45^\mathrm{s}$ & 0.019 & $-53\degr50\arcmin47.84\arcsec$ & 0.025 & $2.59\pm0.04$ & $1.56\pm0.06$ & $-1.0\pm0.2$ \\
G327.1070+00.4933 & Gaussian & $15^\mathrm{h}47^\mathrm{m}33.11^\mathrm{s}$ & 0.010 & $-53\degr53\arcmin56.18\arcsec$ & 0.013 & $2.36\pm0.03$ & $1.39\pm0.03$ & $-1.1\pm0.1$ \\
G327.1160+00.4966 & Gaussian & $15^\mathrm{h}47^\mathrm{m}35.15^\mathrm{s}$ & 0.030 & $-53\degr53\arcmin26.52\arcsec$ & 0.040 & $0.45\pm0.02$ & $0.46\pm0.03$ & $0.0\pm0.4$ \\
G327.1306+00.5258 & Gaussian & $15^\mathrm{h}47^\mathrm{m}32.48^\mathrm{s}$ & 0.001 & $-53\degr51\arcmin31.69\arcsec$ & 0.002 & $21.60\pm0.05$ & $18.40\pm0.05$ & $-0.3\pm0.1$ \\
G327.1761+00.5137 & Gaussian & $15^\mathrm{h}47^\mathrm{m}50.05^\mathrm{s}$ & 0.003 & $-53\degr50\arcmin24.45\arcsec$ & 0.005 & $20.30\pm0.06$ & $19.80\pm0.14$ & $-0.1\pm0.1$ \\
G331.3546+01.0638 & Extended & $16^\mathrm{h}06^\mathrm{m}23.25^\mathrm{s}$ & 0.100 & $-50\degr43\arcmin28.83\arcsec$ & 0.100 & $78.10\pm0.69$ & $86.20\pm3.21$ & $0.2\pm0.2$ \\
G331.5131-00.1020 & Gaussian & $16^\mathrm{h}12^\mathrm{m}09.99^\mathrm{s}$ & 0.006 & $-51\degr28\arcmin37.67\arcsec$ & 0.005 & $128.00\pm2.20$ & $170.00\pm1.20$ & $0.6\pm0.1$ \\
G331.5582-00.1206 & Gaussian & $16^\mathrm{h}12^\mathrm{m}27.27^\mathrm{s}$ & 0.031 & $-51\degr27\arcmin32.64\arcsec$ & 0.026 & $164.00\pm8.50$ & $103.00\pm4.00$ & $-0.9\pm0.3$ \\
G331.5717-00.0498 & Gaussian & $16^\mathrm{h}12^\mathrm{m}12.90^\mathrm{s}$ & 0.124 & $-51\degr23\arcmin51.40\arcsec$ & 0.104 & $7.41\pm0.98$ & $3.33\pm0.51$ & $-1.6\pm0.9$ \\
G332.0625-00.4061 & Gaussian & $16^\mathrm{h}16^\mathrm{m}03.96^\mathrm{s}$ & 0.012 & $-51\degr19\arcmin05.31\arcsec$ & 0.018 & $2.23\pm0.09$ & $1.93\pm0.05$ & $-0.3\pm0.2$ \\
G333.0145-00.4438 & Extended & $16^\mathrm{h}20^\mathrm{m}33.19^\mathrm{s}$ & 0.150 & $-50\degr40\arcmin41.26\arcsec$ & 0.150 & $101.00\pm14.50$ & $101.00\pm4.12$ & $0.0\pm0.7$ \\
G338.9145+00.5470 & Gaussian & $16^\mathrm{h}40^\mathrm{m}33.58^\mathrm{s}$ & 0.038 & $-45\degr42\arcmin28.12\arcsec$ & 0.084 & $3.44\pm0.16$ & $3.35\pm0.26$ & $-0.1\pm0.4$ \\
G338.9195+00.5566 & Gaussian & $16^\mathrm{h}40^\mathrm{m}32.24^\mathrm{s}$ & 0.055 & $-45\degr41\arcmin51.92\arcsec$ & 0.123 & $0.80\pm0.07$ & $0.81\pm0.09$ & $0.0\pm0.7$ \\
G338.9217+00.6233 & Extended & $16^\mathrm{h}40^\mathrm{m}15.52^\mathrm{s}$ & 0.150 & $-45\degr39\arcmin02.91\arcsec$ & 0.150 & $214.00\pm5.15$ & $249.00\pm5.75$ & $0.3\pm0.2$ \\
G338.9237+00.5618 & Gaussian & $16^\mathrm{h}40^\mathrm{m}31.90^\mathrm{s}$ & 0.066 & $-45\degr41\arcmin25.09\arcsec$ & 0.146 & $1.55\pm0.20$ & $2.76\pm0.38$ & $1.2\pm0.9$ \\
G338.9249+00.5563 & Gaussian & $16^\mathrm{h}40^\mathrm{m}33.55^\mathrm{s}$ & 0.018 & $-45\degr41\arcmin37.82\arcsec$ & 0.040 & $1.35\pm0.10$ & $1.81\pm0.07$ & $0.6\pm0.4$ \\
G338.9313+00.5296 & Gaussian & $16^\mathrm{h}40^\mathrm{m}41.90^\mathrm{s}$ & 0.045 & $-45\degr42\arcmin24.54\arcsec$ & 0.101 & $0.74\pm0.07$ & $0.76\pm0.07$ & $0.1\pm0.6$ \\
G339.5664-00.0866 & Gaussian & $16^\mathrm{h}45^\mathrm{m}44.61^\mathrm{s}$ & 0.007 & $-45\degr37\arcmin55.78\arcsec$ & 0.014 & $24.60\pm0.26$ & $20.50\pm0.34$ & $-0.4\pm0.1$ \\
G339.5878-00.0796 & Gaussian & $16^\mathrm{h}45^\mathrm{m}47.67^\mathrm{s}$ & 0.053 & $-45\degr36\arcmin41.25\arcsec$ & 0.100 & $0.94\pm0.08$ & $0.55\pm0.06$ & $-1.1\pm0.7$ \\
G339.6197-00.0994 & Gaussian & $16^\mathrm{h}45^\mathrm{m}59.96^\mathrm{s}$ & 0.020 & $-45\degr36\arcmin00.43\arcsec$ & 0.038 & $0.72\pm0.04$ & $0.72\pm0.03$ & $0.0\pm0.3$ \\
G339.6534-00.1602 & Gaussian & $16^\mathrm{h}46^\mathrm{m}23.31^\mathrm{s}$ & 0.061 & $-45\degr36\arcmin50.46\arcsec$ & 0.115 & $0.93\pm0.08$ & $0.76\pm0.10$ & $-0.4\pm0.8$ \\
G339.9059-01.2753 & Gaussian & $16^\mathrm{h}52^\mathrm{m}13.90^\mathrm{s}$ & 0.062 & $-46\degr08\arcmin11.26\arcsec$ & 0.135 & $0.24\pm0.07$ & $0.26\pm0.03$ & $0.2\pm1.4$ \\
G343.4717-00.5152 & Gaussian & $17^\mathrm{h}01^\mathrm{m}23.46^\mathrm{s}$ & 0.019 & $-42\degr52\arcmin36.32\arcsec$ & 0.035 & $1.81\pm0.03$ & $1.24\pm0.05$ & $-0.8\pm0.2$ \\
G343.5530-00.4503 & Gaussian & $17^\mathrm{h}01^\mathrm{m}23.03^\mathrm{s}$ & 0.024 & $-42\degr46\arcmin21.81\arcsec$ & 0.045 & $17.10\pm0.24$ & $13.00\pm0.61$ & $-0.6\pm0.2$ \\
G343.5539-00.4565 & Gaussian & $17^\mathrm{h}01^\mathrm{m}24.81^\mathrm{s}$ & 0.046 & $-42\degr46\arcmin33.02\arcsec$ & 0.087 & $8.88\pm0.27$ & $6.39\pm0.58$ & $-0.7\pm0.5$ \\
G345.4881+00.3148 & Extended & $17^\mathrm{h}04^\mathrm{m}28.04^\mathrm{s}$ & 0.100 & $-40\degr46\arcmin24.84\arcsec$ & 0.100 & $2330.00\pm31.40$ & $1690.00\pm51.90$ & $-0.7\pm0.2$ \\
\hline
\end{tabular}
\end{scriptsize}
\label{tab:ExtraSources}
\end{table}

\clearpage
\twocolumn

\section{Discussion of individual targets}
\subsection{Jets, candidates and a disc wind}
\subsubsection*{G010.8411-02.5919} 
Otherwise known as the Herbig-Haro object, HH 80-81, an ouflow axis with a P.A. of $\sim20\degr$ and spectral index for the central source of $\alpha\sim-0.3$ has been observed \citep{Marti1993}. Polarization of the radio emission is also present \citep{CarrascoGonzalezMagJet2010} inferring a magnetic field aligned with the jet axis and with a strength of $\sim0.2\,\mathrm{mG}$. At 7mm, a disc with an orientation of $\sim-20\degr$ \citep{Linz2004} is seen as well as an SiO bipolar outflow whose axis lies at a P.A. of $\sim70\degr$ and a monopolar SiO outflow with  $\theta_\mathrm{PA}\sim-20\degr$ \citep{FernandezLopez2011}.

Our observations at 17 and 22.8 GHz derive $\alpha=0.23\pm0.14$ and $\gamma=-0.98\pm0.63$ with water maser emission detected to the ENE (in direction of bipolar SiO outflow). Considering the history of observations towards this object, we can classify it as an unambiguous ionized jet.

\subsubsection*{G012.9090-00.2607}
More commonly known by the alias, W33A previous radio observations derive a spectral index for the central source of $\alpha\sim1.5$ \citep[from $8$ to $15\,\mathrm{GHz}$,][]{Rengarajan1996}. Q-band observations by \citet{VanDerTak2005} resolved W33A into 3 sources (Q1,Q2 and Q3) with a mean separation of $\sim0.8\arcsec$. \citet{GalvanMadrid2010} resolved two previously established mm-cores (MM1 and MM2) $\sim0.1$pc apart, into multiple individual mm-sources. One ESE-WNW outflow centred on MM1-main (Q1) and a NNE-SSW outflow centred on the mm-core MM2. \citet{davies2010a} used resolved Br$\gamma$ emission observations, to infer a fast bipolar wind aligned to the large scale outflow from Q1/MM1-main, with complimentary H$_2$ 2.12$\micron$ emission in the same orientation out to 1$\arcsec$. A rotating disc structure perpendicular to the outflow was also detected in CO, orbiting a central mass of $10^{+9}_{-5}M_\odot$.

Water maser emission is detected coincident with the central continuum ($\sim550$AU SE of MM1-main) source, and $\sim5\arcsec$ to the south, possibly due to collisional pumping from a rapidly moving outflow/jet. By splitting the 17 and 22.8 GHz data into two sub-bands each, we can decompose the emission into 3 components and derive $\alpha = 1.37\pm0.32$ for the emission coincident with Q1, $\alpha>0.36\pm0.47$ for that coincident with Q2 and $\alpha>0.19\pm0.51$ for that coincident with Q3. Without a value for $\gamma$ for the jet this object must be classified as a jet candidate.

\subsubsection*{G014.9958-00.6732}
Also known as the Kleinmann-Wright object \citep{Kleinmann1973}. Previous observations in the near infrared (I-band, 0.81$\micron$, \citet{Chini2004}) resolved the KW-object into 2 separate early B-type stars (KW-1 and KW-2) separated by $\sim2600$AU. Subsequent SED fitting by \citet{Povich2009}, concluded that the object KW-1 was an early Herbig Be star, while KW-2 remained unresolved, yet coincident with X-ray emission. Radio data at 5.0 and 8.5GHz by \citet{RodriguezM172012} determined a spectral index of $\alpha=0.9\pm0.6$ (JVLA number 3) for which the source remained unresolved.

An unresolved point source was detected at 17 and 22.8 GHz coincident with JVLA number 3 from \citet{RodriguezM172012}. Including the flux recorded at 8.46GHz by \citet{RodriguezM172012} yields a spectral index of $\alpha=1.11\pm0.23$, indicative of thermal free-free emission. The emission is offset from the nearest source (KW-1) by $\sim0.88\arcsec$ or $1700$AU. Likely this indicates a third component absent at the NIR I-band due to heavy extinction, and therefore more embedded and younger. Radio fluxes recorded at 22.8 GHz are $\sim50$ times less than that expected from an optically thin HII region of the same bolometric luminosity \citep{Davies2011}. Without a definite outflow or position angle, this is classified as a jet candidate.

\subsubsection*{G251.2337-01.9535}
Both Br$\gamma$ and [FeII] emission are present in NIR spectra (Lumsden et al., in prep.) towards this MYSO.

From our observations, we derive a spectral index of $\alpha=0.08\pm0.06$ for the central object at each of the 4 observing frequencies, however it remained unresolved at 3 out of the 4 frequencies. This spectral index suggests a highly collimated ionized jet from the central MYSO or an optically thin HII region, however it is under-luminous by a factor of 5 at 22.8 GHz compared to the flux expected from optically thin HII regions of the same luminosity (which is well constrained given the isolated nature of the source in the infrared). There are hints of more structure to the NNW though this may be an artefact of the deconvolution process. Confusing NIR spectra and under-luminosity of just less than 20$\%$ classifies this as a jet candidate.

\subsubsection*{G254.0491-00.5615}
G254.0491-00.5615 looks like a point source at both 10$\micron$ and K-band (with possibly a weak nebula surrounding it). \citet{Navarete2015} detect 2.12$\micron$ H$_2$ emission at a position angle of $310\degr$. NIR spectra (Lumsden et al., in prep.) show no Br$\gamma$ emission, ruling out a HII region.

The 17 GHz data was poor and affected by side-lobes with resulting derived (between 5.5 and 9 GHz) values of $\alpha=0.3\pm0.3$ and $\gamma=-0.9\pm0.2$ is found consistent with thermal emission from a collimated jet. Deconvolved position angles at 9 GHz are parallel with the 2.12$\micron$ H$_2$ emission. This leads to the classification of this source as an ionized jet.

\subsubsection*{G254.0548-00.0961}
Optical knots with velocities of $\sim570$kms$^{-1}$ in the ENE-WSW direction (PA$\sim60\degr$) have been identified \citep{Neckel1995} hence the alias, HH 217. \citet{Felli1998} observed unresolved, free-free emission (8.4 and 15.0GHz), and interpolated a spectral index of $\alpha=0.3\pm0.5$. Emission at the [FeII] line is seen in NIR spectra (Lumsden et al., in prep.).

Alignment of the central (thermal jet) and west (non-thermal) lobes is parallel to the previously established Herbig-Haro outflow axis. We calculate $\alpha=0.19\pm0.20$, suggestive of collimated jet, therefore this is classified as a thermal jet with a non-thermal lobe.

\subsubsection*{G263.2283+01.5712}
Ks band observations by \citet{Burkert2000} show a bipolar reflection nebula centred on their source 25 (Figure 1e in the reference) interpreted as the evacuated bipolar cavities emitting at NIR wavelengths after heating by the central source (the NE cavity is oriented towards us). High extinction at angles perpendicular to the cavities, and general morphology of the emission, are suggestive of a circumstellar accretion disc. From inspection of those images (figure 3b in \citet{Burkert2000}), an outflow axis with a P.A.$\sim130\degr$ is inferred. 

The radio intensity contour maps presented in figure \ref{fig:ContourPlots}, show a confirmed radio detection coincident with source 25 of \citet{Burkert2000}. The derived spectral index is $\alpha = 0.19\pm0.24$ indicating thermal free-free emission. From the models of \citet{Reynolds1986}, this suggests a thermal, collimated jet of ionized material. Deconvolved position angle at 17 GHz of P.A.$=147\pm37\degr$ is aligned with the outflow cavities. On the basis of the radio emission being parallel with the cavities and perpendicular to the disc, we classify this as an ionized jet.

\subsubsection*{G263.7434+00.1161}
`Complex and intense' H$_2$ emission at 2.12$\micron$ \citep{deLuca2007} is seen with no discernable orientation. \citet{Thi2010} inferred a jet-bow shock model and wide-angle wind model to explain their observations of both $^{12}\mathrm{CO}$($1\leftarrow2$) and $^{13}\mathrm{CO}$($1\leftarrow2$) which both had large turbulent widths ($10-12$kms$^{-1}$).

\begin{figure}
\includegraphics[width=84mm]{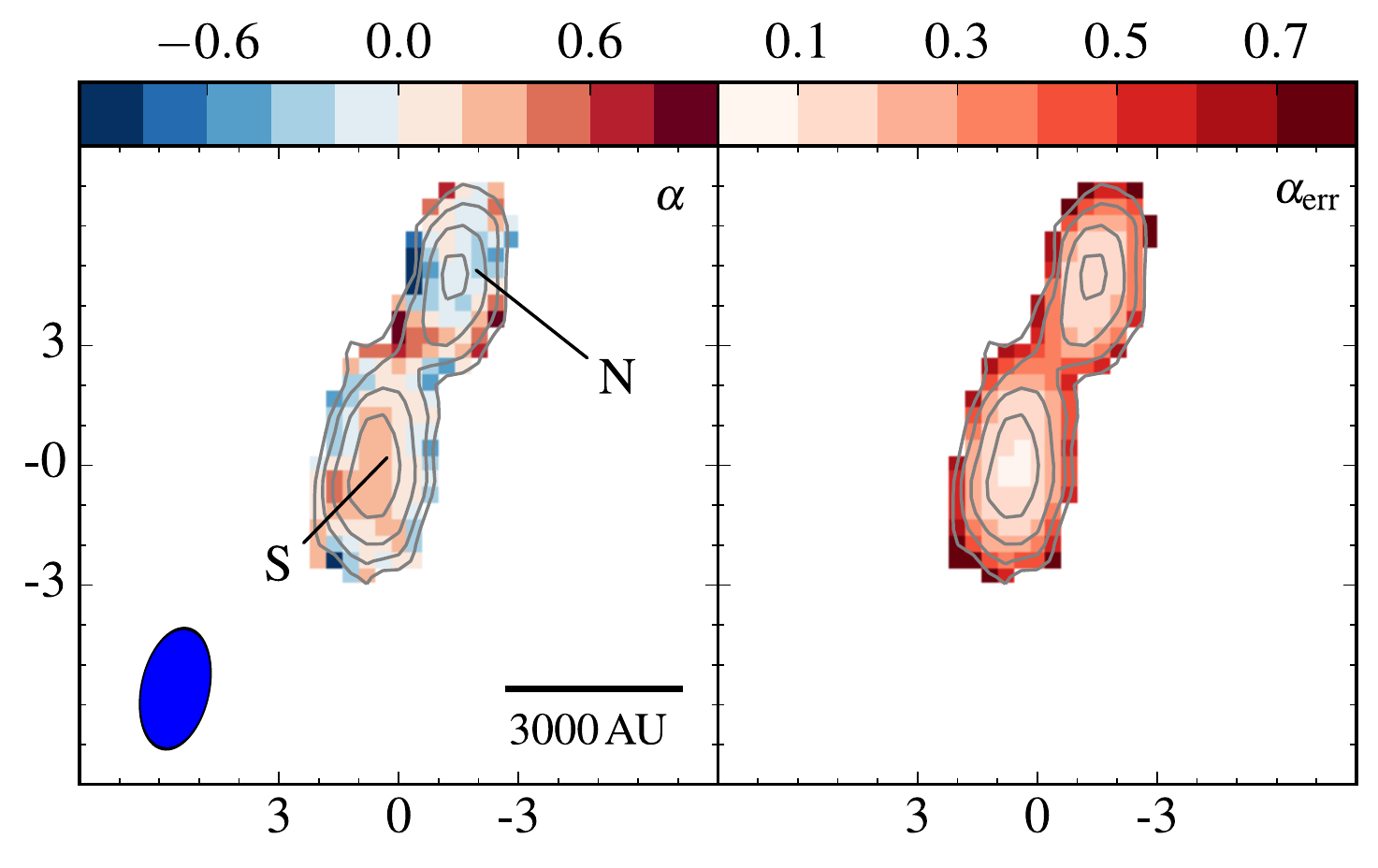}
\caption{A map of the spectral index (between 5.5 and 9GHz) for G263.7434+00.1161 (left) and the associated errors in the spectral index (right). The restoring beams (blue) used for both the 5.5 GHz and 9 GHz data were identical, in the production of the spectral index maps. Grey contours represent the flux densities at 9 GHz starting at 8.04$\times10^{-5}$Jy/beam ($4\sigma$), and scaling by a factor of $\sqrt{3}$ for each subsequent contour.}
\label{fig:G263_7434_SPIX}
\end{figure}

Our observations of G263.7434+00.1161 reveal two distinct lobes of emission present (separated by $\sim3500$AU for a distance of 0.7kpc) at the lower 3 frequencies (northern lobe) and one detected at all frequencies (southern lobe), for which we find spectral indices of $\alpha = -0.45 \pm 0.31$ and $\alpha = 0.39 \pm 0.18$ respectively. We have also constructed a map of the spectral index variation across the source in Figure \ref{fig:G263_7434_SPIX}, reflecting this thermal/non-thermal emission distribution. For $\gamma$, we find a value o $-1.01 \pm 0.24$ for the southern lobe. Given the previous observational history and values for $\alpha/\gamma$, we classify this as an ionized jet with a non-thermal lobe.

\subsubsection*{G263.7759-00.4281}
\citet{Giannini2005} observed the region looking for 2.12$\micron$ H$_2$ emission, and identified two jets with NW-SE and ENE-WSW axes, with the first centred on G263.7759-00.4281 (source \#57 in their paper) and the other a few tenths of an arcsecond to the NW of \#57 (seen as an elongation in the broadband NIR images). Diffraction limited MIR imaging \citep{Wheelwright2012} shows elongation of the MIR emission to the NW-SE, interpreted as emission from the cavity walls (also seen in 2MASS and GLIMPSE imagery). Asymmetry in the images supports the notion of multiplicity, with the second YSO (driving the ENE-WSW outflow) positioned to the NW of G263.7759-00.4281. IRAC GLIMPSE imaging shows 4 knots of excessive 4.5$\micron$ emission along an axis at a position angle of $155\degr$, attributable to the jet \citep{Giannini2013}. 

Our results show a compact source  with extended emission to the NW at a rough position angle of $\sim140\degr$, parallel to the outflow axis seen in H$_2$ emission. Values of $\alpha=0.52\pm$0.39 and -0.78$\pm$0.80 are found for the central and NW components respectively. The likelihood that the NW source represents a second source is unlikely due to the probable non-thermal nature of its spectrum. Considering the observational history of this object, we believe this is good evidence for a jet with a non-thermal lobe.

\subsubsection*{G265.1438+01.4548}
\citet{Wheelwright2012} observed this object using diffraction limited mid-infrared imaging and based upon SED modelling derived a low inclination angle of $32\degr$, a mass loss rate, $2\times10^{-4}\,\mathrm{M}_\odot\,\mathrm{yr}^{-1}$ and an opening angle of $\theta_{\mathrm{o}}=25\degr$. Considering the symmetrical morphology of the mid-infrared emission, the low inclination angle may in fact be head on to the outflowing material, within the constraints of the SED fitting procedure.

As for radio observations, this object was observed at 17 and 22.8 GHz and displayed MASER emission $\sim6\arcsec$ to the north of the source. Imaging of the continuum emission discarded baselines with a length of $<22\mathrm{k}\lambda$ due to the presence of partially resolved-out, strong and extended emission to the north east, the side lobes from which drowned out the relatively weak flux of the target. At the higher frequency band a relatively simple point like morphology is seen, while at 17 GHz the source is elongated to the NW. We derive a spectral index of $\alpha=0.5\pm0.4$) and $\theta_\mathrm{PA}$ at 17 GHz of $51\pm34\degr$. At 22.8 GHz, we find the central source is over-luminous by a factor of $\sim4$ compared to that expected of an optically thin HII region \citep{Davies2011}. These observations support the picture of both a disc wind and also a conical, non-collimated jet. It is worth noting that mid-infrared emission seen by \citet{Wheelwright2012} was symmetrical and interpreted as looking directly down the ionized flow. This favours the disc-wind picture, though observations at both sub-mm and mm would help eliminate one of these possibilities.

\subsubsection*{G286.2086+00.1694}
GLIMPSE imagery show the MSX point source to be a combination of two reddened infrared sources separated by $\sim8\arcsec$. TIMMI2 $10\micron$ emission is seen towards the SE IR source, with weaker 10$\micron$ emission to the NW source. 

Radio observations show radio emission associated to the NW IR source, but not the SE source. The emission is comprised of two components (B and B2), with $\alpha=0.78\pm0.61$ ($\gamma=-0.6\pm1.4$) and $\alpha<-0.27$ respectively. Due to large errors and a lack of other observations we classify this as a jet candidate.


\subsubsection*{G298.2620+00.7394}
GLIMPSE observations reveal two points sources within 2$\arcsec$ of eachother, with possible extended emission to the south. A $^{13}$CO (J$\,2\rightarrow1$) outflow was seen by \citet{Henning2000}, oriented NNW (blue lobe) to SSE. A spectral type for the central MYSO of B0.5 was inferred. \citet{Walsh1998} also detected a methanol MASER at 12$^h$11$^m$47.8$^s$, -61$\degr$46$\arcmin$19$\arcsec$ (J2000), slightly offset from an extended green object (EGO) at 12$^h$11$^m$47.7$^s$, -61$\degr$46$\arcmin$21$\arcsec$ \citep{Cyganowski2008}. 

A single lobe of emission with $\alpha=0.79\pm0.20$ and $\gamma=-1.09\pm0.33$ is detected at all frequencies. An average position angle of $(92\pm28)\degr$ is also deduced.  The spectral index of the emission can be modelled by a standard, bi-conical jet/wind \citep{Reynolds1986}. Considering the shocked EGO emission and radio emission perpendicular to the outflow, this is classified as a disc wind.

\subsubsection*{G310.0135+00.3892}
\citet{Kraus2010} used near-infrared interferometry to observe a hot, dusty, compact ($13 \mathrm{AU} \times 19 \mathrm{AU}$) disc with $\theta_\mathrm{PA}=114\degr$. Sub-mm emission from a collimated (opening angle $\sim 6\degr$) bipolar outflow \citep{Kraus2010} is found perpendicular to the disc. Furthermore, investigations of IRAC images showed two lobes of excess emission at $4.5\micron$ separated by $7\arcmin$ (indicative of shocked molecular hydrogen) and aligned with the outflow axis. Corroborating these findings, diffraction limited mid-infrared imaging at $20 \micron$ by \citet{Wheelwright2012} suggested that the dominant emission at $20 \micron$ was due to the presence of cavities evacuated by outflows (along a SW-NE axis). \citet{Ilee2013} observed CO bandhead emission indicative of entrained envelope material in an outflow, and subsequent modelling found temperature gradients consistent with \citet{Kraus2010} and with a flared, irradiated disc around a MYSO with $M_\star=21.8M_\odot$. Broad linewidths (25.5kms$^{-1}$) in the CO are also present, as well as high-velocity wings attached to the P-cygni profile of the Br$\gamma$ line \citep{Stecklum2012}, suggestive of an ionized wind. Most recently \citet{Caratti2015} observed the H$_2$ 2.21 $\micron$ transition, detecting lobes of emission spread over 6.9pc at a position angle of $\sim206\degr$, coincident with the SE--NW outflow.

\begin{figure}
\includegraphics[width=84mm]{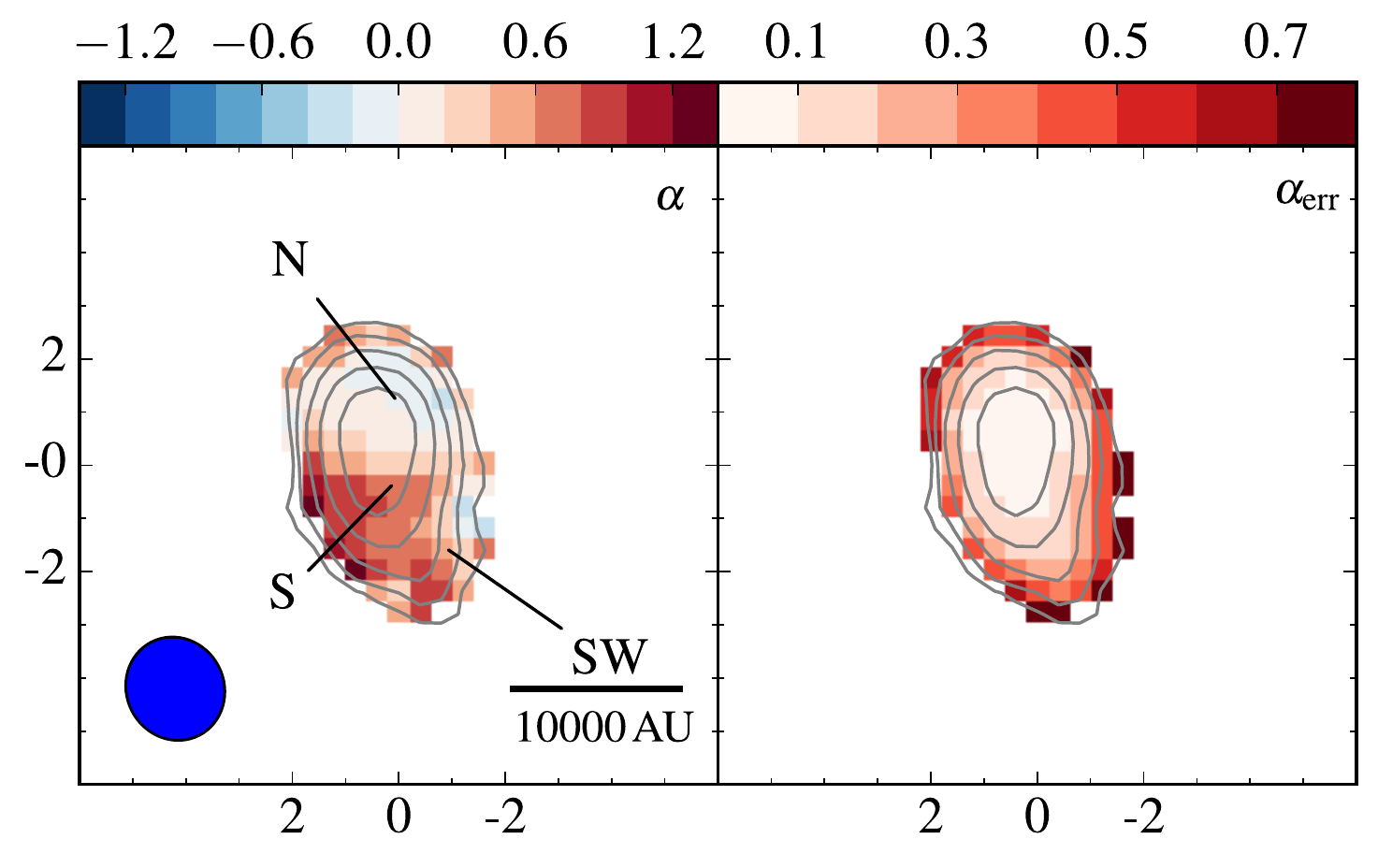}
\caption{A map of the spectral index (between 5.5 and 9GHz) for G310.0135+00.3892 (left) and the associated errors in the spectral index (right). The restoring beams (blue) used for both the 5.5 GHz and 9 GHz data were identical, in the production of the spectral index maps. Grey contours represent the flux densities at 9 GHz starting at 5.64$\times10^{-5}$Jy/beam ($4\sigma$), and scaling by a factor of $\sqrt{3}$ for each subsequent contour.}
\label{fig:G310_0135_SPIX}
\end{figure}

Two distinct lobes of emission are detected at each frequency. The southern lobe is centred on the IRAS point source, whereas the northern lobe is offset to the north by $\sim2\arcsec$. There is also an additional component to the SW, seen at 5.5 GHz as a slight extension to the southern lobe, and at 9 GHz as distinct emission at a position angle of $\sim215\degr$ and separated by 1.5$\arcsec$ ($\sim4900$AU) from the southern lobe. Spectral indices of $\alpha_N=-0.17\pm0.12$ and $\alpha_S=1.27\pm0.18$ are derived for the north and south lobes respectively. A spectral index for the south west component of $\alpha_{\mathrm{SW}}\sim0.7$ is derived between 5.5 and 9 GHz. This SW component is also parallel with the observed outflow axis, and perpendicular to the disc inferred by \citet{Kraus2010}, thus we conclude this to be part of the jet. Corroborating this scenario, the spectral index map in Figure \ref{fig:G310_0135_SPIX} shows an orientation for the thermal emission along the outflow axis. The southern component centred on the accretion disc, based on its spectral index, is likely to be a pressure confined ionized jet of material imaged close to the accretion disc. We interpret the northern source as a working surface located in the cavities upon which the edges of the ionized jet impinge upon. Thus we classify this as a jet with lobes.

\subsubsection*{G310.1420+00.7583A}
\begin{figure}
\includegraphics[width=84mm]{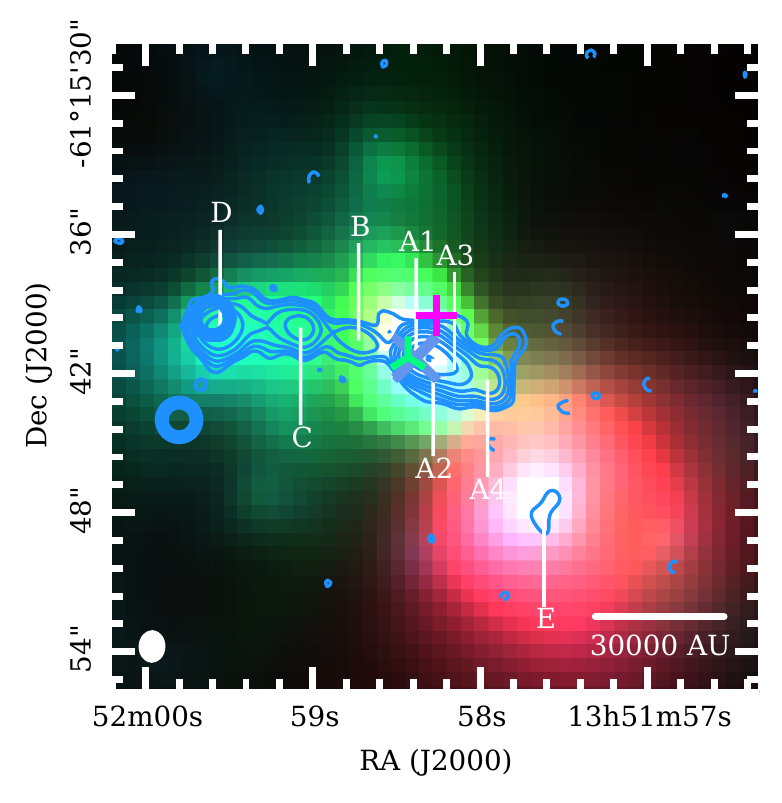}
\caption{The 9GHz radio contours overlaid on an RGB (8.0$\micron$, 4.5$\micron$ and 3.6$\micron$) composite image of existing GLIMPSE data for G310.1420+00.7583A. Contours step up from 4$\sigma$ by a factor of $\sqrt{3}$ per level and negative (-4$\sigma$) contours are dashed. Other annotations have the same meanings as in Figure \ref{fig:ContourPlots}.}
\label{fig:G310_1420_RGBplot}
\end{figure}

This source is associated to IRAS 13484--6100 and is in close proximity ($\sim8\arcsec$ offset) to a cometary UCHII (G310.1420+00.7583B). Different masing species (OH, H$_2$O and CH$_3$OH) have been detected towards this source (\citet{Walsh1998}, \citet{Urquhart2009} and \citet{Green2012} respectively). \citet{Caratti2015} observed two knots of H$_2$ emission (blue circles in Figure \ref{fig:G310_1420_RGBplot}), one of which displayed Br$\gamma$ emission indicative of strong dissociative shocks (`knot 1') where the shock velocities exceeded 90 kms$^{-1}$ within a medium of density $\sim10^5$cm$^{-3}$. Jet properties inferred included a length of 0.4pc, precession of 17$\degr$ and electron density of $n_e=(4\pm1)\times10^4$cm$^{-3}$. An extended green object was detected offset to knot 1 by $\sim16\arcsec$ at a PA of 57$\degr$ \citep{Cyganowski2008}. 

\begin{figure}
\includegraphics[width=84mm]{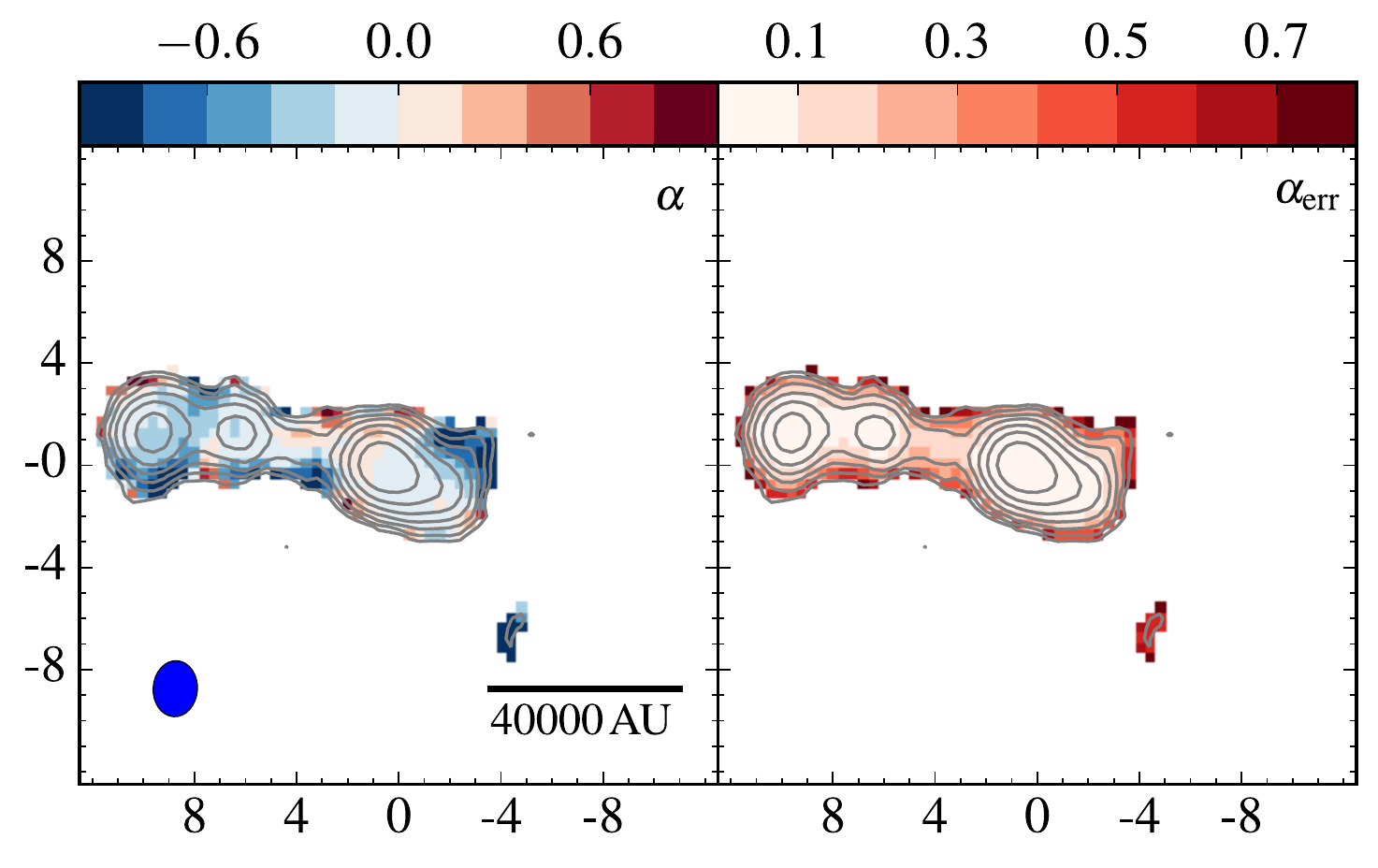}
\caption{A map of the spectral index (between 5.5 and 9GHz) for G310.1420+00.7583A (left) and the associated errors in the spectral index (right). The restoring beams (blue) used for both the 5.5 GHz and 9GHz data were identical, in the production of the spectral index maps. Grey contours represent the flux densities at 9GHz starting at 8.0$\times10^{-5}$Jy/beam ($5\sigma$), and scaling by a factor of $\sqrt{5}$ for each subsequent contour.}
\label{fig:G310_1420_SPIX}
\end{figure}

Morphologically, the emission forms a string of at least 7 lobes whose axis lies at a position angle of $\sim77\degr$, with the eastern lobes of emission becoming weaker with increasing frequency. The furthest extent of the eastern emission is coincident with H$_2$ emission \citep[knot 1 from][]{Caratti2015}. The source centred on the observed water masers has a thermal, slightly positive spectral index of $\alpha=0.18\pm0.09$. The rest of the derived parameters are shown in table \ref{tab:Jets+LobeSizes} and due to the complex morphological nature of the target, we plot a map of the spectral index in Figure \ref{fig:G310_1420_SPIX}. This is indicative of thermal emission processes for lobes A1 and B. Component D becomes resolved at the upper two frequencies, and possibly is comprised of multiple sources. Considering the orientation of the two knots of H$_2$ emission with the apparent orientation of the resolved emission of D (especially at 17 GHz), this complicates the picture. From the observations presented here, we believe this to be an ionized jet displaying both thermal and non-thermal emission.

\subsubsection*{G313.7654-00.8620}
Associated to the IRAS point source IRAS 14212--6131, this object has a $^{13}$CO outflow detection \citep{Urquhart2007CO} and masers (OH and methanol). \citet{Caratti2015} detected 4 knots of H$_2$ emission which, if tracing a jet, show evidence of a precession in the jet's axis of $32\degr$. Knot 1 (closest to the MSX point source) has an inferred electron density of $(1\pm0.5)\times10^4 \mathrm{cm}^{-3}$, while knot 4 (most distant) displays Br$\gamma$ emission indicative of strong J-type shocks with a shock speed of $\sim60\mathrm{kms}^{-1}$. Currently the jet axis is defined at a position angle of 125$\degr$, with a length (on one side) of 1.4pc. GLIMPSE images show diffuse emission in the direction of the H$_2$ knot 1, an extended HII region to the south and two point sources within 5$\arcsec$ of the MSX point source position.

\begin{figure}
\includegraphics[width=84mm]{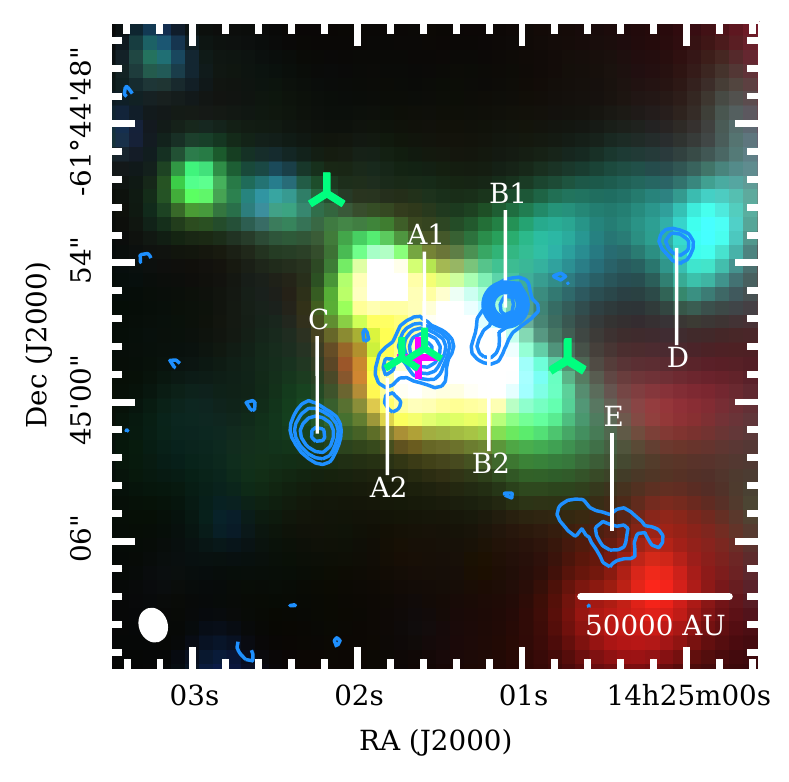}
\caption{The 9GHz radio contours overlaid on an RGB (8.0$\micron$, 4.5$\micron$ and 3.6$\micron$) composite image of existing GLIMPSE data for G313.7654-00.8620. Contours step up from 3$\sigma$ (where $\sigma=1.58\times10^{-5}$Jy/beam by a factor of $\sqrt{3}$ per level.}
\label{fig:G313_7654_RGBplot}
\end{figure}

\begin{figure}
\includegraphics[width=84mm]{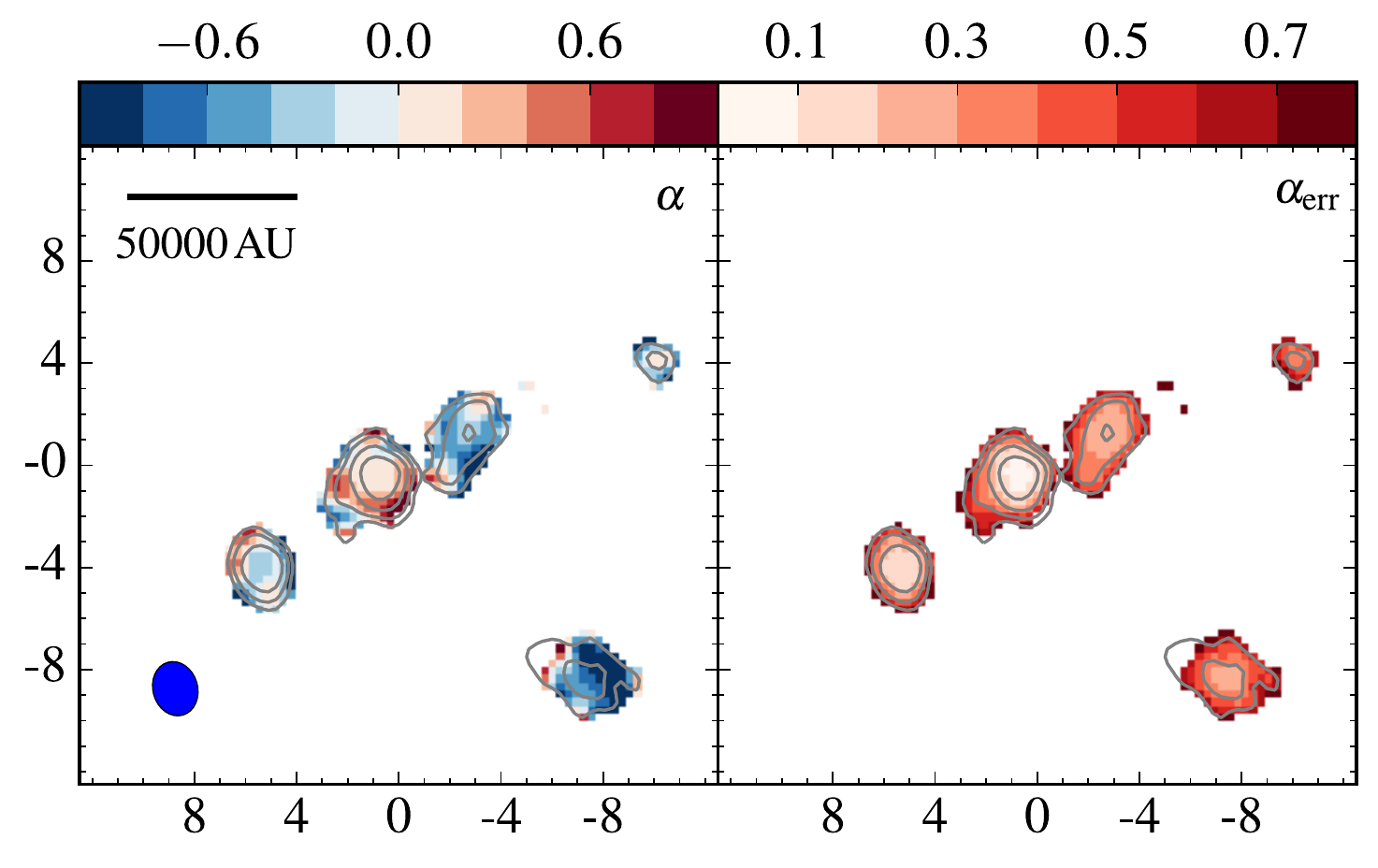}
\caption{A map of the spectral index (between 5.5 and 9 GHz) of G313.7654-00.8620 (left) and the map of errors in the spectral index (right). Grey contours represent the flux densities at 9GHz starting at 7.40$\times10^{-5}$Jy/beam ($5\sigma$), and scaling by a factor of $\sqrt{5}$ for each subsequent contour.}
\label{fig:G313_7654_SPIX}
\end{figure}

Our observations show at least 7 distinct lobes of emission at both observing frequencies. Lobe B1 coincides with a H$_2$ emission knot \citep{Caratti2015}, while A1 is coincident with both methanol (Purser et al., in prep.) and hydroxyl maser emission \citep{Caswell1998}. The NW string of emission (i.e. B1/B2 and D) coincide with extended MIR emission potentially tracing outflow cavities (see Figure \ref{fig:G313_7654_RGBplot}). All emission (apart from E) is sited in a linear distribution with the brightest emission (A1) being co-located with the MSX point source's peak position. A large range of spectral indices are present as shown in Figure \ref{fig:G313_7654_SPIX} with spectral indices of 0.3, -0.7, -0.96, 1.9, 0.04, -0.32 and 0.29 calculated for lobes A1, A2, B1, B2, C, D and E respectively. B1 has the steepest negative spectral index and is coincident with an H$_2$ knot \citep[knot 1 in][]{Caratti2015}. The current position angle of this jet from our observations (from the alignment of A1, A2 and B1) is 121$\degr$, close to that derived by \citet{Caratti2015}. We classify this as an ionized jet with lobes.

\subsubsection*{G318.9480-00.1969A}
\citet{Lee2001} detected H$_2$ line emission with two H$_2$ knots present (linearly distibuted with the MYSO) separated from the target by $\sim15\arcsec$ to the WNW and ESE. \citet{deBuizer2003} also found 5 knots oriented on the same axis. Methanol masers are distributed in a linear fashion and offset from the centre of 11.7$\micron$ emission which is elongated in the direction of the knots (PA$\sim145\degr$), as well as an SiO outflow \citep[blue lobe to the WNW, red to the ESE,][] {deBuizer2009}. Considering the apparent misalignment of the methanol masers with the outflow, the hypothesis of the methanol masers residing in the disc was rejected by \citet{deBuizer2009}. However, considering that discs can span up to 30000AU in some cases (see the review by \citet{Cesaroni2007}), with a flared cross-section, an alternate interpretation is that the masers are residing in the flared outer region of a disc around 1500-2000AU from the central object.

With the observations presented in this work (figure \ref{fig:ContourPlots}), radio continuum from this object is detected at all 4 observing frequencies. The emission is point like and unresolved, offset from the 11.7$\micron$ emission by $1\arcsec$ (3000AU) to the ESE. With a derived spectral index of $\alpha=0.69\pm0.22$, this fits Reynold's model of a conical, non-collimated jet. Lower limits on the deconvolved sizes infer a value for $\gamma>-0.9$, while the deconvolved position angle of $\sim80\degr$ (at 22.8 GHz) is parallel with the observed H$_2$ emission. We dismiss the possibility of a HII region on the basis of the radio luminosity (at 22.8 GHz) being 1.4$\%$ of that expected for $\mathrm{L}_\mathrm{Bol}=1\times10^4\,\mathrm{AU}$. However, without more information on the deconvolved properties, we must classify this as a jet/disc wind.

\subsubsection*{G326.6618+00.5207}
The 1.2mm emission in this region \citep{Miettinen2006} revealed a complex large-scale morphology. This complexity continues on the smaller core/clump scale with IRAS 15412--5359 being a compact site of massive star formation surrounded by an ionized region powered by several massive stars, one 09V and several B0V type \citep{BarresDeAl2006}. \citet{Ellingsen2006} detected 6.7GHz methanol maser activity centred on the continuum emission presented in this work (see Figure \ref{fig:ContourPlots}).

We derive $\alpha=1.0\pm0.6$ (between 17 and 22.8 GHz), the emission for which is resolved with an angular size of $(0.51\pm0.29)\arcsec\times(0.10\pm0.14)\arcsec)$ at 22.8 GHz and position angle of $153\degr\pm20\degr$. It is also coincident with water maser emission detected in the highest frequency band. A wide range of scenarios are compatible with the relatively large errors on the spectral index. The elongation of the radio emission hints at elongated morphology for the deconvolved emission, though the errors are too large to conclude this. At 22.8 GHz, the flux is 0.78\% that expected from the bolometric flux assuming a HII region. On this basis, we classify it as a jet candidate.

\subsubsection*{G327.1192+00.5103}
\citet{phillips1998} detected a weak 0.5mJy source, which was classified as a HII region powered by a B1 type star (though this is underluminous to the flux expected for L$_\mathrm{Bol.}=3.7\times10^4\mathrm{L}_\odot$ by a factor of $\sim500$). Previously, both methanol and hydroxyl masers have been detected close to the MSX point source position with the methanol masers aligned in a linear arrangement with a position angle of $\sim80\degr$ \citep{Walsh1998}. An extended green object is detected at 15$^\mathrm{h}$45$^\mathrm{m}$32.7$^\mathrm{s}$ -53$\degr$52$\arcmin$39.4$\arcsec$ and is listed as a ``likely MYSO candidate" by \citet{Cyganowski2008}. Br$\gamma$ emission is absent from NIR spectra (Lumsden et al., in prep.).

A single lobe with $\alpha=0.16\pm0.09$ is detected at all frequencies with water masers $\sim1\arcsec$ to the NW. All species of maser are aligned in a roughly linear configuration with a position angle of $\sim125\degr$. At 22.8 GHz, there are hints of emission detected toward the water masers at the 3$\sigma$ level. The source is underluminous at radio frequencies compared to the bolometric flux (0.18\% of that expected for a HII region). Since it cannot be a HII region, or a disc wind ($\alpha\neq0.6$), we classify this as an ionized jet.

\subsubsection*{G331.3576+01.0626}

We derive $\alpha=0.47\pm0.09$ for the single component, typical of a collimated jet \citep{Reynolds1986}. The presence of a strong HII region 25$\arcsec$ to the West necessitated the neglect of baselines shorter than $40\,\mathrm{k\lambda}$ at 5.5 and 9 GHz. Deconvolved dimensions at 22.8 GHz show a spherical structure (with admittedly large errors), and consequent fitting with a model for a homogeneous HII region yields an emission measure of 2.4$\times10^8\mathrm{pccm}^{-6}$ over an area of $\sim0.08\arcsec^2$ or $400$AU$^2$, however this is poorly fitted and radio flux at 22.8 GHz is underluminous by a factor $\sim17$ compared to the bolometric luminosity. Considering the lack of historical observations, we can only classify this object as a jet candidate.

\subsubsection*{G332.0939-00.4206}
One of the brightest objects in our sample with L$_{\mathrm{bol}}=9.3\times10^4\mathrm{L}_\odot$, it shows a slight extension in K-band in the SE/NW direction (2MASS images). In GLIMPSE imagery it is heavily saturated in all filters, with extended red emission still discernible to the NE. Weak Br$\gamma$ is present in NIR spectra (Lumsden et al., in prep.) as well as faint CO bandhead emission, too weak to be fitted \citep{Ilee2013}. A methanol maser has been detected close ($<1\arcsec$) to the source position \citep{Ellingsen2006} also.

Between 5.5/9 GHz and 17/22.8 GHz, $\alpha=0.9\pm0.5$ and $\alpha=0.6\pm0.3$ are derived respectively, while across all frequencies we derive $\alpha=1.80\pm0.14$. We believe the steep overall index when compared to the values for $\alpha$ calculated across the two spectral bands indicate significant variability between the two epochs. In this case the fluxes at 17 and 22.8 GHz are overluminous by a factor of $\sim2.7$, when compared to that expected from the fluxes at 5.5 and 9 GHz (using $\alpha=0.9\pm0.5$). Comparing the bolometric flux to the radio flux at 22.8 GHz indicates the radio flux at 22.8 GHz to be 0.17\% of that expected. Together with faint CO bandhead emission indicating an accretion disc, this indicates a ionized jet.

\subsubsection*{G332.9868-00.4871}
CO bandhead emission attributable to an accretion disc around a central source of 16.6M$_\odot$ was detected by \citet{Ilee2013}. Diffraction limited MIR observations \citep{Wheelwright2012} showed a spherically symmetric and compact morphology. NIR observations revealed weak Br$\gamma$ emission also (Lumsden et al., in prep.).

This source has a confirmed detection at all four observing frequencies, and was too compact to be deconvolved. The radio spectrum itself displays possible evidence for a turnover at higher frequencies, and when fitted with a homogeneous HII region model, an emission measure of ($1.17\pm0.28)\times10^9$pccm$^{-6}$, a size of ($\sim330$AU) and $\mathrm{n_e}=(8\pm1)\times10^5\,\mathrm{cm^{-3}}$ is derived. Fitting with a simple power-law yield a spectral index of $\alpha=1.27\pm0.14$. Optically thin radio flux (as fitted by simple HII models) is 18\% of that expected from the bolometric flux, which works against the HII region picture. Thus we classify this object as a jet candidate on account of uncertainty in deconvolved dimensions and orientations.

\subsubsection*{G338.9196+00.5495}
In K-band 2MASS images, an infrared point source is present with a tail of diffuse emission to the SE which is not detected at longer infrared wavelengths. The MSX source position itself is coincident with a 4.5$\micron$ excess (EGO) indicative of shocks \citep{Cyganowski2008} and outflow activity.

The detection of multiple, bright, extended HII regions meant image quality was reduced at the lower frequencies due to residual side lobes. By neglecting baselines $<30\mathrm{k\lambda}$, a spectral index of $\sim1$ is derived. Water maser activity is detected, offset 1$\arcsec$ ($\sim5000$AU) to the south of the free-free emission. At 22.8 GHz, the radio flux is a factor of $\sim800$ times weaker than that expected from an optically thin HII region. The presence of shocks in the MIR, and weak radio flux would classify this as a jet, however without reliable estimates for $\gamma$ or direct outflow observations, we must classify this as a candidate.

\subsubsection*{G339.6221-00.1209}
Diffraction-limited mid-infrared imaging of this object by \citet{Wheelwright2012} showed a bi-polar, cometary morphology whose major axis is oriented NE-SW. Modelled with a high bolometric luminosity ($L_{\mathrm{bol}}=5.2\times10^5L_\odot$) in this reference, yet with a radio-quiet profile, this object was postulated to be a weak, distant, extended HII region. However recent revisions of its bolometric luminosity (see table \ref{tab:targets}) fit in more with an MYSO picture. With this in mind, the mid-infrared emission may trace the warm dust in the outflow cavities assuming a single MYSO at the MSX point source position. \citet{Navarete2015} observed 2.12$\micron$ H$_2$ emission coincident with the source, indicative of shock emission from outflows.

Observations of G339.6221-00.1209 at 5.5 and 9 GHz show the detection of two components of free-free emission in the (robustness of 0.5) clean maps. The east component has a thermal, flat spectral index of $\alpha=0$ and coincides with OH maser emission whilst offset from a string of methanol maser spots residing 2$\arcsec$ to the east of the lobe's centroid. The west lobe of emission is relatively weak and has a negative spectral index of $\alpha=-1$. Because of interference from strong HII regions in the area, re-imaging was performed neglecting shorter baselines ($<12.5\,\mathrm{k\lambda}$) and a robustness of -1 to suppress residual side lobes. This yielded $\alpha=-0.17\pm0.26$ and $\gamma=-0.33\pm0.25$ for E, however W was not detected. From Equations \ref{eq:reynoldsAlphaBeta} and \ref{eq:reynoldsGammaBeta}, these are incompatible assuming an ionized jet picture. However, considering the deconvolved position angle is roughly oriented with the major axis of the beam, deconvolved dimensions may be compromised by the low signal to noise of the target combined with the elongated beam. Flux densities of the source are more certain and are $\sim110$ times too weak compared to the flux expected of an optically thin HII region of the same bolometric luminosity. Although the observations in the literature show jet activity in the area, our radio images cannot definitely classify this source as an ionized jet, only as a candidate.

\begin{figure}
\includegraphics[width=84mm]{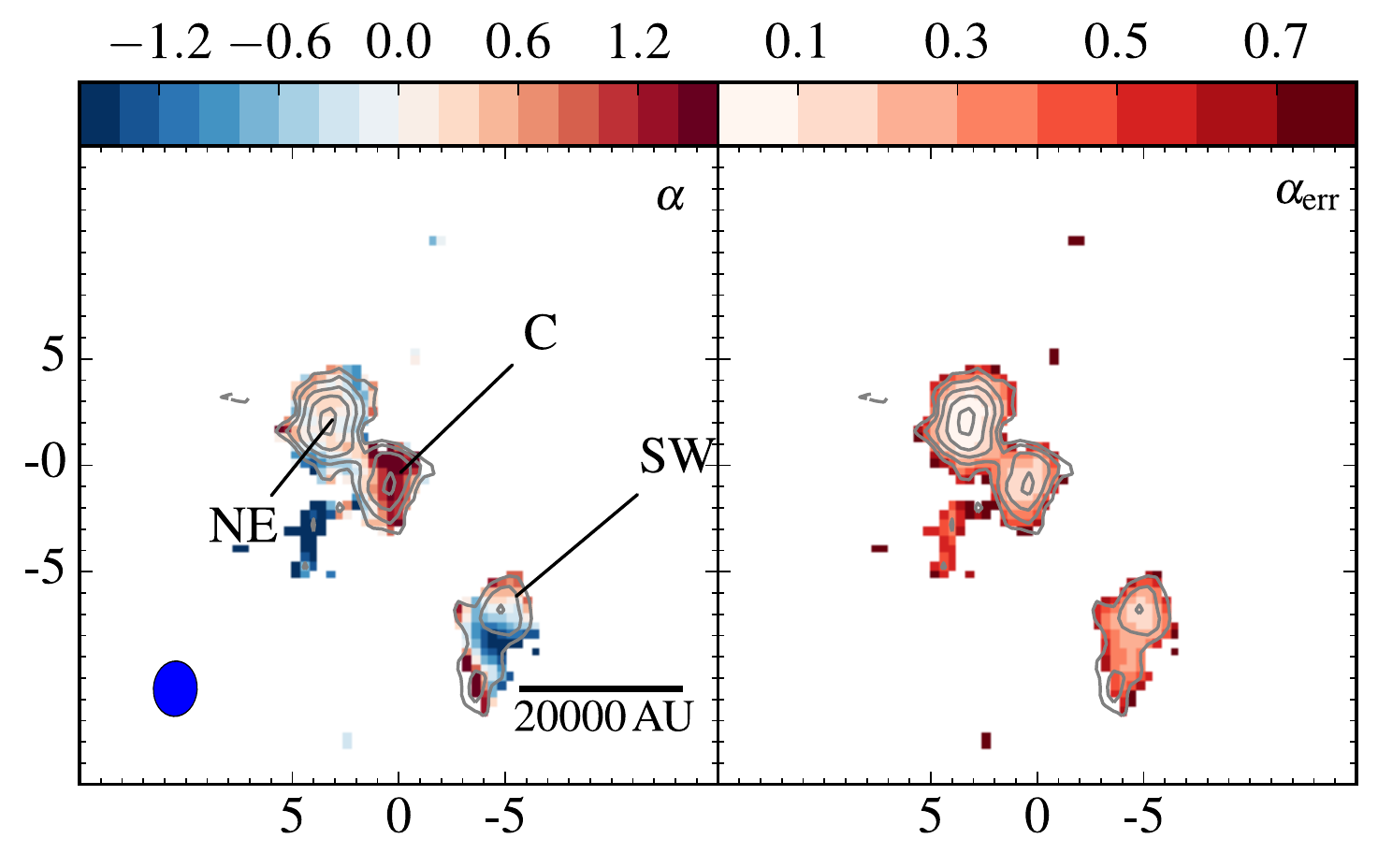}
\caption{A map of the spectral index (between 5.5 and 9GHz) for G339.8838--01.2588 (left) and the associated errors in the spectral index (right). The restoring beams (blue) used for both the 5.5 GHz and 9GHz data were identical, in the production of the spectral index maps. Grey contours represent the flux densities at 5.5 GHz starting at 1.40$\times10^{-4}$Jy/beam ($4\sigma$), and scaling by a factor of $\sqrt{3}$ for each subsequent contour.}
\label{fig:G339_8838_SPIX}
\end{figure}

\subsubsection*{G339.8838--01.2588}
Associated to the object, IRAS 16484--4603, this MYSO has an extensive observational history. \citet{Ellingsen1996} observed elongated, ionized emission at a position angle of $-45^{\degr}$, which within astrometrical errors, was coincident with the position of, and perpendicular to the distribution of, multiple, linearly-distributed (both spatially and in velocity) CH$_3$OH MASER spots. \citet{Stecklum1998} also observed 10$\micron$ emission elongated at the same position angle as the MASERs, leading to an interpretation of a disc/jet model with irradiated MASERs in Keplerian rotation, embedded in the circumstellar disc. However, further observations at 10$\micron$ and 18$\micron$ by \citet{deBuizer2002} resolved the elongated MIR emission into three sources, whose temperature profile peaked with the radio emission spatially, however the resolved point source, 1B, was offset $\sim0.4\arcsec$ to the SE, and coincident with the outflow axis of the radio knots from \citet{Ellingsen1996}. \citet{deBuizer2002} further concluded that there are two sources at work, one massive, unobscured star with an extended HII region (partially resolved out in existing radio data) and a second, which is embedded and centred on the previously detected radio emission and also responsible for outflow activity in the region. 

The nature of the observations in this paper preclude the detection of extended emission on the scales discussed in the above references (some extended emission is partially resolved out at 5.5 GHz). Therefore it is safe to assume that any flux imaged comes from free-free emission associated to compact sources (i.e. radio lobes/central jets/HCHIIs). 

From the radio flux contour maps in figure \ref{fig:ContourPlots}, 3 compact components are present through all 4 observing frequencies. A range of spectral indices are seen over these components as illustrated in Figure \ref{fig:G339_8838_SPIX}. Calculated values for $\alpha$ and $\gamma$ (component C) are $0.77\pm0.04$ and $\gamma=-1.0\pm0.5$ respectively indicate either a conical, thermal jet undergoing some degree of recombination and/or acceleration in the flow \citep{Reynolds1986} or an optically thick HCHII. At 5.5 GHz, the central source also becomes slightly extended. Considering the turn-off at higher frequencies, it is possible that this represents a young, optically thick HCHII region of small spatial extent, as well as an ionized jet, however it is under-luminous by a factor of $\sim1800$ to that expected from the bolometric luminosity, supporting the ionised jet picture which is our final classification.

\subsubsection*{G340.0543-00.2437A}
Though observations of this region are sparse in the literature, this is none-the-less a very active region with star formation at a variety of stages. Within 10$\arcsec$ there are 5 separate objects registered as HII regions or YSOs in the RMS database with G340.0543-00.2437D overlapping the extended green object determined from GLIMPSE.

Due to the large fluxes and extended emission of G340.0543-00.2437E (the UCHII in figure \ref{fig:ContourPlots}), robustness of the images was reduced to -1 to counter its strong sidelobes. Noise in the images is still relatively high as a result of this but also due to partially resolved out extended emission in the general area. As a result, fitting of a Gaussian to the image plane was ineffectual and a spectral index of $\alpha=0.4\pm0.7$ is calculated from the integrated flux within the 3$\sigma$ contours of G340.0543-00.2437D. The radio emission coincides with the EGO and is probably associated with outflow activity because of its coincidence with this 4.5$\micron$ excess and underluminosity compared to HII regions of similar luminosities. Through a lack of observations in the literature as well the results presented here, we classify this object as a jet candidate.

\begin{figure}
\includegraphics[width=84mm]{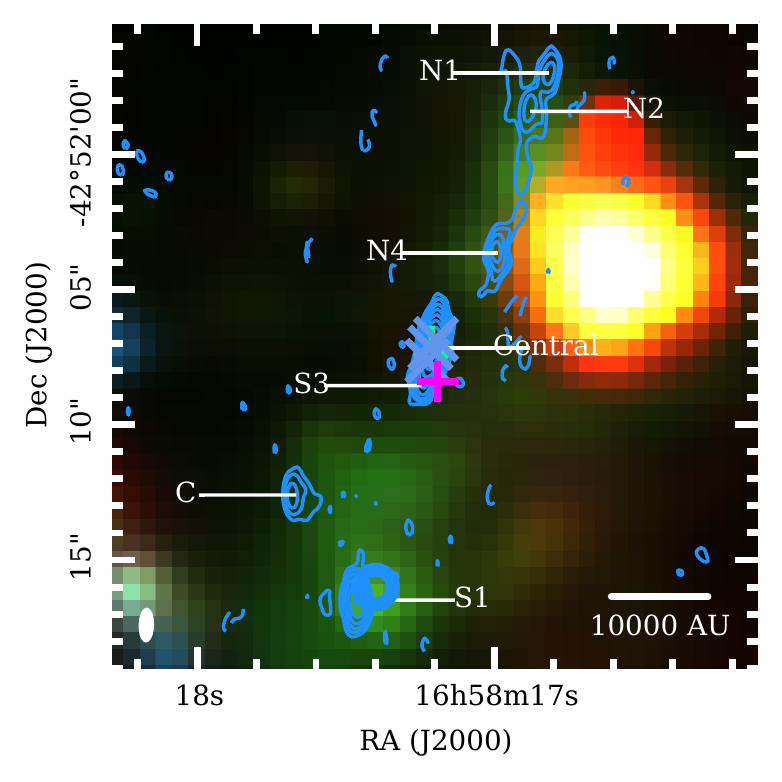}
\caption{The 17 GHz radio contours overlaid on an RGB (8.0$\micron$, 4.5$\micron$ and 3.6$\micron$) composite image of existing GLIMPSE data for G343.1261-00.0623. Contours step up from 5$\sigma$ by a factor of $\sqrt{3}$ per level. Arrows represent the proper motions derived from our data and that found in \citet{Rodriguez2005}, who's lengths are relative and in proportion.}
\label{fig:G343_1261_RGBplot}
\end{figure}

\subsubsection*{G343.1261-00.0623}
Associated to the object IRAS 16547--4247, this source has an extensive observational history. Most recently, cycle-0 ALMA observations at 880$\micron$ have revealed two dust cores separated by $\sim5000\mathrm{AU}$, Core A (closest to IR source) and Core B \citep[masses 15$M_\odot$ and 12$M_\odot$ respectively,][]{Higuchi2015}. The same set of observations showed two, perpendicular, bi-polar molecular outflows in $^{12}$CO(3-2) (one NNE-SSW, the other WNW-ESE for the red-blue lobe alignment axis), which are both separate again to a larger-scale bi-polar outflow detected with the Atacama Pathfinder EXperiment \citep{Garay2007} with a N-S alignment. This larger scale outflow is slightly misaligned to the previously established radio jet of \citet{Rodriguez2005} which is comprised of at least 9 components, the central thermal jet being deduced to have a spectral index of $\alpha=0.33\pm0.05$ (between 8.46 and 14.9 GHz). The thermal jet also displays evidence for precession \citep{Rodriguez2008} which is likely caused by the proximity of the two cores. Other ALMA observations of spectral lines associated to CH$_3$SH and CH$_3$OH \citep{Zapata2015} reveal a compact disc and display Keplerian rotation profiles (along a position angle of $\sim40\degr$, perpendicular to the free-free thermal jet). These findings lead to a conclusion of a central MYSO with a mass of $\sim20$M$_\odot$ and disc mass of $\sim6$M$_\odot$ , calculated from concurrent observations of 0.86mm continuum emission.

\begin{figure}
\includegraphics[width=84mm]{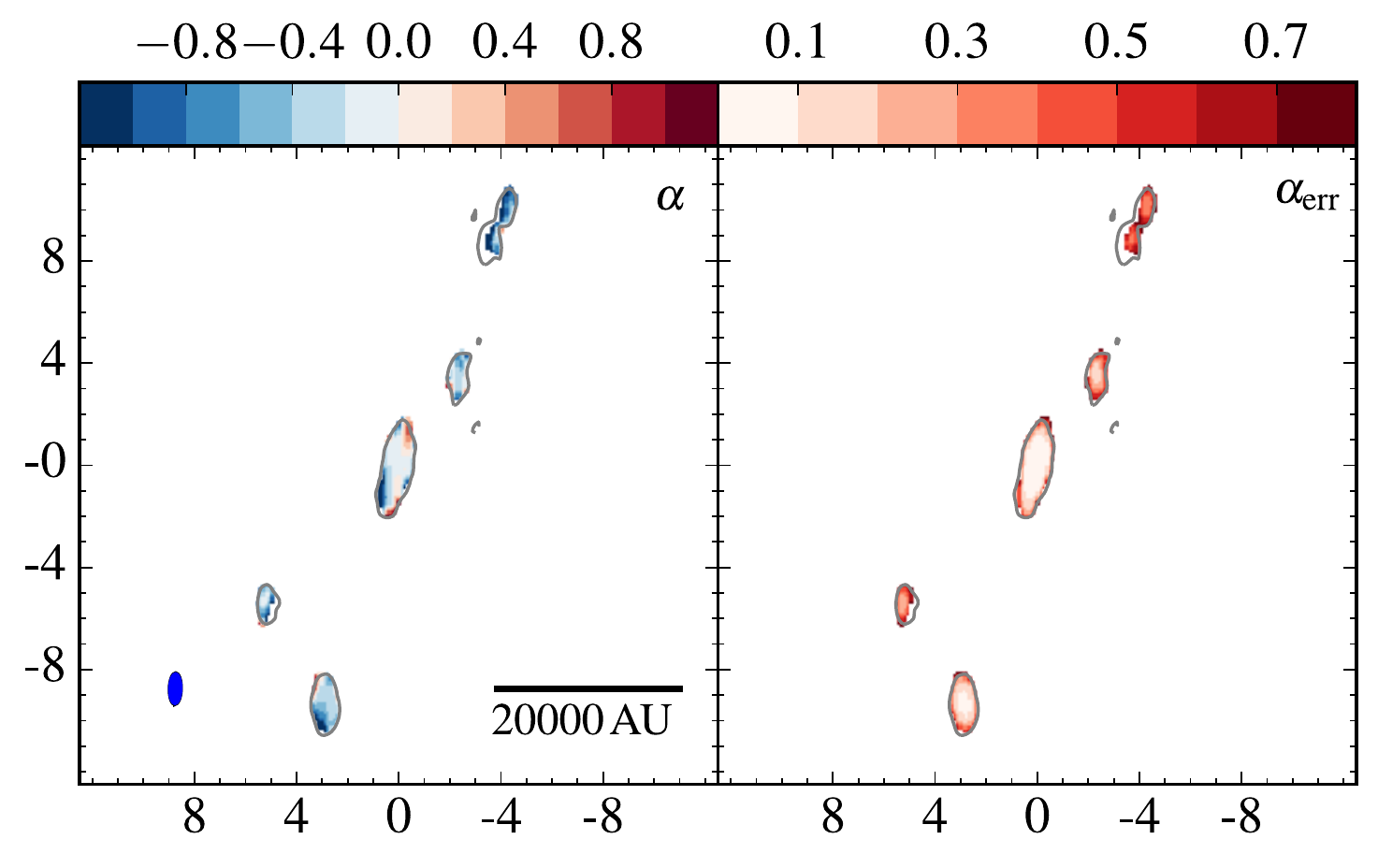}
\caption{A map of the spectral index (between 17 and 22.8 GHz) for G343.1261-00.0623 (left) and the map of errors in the spectral index (right). The grey contour represents a flux density, at 17 GHz, of 3.03$\times10^{-4}$Jy/beam ($5\sigma$). For clarity, no further contouring has been included.}
\label{fig:G343_1261_SPIX}
\end{figure}

Taking the terminology from \citet{Rodriguez2005}, we detect lobes of emission towards the positions of components C, N2, N1, S-1 and the central source. Source B was not detected but is known to be time variable. New lobes of emission are detected, dubbed N-4 and S-3, of which the first was detected by \citet{Rodriguez2005} but not named. At 17 GHz there is some weak, diffuse emission connecting the northern lobes which is either resolved out or too weak at 22.8 GHz to be detected. The central component appears to be comprised of multiple components for which a 2-component model best fits the 17 GHz and 3-component model best fits the 22.8 GHz data. For consistency, spectral indices derived were taken from the 2-component model (S-3 and central lobes) at each frequency, however flux is left over at 22.8 GHz (with an integrated value of $\sim$1mJy) in the residual map which cannot be fitted without an extra component. Spectral indices derived are -0.1 for the central source, and -1.2 for the second component, indicating thermal and non-thermal emission respectively, though the reliability of these is unknown requiring higher resolutions. Considering the relatively large errors at 22.8 GHz, these spectral indices have inherently large errors. For the other lobes, spectral indices of -1.4, -0.45, -2.5, -2.1 and -0.7 are derived for lobes C, S-1, N-1, N-2 and N-4 respectively. Because of the diffuse emission resolved out at 22.8 GHz, the spectral indices for the northern components and component C should be taken as lower limits. Lobe S-1 has a spectral index consistent with that found by \citet{Rodriguez2005}. With expected fluxes from \citet{Rodriguez2005} at 17 and 22.8 GHz of 10.9 and 12.3mJy respectively, and measured fluxes of 16.8 and 16.3mJy, the central source appears to display a degree of time variability. Water MASER emission is also detected across a broad range of channels, with the MASER spots arranged in a linear pattern parallel to the major axis of the central component. This would indicate they are excited in the outflow cavities rather than the accretion disc. This source is classified as a jet with lobes.

\subsubsection*{G343.5213-00.5171}
Modelling of the SED profile by \citet{Wheelwright2012} suggested an MYSO inclined 57$\degr$ to the line of sight with a mass loss rate of $2.5\times10^{-4}\,\mathrm{M_\odot\,yr^{-1}}$ and opening angle of 10$\degr$, however the confused nature of the source detrimentally influenced the reliability of the fit. No other significant previous observations of this object are present in the literature other than an extended green object detected by \citet{Cyganowski2008} offset 30$\arcsec$ to the north. Whether this is directly related to this particular source, and not the star formation region in general, is unknown. No Br$\gamma$ emission is seen, however [FeII] emission is present (Lumsden et al., in prep.).

The detection of 2 components at lower frequencies, and 1 component at higher frequencies are made for this source. At the risk of inconsistency, the robustness parameter was increased to 4 at 22.8 GHz due to the point like nature of the source and also the very low signal to noise ratio present at this frequency. This allowed a more reliable fit to the spectral index for the west component of $\alpha=-0.05\pm0.15$. The east component yielded a spectral index of $\sim$0 for the two frequencies it was detected at. The SW extension of the emission at 5.5 GHz, and west/east lobes of emission are coincident with the MIR emission in figure 3K of \citet{Wheelwright2012}. The possibility of both components representing separate MYSOs is considered, which would mean the accepted bolometric luminosity would be split between the two, potentially explaining the lower than expected fluxes for the main component. However, no Br$\gamma$ emission is seen (Lumsden et al., in prep.), ruling out the possibility of a HII region. As such, we classify this as a jet candidate.

\subsubsection*{G345.4938+01.4677}
A paper by \citet{Guzman2010} detected 5 distinct lobes of thermal free-free emission associated to this object, and promoted an episodic accretion/ejection scenario. The central source was inferred to be an ionized jet with a spectral index of $\alpha=0.85\pm0.15$, and a position angle for the jet of $\sim99\degr$. Emission in the inner/outer lobes was optically thin with flat spectral indices of $\alpha\sim-0.1$, and was associated to excess emission at 4.5$\micron$ due to shocks in the flow. K$_\mathrm{s}$ band images show enhanced emission, as a result of the excitation of the 2.12$\micron$ line, parallel to the jet on both sides of the source, and also perpendicular to it with a N-S alignment. The N-S emission displayed a morphology similar to that seen in the bow-shocks of Herbig-Haro objects, suggesting a second outflow in the region. Emission at 10$\micron$ \citep{MottramRMS2007} coincided with the I-E lobe (terminology as in \citet{Guzman2010}) and central source, with the prior being associated to warm dust in the outflow cavity. Follow up observations of the lines CO $J=3\rightarrow2$, $6\rightarrow5$ and $7\rightarrow6$ (with APEX) showed a quadrupolar morphology (more distinctive in the two higher excitation temperature transitions), with one red-blue lobe pair aligned with the radio jet and the other aligned with the N-S K$_\mathrm{s}$ band emission \citep{Guzman2011}. The velocity range of the outflows was $\sim60$kms$^{-1}$, with collimation factors of $\sim1$ and 2 for the N-S (P.A. of 7$\degr$) and SE-NW (P.A. of 107$\degr$) outflows respectively, as expected with MYSOs. GLIMPSE images show the walls of an outflow cavity aligned with the eastern jet.

The radio lobes detected in previous works are all detected at 17 GHz, however at 22.8 GHz only the central component and components I-E and I-W are detected. The jet is aligned with the outflow cavities seen in GLIMPSE imagery. Water maser activity is detected towards at least 5 separate sites in the field of view, and is probably associated to radiative processes on material in the west outflow cavity walls. The spectral index of the central source (derived from our observations alone) is $\alpha\sim0.8$, in close agreement with \citet{Guzman2010}. However the fluxes recorded at both frequencies are $\sim30\%$ lower than expected, suggesting the central source's flux may be time variable (absolute flux calibration errors of $<10\%$ are expected with the ATCA). We classify this source as a jet with lobes.

\subsubsection*{G345.5043+00.3480}
Outflow activity is detected in MOPRA observations of HCO$^+$, however it is relatively uncollimated with a rough S-N (red-blue) configuration centred on G345.5043+00.3480 \citep{Yu2014}. 

Observations detect a radio source offset by less than 1$\arcsec$ to the NE of the phase centre (at all frequencies), as well as an extremely strong HII region located $\sim130\arcsec$ to the SE (RMS survey designation, G345.4881+00.3148). Due to the limited dynamic range of the observations, strong side lobes were present at the lower two frequencies, which interfered with the deconvolution of the target of interest. A robustness of -1 was therefore used to suppress the side lobes at as well as neglecting baselines $<40\,\mathrm{k \lambda}$ at the cost of higher noise levels in the resultant images. Large errors on the deconvolved size at 17 GHz prohibit a power-law index for the deconvolved major axis being derived, however a spectral index of $\alpha=0.92\pm0.10$ is calculated, indicative of thermal free-free emission. At 22.8 GHz, water maser emission is detected in two spots, one coincident with and the second offset $\sim3\arcsec$ to the SE (previously seen by \citet{Forster1989}) from, the target. It is underluminous at 22.8 GHz by a factor of $\sim3000$, however without more information on the behaviour of the major axis with frequency, we can only classify this object as jet candidate.

\subsection{`Embedded' HCHII regions}
\label{sec:CoreHIIRegions}
From the 14 HII regions which were identified in our sample, three displayed a compact, optically thick core, centred on the MSX point source position. These cores were embedded in an optically thin region of extended emission and generally qualify for HCHII regions, as opposed to jets, based on their radio fluxes and high-frequency turnovers present in their spectra (jet turnover frequencies are likely $\nu_\mathrm{m}>50\mathrm{GHz}$). All of these examples are discussed below in further detail. 

\subsubsection*{G274.0649-01.1460A}
Our observations show a compact, bipolar HII region with a bright core which becomes more prominent at 17 and 22.8 GHz. From the morphology alone, this object is similar to bipolar HCHIIs like NGC 7538 IRS1 (\citet{Campbell1984}; \citet{Sandell2009}). It is thought that these objects may be a transitional phase between MYSOs and an expanding HII region, as the ionized front moves along the density gradient of the evacuated cavities previously cleared by outflows and jets (which may still be active), to produce the bipolar morphology. However considering the distance-luminosity expected at 22.8 GHz ($\sim25\,\mathrm{mJy\,kpc^2}$), G274.0649-01.1460A is $\sim4$ times too luminous suggesting either other ionization processes may be responsible for the flux observed or simply that the bolometric luminosity has been underestimated due to the complex environment. The fluxes of the core were fitted with a Gaussian profile and had a spectral index (between 17 and 22.8 GHz) of $\alpha=0.81\pm0.81$, for a robustness of 0.5. However, the extended emission affected the fitting of the central component as they could not be separated effectively. Therefore, we re-imaged the object using a robustness of -1 at 9 (neglecting baselines $<50\mathrm{k}\lambda$), 17 and 22.8 GHz (5.5 GHz imaging was ineffective at separating the extended and compact components), yielding a more reliable value for the spectral index of $\alpha=0.37\pm0.17$ ($\gamma$ could not be determined as the central component remained unresolved at all frequencies). It is worth noting there was no sign of a turnover and the fluxes remained within errors for a standard power-law profile. The spectral index of the emission as a whole (between 5.5 and 9GHz since much emission is resolved out at the upper two frequencies) is $\alpha\sim-0.1$, indicative of optically thin free-free emission from a HII region. 

Two other HII regions were detected within 20$\arcsec$ of the central source, G274.0649-01.1460B (most southerly in Figure \ref{fig:ContourPlots}) and G274.0649-01.1460C. These two HII region have spectral indices of $\alpha\sim0.1$ and $\alpha=-0.06\pm0.06$ for G274.0649-01.1460B (2 frequencies) and G274.0649-01.1460B (4 frequencies) respectively, both indicative of optically thin emission from a HII region. The reason for using differing numbers of frequencies for the derivation of spectral indices was based upon whether or not extended emission was becoming resolved out (as in A/B's case) or not (as in C's case). Based upon the 9 GHz flux densities, emission measures of 1.47$\times10^6\,\mathrm{pc\,cm}^{-6}$, 2.30$\times10^6\,\mathrm{pc\,cm}^{-6}$ and 2.83$\times10^6\,\mathrm{pc\,cm}^{-6}$ are derived for A, B and C respectively. All of these regions would be powered by a B0.5 ZAMS type star \citep{Davies2011}. The compact core seen towards the target is of particular interest, and its exact nature remains undetermined. Considering the spectral index of the extended emission, and bipolar morphology, a young HCHII region expanding down the outflow cavities previously cleared by jet activity is likely.

\subsubsection*{G305.5610+00.0124}
TIMMI2 observations of this source \citep{MottramRMS2007} show bright, compact 10.4$\micron$ emission, coincident with the MSX point source position. Faint, extended emission is also seen to the north and east of the bright source. GLIMPSE imaging is saturated, while 2MASS shows a reddened point source with extended faint emission in the background. 

Flux density contour maps presented in Figure \ref{fig:ContourPlots} show a complicated picture. An apparent UCHII is present, which becomes resolved out at higher frequencies. One compact component coincident with the MSX point source is persistent at all frequencies and likely represents the position of the MYSO. Undoubtedly this source displays characteristics of multiplicity from the distribution of ionized material since the morphology does not fit in with those expected for a HII region powered by a single source. The water maser spots detected are located on opposite sides of the UCHII region, suggesting excitation via shocks at the ionization front of the overall expanding HII region. The compact source at the phase centre of the observations holds a spectral index of $\alpha\sim0.8$, which was derived from the upper two frequencies where the extended emission had been resolved out enough to enable fitting by a Gaussian profile. However this spectral index will have large inherent errors due to the morphology of the region itself. The spectral index plot in  The north east emission has a spectral index of $\alpha=-0.14\pm0.06$ from the 5.5, 9.0 and 17 GHz (using a robustness of 0.5) data. If this were a separate, optically thin HII region, we derive an emission measure of $5.44\times10^6\,\mathrm{pc\,cm}^{-6}$, a Lyman continuum flux of 2.27$\times10^{46}\,\mathrm{s}^{-1}$ and a ZAMS spectral type of B0.5 \citep{Davies2011}. Using the deconvolved size at 9 GHz, the size of the north east component is found to be 0.03pc, which in combination with the other derived physical parameters is suggestive of an UCHII region.

The central, compact source could represent an ionized jet (considering a radio under-luminosity of 25$\%$ that expected from the bolometric luminosity) or an early-stage HCHII region, embedded in extended HII regions from other young massive stars in the environment. Deconvolution at 22.8 GHz of the central source yields a source size of (0.48$\pm$0.02)$\arcsec$ $\times$(0.17$\pm$0.08)$\arcsec$, oriented at a position angle of $(171\pm4)\degr$. This position angle is oriented towards the extended emission to the SE (which is apparent at all observing frequencies) and NW, suggesting a density gradient in these directions. Considering the optically thick nature of the emission and only slight under-luminosity, it is most likely this core is in fact an embedded HCHII region.

\subsubsection*{G317.4298-00.5612}
\citet{Guzman2012} observed this object at similar frequencies to this paper and concluded, from radio spectral analysis, that this was a HCHII with an emission measure of 1.9$\times10^8$pccm$^{-6}$, an effective area of 0.08 arcsec$^2$ and powered by an O6/5.5 ZAMS type star. Morphologically, two lobes of radio emission were detected, one centred on the MSX source position and the other offset $\sim4\arcsec$ to the SE. The second source was concluded to be an optically thin UCHII region powered by a O9.5 type ZAMS star. 

This source was observed at 5.5 and 9.0 GHz, for which a central compact source and diffuse emission to its south east are detected. At 9 GHz, the diffuse emission looks partially resolved out by the smaller synthesised beam, and at both frequencies the central compact source is resolved. Fitting a Gaussian profile to the central source in the image domain yields a spectral index of $\alpha=0.9$, indicative of thermal free-free emission, and a power law constant of $\gamma=-1.1$ for the variation of the deconvolved major axis with frequency. It is important to note that the deconvolved dimensions may be affected by the diffuse emission to the south east. Since we do not have observations at higher frequencies, it is hard to tell whether this is an HCHII region approaching its turnover frequency in our observations, or an extremely bright ionized jet. The deconvolved size at 9.0 GHz of $0.45\arcsec\times0.27\arcsec$ (or $0.024\mathrm{pc}\times0.005\mathrm{pc}$ for a distance of 14.2kpc) and emission measure of 5.57$\times10^8\,\mathrm{pc\,cm^{-6}}$ are suggestive of a HCHII/UCHII region around a B0 type ZAMS star. The calculated emission measure assumed an optically thin regime at 9GHz and is thus a lower limit. The deconvolved position angles of $136.0\degr\pm2.6\degr$ and $136.6\degr\pm1.6\degr$ at 5.5 and 9 GHz respectively, align the major axis with the diffuse emission to the south east, approximately bisecting it. One interpretation of these results would be a MYSO with a HCHII region which is expanding down the evacuated cavities in the circumstellar envelope (explaining the elongated deconvolved dimensions). The diffuse emission to the south east is possibly the expansion of the HII region down these outflow cavities, which possess relatively wide opening angles. However, considering the scale of the extended emission ($\sim0.65pc$) this is unlikely. Instead this is probably a separate, classical HII region. In this case, we derive an emission measure (using 5.5 GHz images, and assuming this is in the optically thin regime) of $(1.26\pm0.09)\times10^5\,\mathrm{pc\,cm}^{-6}$, a derived Lyman flux of $(8.79\pm1.36)\times10^{47}\,\mathrm{s}^{-1}$. These are indicative of a classical HII region powered by an O9.5 type ZAMS star \citep{Davies2011} with $\mathrm{L_{Bol}}~4\times10^4\,\mathrm{L_\odot}$. Comparing the combination of the two massive stars contributing to the bolometric luminosity ($3.5\sim10^5\mathrm{L}_\odot$), there is a distinct under-luminosity ($\sim7\%$ of that expected) lending some support to the core being in the jet phase. However since the core is still optically thick and extremely bright compared to the rest of the jets, we classify it to be a HCHII.

\subsection{A curious case - G301.1364-00.2249}
\label{sec:G301.1364}
\citet{Walsh1998} detect the radio emission from two UCHII regions associated with IRAS 12326-6245, both coincident with excess 4.5$\mu m$ emission/EGOs in GLIMPSE RGB images. Both UCHII regions were inferred to be powered by B0 type stars, which are both members of a deeply embedded proto-cluster, with a NE-SW outflow detected in molecular line maps (though the 23$\arcsec$ beam was too coarse to resolve the origin). Inferred parameters for the outflow include mass loss rates of $\dot{M}\sim0.02$M$_\odot$yr$^{-1}$ and a momentum rate of $F\sim0.15$M$_\odot\mathrm{km\,s}^{-1}$yr$^{-1}$ \citep[from][where it was named MIR1]{Henning2000b}. H$_2$ (2.12$\micron$) emission is detected to the SE and W of G301.1364-00.2249 and loosely associated with the other UCHII. A 6.7 GHz methanol maser with a peak velocity of $-39.6\,\mathrm{km\,s}^{-1}$ is also detected coincident with the southern UCHII (Urquhart, private communication). Class I methanol maser emission (36 and 44 GHz) is also detected coincident with the southern HII region, displaying a `remarkable' absorption feature \citep{Voronkov2014}.
\begin{figure}
\includegraphics[width=84mm]{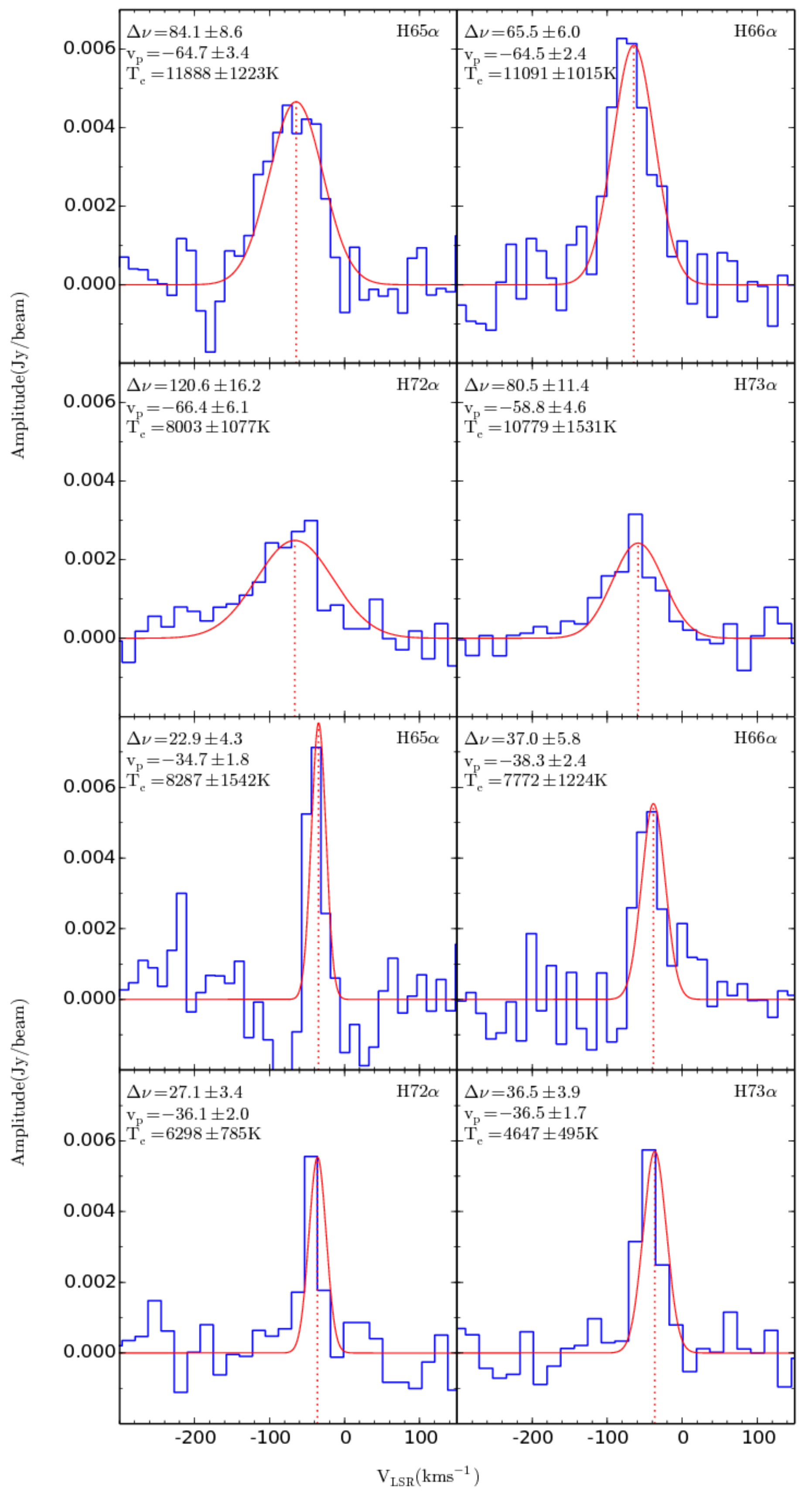}
\caption{A spectral plot of the hydrogen recombination lines detected during our observations of G301.1364-00.2235 (bottom 4 plots) and G301.1364-00.2249 (top 4 plots). The transition name is annotated in the top right of each plot, while fitted parameters are annotated in the top left, where $\Delta\nu$ and v$_\mathrm{p}$ are the line-width and peak velocity respectively (both in $\mathrm{km\,s}^{-1}$). The fitted Gaussian profile is indicated by the solid, red line, while the peak velocity is illustrated by the dotted, red line.}
\label{fig:G301_1364_HIIRecombLines}
\end{figure}

Our observations at 17 and 22.8 GHz show the presence of two apparent HII regions, G301.1364-00.2249 (phase centre) and G301.1364-00.2235 ($\sim7\arcsec$ north). Spectral indices of $\alpha=1.0\pm0.1$ and $\alpha = 0.3\pm0.2$ are calculated between the two observed frequencies, for the target and HII region to the north respectively. At 22.8 GHz, extensive water maser activity is detected, coincident with each HII region, to the south of the target and in at least 8 other spots spread between the HII regions. Archival ATCA data at 2, 5.5, 9, 33 and 46 GHz was reduced (taken on the 29th January 2011 for the 2, 5.5 and 9 GHz data and 16th August 2009 otherwise) providing a dataset from which (for G301.1364-00.2235) we calculate an emission measure of (7.91$\pm$1.29)$\times10^7\,\mathrm{pc\,cm}^{-6}$, a size of 1.25$\arcsec\times$0.45$\arcsec$ (or a diameter, assuming spherical morphology, of $0.03\pm0.01$pc, for a distance of 4.3kpc), and an inferred ZAMS spectral type of B0. This classifies it as a UCHII. Given the deconvolved size for G301.1364-00.2249 of 0.67$\arcsec\times0.36\arcsec$ (2800$\times1500$AU), and derived emission measure of (1.19$\pm0.17$)$\times10^9pc\,cm^{-6}$, this can be classified as an elongated (potentially bipolar at higher resolutions) HCHII region. At 22.8 GHz, there is also a (relatively) weak extension to the emission of G301.1364-00.2249 to the NW, with an integrated flux of 21.4$\pm1.7$mJy, a deconvolved size of 0.86$\arcsec\times0.18\arcsec$ and a position angle of $150\pm3\degr$.

Figure \ref{fig:G301_1364_HIIRecombLines} shows four hydrogen recombination lines detected during observations of this region (using the pixels with values greater than 5$\sigma$ in each channel, within a bounding box for each HII region). These radio recombination lines (RRLs) display broad line-widths in excess of $60\,\mathrm{km\,s}^{-1}$ for G301.1364-00.2249 and narrower line-widths of $\sim30\,\mathrm{km\,s}^{-1}$ for G301.1364-00.2235, further clarifying their status as HCHII and UCHII regions respectively. The peak velocities for the two HII regions do, however, differ considerably. The southern HCHII has RRLs which consistently peak at $\sim-64\mathrm{km\,s}^{-1}$, while the northern UCHII data peaks at $\sim-36\mathrm{km\,s}^{-1}$. To have members of the same clump with velocities differing by $>10\mathrm{km\,s}^{-1}$ is unusual. For this clump (M$\sim6000$M$_\odot$), the virial velocity at a separation of 2.2$\arcsec$ for a distance of 4.3kpc is $\sim24\mathrm{km\,s}^{-1}$, and since the velocity difference is of this order, a scenario whereby this source was recently ejected from the cluster is possible. From the archived data, hydrogen recombination lines at 34.6GHz (H57$\alpha$) and 45.5 GHz (H52$\alpha$) were also analysed. Both the H52$\alpha$ and H57$\alpha$ RRLs showed a similar profile to those lines in the 17 and 22.8 GHz data. Many individual channels were pre-flagged by the CABB correlator during these observations, some of which lie on the H52$\alpha$ RRL itself, however the fitted data adheres closely to a Gaussian. 

The clear offset in the V$_\mathrm{LSR}$ cannot be explained by the observations presented here. A possible, speculative explanation for the observations would be an optically thick HCHII region expanding down previously evacuated outflow cavities which are oriented towards the observer. This would explain both the classical HII region spectral profile (i.e. with an optically thick spectral index of $\sim2$ rather than $\sim1$ typical of HCHII regions) and broad linewidths offset from the LSR velocity of the source itself \citep{Lizano2008}. Further observations of RRLs at higher spectral resolutions would help to definitively resolve this situation.

\onecolumn

\section{Figures}
\begin{figure*}
\caption{The radio flux density contour plots of all detected objects in our original sample. The frequency of each is indicated in the top right corner of each subplot, and a scale bar is indicated in the bottom right for distances assumed by the RMS survey \citep{Lumsden2013}. Blue crosses (x) indicate water masers detected in this work, red crosses (+) indicate OH masers \citep{Caswell1998}, green crosses (three vertices) represent 6.7 GHz methanol masers \citep{Walsh1998}, and blue circles represent H$_2$ knots \citep{Caratti2015}. The synthesised beam is indicated in the bottom left of each plot and contours start at -3, 3, 5 and 7 times the image RMS noise and range up to the peak of the image in equal intervals.}

\includegraphics[width=0.81\textwidth]{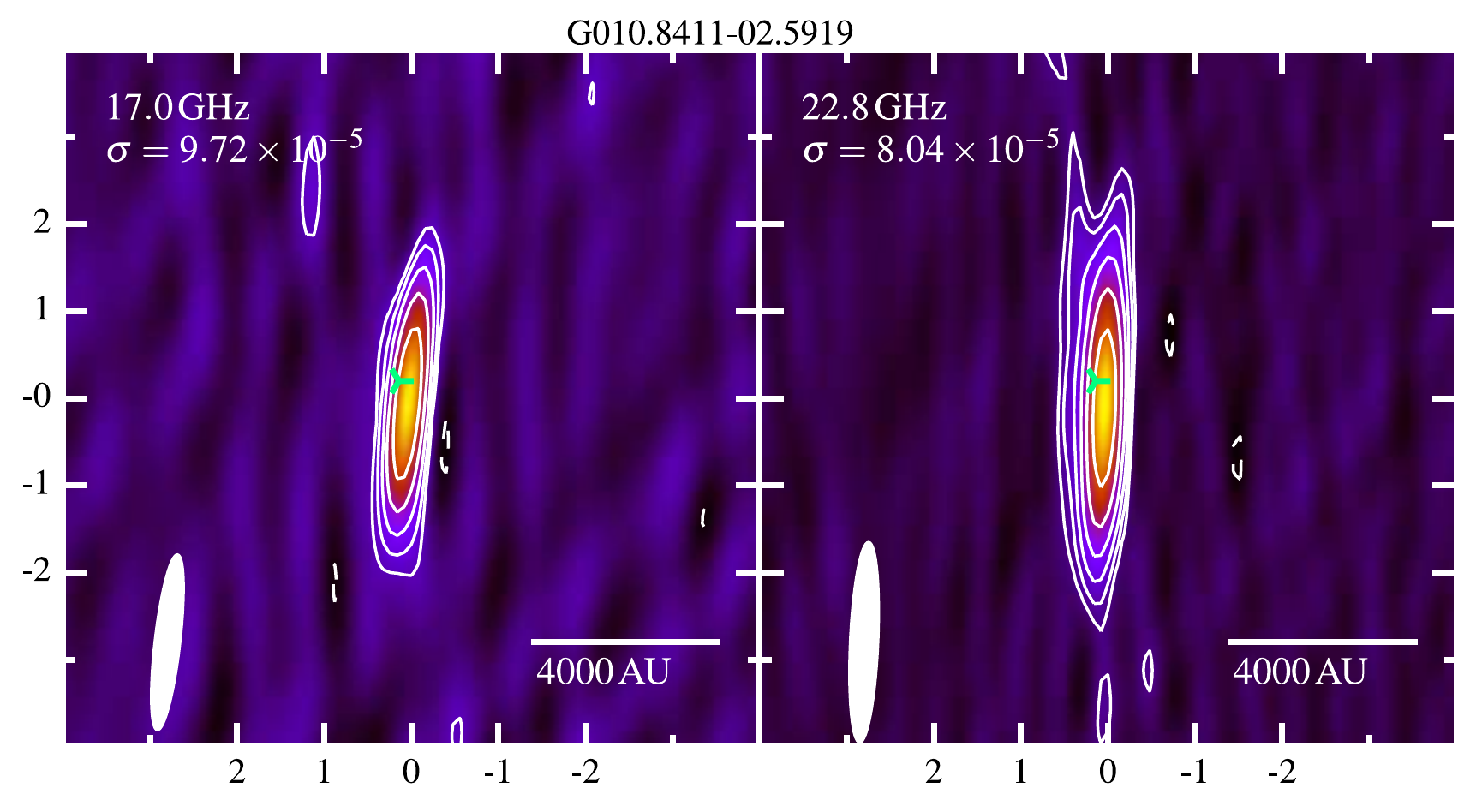}\\
\label{fig:ContourPlots}
\end{figure*}
\begin{figure*}
\includegraphics[width=0.81\textwidth]{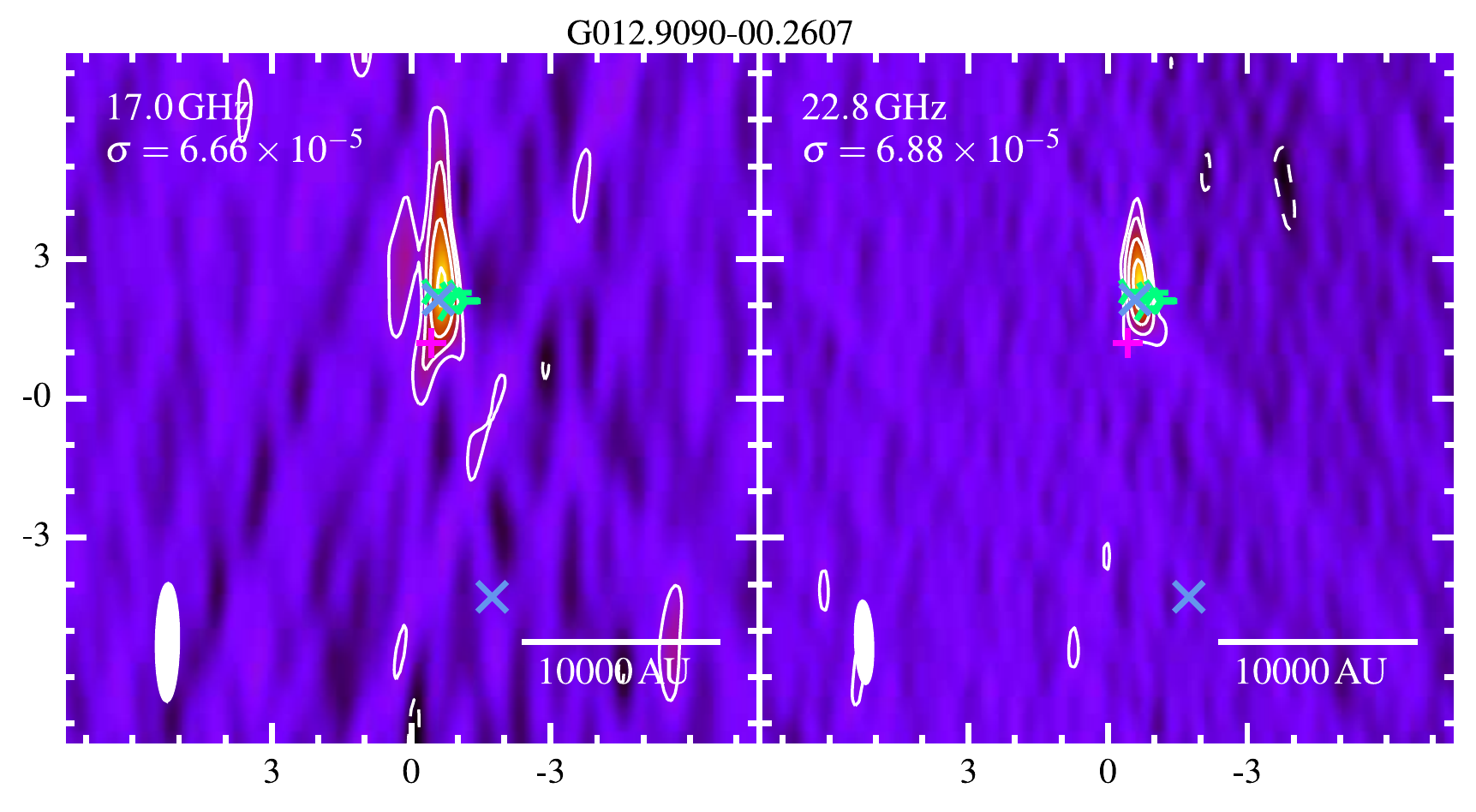}\\
\end{figure*}
\clearpage

\begin{figure*}
\includegraphics[width=0.81\textwidth]{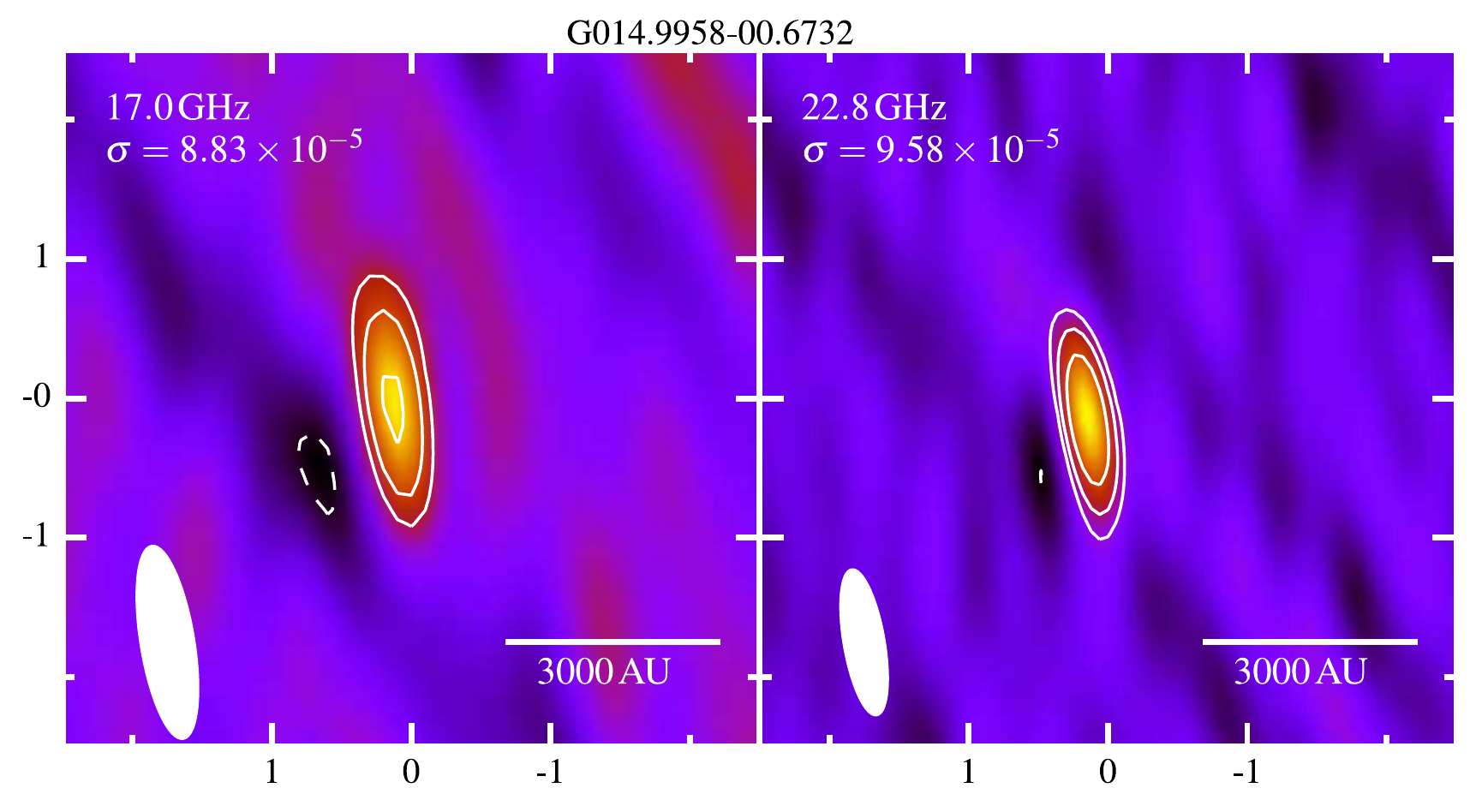}\\
\end{figure*}
\begin{figure*}
\includegraphics[width=0.81\textwidth]{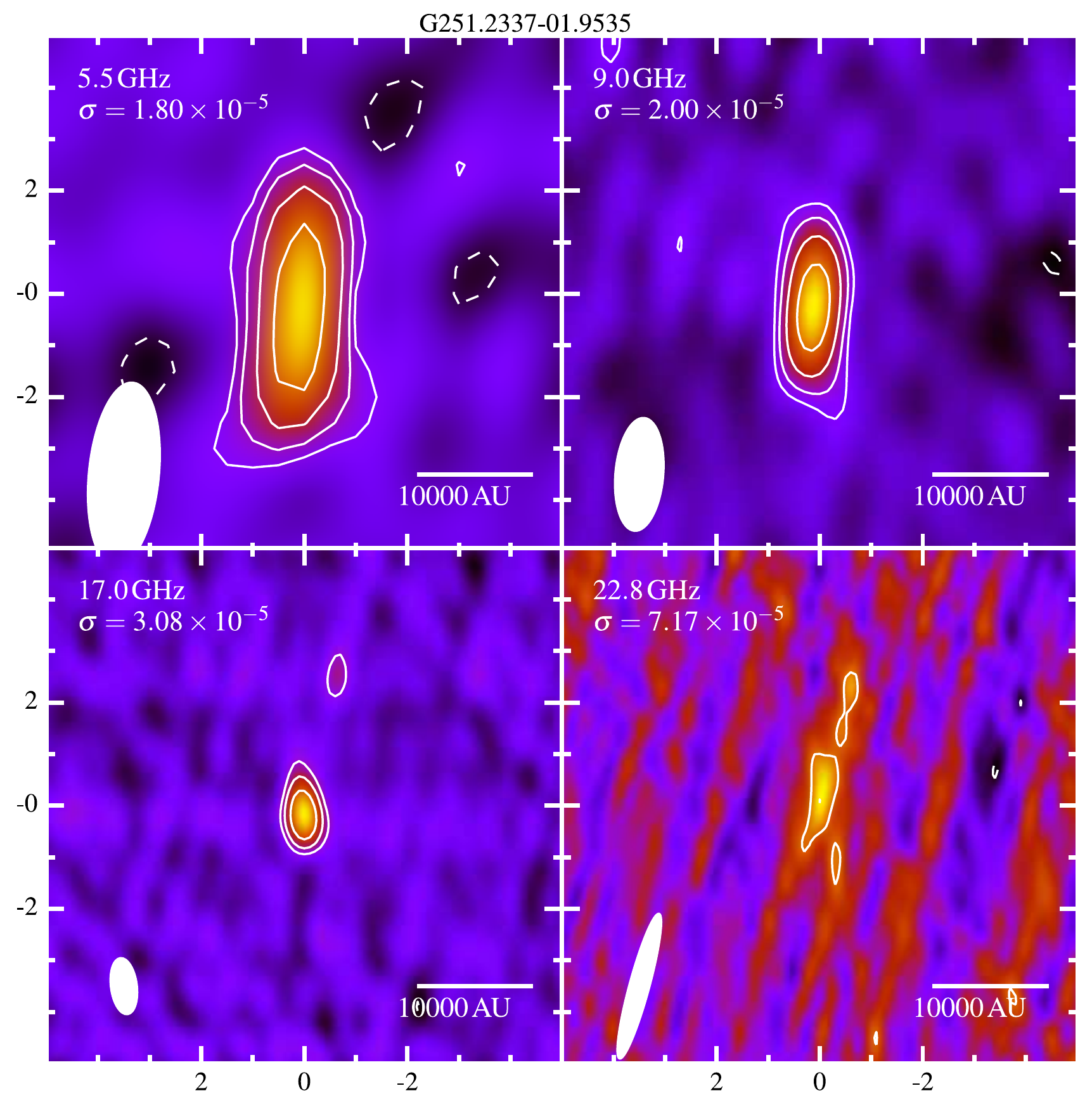}\\
\contcaption{}
\end{figure*}
\clearpage

\begin{figure*}
\includegraphics[width=0.81\textwidth]{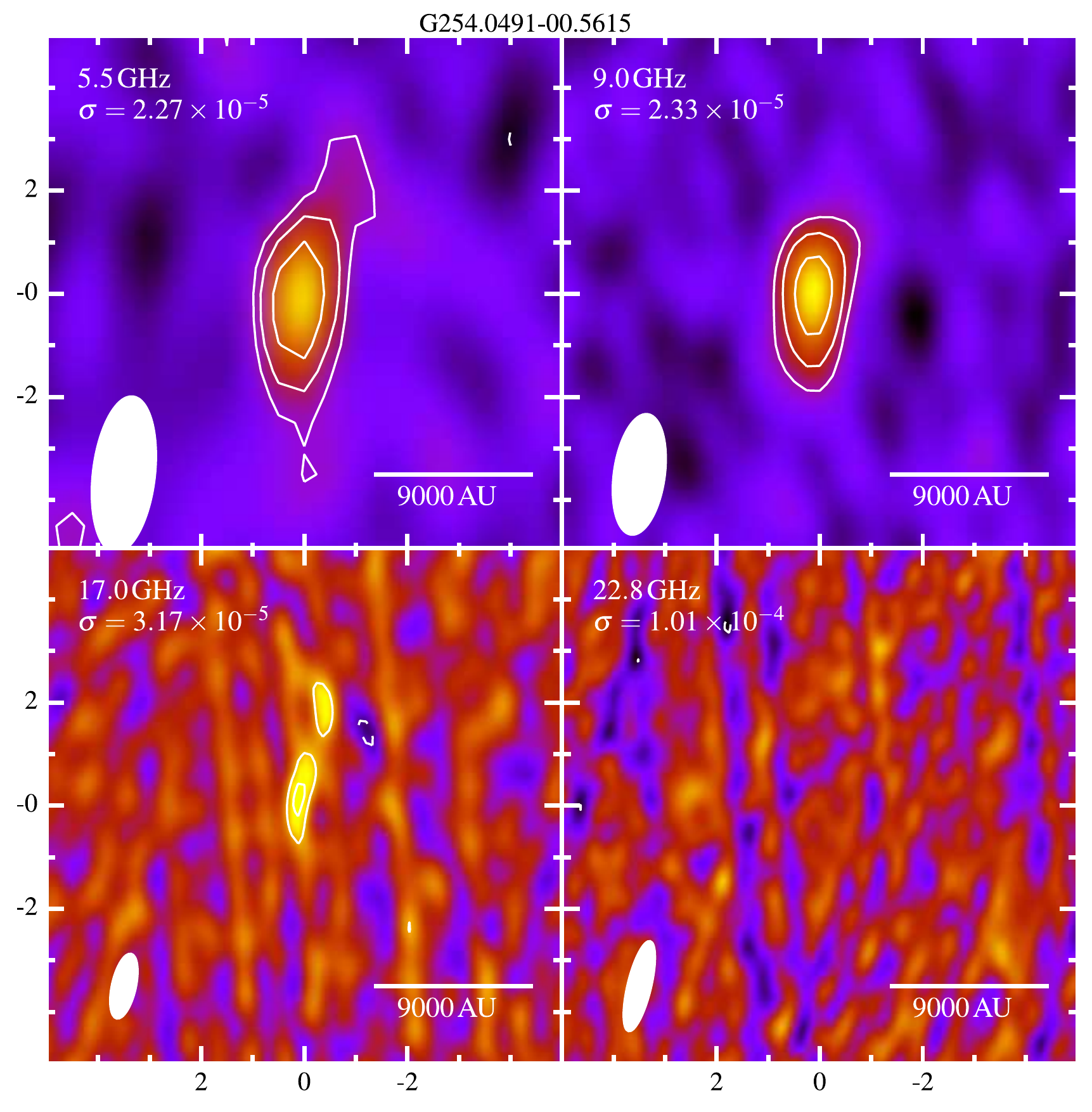}\\
\contcaption{}
\end{figure*}
\clearpage

\begin{figure*}
\includegraphics[width=0.81\textwidth]{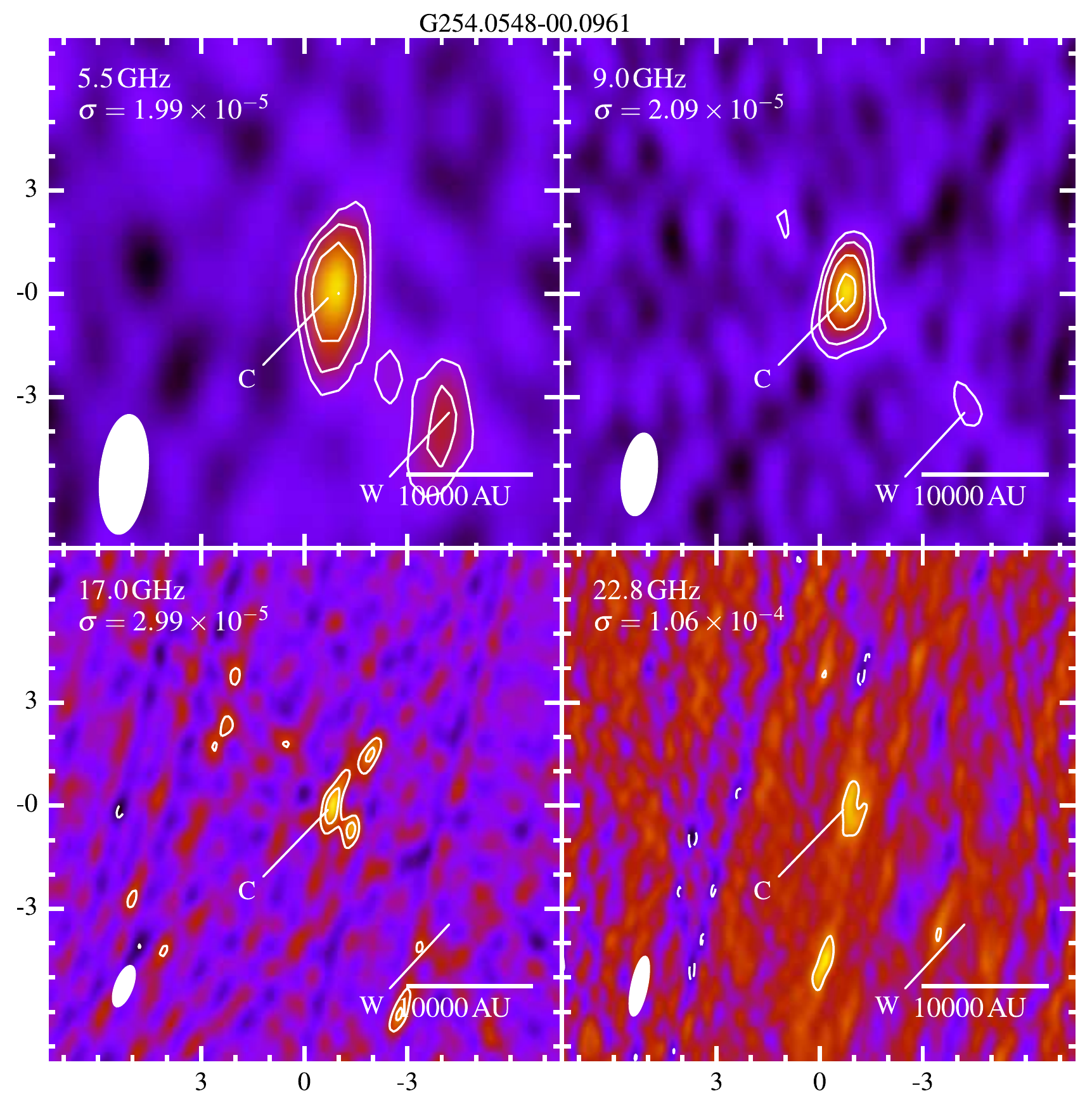}\\
\contcaption{}
\end{figure*}
\clearpage

\begin{figure*}
\includegraphics[width=0.81\textwidth]{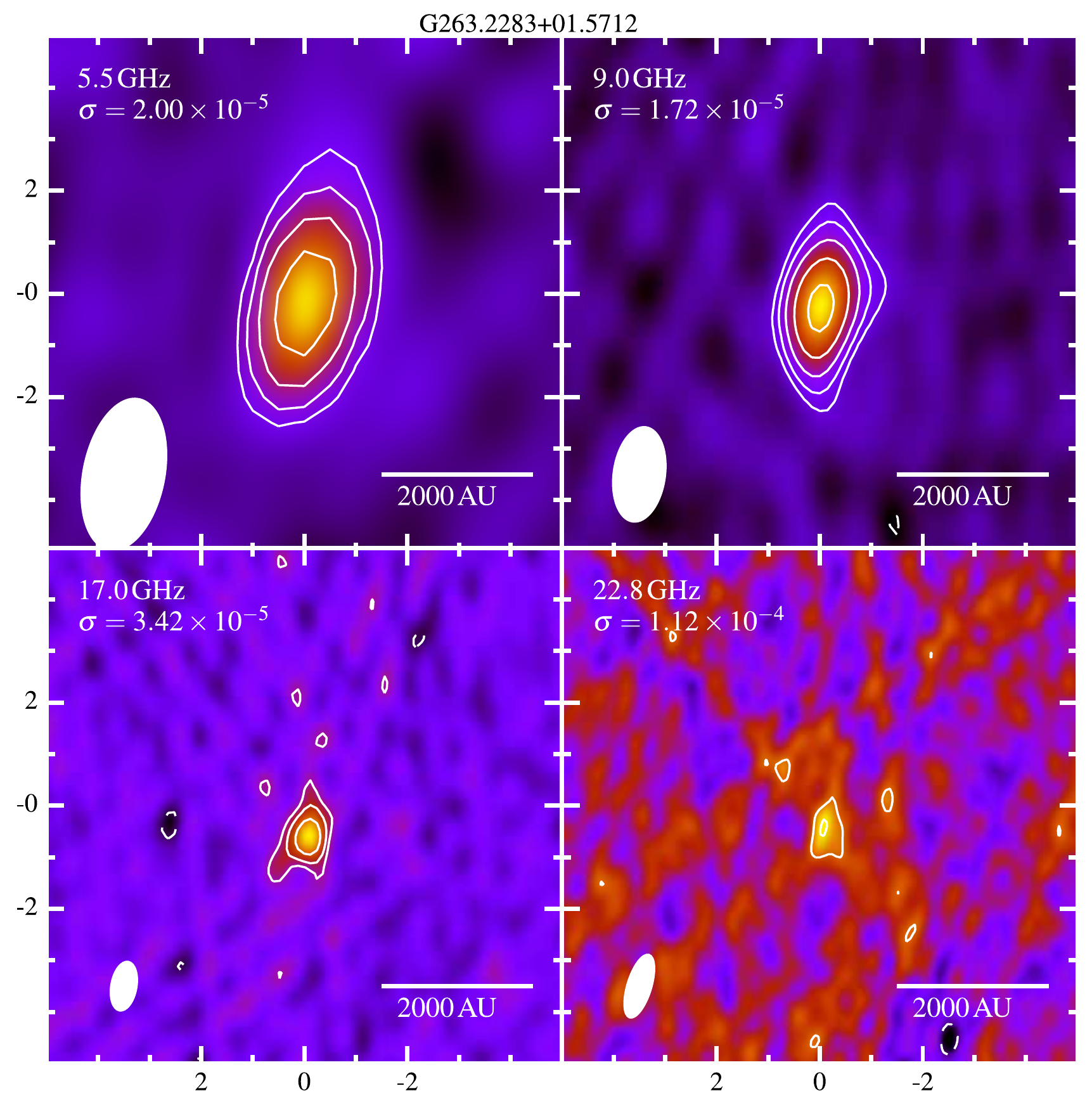}\\
\contcaption{}
\end{figure*}
\clearpage

\begin{figure*}
\includegraphics[width=0.81\textwidth]{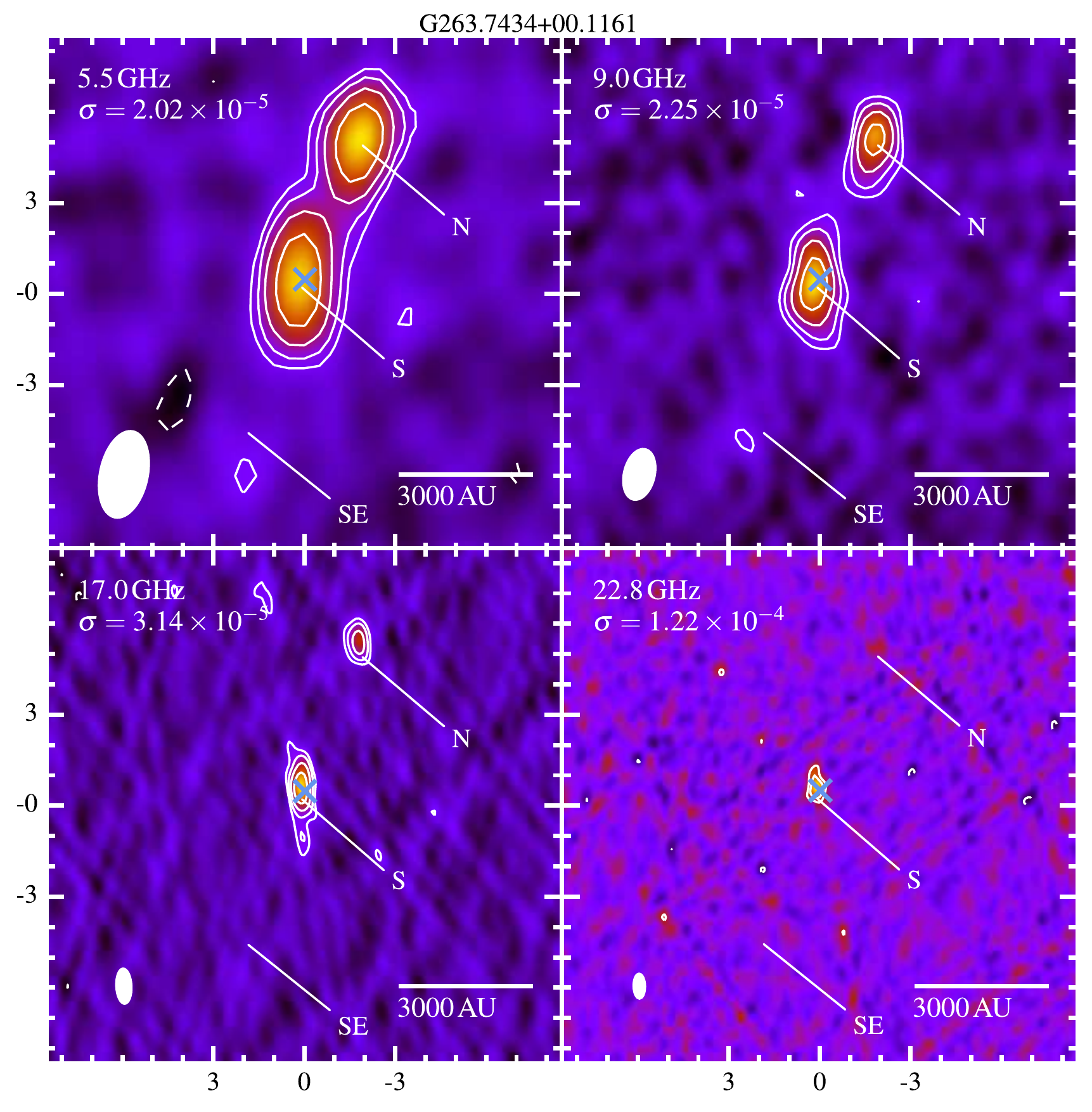}\\
\end{figure*}
\begin{figure*}
\includegraphics[width=0.81\textwidth]{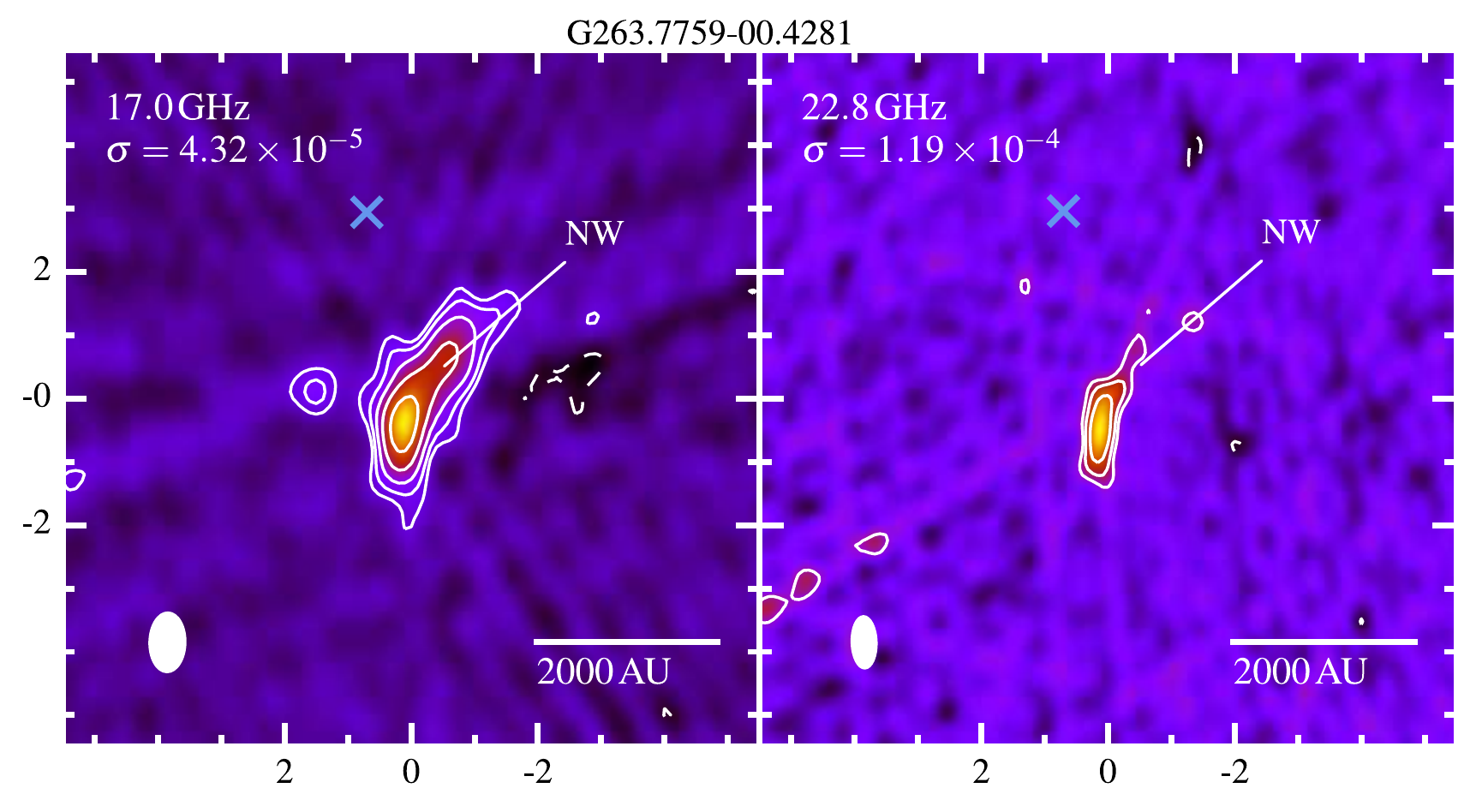}\\
\contcaption{}
\end{figure*}
\clearpage

\begin{figure*}
\includegraphics[width=0.81\textwidth]{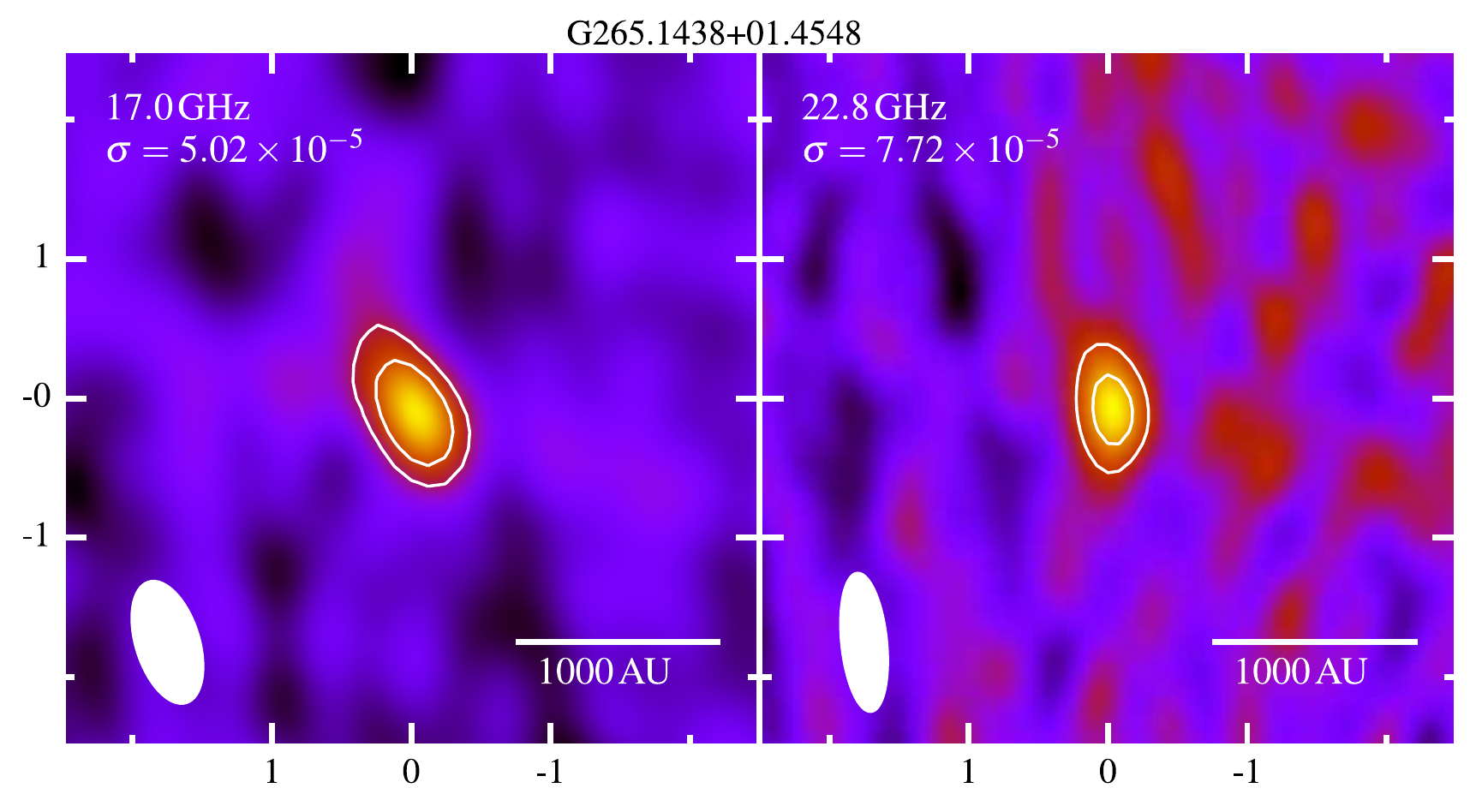}\\
\end{figure*}
\begin{figure*}
\includegraphics[width=0.81\textwidth]{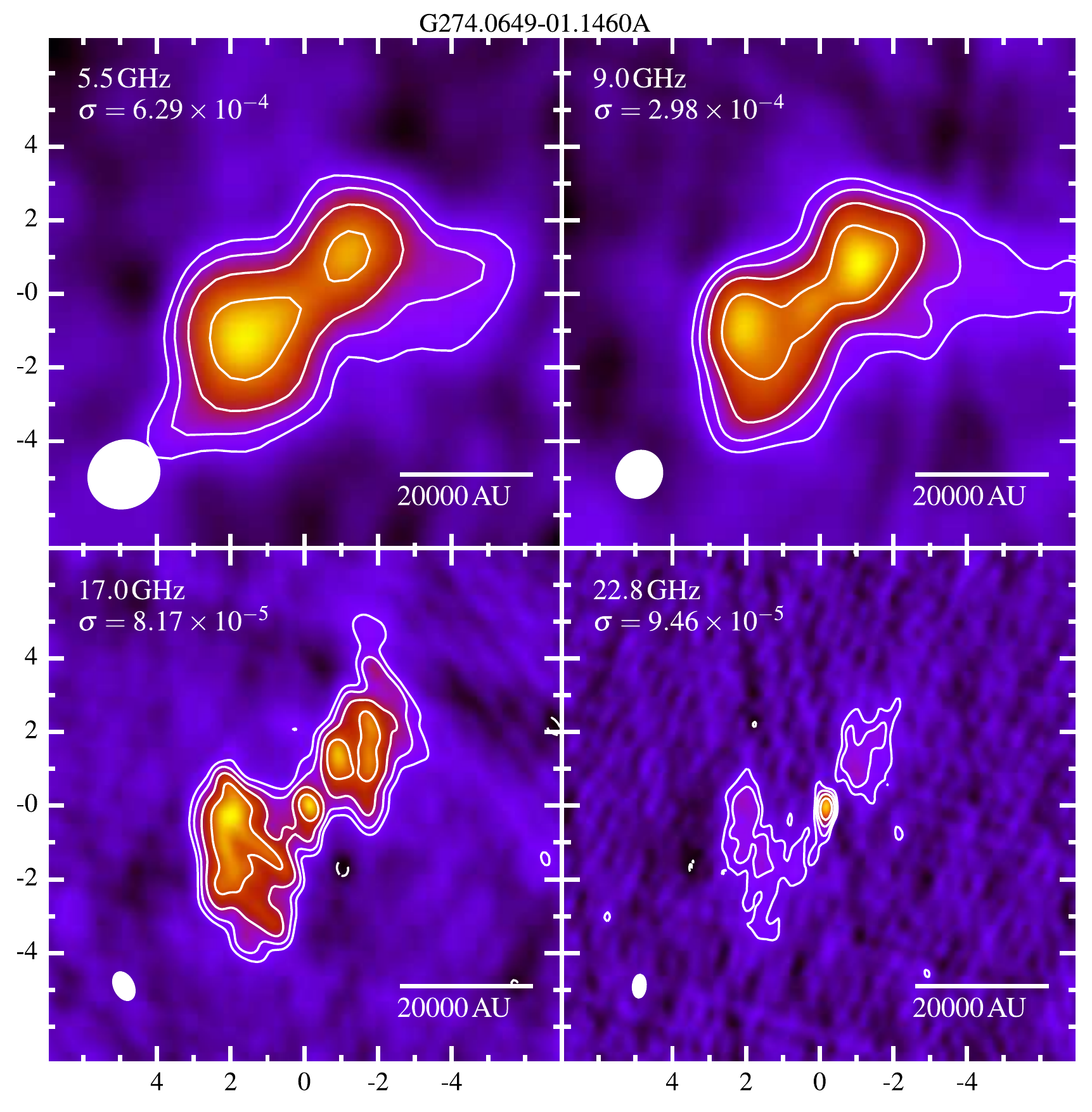}\\
\contcaption{}
\end{figure*}
\clearpage

\begin{figure*}
\includegraphics[width=0.81\textwidth]{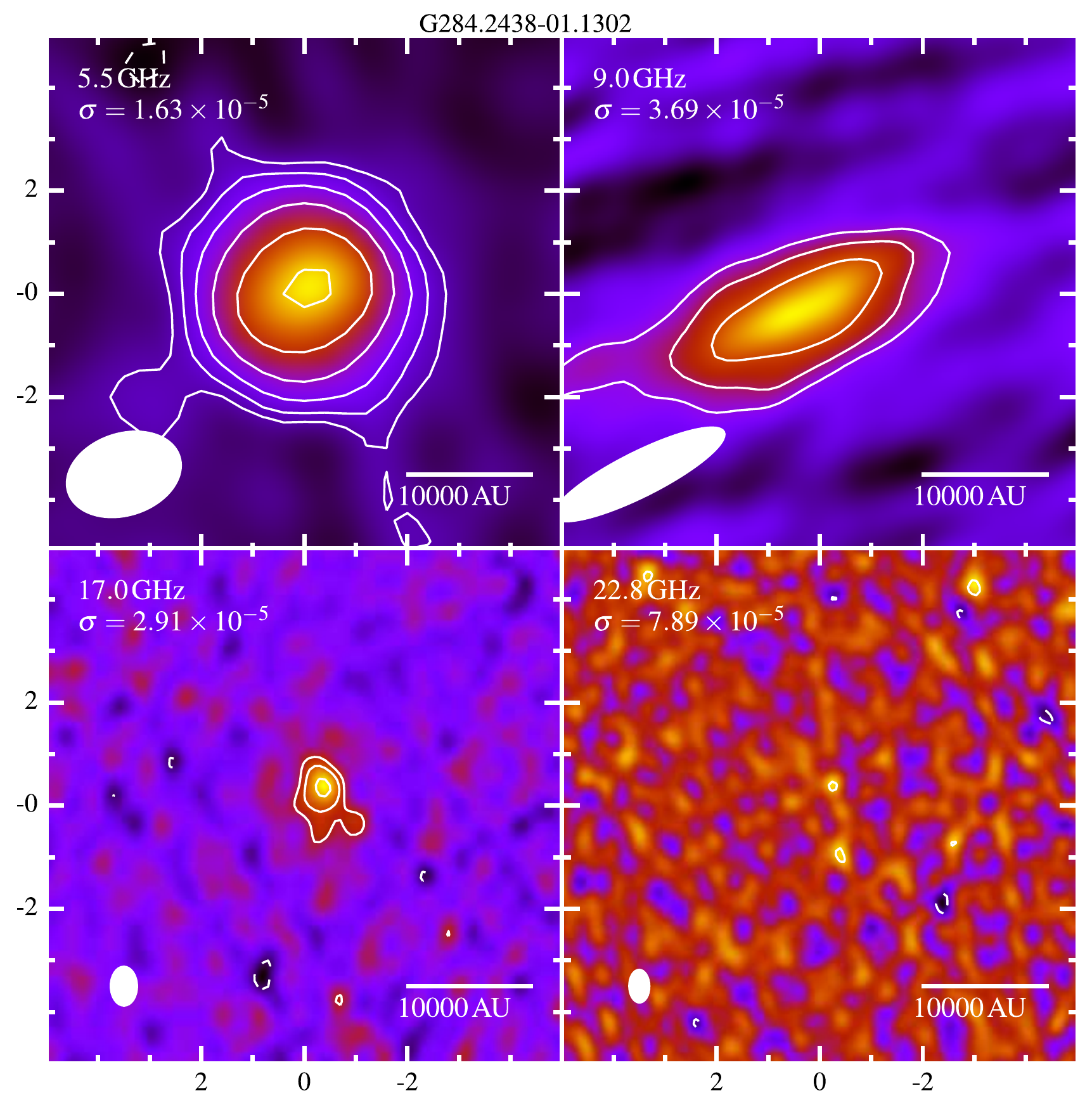}\\
\end{figure*}
\begin{figure*}
\includegraphics[width=0.81\textwidth]{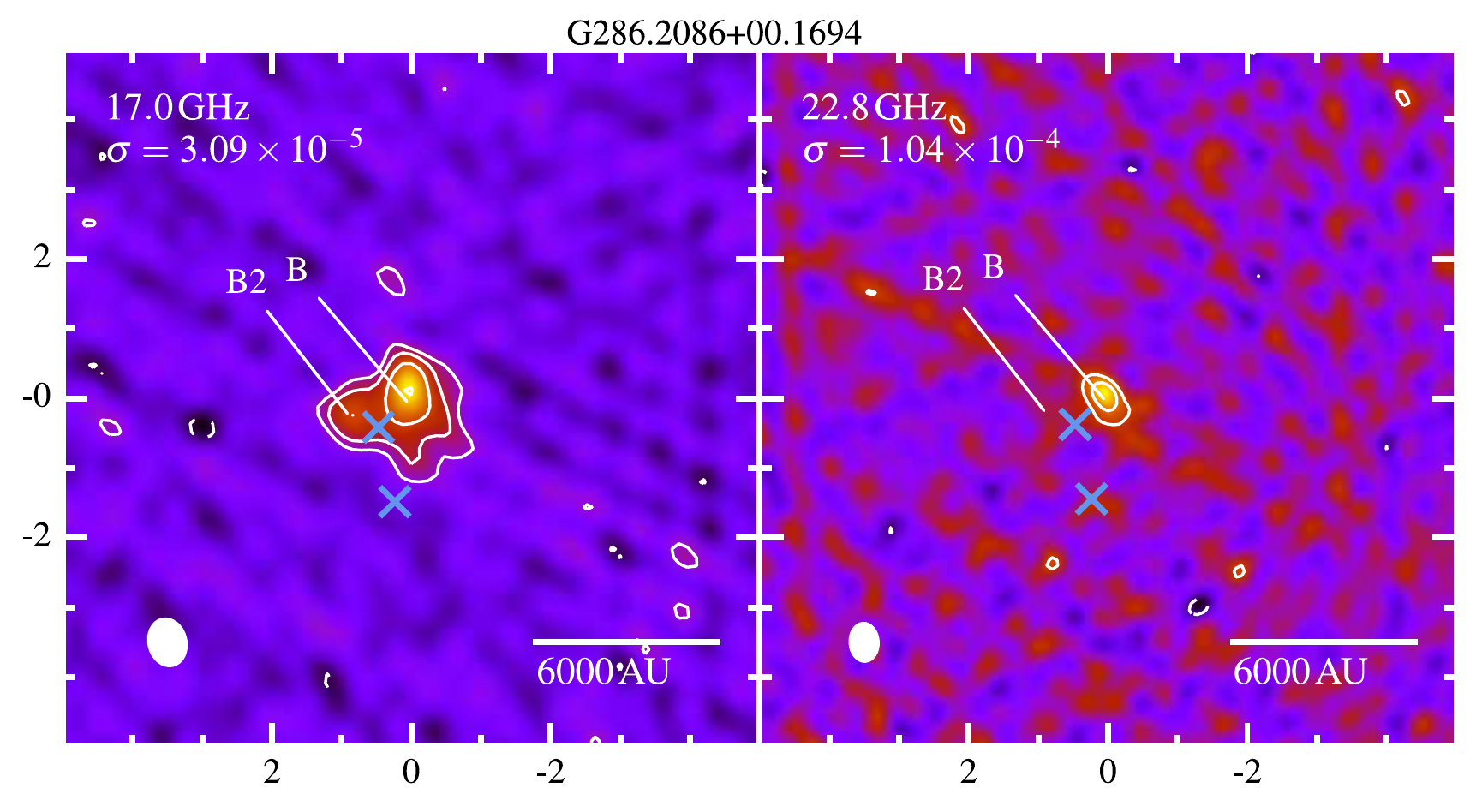}\\
\contcaption{}
\end{figure*}
\clearpage

\begin{figure*}
\includegraphics[width=0.81\textwidth]{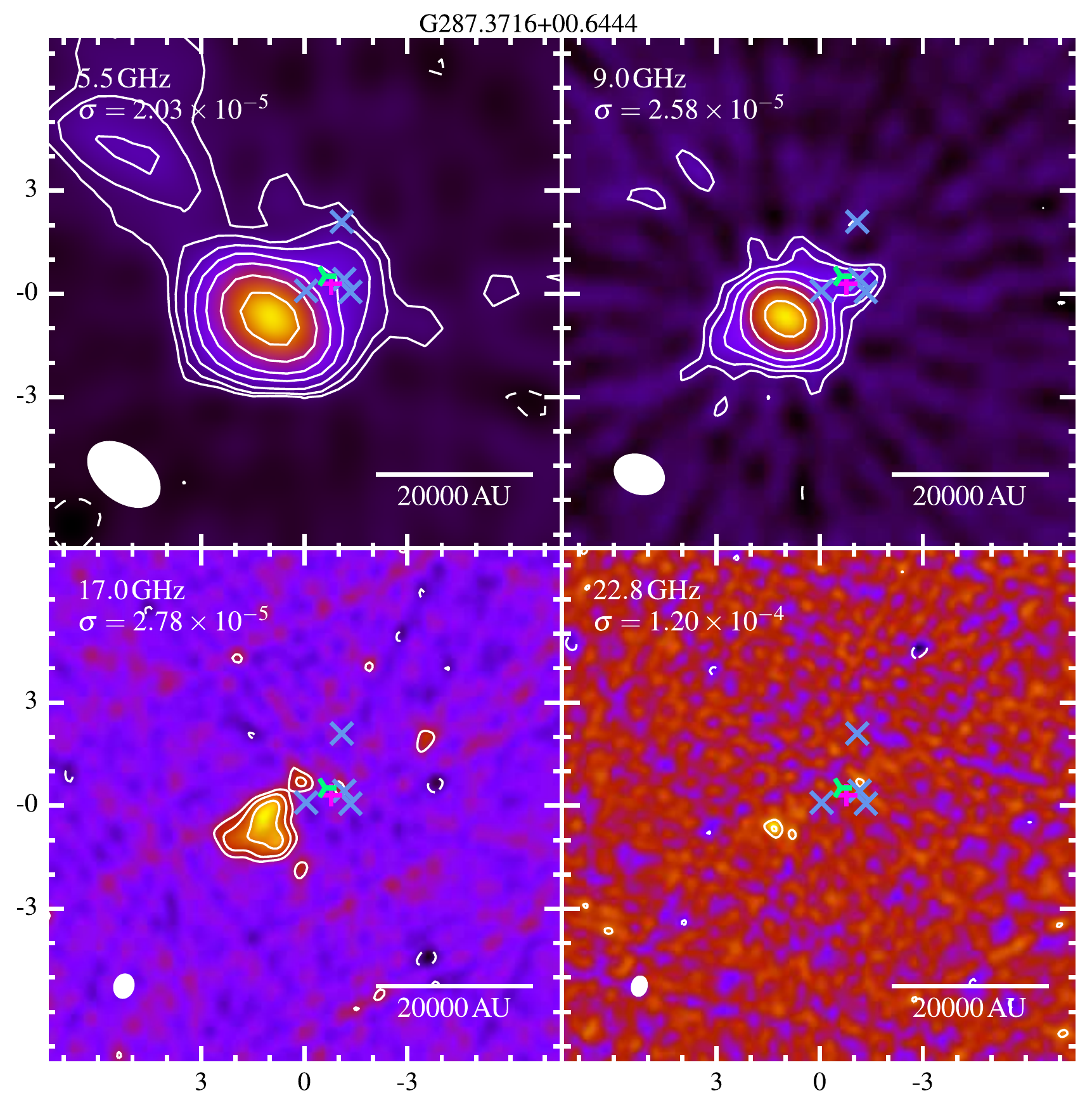}\\
\contcaption{}
\end{figure*}
\clearpage

\begin{figure*}
\includegraphics[width=0.81\textwidth]{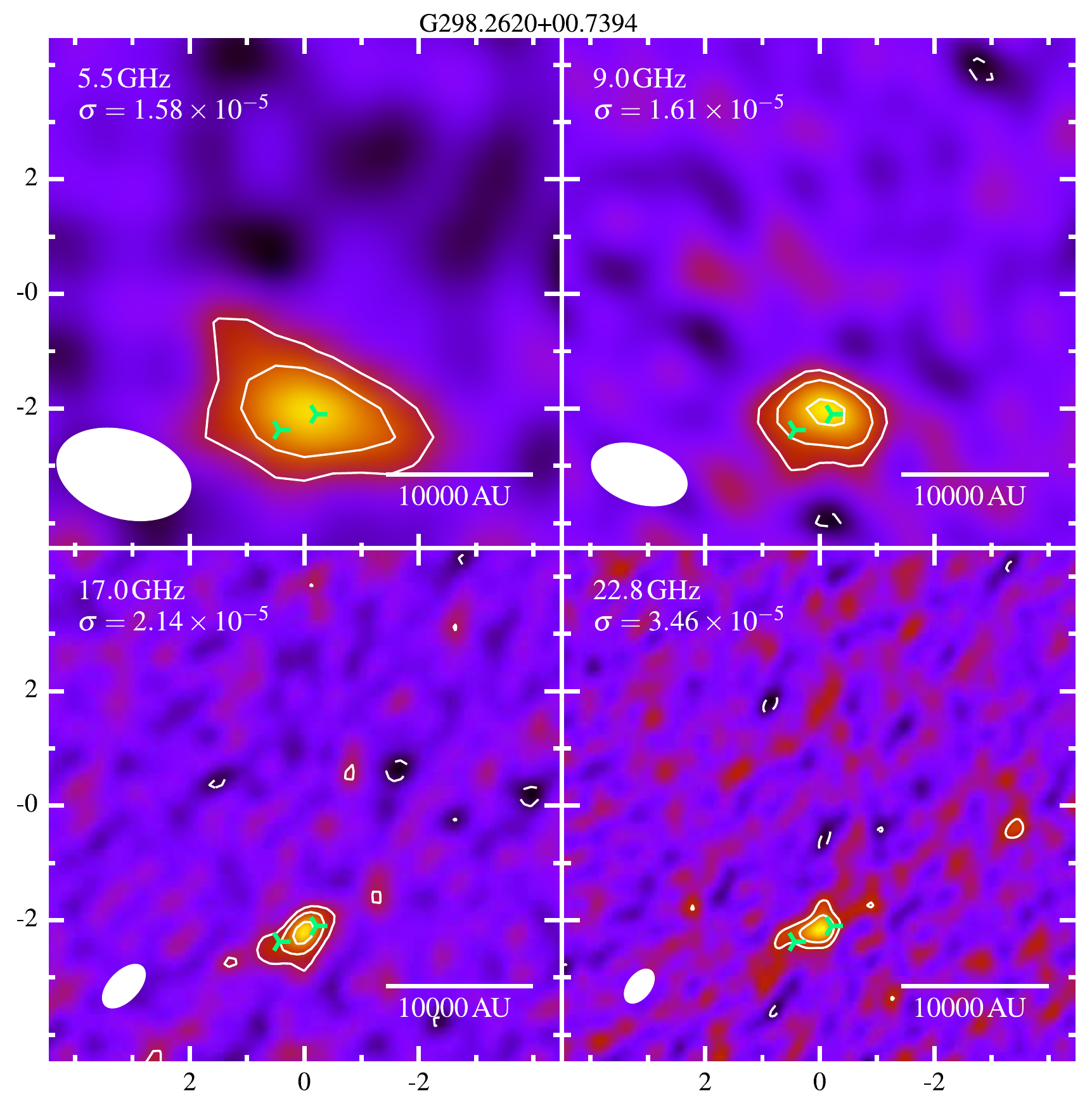}\\
\end{figure*}
\begin{figure*}
\includegraphics[width=0.81\textwidth]{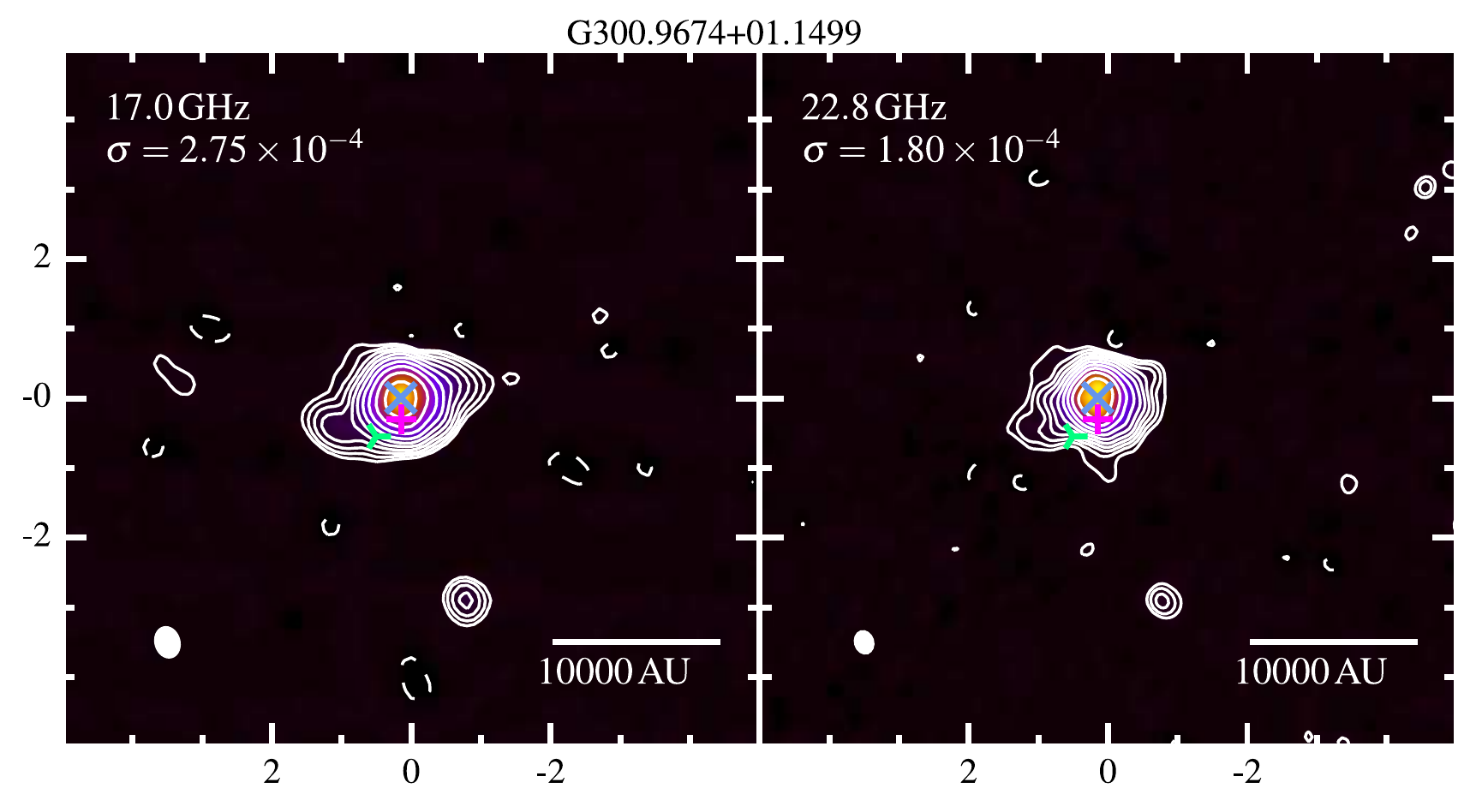}\\
\contcaption{}
\end{figure*}
\clearpage

\begin{figure*}
\includegraphics[width=0.81\textwidth]{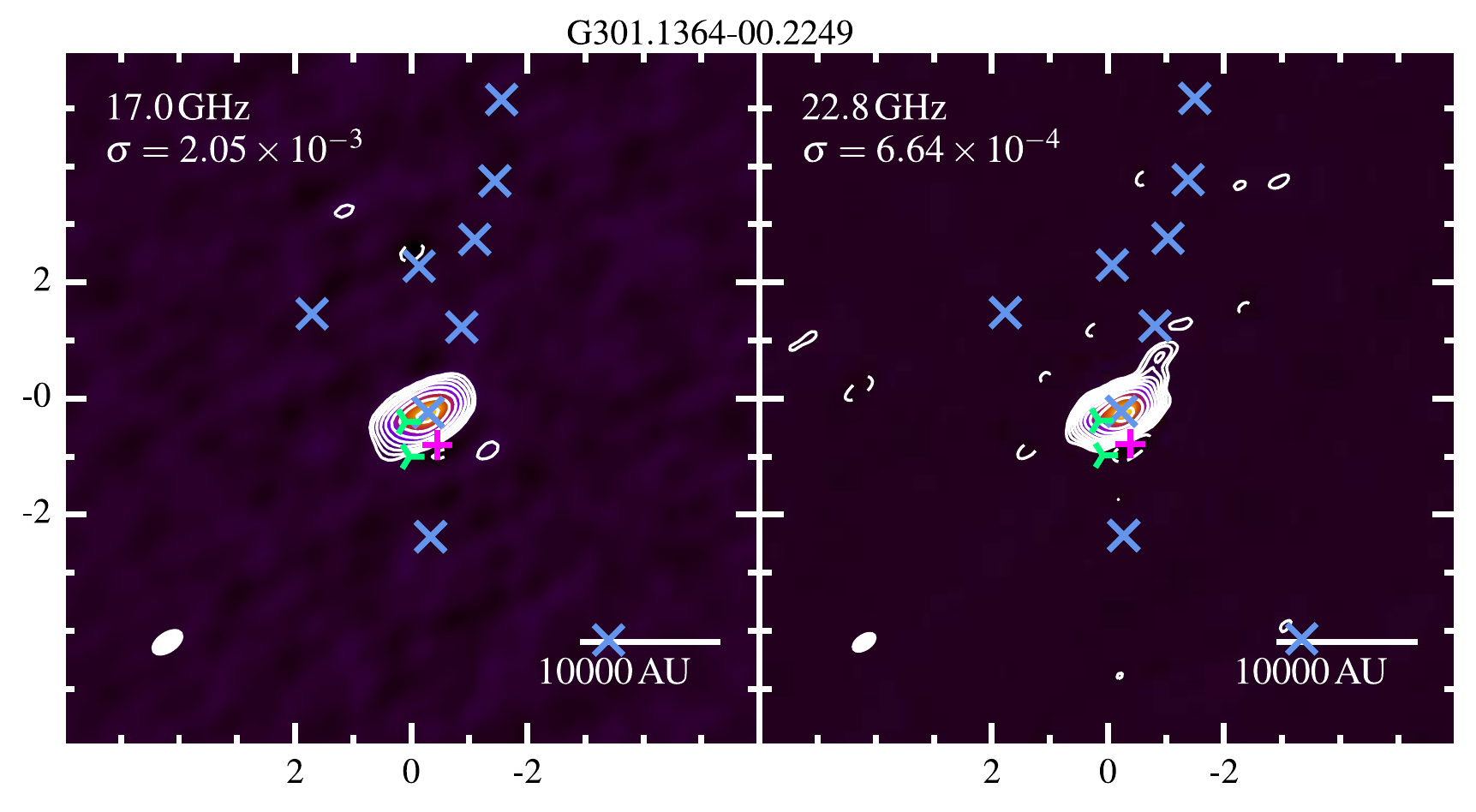}\\
\end{figure*}
\begin{figure*}
\includegraphics[width=0.81\textwidth]{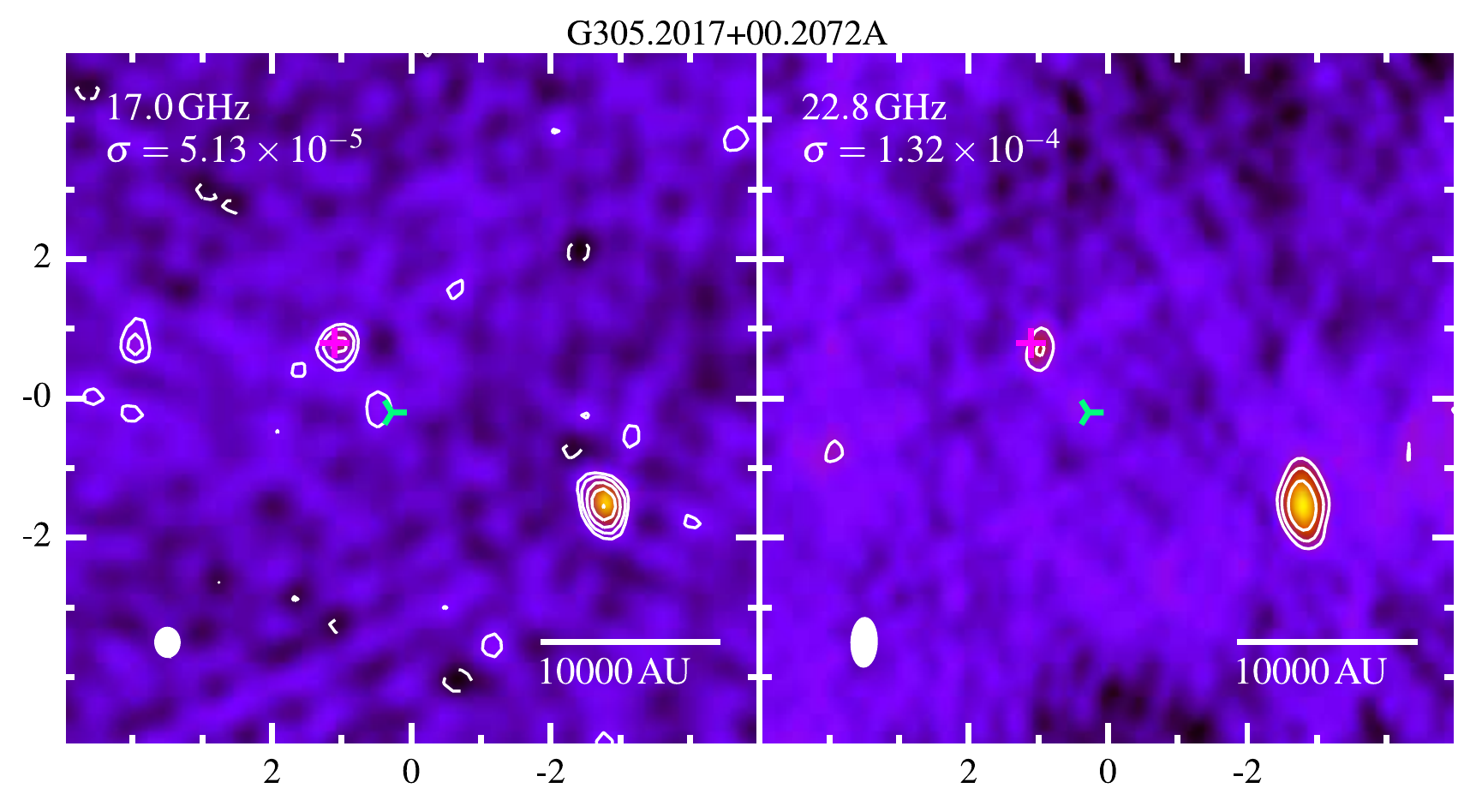}\\
\contcaption{}
\end{figure*}
\clearpage

\begin{figure*}
\includegraphics[width=0.81\textwidth]{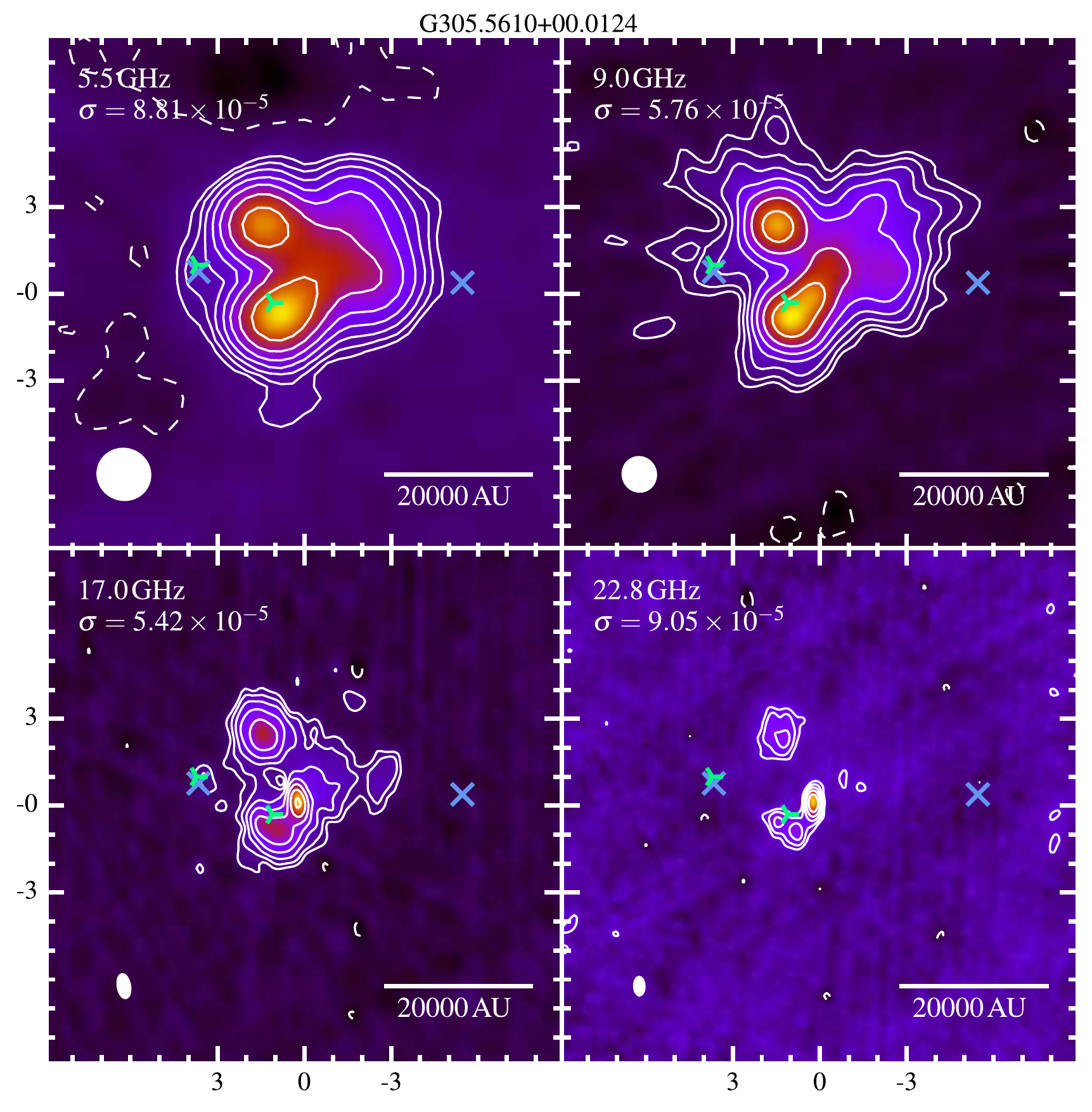}\\
\end{figure*}
\begin{figure*}
\includegraphics[width=0.81\textwidth]{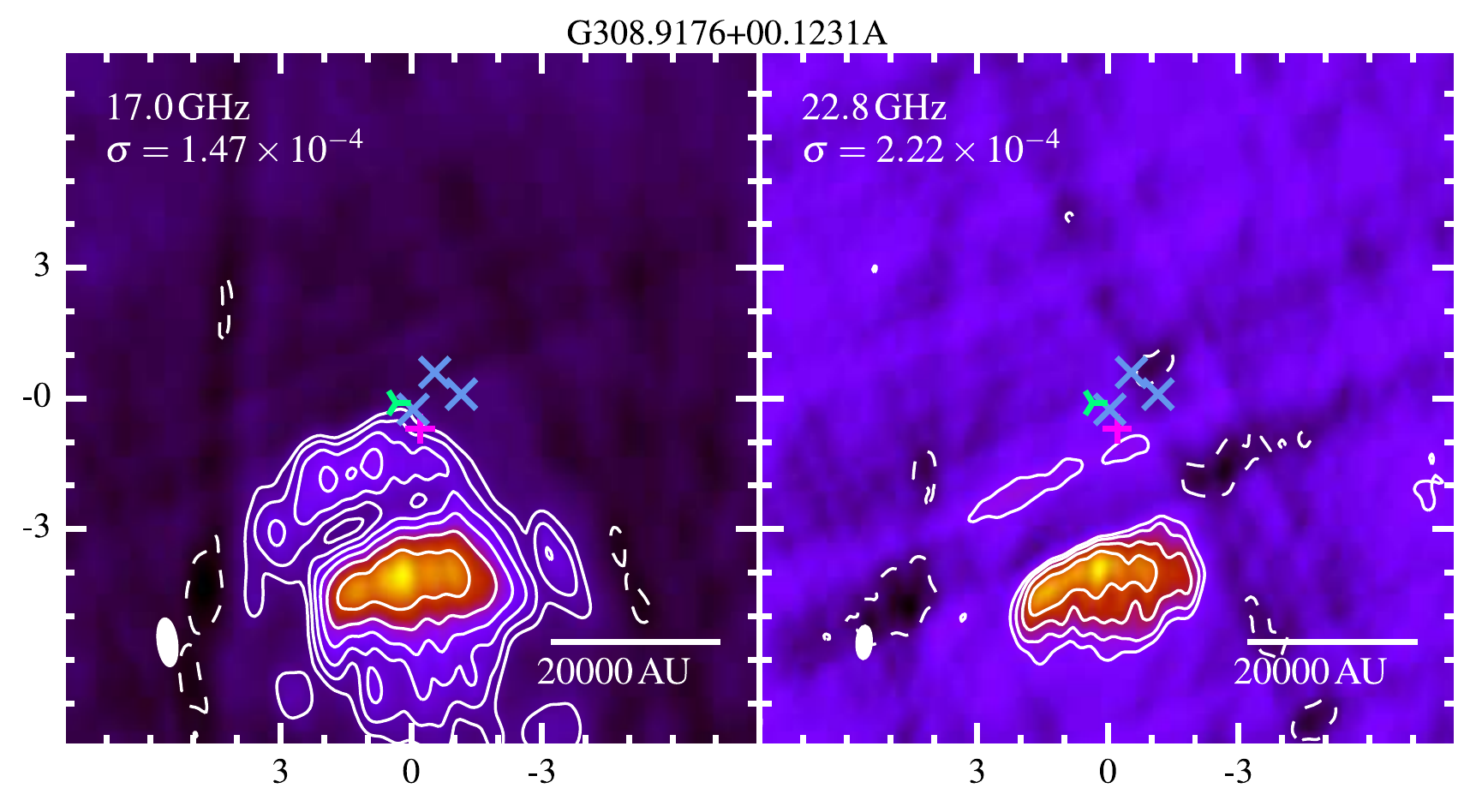}\\
\contcaption{}
\end{figure*}

\clearpage
\begin{figure*}
\includegraphics[width=0.81\textwidth]{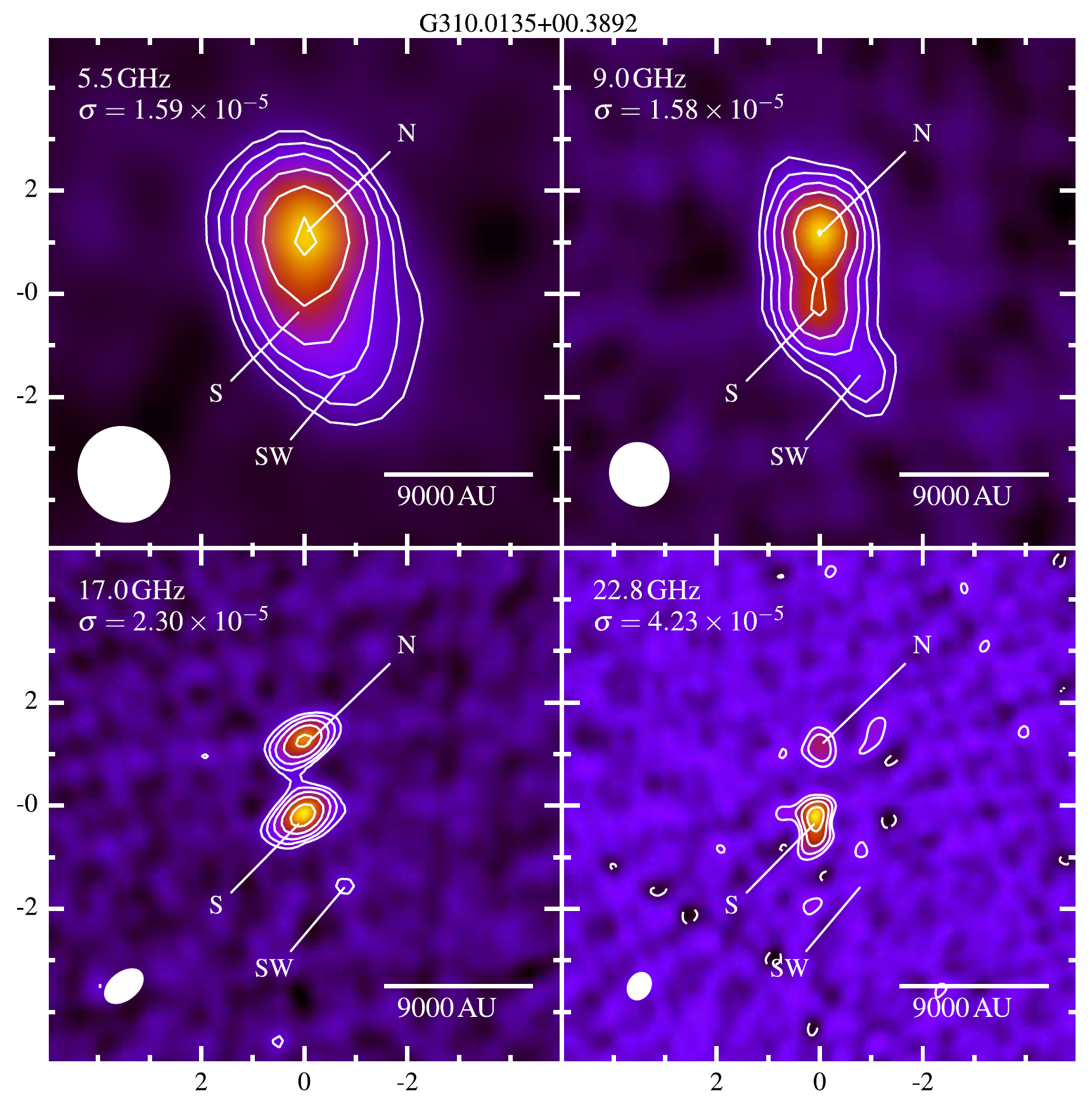}\\
\contcaption{}
\end{figure*}
\clearpage

\begin{figure*}
\includegraphics[width=0.81\textwidth]{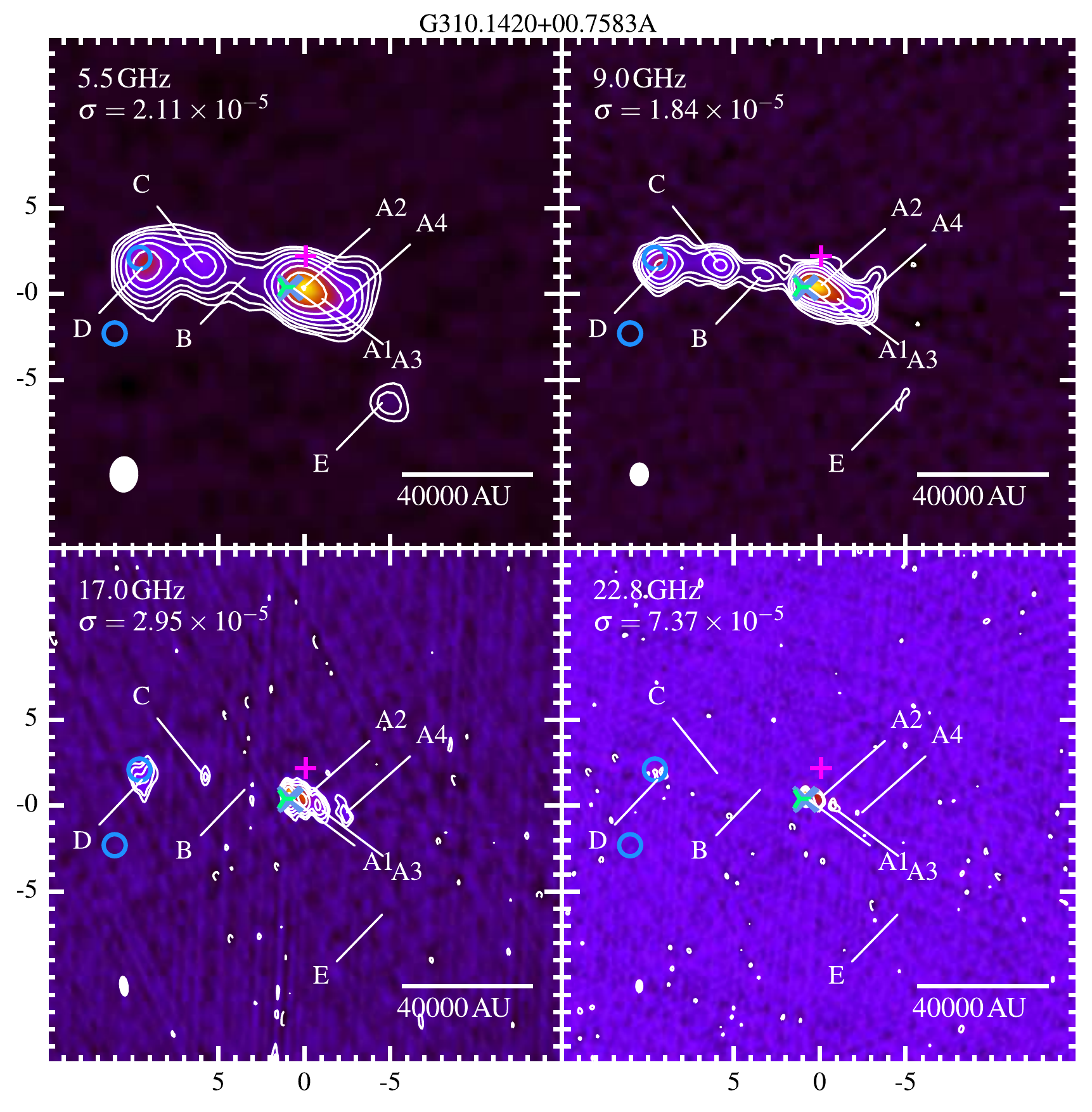}\\
\end{figure*}
\begin{figure*}
\includegraphics[width=0.81\textwidth]{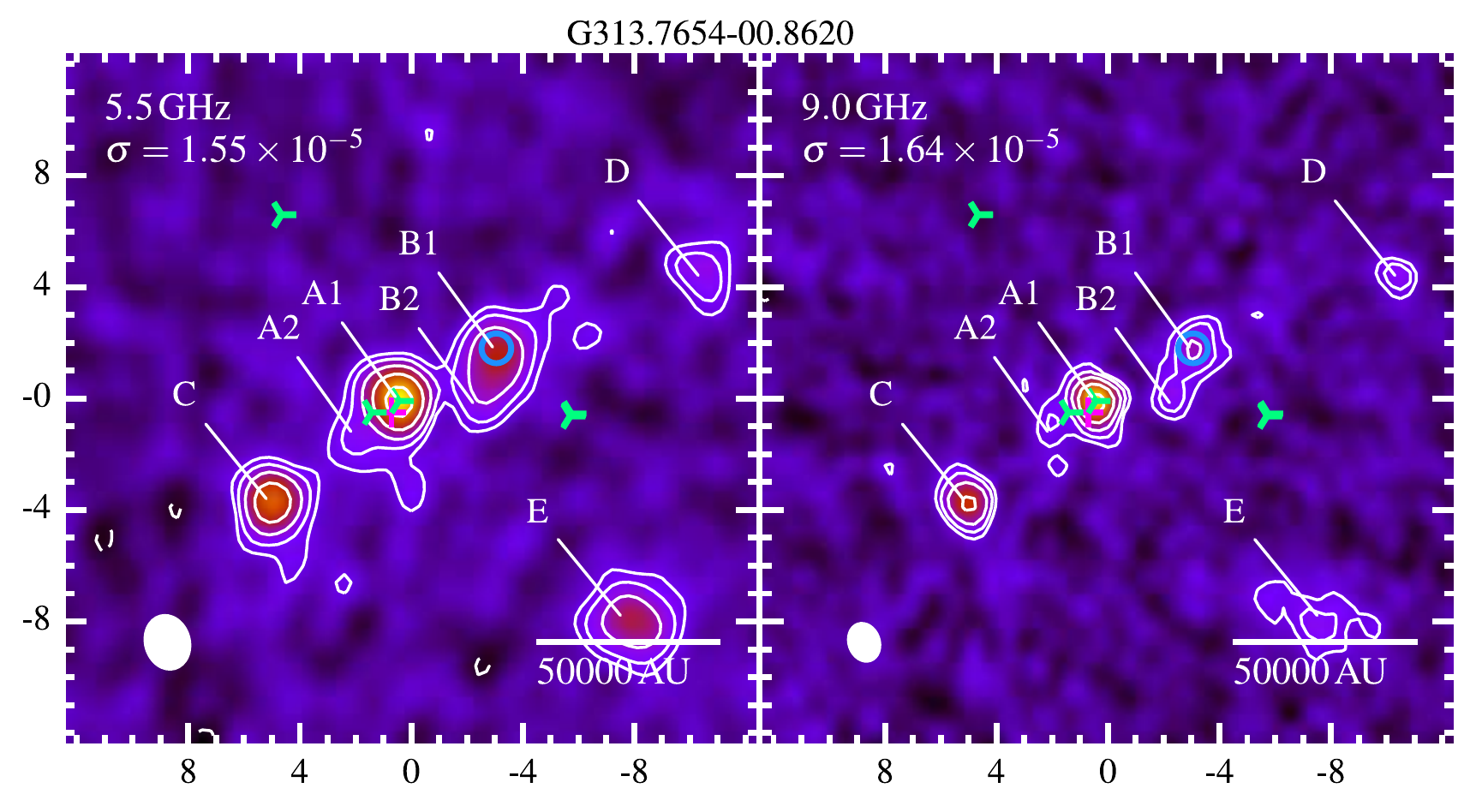}\\
\contcaption{}
\end{figure*}
\clearpage

\begin{figure*}
\includegraphics[width=0.81\textwidth]{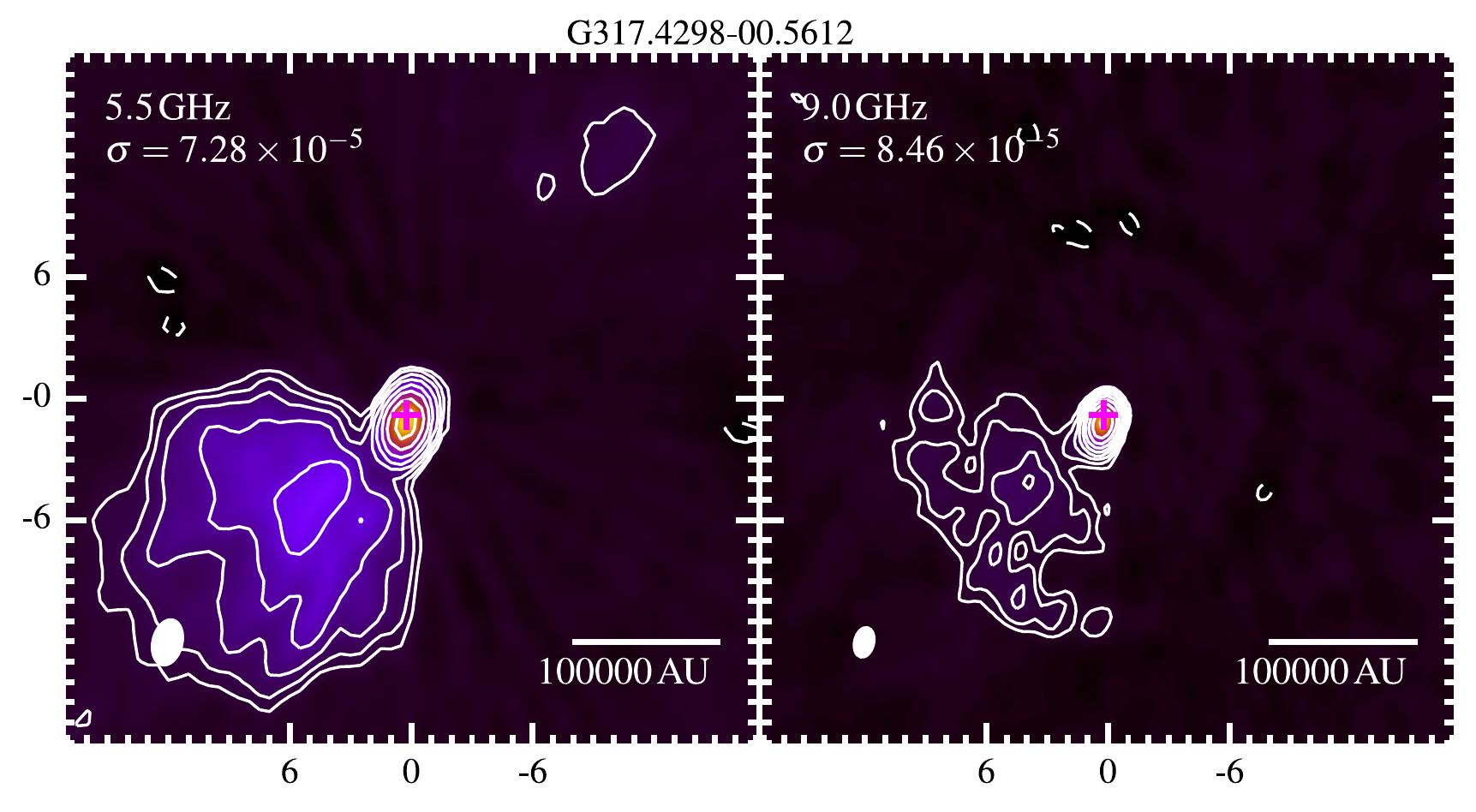}\\
\end{figure*}
\begin{figure*}
\includegraphics[width=0.81\textwidth]{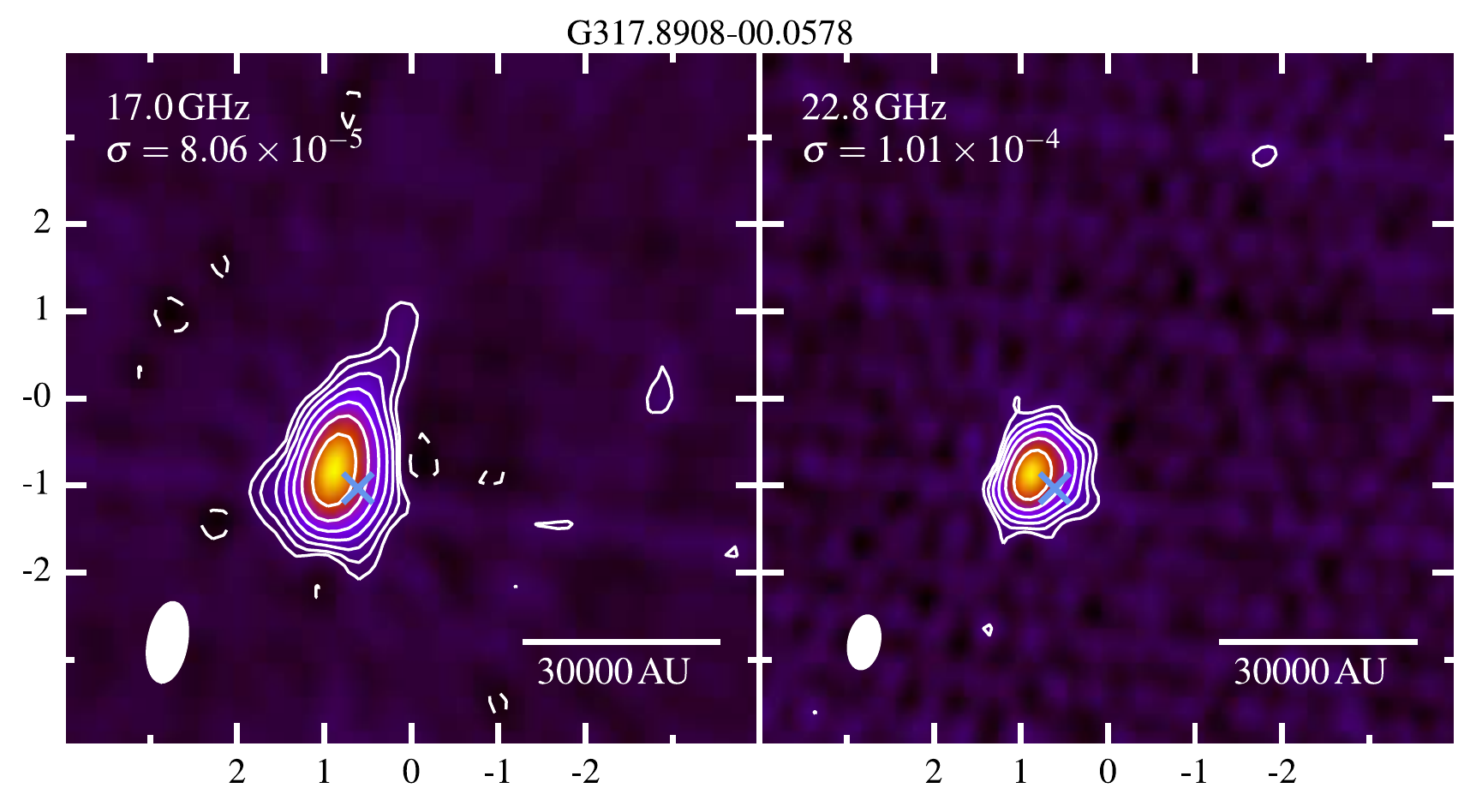}\\
\contcaption{}
\end{figure*}
\clearpage

\begin{figure*}
\includegraphics[width=0.81\textwidth]{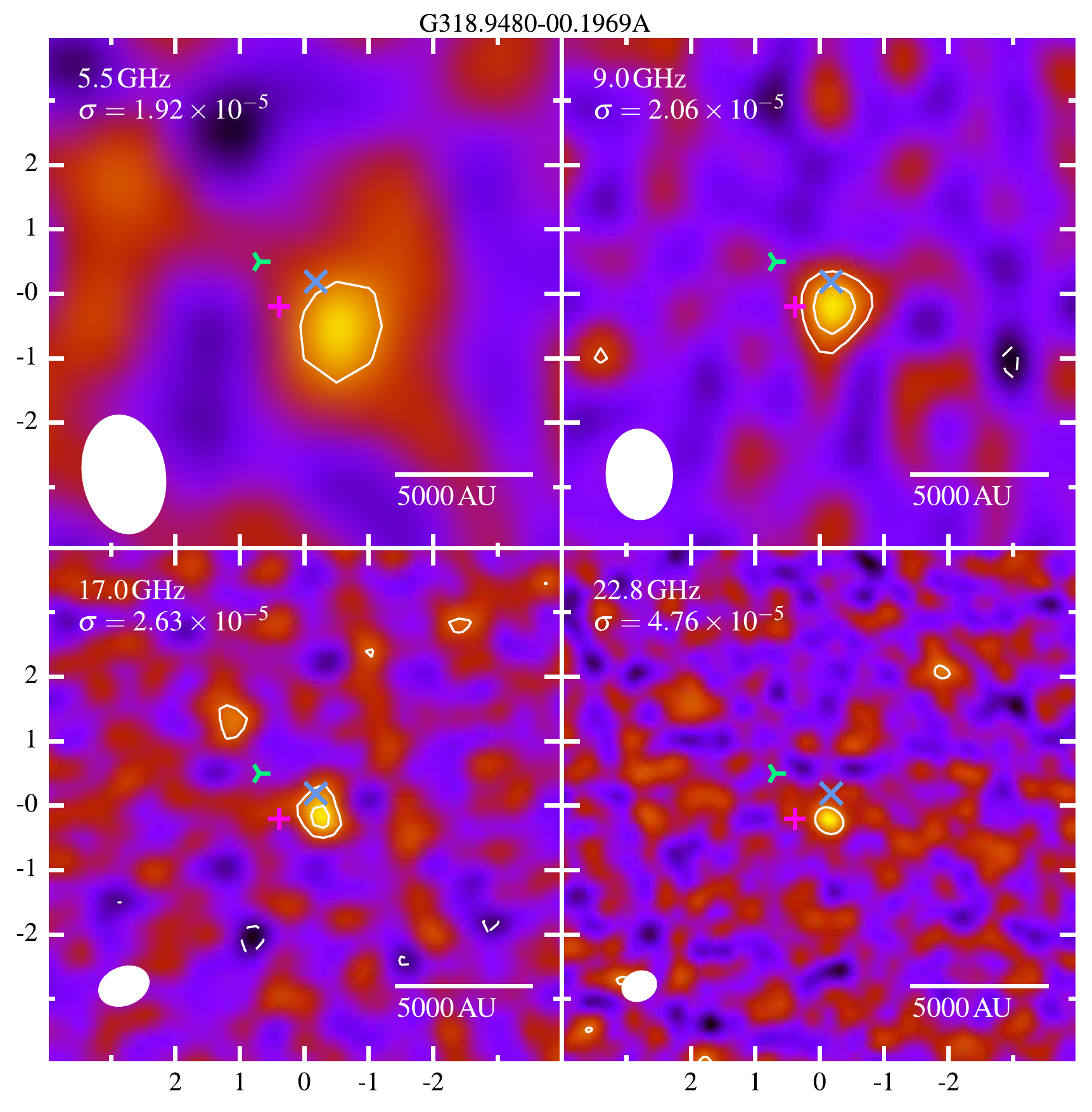}\\
\contcaption{}
\end{figure*}
\clearpage

\begin{figure*}
\includegraphics[width=0.81\textwidth]{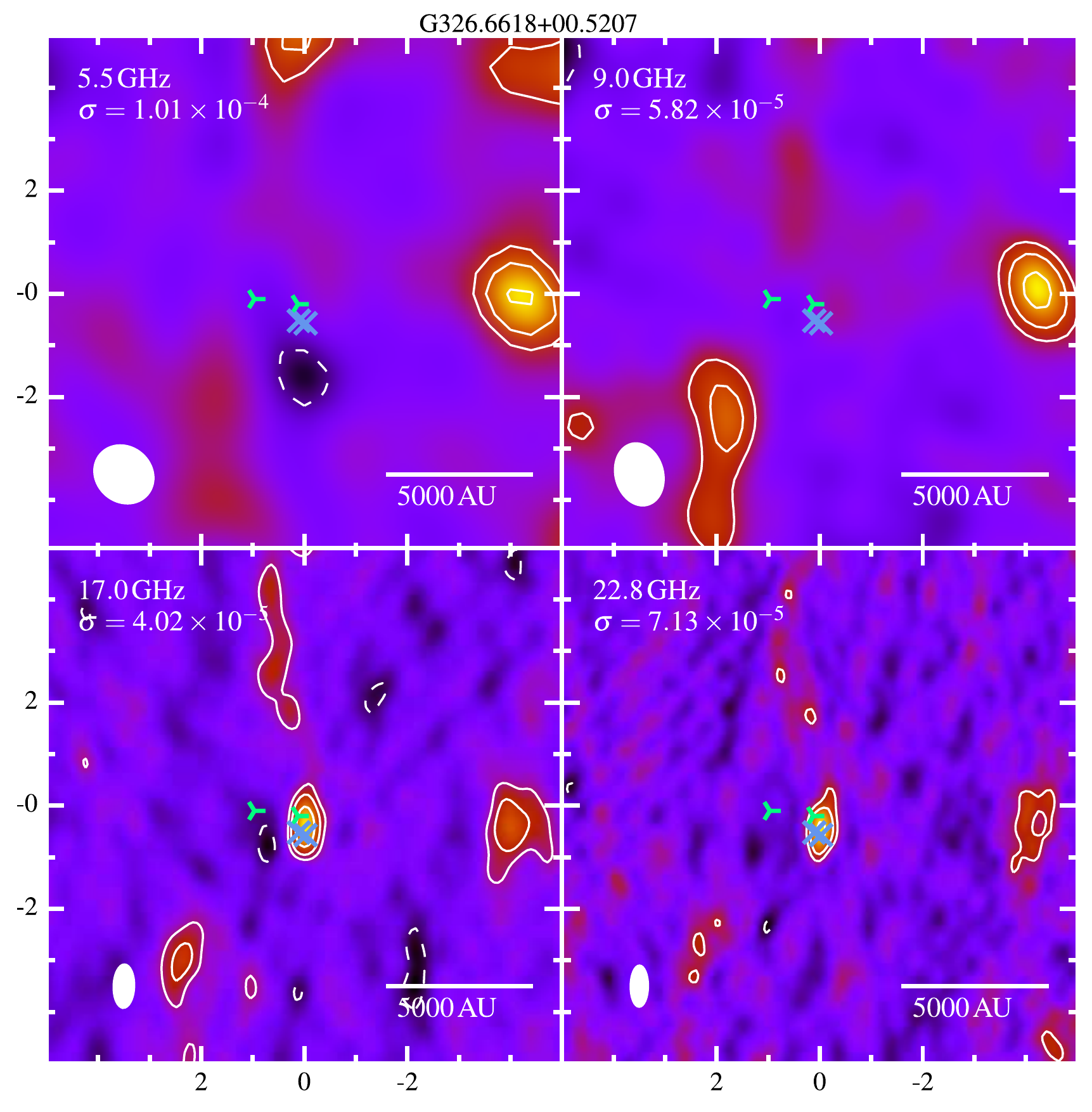}\\
\contcaption{}
\end{figure*}
\clearpage

\begin{figure*}
\includegraphics[width=0.81\textwidth]{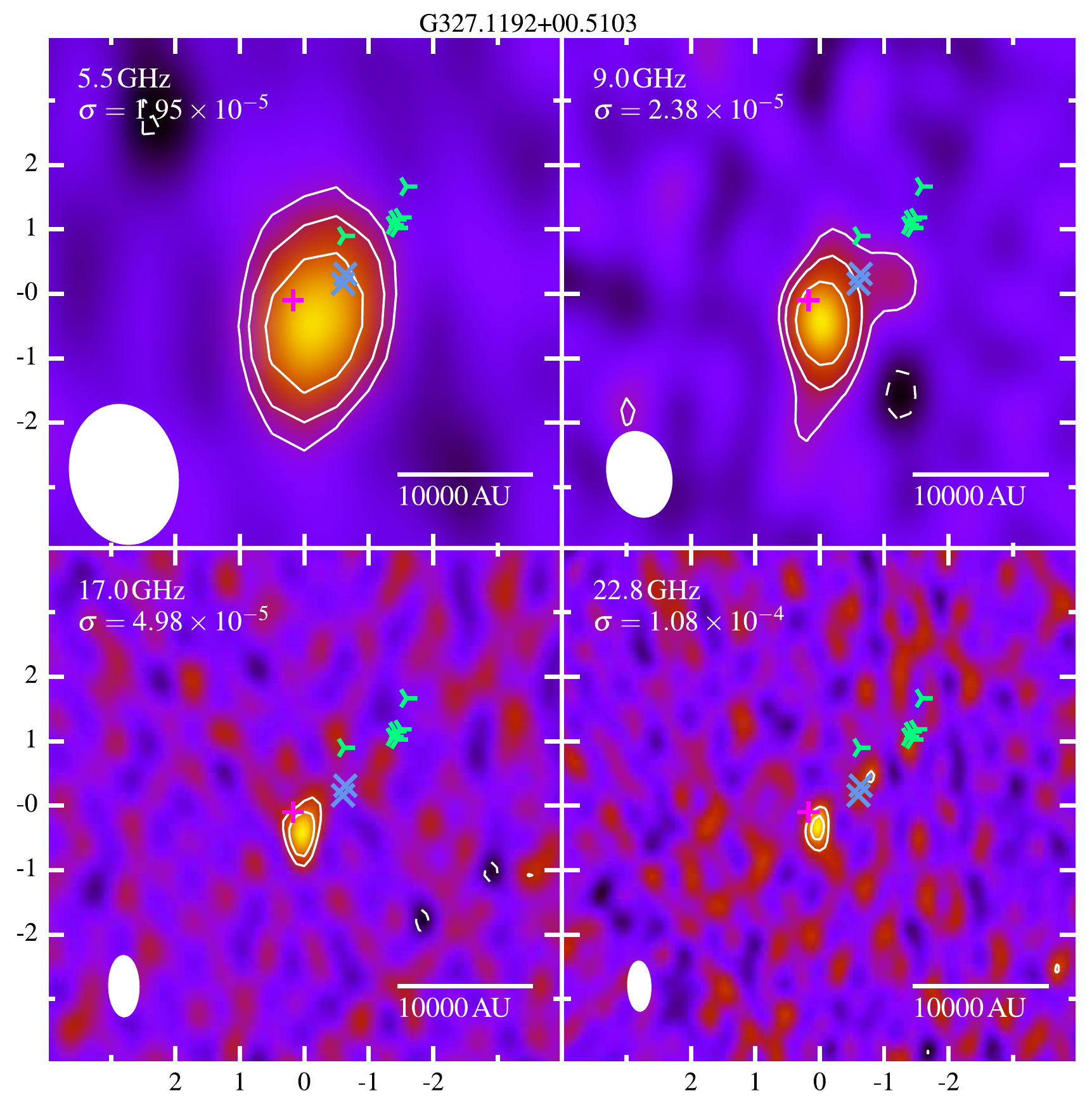}\\
\contcaption{}
\end{figure*}
\clearpage

\begin{figure*}
\includegraphics[width=0.81\textwidth]{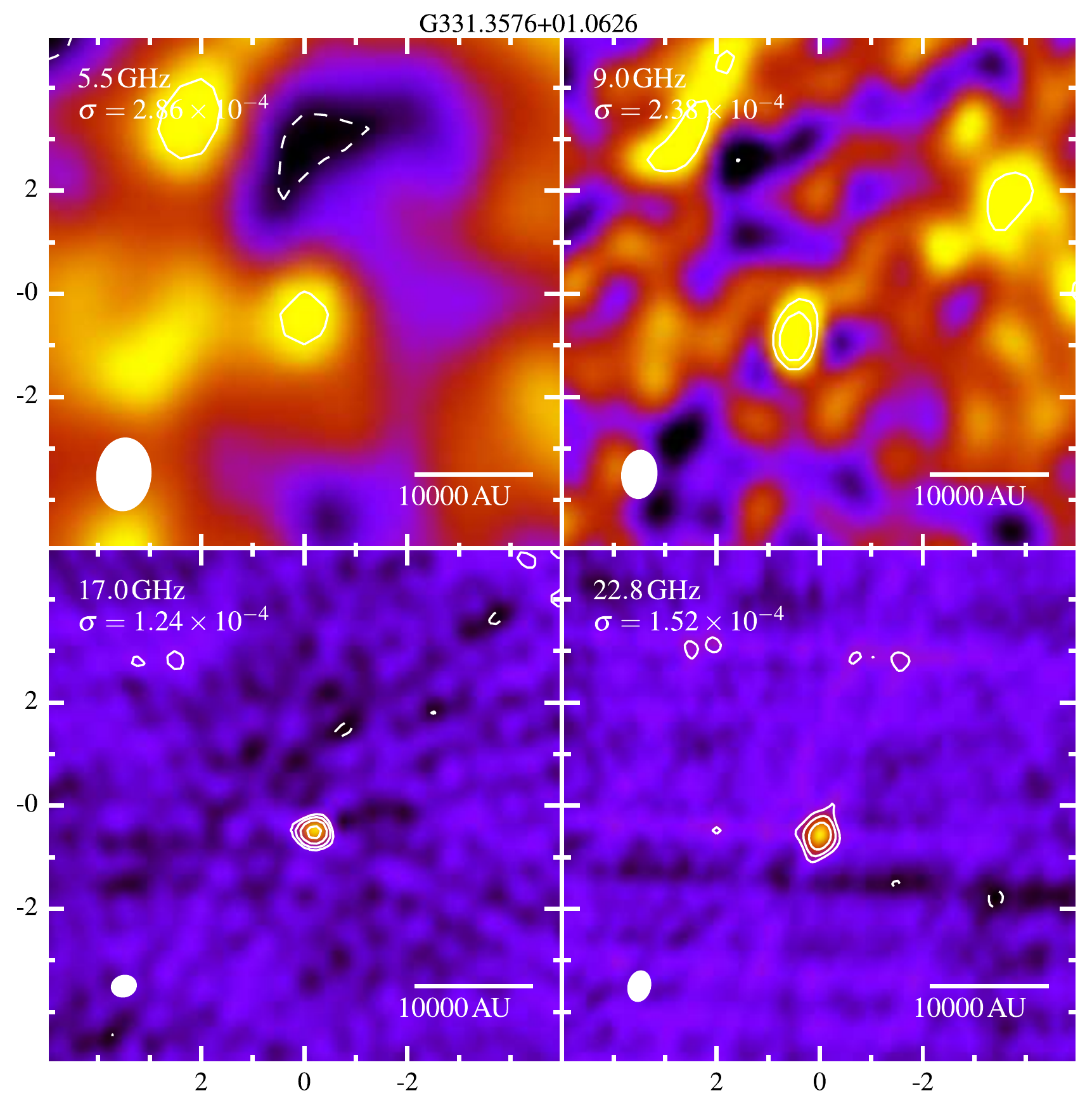}\\
\end{figure*}
\begin{figure*}
\includegraphics[width=0.81\textwidth]{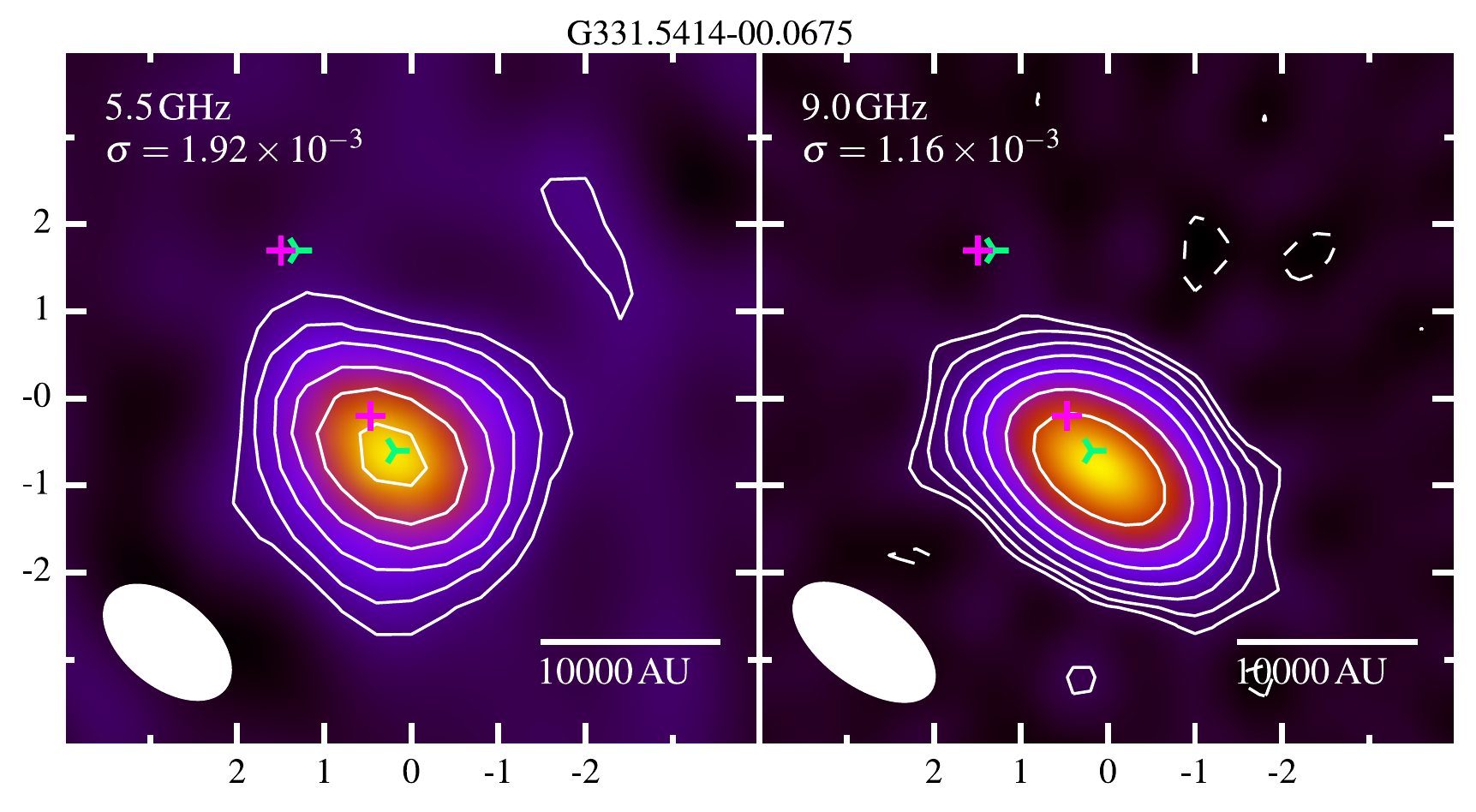}\\
\contcaption{}
\end{figure*}
\clearpage

\begin{figure*}
\includegraphics[width=0.81\textwidth]{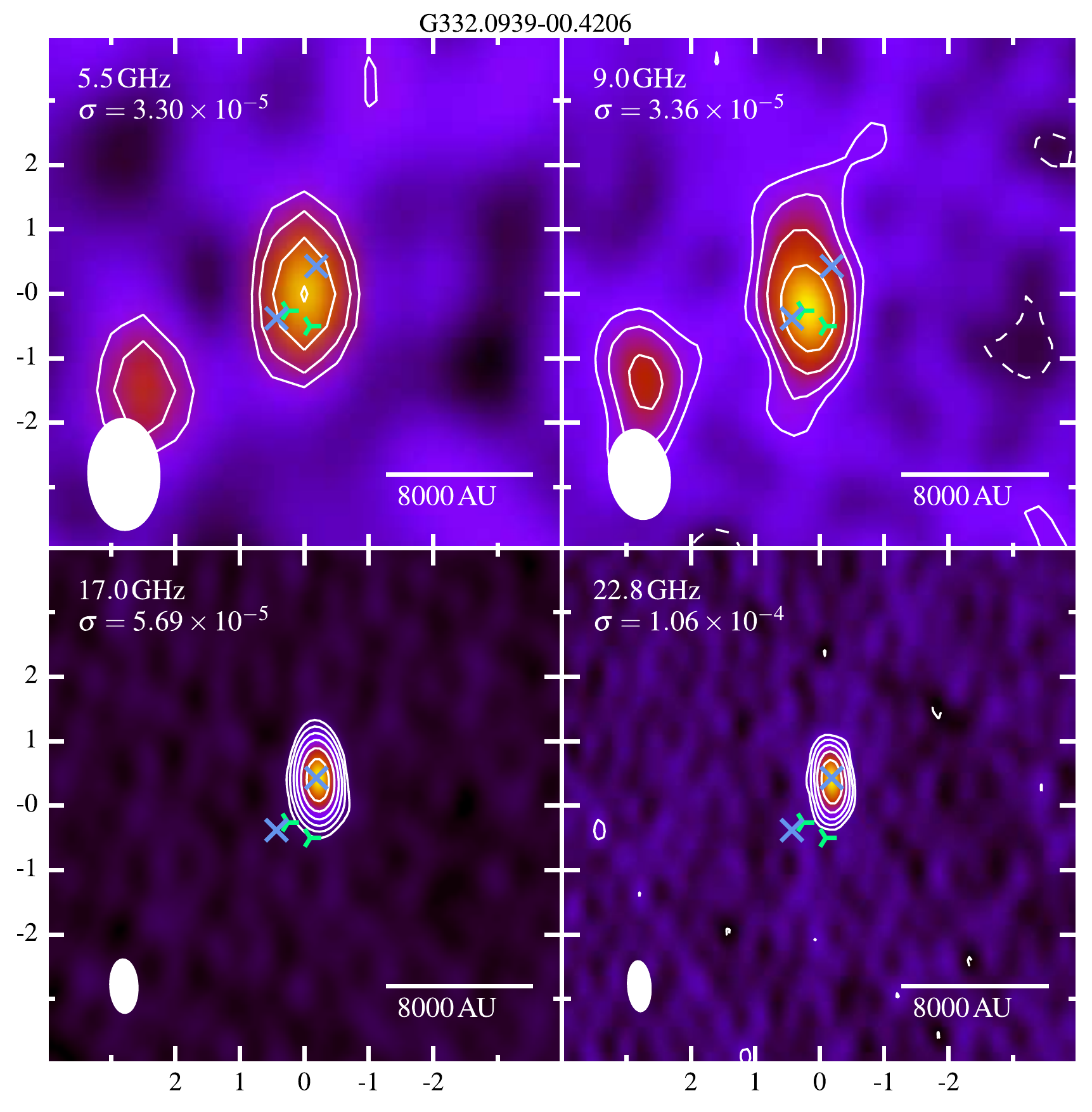}\\
\end{figure*}
\begin{figure*}
\includegraphics[width=0.81\textwidth]{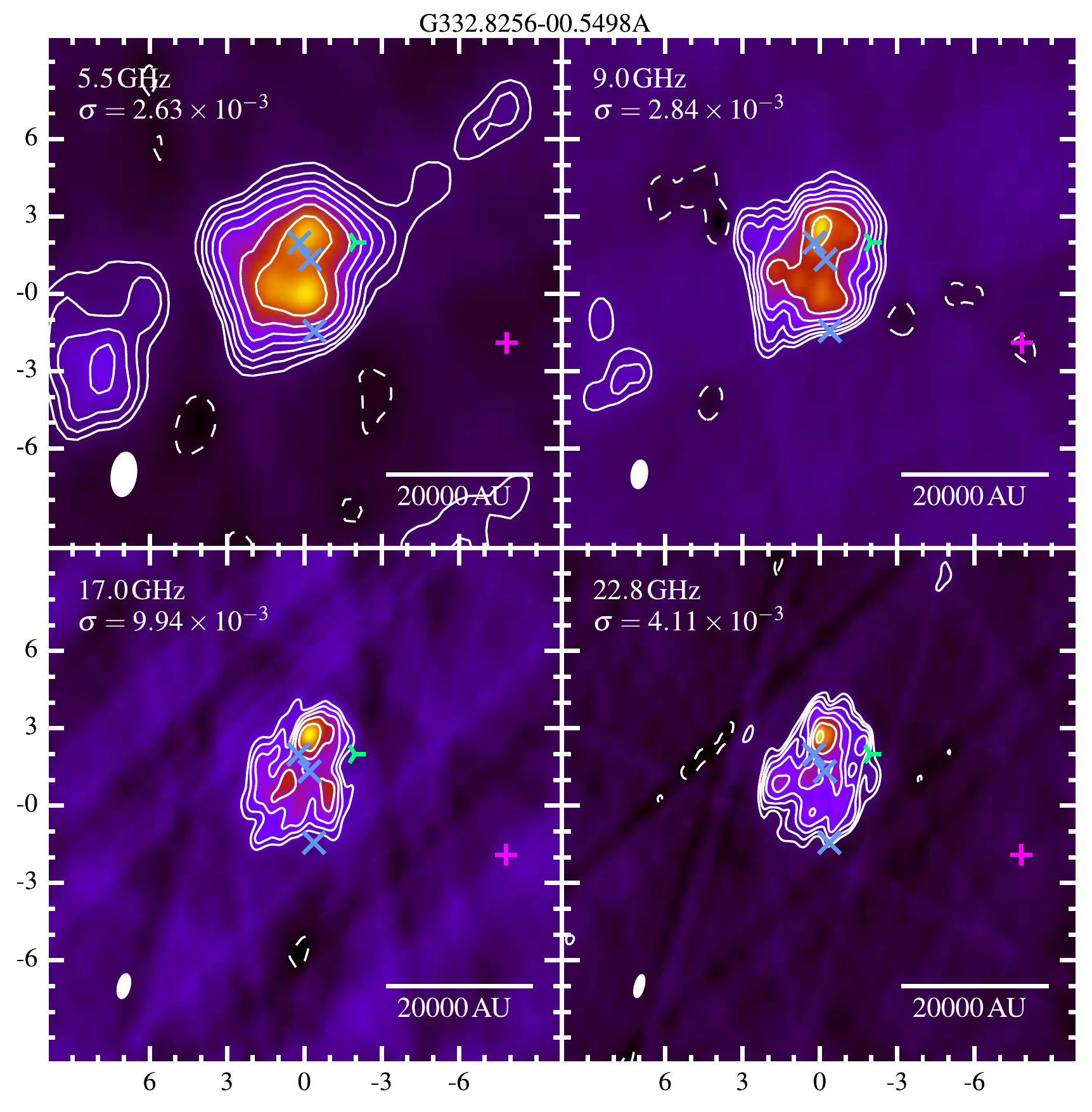}\\
\contcaption{}
\end{figure*}
\clearpage

\begin{figure*}
\includegraphics[width=0.81\textwidth]{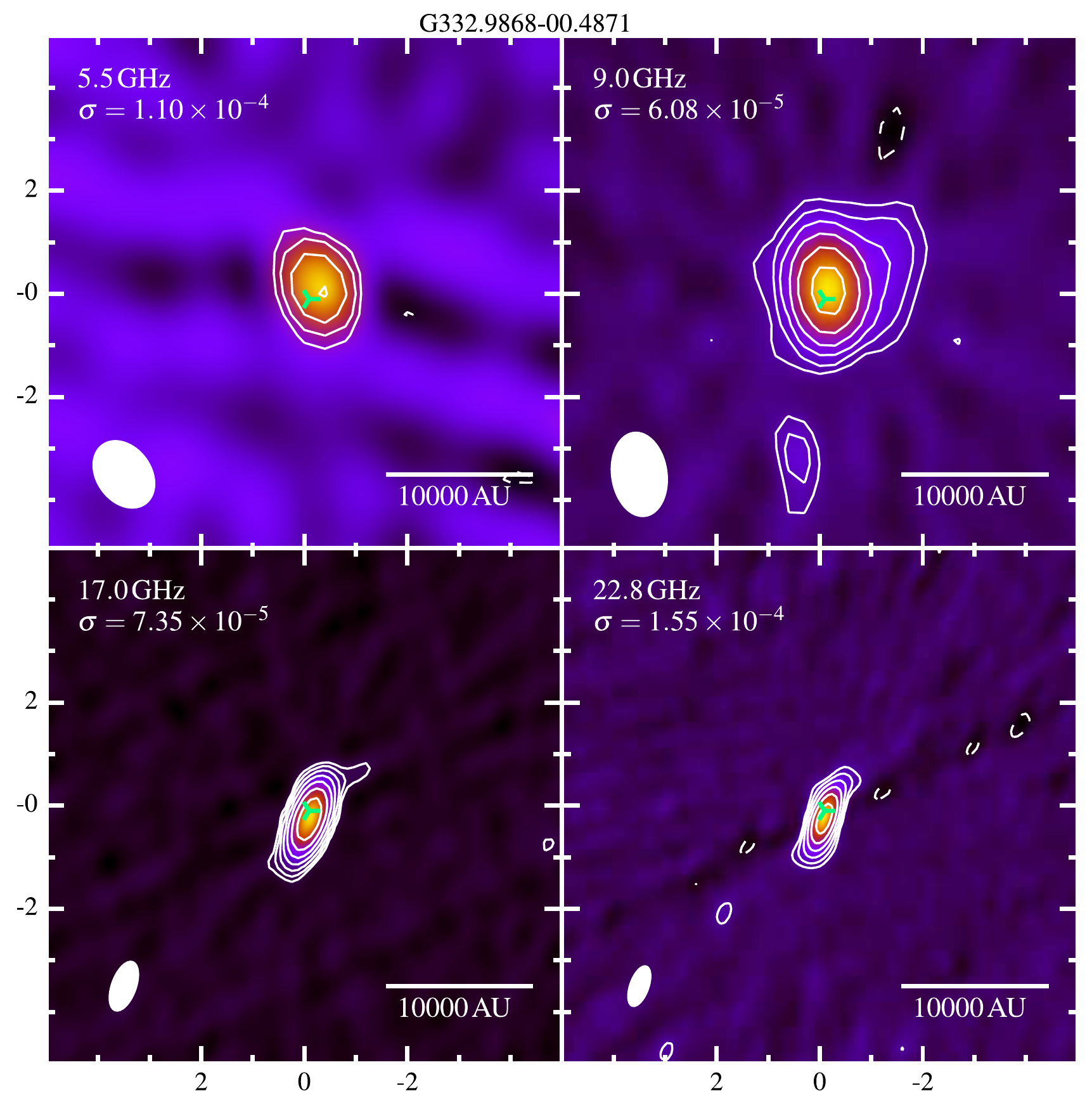}\\
\end{figure*}
\begin{figure*}
\includegraphics[width=0.81\textwidth]{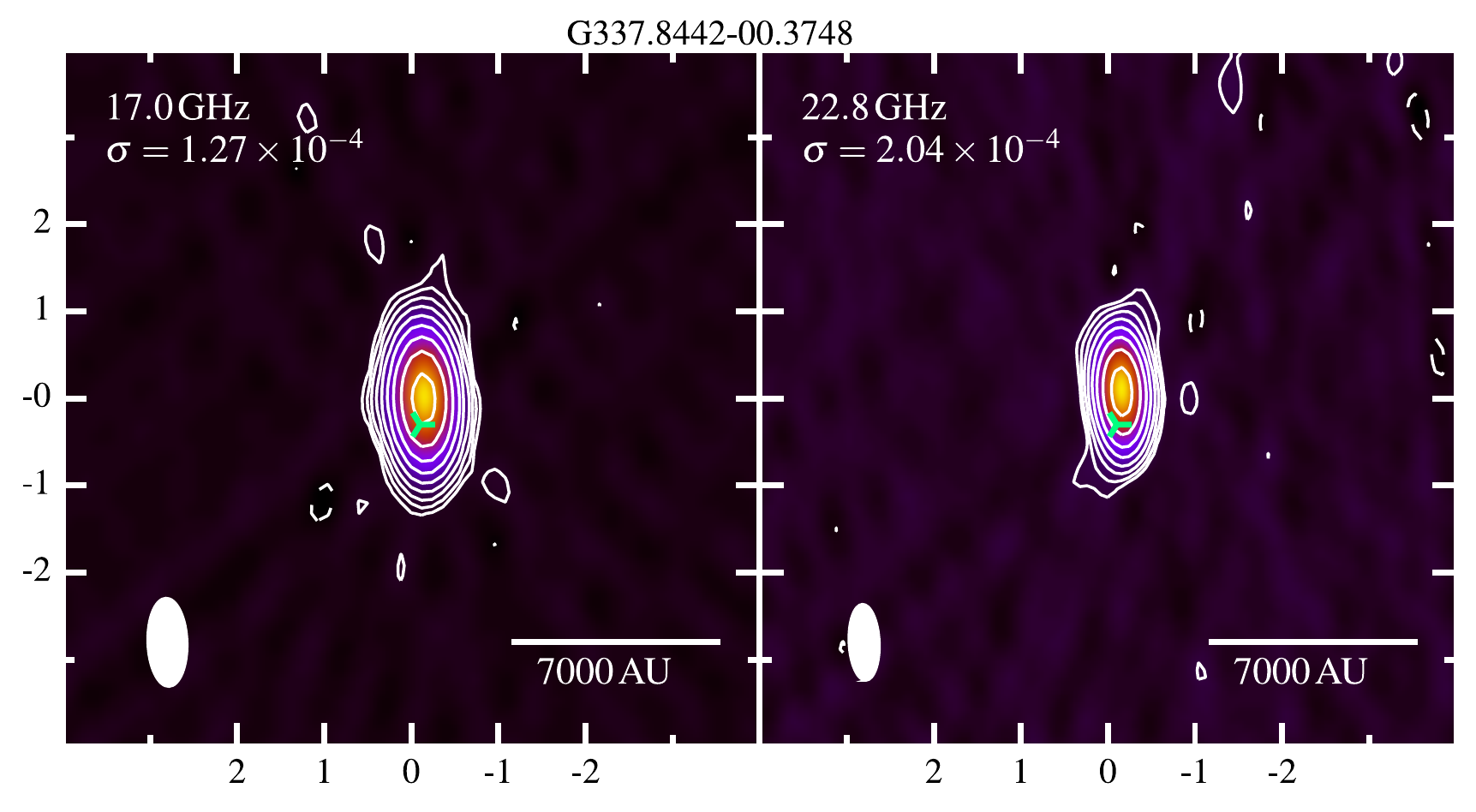}\\
\contcaption{}
\end{figure*}
\clearpage

\begin{figure*}
\includegraphics[width=0.81\textwidth]{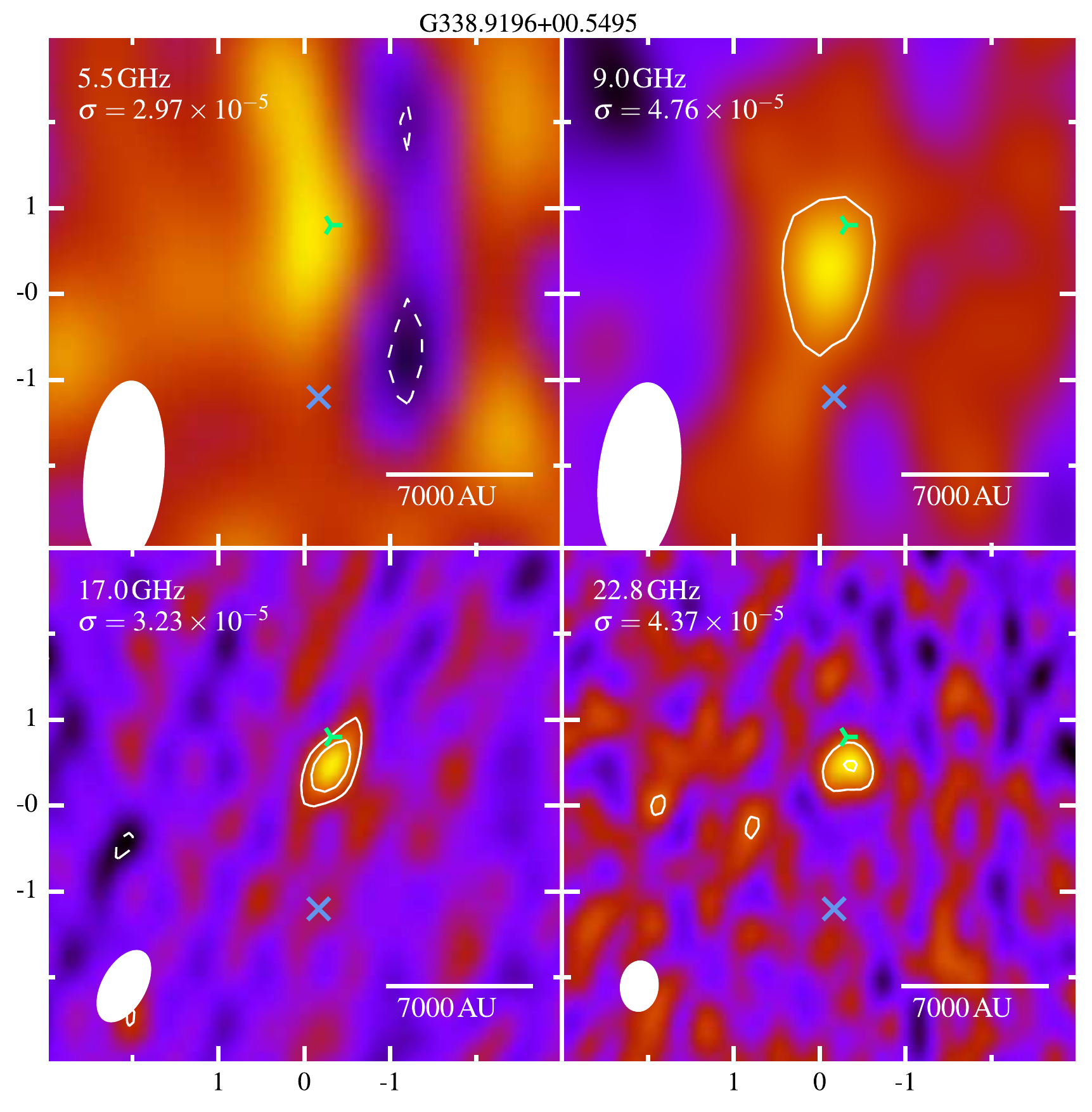}\\
\end{figure*}
\begin{figure*}
\includegraphics[width=0.81\textwidth]{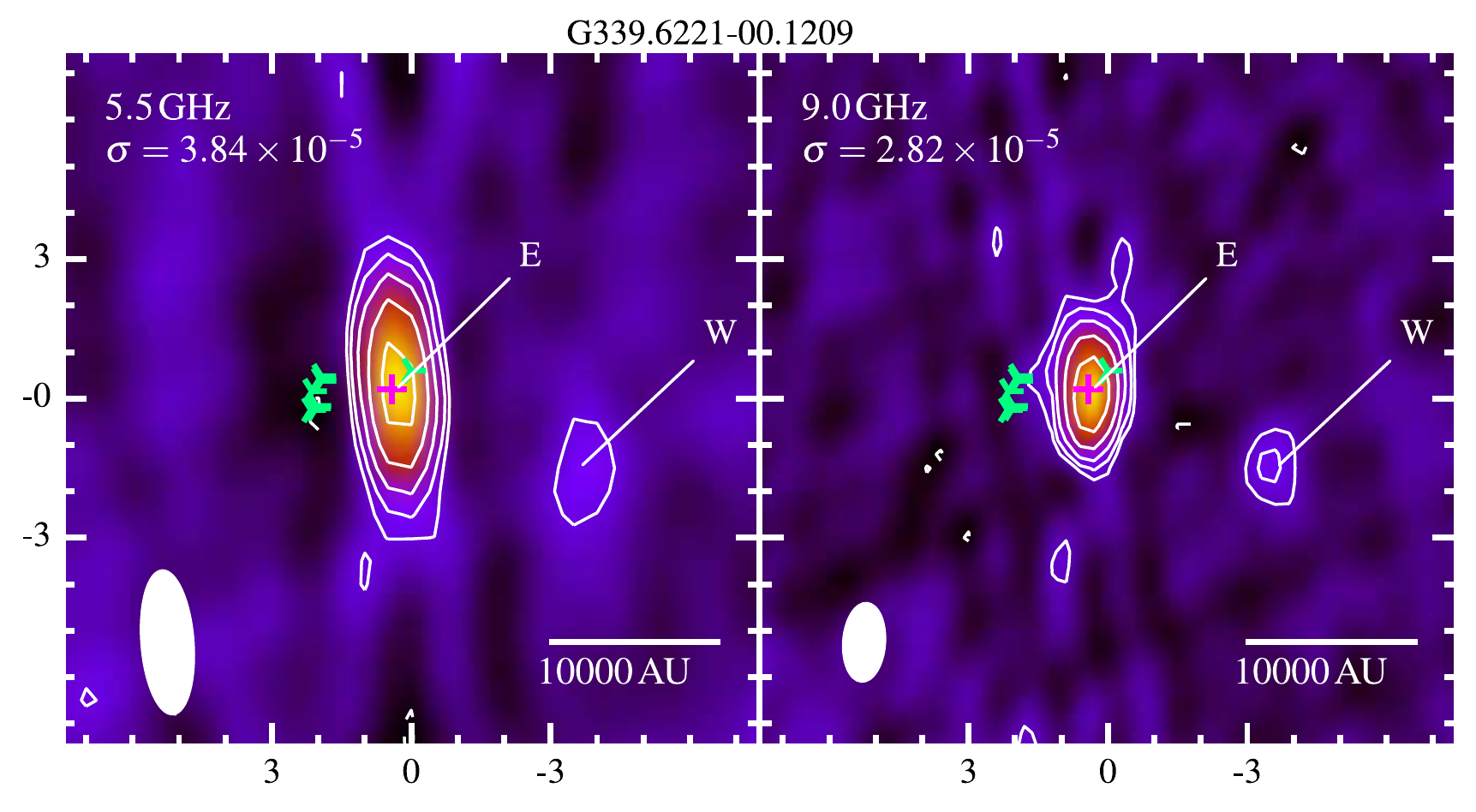}\\
\contcaption{}
\end{figure*}
\clearpage

\begin{figure*}
\includegraphics[width=0.81\textwidth]{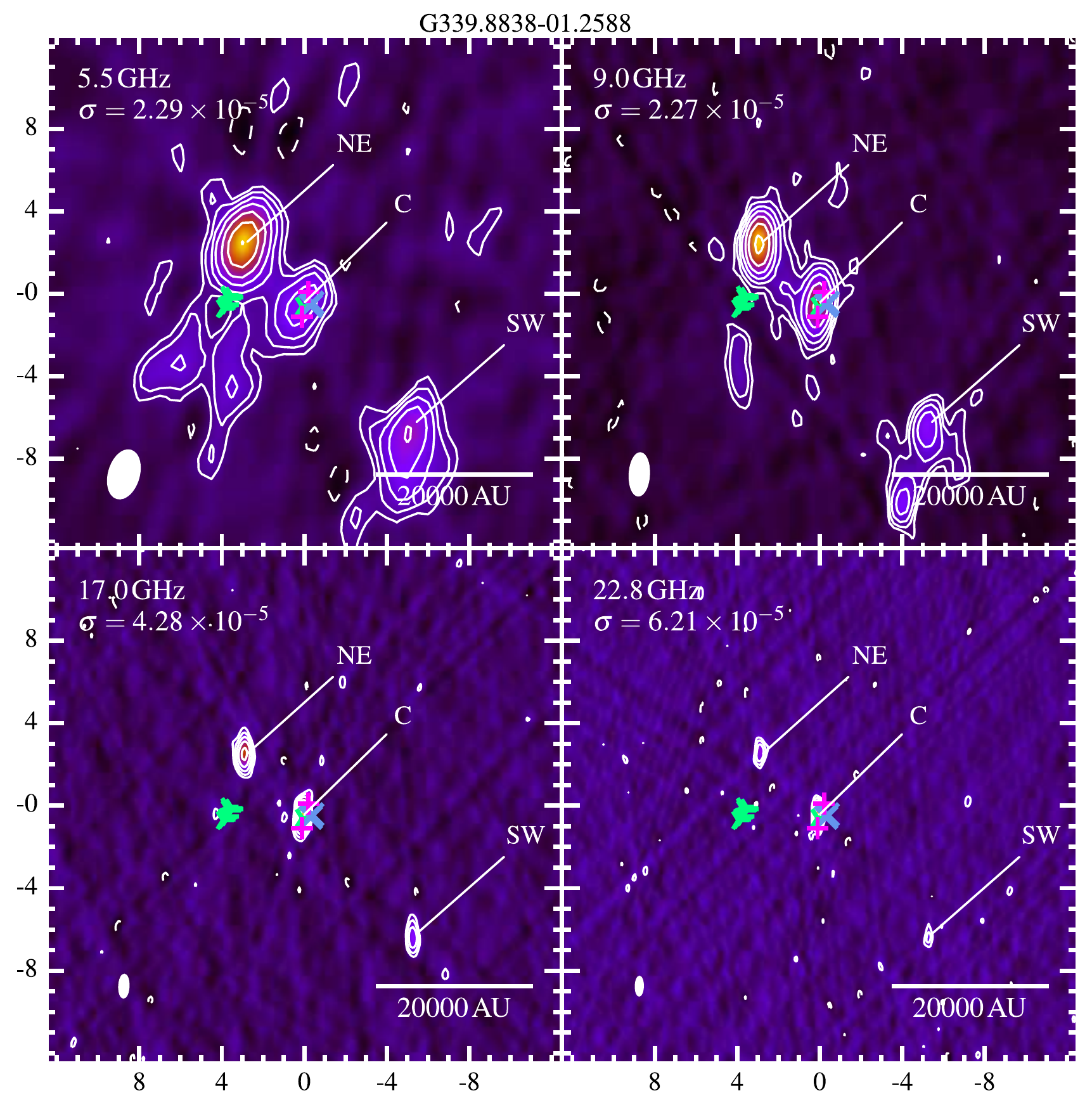}\\
\end{figure*}
\begin{figure*}
\includegraphics[width=0.81\textwidth]{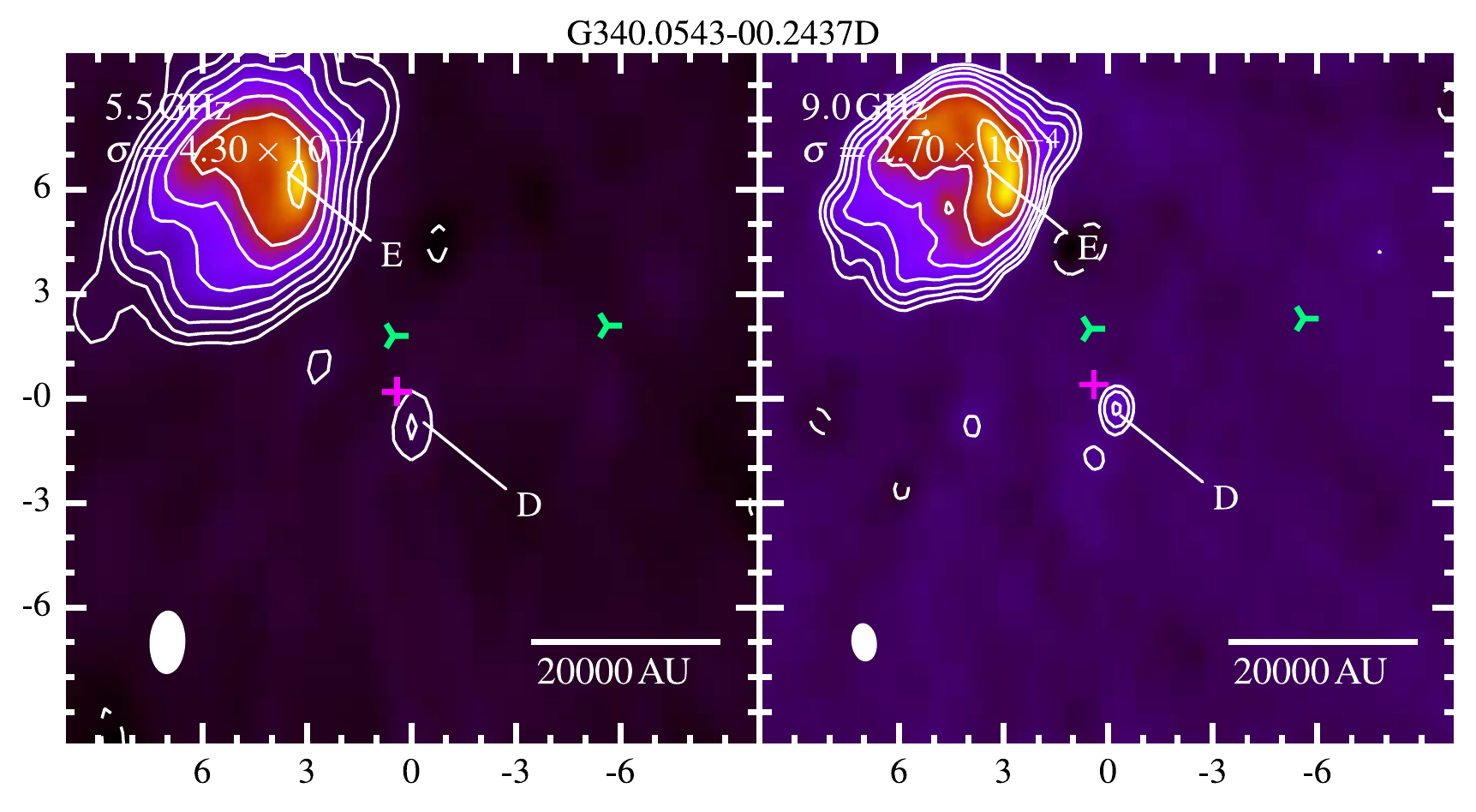}\\
\contcaption{}
\end{figure*}
\clearpage

\begin{figure*}
\includegraphics[width=0.81\textwidth]{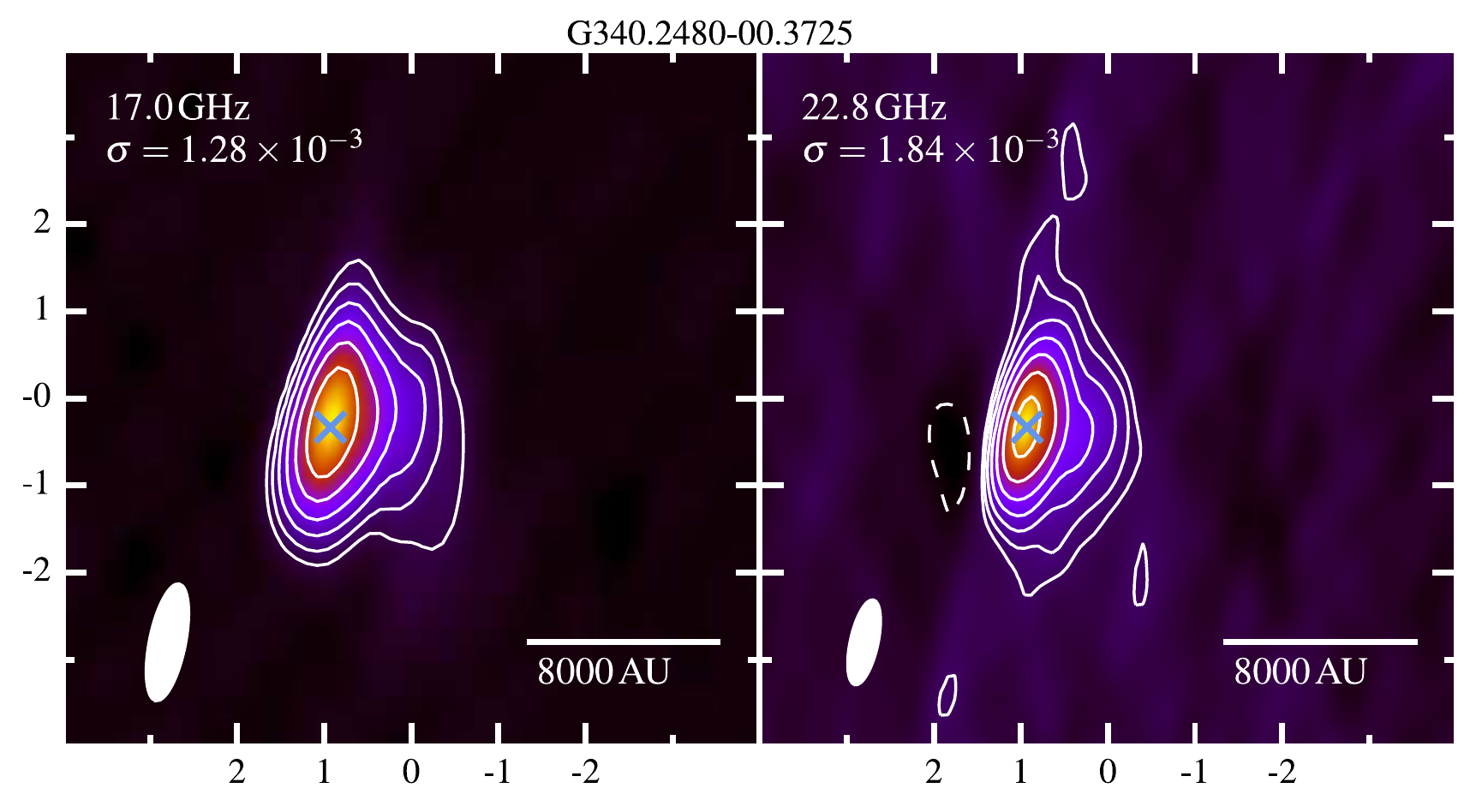}\\
\end{figure*}
\begin{figure*}
\includegraphics[width=0.81\textwidth]{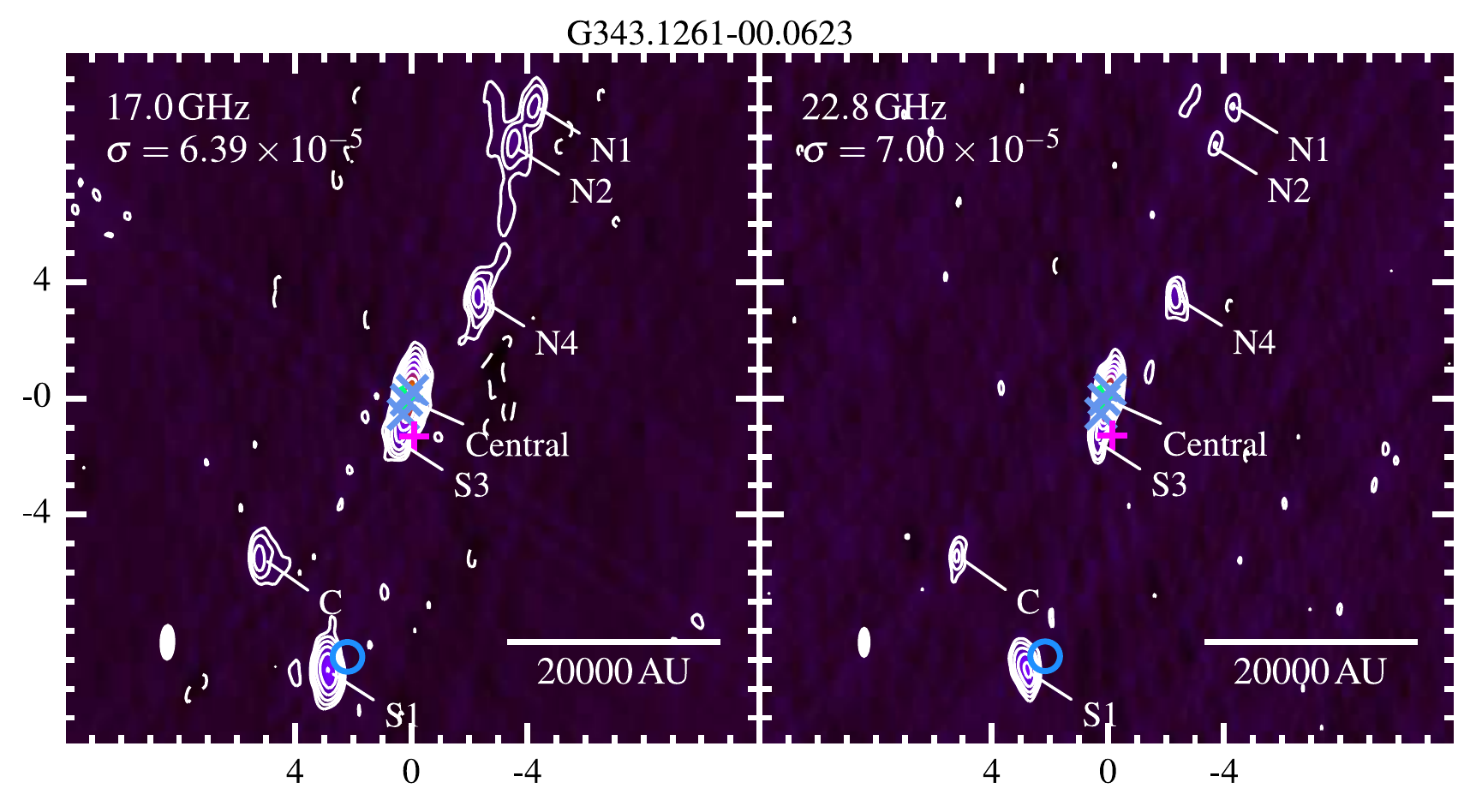}\\
\contcaption{}
\end{figure*}
\clearpage

\begin{figure*}
\includegraphics[width=0.81\textwidth]{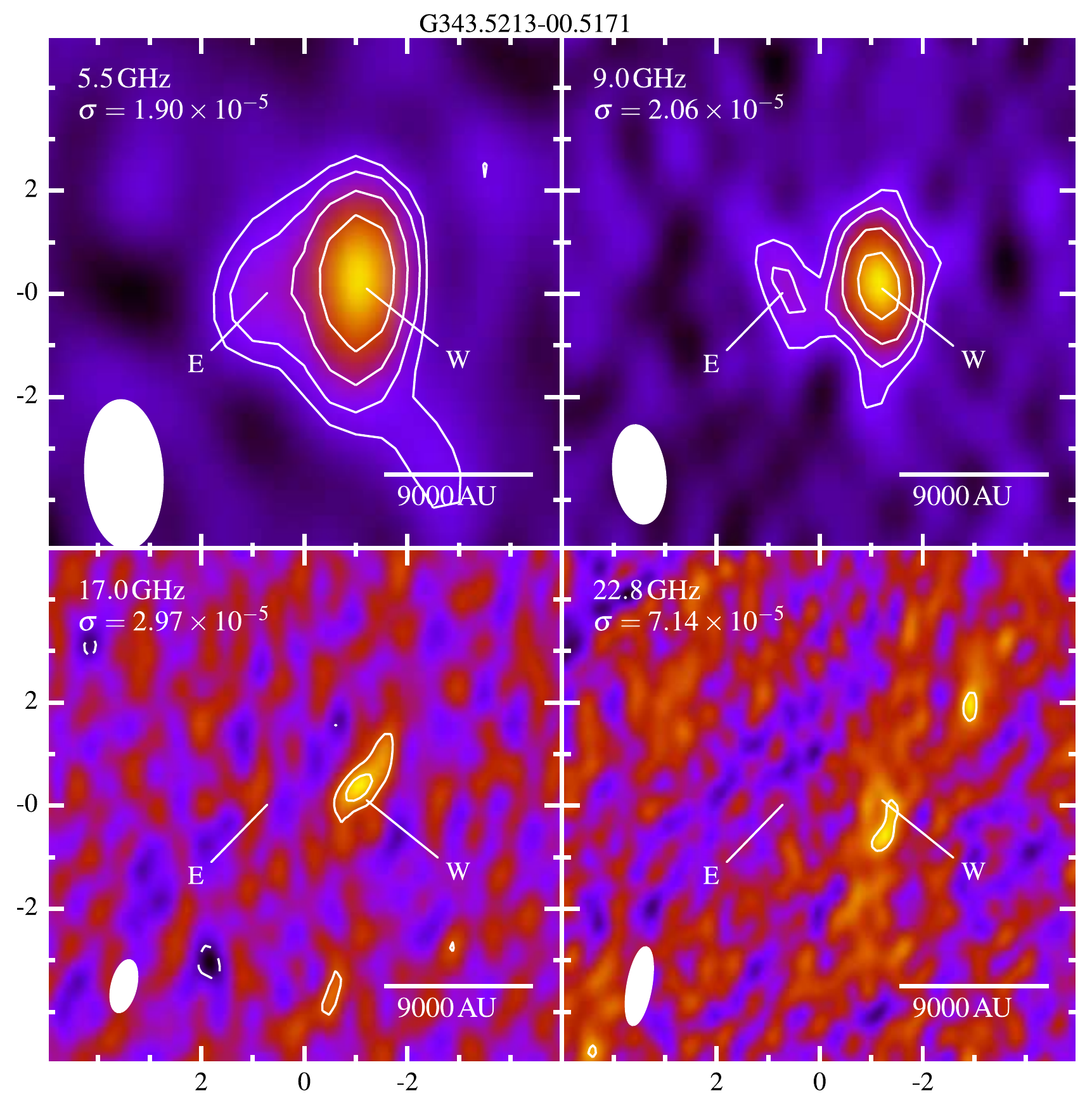}\\
\end{figure*}
\begin{figure*}
\includegraphics[width=0.81\textwidth]{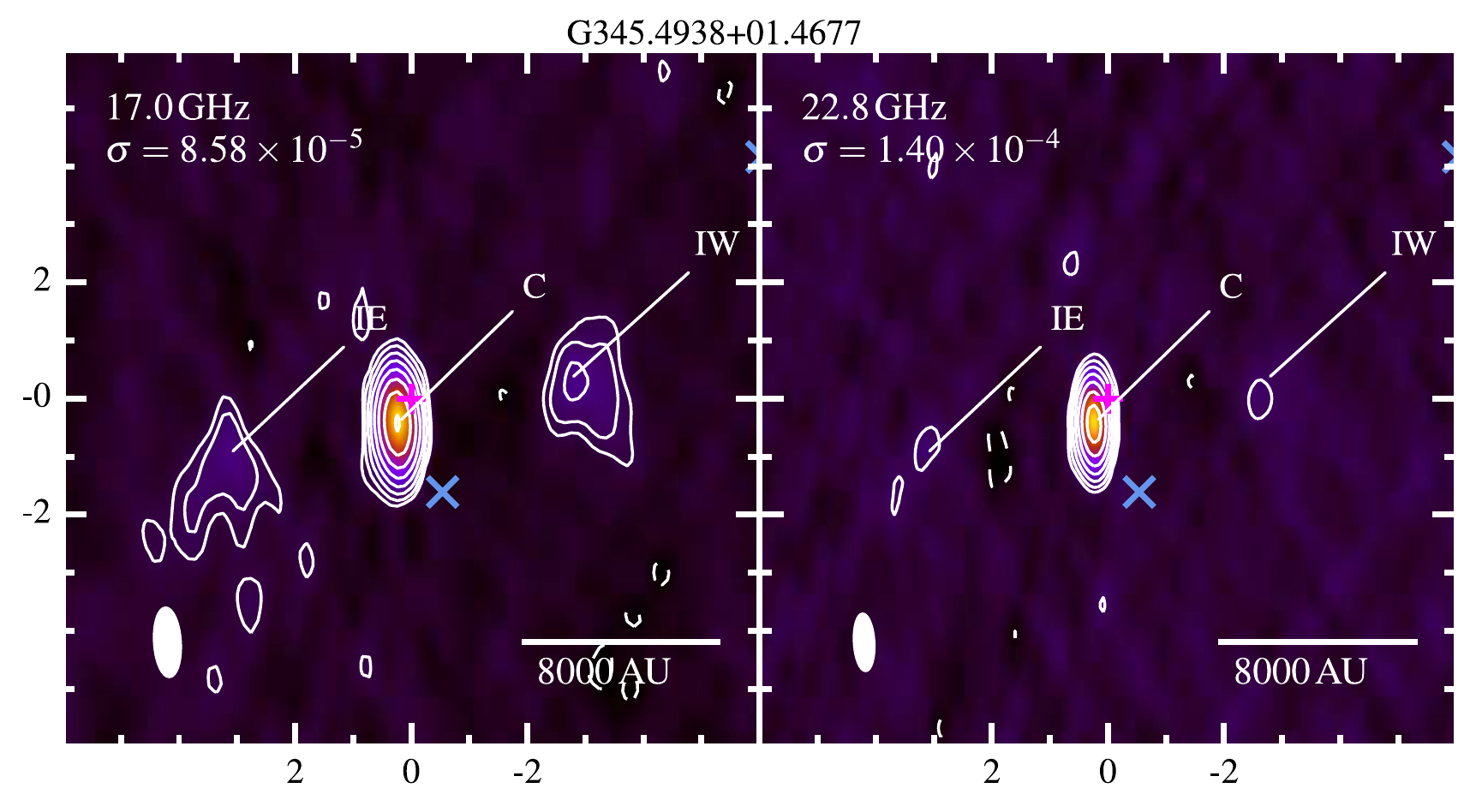}\\
\contcaption{}
\end{figure*}
\clearpage

\begin{figure*}
\includegraphics[width=0.81\textwidth]{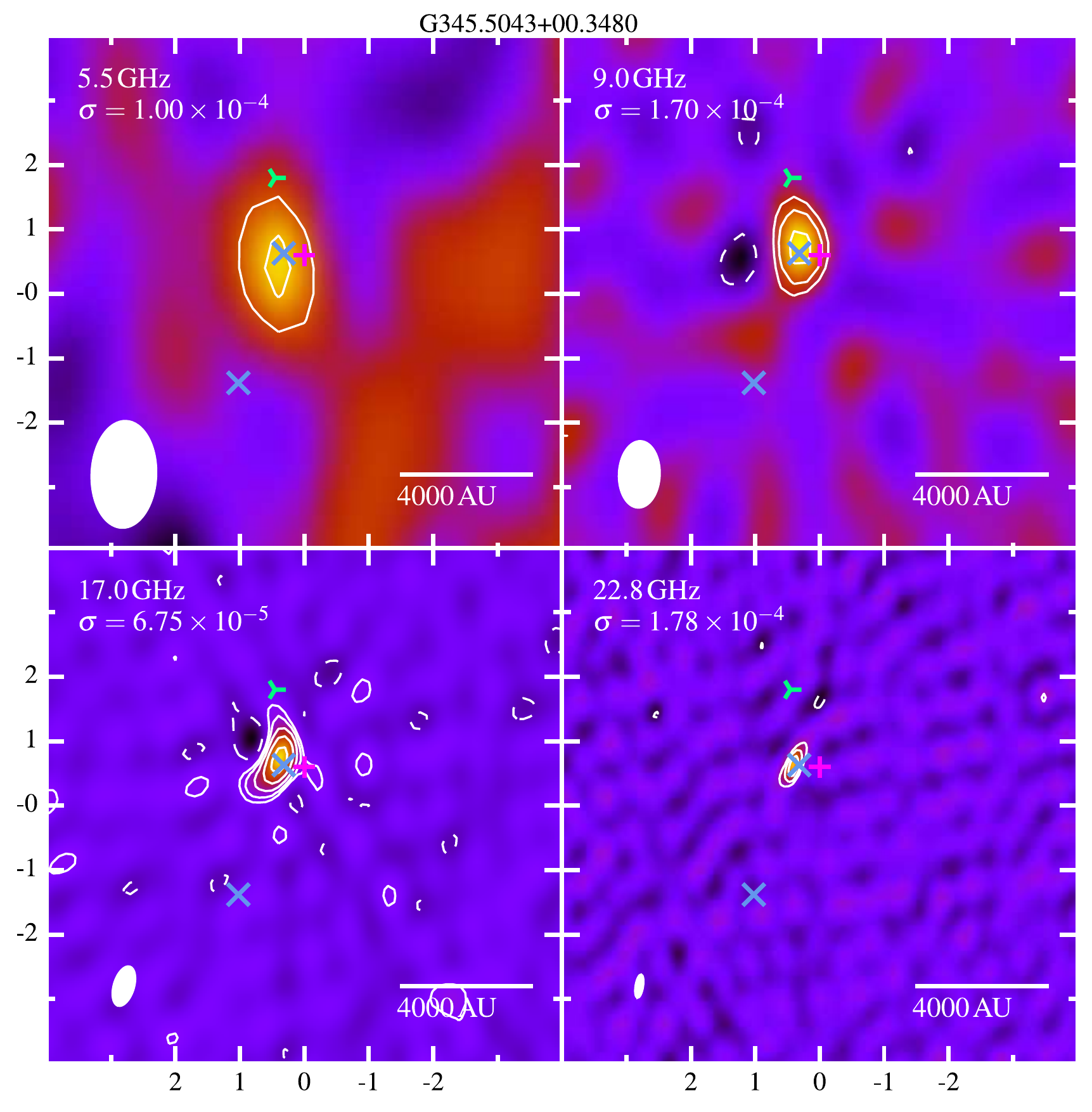}\\
\contcaption{}
\end{figure*}	
\clearpage

\label{lastpage}
\begin{figure*}
\includegraphics[width=0.81\textwidth]{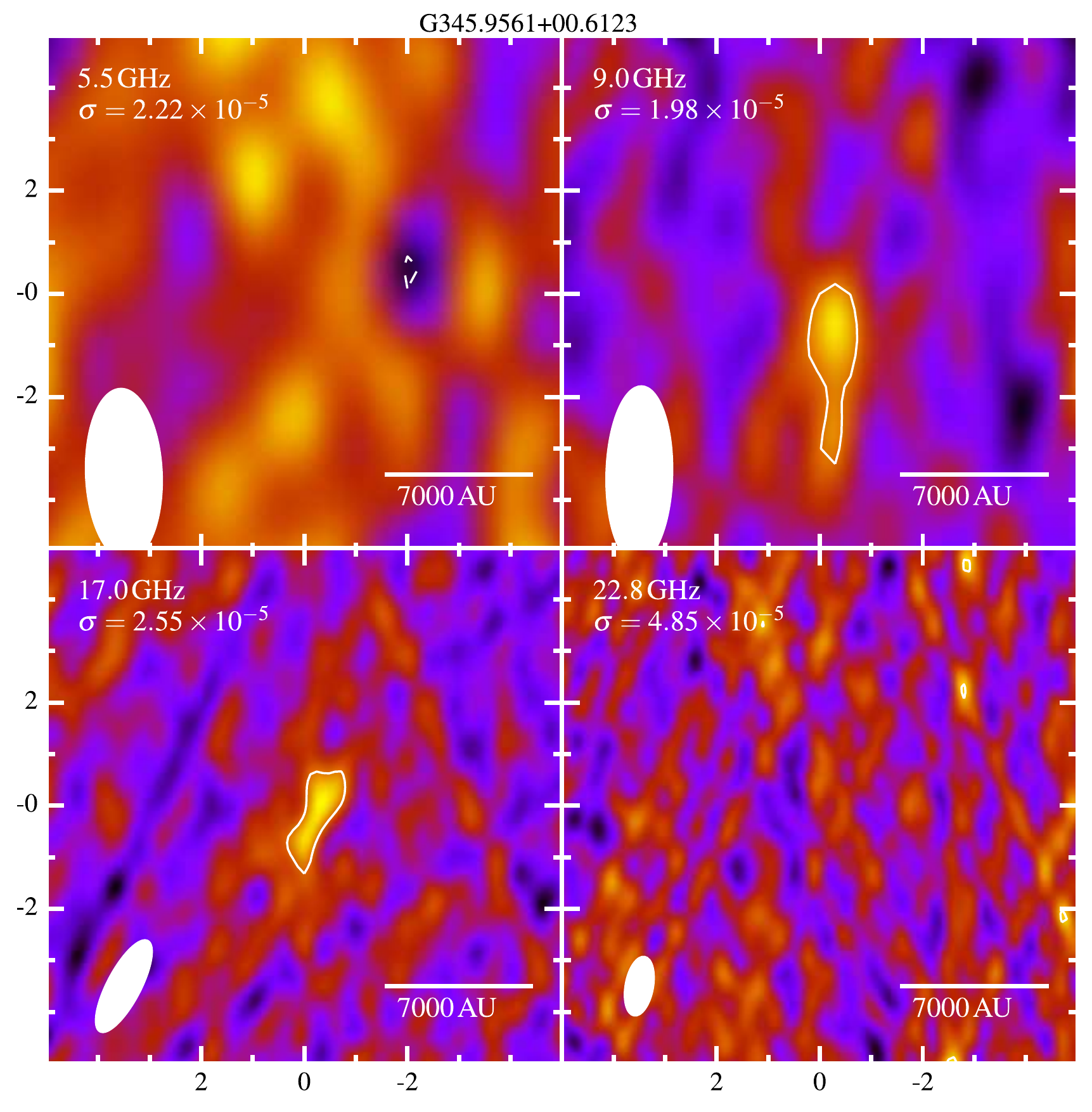}\\
\end{figure*}
\begin{figure*}
\includegraphics[width=0.81\textwidth]{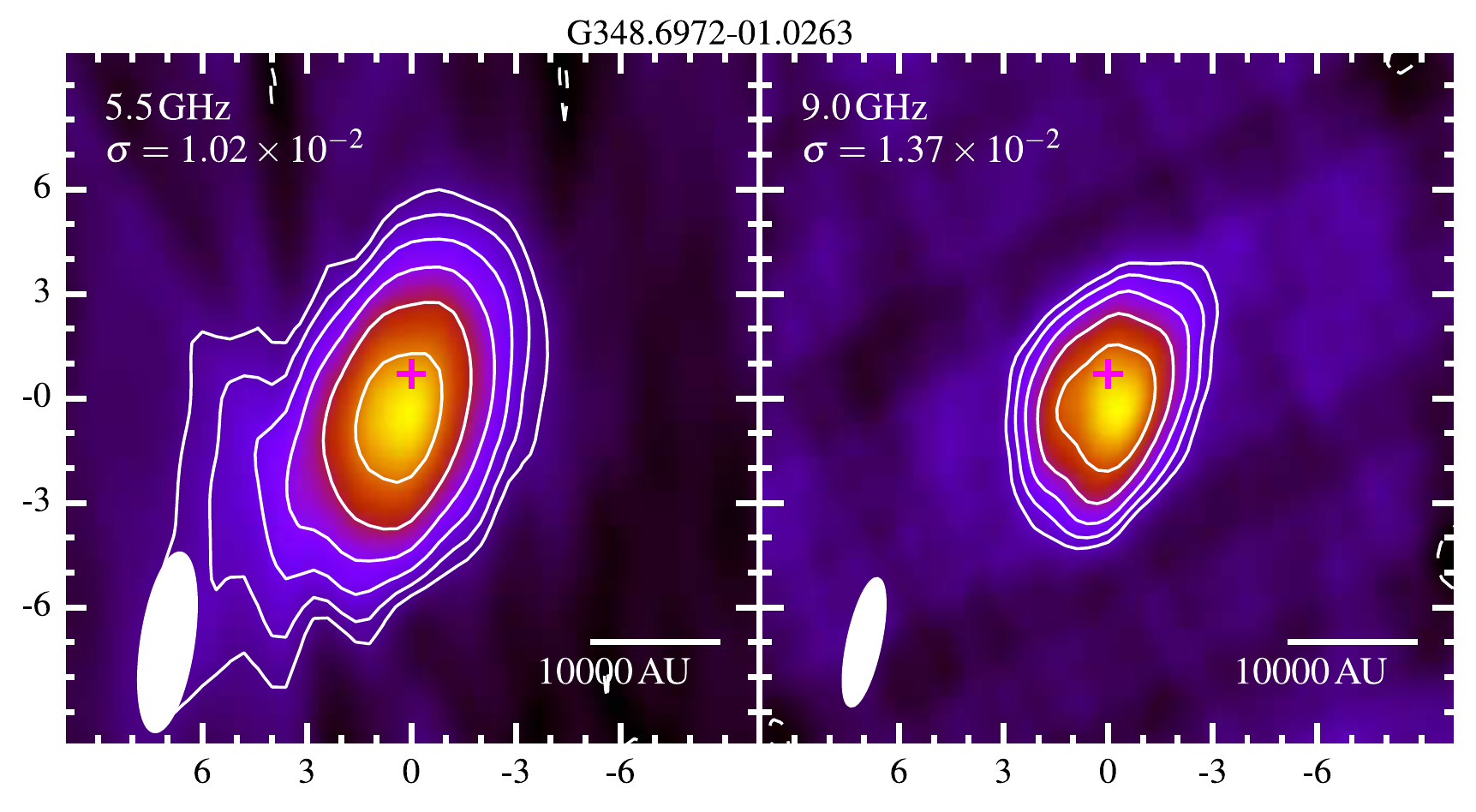}\\
\contcaption{}
\end{figure*}

\end{document}